%% file: thesis.tex
\documentclass[11pt,twoside,openright]{report}
\usepackage{./macros}

\begin{document}

\begin{fmffile}{./thesis_}
\fmfset{dash_len}{\feynmflen}
\fmfset{wiggly_len}{dash_len}
\fmfset{arrow_len}{1.5dash_len}
\fmfset{arrow_ang}{20}

\begin{titlepage}
\noindent
\begin{flushright}
\large
BONN-IR-98-06
\end{flushright}

\vspace*{1cm}

\begin{center}
{\huge \bf
  \input{./title}
}
\vspace{2cm}

{\Large
 Ralf Pantf\"order\footnote {Address: Neue Maastrichter Stra{\ss}e 2,
 D-50672 K\"oln, Germany}
}

\vspace{.5cm}

{\large
\textit {Physikalisches Institut der Universit\"at Bonn} \\
\textit {Nu{\ss}allee 12, 53115 Bonn, Germany}
}

\vspace{3cm}

\textbf {\large Abstract}

\end{center}

\input{./abstract}

\end{titlepage}

\setcounter{page}{0}
\clearpage{\thispagestyle{empty}\cleardoublepage}

\clearpage{\pagestyle{empty}\clearpage}
\pagenumbering{roman}
\tableofcontents

\clearpage{\pagestyle{empty}\cleardoublepage}
\chapter{Introduction}
\pagenumbering{arabic}

\section{History}

The Gerasimov-Drell-Hearn (GDH) sum rule has been found in 1965 by
Sergo Gerasimov \cite {Gerasimov65}, and then independently by Sidney
Drell and Anthony Hearn \cite {Drell66} in 1966.  The Russian original
of Gerasimov's paper was published in the October 1965 issue of \emph
{Yadernaya Fizika;} the translation in \emph {Soviet Journal of
Nuclear Physics} appeared in April 1966, a month before the letter by
Drell and Hearn was published in \emph {Physical Review Letters.} This
strict time ordering is why I call it the GDH sum rule and \emph {not}
DHG or even DH sum rule -- names that sadly won't disappear even from
the latest literature.  By the way: There has been a third independent
discovery of the sum rule.  On May 20th 1966, the editors of \emph
{Progress of Theoretical Physics} received a letter by Hosoda and
Yamamoto \cite {Hosoda66a}, who, besides, were the first to utilize a
current-algebra technique for the derivation.

The most remarkable aspect of sum rules is that they relate particle
properties that appear to be completely independent at first sight: some \emph
{low-energy} limit is expressed in terms of an integral that runs over
\emph {all energies}.  In particular, the GDH sum rule
\begin{equation} \label{GDH1}
 \boxed{ -2\pi\mu_\ta^2 = \int_{\nu_0}^\infty\! \frac{\td\nu}\nu
  \bigl( \sigma_{1/2}(\nu) - \sigma_{3/2}(\nu) \bigr) }
\end{equation}
relates the anomalous magnetic moment $\mu_\ta$ of the nucleon (proton
or neutron) to an energy-weighted integral of the difference
$\sigma_{1/2}(\nu)-\sigma_{3/2}(\nu)$ of the nucleon's polarized total
photoabsorption cross sections.  In \Eq {GDH1}, $\nu$ denotes the
lab-frame energy of the photon, $\nu_0$ is the pion-production
threshold, and subscripts 1/2 and 3/2 denote total helicity in the
initial state, i.e., antiparallel and parallel $\gamma$N
polarizations, respectively.

Albeit more than thirty years old today, the GDH sum rule still lacks
a direct experimental check, since both beam and target have to be
polarized and a wide range of photon energies has to be covered, which
presents an enormous challenge.  Presently, an experiment is in
preparation at the Mainz microtron MAMI and -- using the same target
-- at the electron stretcher facility ELSA at Bonn \cite {Anton95}.
MAMI covers the energy range from pion-production threshold up to
roughly 800 MeV, while photon energies from 500 to 3,000 MeV are
generated at ELSA.

For the present, all we have experimentally are multipole analyses of
unpolarized single-pion photoproduction data, from which -- in an
indirect fashion -- estimates of the contribution from low-lying
resonances to the GDH integral can be extracted [\plaincite
{Karliner73}\nocite {Workman92}--\plaincite {Sandorfi94}].  The result
of the latest resonance saturation by Sandorfi, Whisnant, and
Khandaker \cite {Sandorfi94} is presented in Tab.\ \ref {Tab:sat}.
Besides resonance contributions, values for the GDH integral include
non-resonant background and an estimate of the contribution of the
$\pi\pi$N final state, extracted from known \piN/$\pi\pi$N branching
ratios of the resonances under consideration \cite {Karliner73}.  The
latter contribution is indicated in parentheses in Tab.\ \ref
{Tab:sat}.  A more detailed investigation of double-pion production has
been performed by Coersmeier \cite {Coersmeier93}.

Hammer, Drechsel, and Mart \cite {Hammer97} investigated the possible
role of K$\Sigma$ and K$\Lambda$ final states, as well as $\phi$ and
$\eta$ meson production.  They found that kaon production contributes
roughly $-3~\mu$b to the GDH integral both for the proton-neutron
difference and  for the proton-neutron average, and
that the contribution from $\phi$ and $\eta$ production is even
smaller.  Observe that the magnitude of these numbers is way below the
discrepancy indicated in Tab.\ \ref {Tab:sat}.

Although the explicit suppression of higher photon energies by the
integration measure of the GDH sum rule supports the assumption that
single-pion production saturates the sum rule, results of the
saturations must be taken with care, since generally no errors can be
attributed to the omission of resonances and final states.
Nevertheless, a severe discrepancy remains for the proton-neutron difference.
\begin{table}
 \begin{displaymath} \begin{array}{|c|c|c|}
  \hline
  & -2\pi\mu_\ta^2 & \displaystyle \int^{}\!\frac{\td\nu}{\nu}
    \bigl( \sigma_{1/2}(\nu)-\sigma_{3/2}(\nu) \bigr) \\
  & & \text {from resonance saturation} \\
  \hline
  \text{proton} \begin{matrix} \\ \\ \end{matrix}
  & -204~\mub & -289~\mub\quad(-65~\mub) \\
  \text{neutron} \begin{matrix} \\ \\ \end{matrix}
  & -233~\mub & -160~\mub\quad(-35~\mub) \\
  \begin{matrix} \text{proton-neutron}\\ \text{difference} \end{matrix}
  & \hfill 29~\mub & -129~\mub\quad(-30~\mub) \\
  \hline
 \end{array} \end{displaymath}
 \caption[]{
  Resonance saturation of the GDH integral \cite {Sandorfi94}.
  Numbers in parentheses estimate the contribution
  of the $\pi\pi$N final state \cite {Karliner73}.
  \label{Tab:sat}}
\end{table}

The conclusion to be taken now is of course: Something is wrong,
either with the saturation or with the sum rule itself!  As far as the
first possibility is concerned, it may be best to simply await the
result of the MAMI and ELSA experiments instead of worrying too much
about an interim solution.  As for the second variant, there has in
fact been much talk about possible modifications to the GDH sum rule
ever since the late sixties.  Kawarabayashi and Suzuki \cite
{Kawarabayashi66c} and Khare \cite {Khare75} demonstrated that a
modification arises if quarks possess anomalous magnetic moments.
Abarbanel and Goldberger \cite {Abarbanel68} pointed out that if the
nucleon's Compton amplitude exhibits a $J=1$ fixed pole in complex
angular-momentum plane, then the sum rule is endowed with an additional term
proportional to the residue of the pole.  Owing to the lowest-order
consideration of electromagnetic interactions, the fixed pole is not
ruled out from $t$-channel unitarity.  Chang, Liang, and Workman \cite
{Chang94a} claimed that the chiral anomaly brings about a modification
by virtue of an anomalous charge-densities commutator calculated by
Chang and Liang \cite {Chang91,Chang92}.  Ying \cite {Ying96a}
suggested a modification due to ``localized spontaneous breakdown of
electromagnetic gauge symmetry''.

Considering polarized inclusive electroproduction on the nucleon, the
GDH sum rule can be generalized to non-zero values of the photon
virtuality $Q^2$.  At large $Q^2$, it has counterparts in the Bjorken
\cite {Bjorken66} and Ellis-Jaffe \cite {Ellis74} sum rules, which have
been measured at SLAC [\plaincite {Baum83}\nocite
{Anthony93}--\plaincite {Abe97b}] and at CERN \cite
{Ashman89,Adeva93}.  Intermediate momentum transfer ($Q^2=0.5$~GeV$^2$
and 1.2~GeV$^2$) was recently explored in the SLAC E143 experiment
\cite {Abe97a}.  Forthcoming experiments at Jefferson Lab will considerably
refine the data.  The transition from photoabsorption, which is mainly
driven by resonances, to the high-$Q^2$ region, being governed by
perturbative QCD, is an interesting field in itself.

\section{The scope of this thesis}

The aim of the present thesis is a comprehensive treatment of the GDH
sum rule and all possible sources of modifications suggested in the
literature.  A separate chapter is devoted to a discussion of the
$Q^2$ evolution of the sum rule.

\subsubsection{Derivations of the GDH sum rule}
In \Ch {derivations}, different derivations of the GDH sum rule are
presented and discussed in detail: the dispersion-theoretic approach
given in the original publications by Gerasimov \cite {Gerasimov65}
and by Drell and Hearn \cite {Drell66}, the derivation from the
equal-times commutator of electric-charge-density operators \cite
{Hosoda66a,Kawarabayashi66a}, and, finally, the light-cone
current-algebra approach \cite {Dicus72}.  Important relations and
distinguishing marks between different derivations are pointed out.

Particular attention is paid to the infinite-momentum limit \cite
{Pantfoerder98}, which enters the derivation from equal-times current
algebra as a mere conjecture.  In contrast to the existing literature,
the infinite-momentum limit is postponed to the very end of the
derivation presented in \Sect {ETCA} in order to clarify its physical
and mathematical meaning.  Moreover, this procedure opens the prospect
of thoroughly investigating the legitimacy of the infinite-momentum
limit.  The \emph {finite-momentum GDH sum rule}, i.e., the form that
the sum rule takes \emph {before} the infinite-momentum limit is
taken, is discussed in detail.  Understanding the finite-momentum GDH
sum rule is vital for the comprehension of the essentials of
equal-times current algebra and the infinite-momentum limit.

\subsubsection{The GDH sum rule within perturbative models}
\Ch {pert} is devoted to an inspection of theoretical tests of the GDH
sum rule within perturbative models.  In \Sect {GM}, I consider the
simplest model of ``pointlike'' pions and nucleons with pseudoscalar
coupling among themselves and minimal coupling to the electromagnetic
field \cite {Gerasimov75}.  In \Sect {QED}, I discuss QED \cite
{Altarelli72,Tsai75}, while \Sect {WSM} presents the Weinberg-Salam
model of electro-weak interactions of leptons \cite {Altarelli72}.
This ordering implies increasing degree of complexity.

The Weinberg-Salam model is particularly well suited for investigating
the legitimacy of the infinite-momentum limit in presence of an
anomalous-commutator correction to the finite-momentum GDH sum rule.

\subsubsection{Possible sources of modifications}
In \Ch {mod}, I review the proposed sources of modifications of the
GDH sum rule.  \Sect {fixed-pole} is devoted to a discussion of a
possible $J=1$ fixed pole in angular-momentum plane \cite
{Abarbanel68}.  In \Sect {anomcomm}, I discuss the claimed
modification coming from an anomalous charge-density algebra \cite
{Chang94a}, while \Sect {IML} aims at illustrating the significance of the
infinite-momentum limit.  In \Sect {PRP}, the Weinberg-Salam model is
adopted to simultaneously calculate the anomalous charge-density
commutator \emph {and} the effect of the infinite-momentum limit \cite
{Pantfoerder98}.  It is shown that both of these points yield a
non-trivial contribution to the GDH sum rule, in such a way that the
individual modifications cancel exactly.  \Sect {a1} presents a short
treatise on $t$-channel exchange of axial-vector mesons, in order to
further illustrate that the conclusion of \Sect {PRP} is by no means
restricted to the Weinberg-Salam model, but applies to hadrons, too.

The remaining sections of \Ch {mod} are devoted to a critical discussion
of several other claims of possible modifications that can be found in the
literature.

\subsubsection{$Q^2$ evolution of the GDH sum rule}

In \Ch {Q2}, I investigate the transition from polarized
photoabsorption to polarized inclusive electroproduction.  After an
introduction to the notion of virtual photoabsorption, I explain the
physical meaning of different linear combinations of polarized nucleon
structure functions.  \Sect {gen} is devoted to diverse
generalizations of the GDH integral to non-zero photon virtuality.
The significance of structure function $G_2(\nu,Q^2)$ is pointed out.
Since all generalizations coincide both for real photons and in the
scaling limit, non-leading terms in the limits $Q^2\to0$ and
$Q^2\to\infty$ are analyzed and discussed.

In view of resonance saturation of generalized GDH integrals, the
contribution from single-pion production to structure functions and
inclusive cross sections is treated in \Sect {piN}.  It is expressed
in terms of helicity amplitudes, Chew-Goldberger-Low-Nambu amplitudes,
and helicity multipoles.  The isospin decomposition is sketched.

\subsubsection{Tools}
Some quite extensive calculations presented in this thesis have been
performed with the aid of the computer algebra system \textsl
{Mathematica} \cite {Wolfram91}.  Feynman graphs were drawn using the
\LaTeX\ package \texttt {feyn}\textlogo {MF} \cite {Ohl95} available from
any CTAN host (e.g., ftp.dante.de).

\section{Notation and conventions}

Throughout this thesis, units are chosen such that Planck's constant
$\hbar$, the vacuum speed of light $c$, and the permittivity of the
vacuum $\eps_0$ equal unity.  Then the fine-structure constant is
$\alpha = e^2/4\pi$.

Spacetime indices are represented by greek letters $\mu,\nu,\ldots$,
whereas latin letters $i,j,\ldots$ are used for purely spatial
indices.  The metric is $g_{\mu\nu} = \diag({+}{-}{-}{-})$.
Three-vectors are denoted by boldface symbols, e.g.\ $\bx$.  I use a
Euclidian dreibein $\{\be_k\}$ such that $x^k=\bx\ndot\be_k$.  Partial
derivatives are abbreviated as
\begin{equation}
 \partial_\mu = \frac\partial{\partial x^\mu},\quad
 \nabla^i = \frac\partial{\partial x^i} = -\partial^i.
\end{equation}
The four-dimensional Fourier transform reads
\begin{equation}
 \tilde f(k) = \int\!\td^4x\, e^{ik\ndot x} f(x),
\end{equation}
with inverse
\begin{equation}
 f(x) = \int\!\frac{\td^4k}{(2\pi)^4}\, e^{-ik\ndot x} \tilde f(k).
\end{equation}
The Fourier transform of the gradient $i\partial^\mu f(x)$ is $\tilde
f(k)k^\mu$.  For a translationally invariant, local Hilbert-space
operator $A(x)$, this relation can be expressed in the form
\begin{equation} \label{[A,p]}
 i\partial^\mu A(x) = \bigl[ A(x), \hat p^\mu \bigr],
\end{equation}
where $\hat p^\mu$ denotes the energy-momentum operator.  \Eq {[A,p]}
corresponds to the convention that the Fourier transformed operator
$\tilde A(k)$ maps a \emph {bra} state $\langle\tX|$ of given
four-momentum $p_\tX$ onto a state $\langle\tX|\tilde A(k)$ having
four-momentum $p_\tX+k$.  The integral form of \Eq {[A,p]} reads
\begin{equation} \label{A(x)}
 A(x) = e^{i\hat p\ndot x} A(0) e^{-i\hat p\ndot x}.
\end{equation}
Sandwiched between states $|\tX\rangle$ and $|\tY\rangle$ with
definite momenta $p_\tX$ and $p_\tY$, \Eq {A(x)} reduces to
\begin{equation} \label{trans}
 \me {\tY}{A(x)}{\tX} = \me {\tY}{A(0)}{\tX}\, e^{i(p_{\tY}-p_{\tX})\ndot x}.
\end{equation}

In the theory of current commutators, I frequently deal with derivatives
of delta functions, particularly the gradient of a three-dimensional
delta function, which is defined by
\begin{equation}
 \int\! \td^3y\, \bnabla\delta(\bx-\by)\,\Phi(\by) = \bnabla\Phi(\bx),
\end{equation}
where $\Phi(\bx)$ is a test function.  That is to say, $\bnabla\delta$
may be regarded as a conventional function capable of being partially
integrated and obeying
\begin{equation}
 \bnabla\delta(\bx-\by)
 = \frac\partial{\partial\bx} \delta(\bx-\by)
 = - \frac\partial{\partial\by} \delta(\bx-\by).
\end{equation}
The Heaviside step function is defined by
\begin{equation}
 \theta(t) =
 \begin{cases}
  1 & \text{if } t>0, \\
  0 & \text{if } t<0.
 \end{cases}
\end{equation}
Its derivative is the delta function, $\theta'(t)=\dirac{}(t)$, and its
Fourier transform equals $i/(E+i\eps)$, i.e.,
\begin{equation} \label{theta-Fourier}
 \theta(t) = \frac i{2\pi} \!\int_{-\infty}^{\infty}\! \td E\,
 \frac {e^{-iEt}}{E+i\eps}
\end{equation}
and, inversely,
\begin{equation}
 \frac i{E+i\eps} = \int_0^\infty\! \td t\, e^{iEt}.
\end{equation}

The totally antisymmetric tensors are normalized to
$\eps^{123}=\eps^{0123}=+1$.  For any four-vector $r$, one has the identity
\begin{equation} \label{anti5}
 r^\mu\eps^{\nu\lambda\rho\sigma} = r^\nu\eps^{\mu\lambda\rho\sigma} +
 r^\lambda\eps^{\nu\mu\rho\sigma} + r^\rho\eps^{\nu\lambda\mu\sigma} +
 r^\sigma\eps^{\nu\lambda\rho\mu},
\end{equation}
which follows from the fact that a totally antisymmetric tensor of
fifth or higher rank vanishes in four dimensions.  Furthermore, I
adopt the convention $\gamma_5=i\gamma^0\gamma^1\gamma^2\gamma^3$,
which gives rise to the trace theorem
$\tr(\gamma_5\gamma^\mu\gamma^\nu\gamma^\rho\gamma^\sigma) =
-4i\eps^{\mu\nu\rho\sigma}$.  Dirac sigma matrices are defined by
\begin{equation}
 \sigma^{\mu\nu} = \tfrac i2\, [\gamma^\mu, \gamma^\nu].
\end{equation}

The nucleon's mass, charge (in units of $e$), and anomalous magnetic
moment (in units of the nuclear magneton $e/2M_{\text{p}}$) are denoted by
$M$, $Z$, and $\kappa$, respectively.  Thus,
\begin{subequations}
\begin{align}
 M_{\tp} &= 938 \text{~MeV} & Z_{\tp} &= 1 & \kappa_{\tp} &= 1.793 \\
 M_{\tn} &= 940 \text{~MeV} & Z_{\tn} &= 0 & \kappa_{\tn} &= -1.913
\end{align} \label{MZkappa}%
\end{subequations}
where the small mass difference is usually neglected.  One-nucleon states
$|p,\lambda\rangle$ with four-momentum $p$ and helicity
$\lambda=\pm\frac12$ are normalized covariantly,
\begin{equation} \label{N-norm}
 \langle p', \lambda' | p, \lambda \rangle =
 (2\pi)^3\,2p^0\,\dirac3(\bp' - \bp)\,\delta_{\lambda'\lambda}.
\end{equation}
Relative phases between different-helicity states are fixed according
to Jacob and Wick \cite {Jacob59}.  Spinors $u(p,\lambda)$ are
normalized as
\begin{equation}
 \bar u(p,\lambda') u(p,\lambda) = 2M\,\delta_{\lambda'\lambda}.
\end{equation}
Besides this helicity basis, I occasionally use states $|p,s\rangle$
and spinors $u(p,s)$ with arbitrary spin four-vector $s$ obeying
\begin{align}
 s^2 & = -1, \\*
 p\ndot s & = 0,
\end{align}
and
\begin{equation}
 \bar u(p,s) \gamma^\mu \gamma_5 u(p,s) = 2M s^\mu.
\end{equation}
Moreover, one has that
\begin{equation}
 u(p,s) \bar u(p,s) = (\pslash+M)\, \frac{1+\gamma_5\sslash}2.
\end{equation}
The matrix $u(p,s)\bar{u}(p,s)/2M$ projects onto spin $s$ and positive energy.
Helicity eigenstates are those for which the three-vectors $\bp$ and
$\bs$ are collinear.

For the electron, which is considered in \Sects {QED}, \sect {WSM},
and \sect {PRP}, the nucleon mass $M$ has to be replaced by the
electron mass $m$ in above formulae.

One-pion states $|p_\pi\rangle$, occurring in \Ch {Q2}, are
normalized as
\begin{equation} \label{pi-norm}
 \langle p_\pi' | p_\pi \rangle =
 (2\pi)^3\,2p_\pi^0\,\dirac3(\bp_\pi' - \bp_\pi).
\end{equation}
Pion-nucleon states $|p_\pi;p,\lambda\rangle$ are of course simply products
of pion states $|p_\pi\rangle$ and nucleon states $|p,\lambda\rangle$.
Their normalization is therefore given by the product of \Eqs
{N-norm} and \eq {pi-norm}.

Current densities $J^\mu(x)$ are defined in units of elementary charges per
volume, i.e., for a single charged Dirac field $\psi(x)$ one has
$J^\mu(x) = \bar\psi(x)\gamma^\mu\psi(x)$,
and the physical current density (in proper units) is $eJ^\mu(x)$.

\clearpage{\pagestyle{empty}\cleardoublepage}
\chapter{Derivations of the GDH sum rule}
\label{Ch:derivations}

In this chapter, I present three different derivations of the GDH sum
rule, each of which having its advantages as well as its drawbacks.
It is advisable to comprehend all of them, particularly in view of
possible modifications to the GDH sum rule.  The dispersion-theoretic
derivation was given in the original publications by Gerasimov \cite
{Gerasimov65} and by Drell and Hearn \cite {Drell66}, and it is
comparatively simple. The derivation from equal-times current algebra
has been presented by Hosoda and Yamamoto \cite {Hosoda66a},
Kawarabayashi and Suzuki \cite {Kawarabayashi66a}, and, six years
later, by Pradhan and Khare \cite {Pradhan72}.  (All of these papers
are extremely concise.  Also, I must warn the reader of \Ref
{Pradhan72}, which contains a vast number of errors on less than two
pages.)  Finally, there is a single and not very widely-known
publication by Dicus and Palmer \cite {Dicus72} that presents a
derivation starting from a light-cone commutator of currents.

The current-algebra approaches appear much more involved than the
dispersion-theoretic one, but partly this is due to the fact that in a
way they re-derive the low-energy theorem, which is taken as a premise
to the dispersion-theoretic approach.  The general ideas of current
algebra are treated in much detail in the textbook of Adler and Dashen
\cite {Adler68}, which is supplemented by a nice collection of
reprints of classical papers on this topic.

\section{Dispersion theory}
\label{Sect:DT}

\subsection{Forward Compton amplitude}
\label{Sect:forwComp}

\begin{figure}[tb]
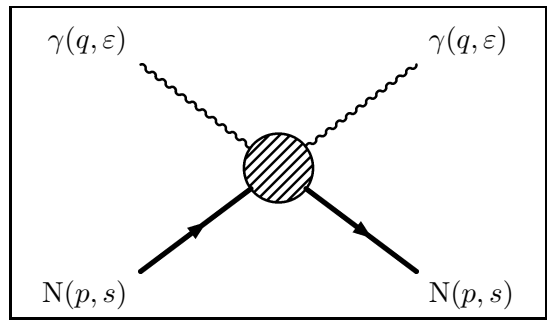

 \begin{displaymath}
  \boxedgraph(5,3){
   \fmfleft{i1,i2} \fmfright{o1,o2}
   \fmflabel{N($p,s$)}{i1} \fmflabel{N($p,s$)}{o1}
   \fmflabel{$\gamma(q,\varepsilon)$}{i2}
   \fmflabel{$\gamma(q,\varepsilon)$}{o2}
   \fmf{fermion,width=thick}{i1,v,o1}
   \fmf{boson}{i2,v,o2}
   \fmfblob{1\ul}{v}
 }
 \end{displaymath}
 \caption[]{
  Forward Compton scattering off the nucleon
  \label{Fig:forw}}
\end{figure}

Consider the forward Compton-scattering process as depicted in \Fig {forw}.%
The corresponding amplitude may be written
\begin{equation}  \label{forw-amp1}
 T^{\mu\nu} = i\!\int\!\td^4x\, e^{iq\ndot x}
  \me {p,s} {\tT J^\mu(x)J^\nu(0)} {p,s},
\end{equation}
where T denotes time ordering.  This definition has no regard to the
actual value of the photon virtuality $q^2$.  In \Ch {Q2}, it will be
utilized for the generalization to virtual Compton scattering, but
presently only the case $q^2=0$ (real photons) is of interest.

I further define the amplitude
\begin{equation}  \label{comp-lab}
 T = \frac{e^2}{8\pi M}\,\eps_\mu^*T^{\mu\nu}\eps_\nu,
\end{equation}
where the factor $e^2$ is due to the fact that the physical current
density (in units of charge per volume) is $eJ^\mu(x)$, and $1/8{\pi}M$
absorbs a kinematic factor from the relation of this amplitude to the
forward differential Compton cross section in the lab frame:
\begin{equation} \label{sigmaComp}
 \left.\frac{\td\sigma}{\td\Omega^{\tlab}}\right|_{\text{forward}}
 = |T|^2.
\end{equation}
The complex conjugation of the polarization vector of the outgoing
photon in \Eq {comp-lab} is needed because circularly polarized
photons will be considered.  Adopting the Coulomb gauge $\eps^0=0$, this
amplitude has the decomposition
\begin{equation}  \label{T-dec}
 T(\nu) = \chi^\dagger\bigl[f_1(\nu) \,\beps^*\ndot\beps
  +i\nu f_2(\nu) \,\bsigma\ndot(\beps^*\ntimes\beps)\bigr]\chi,
\end{equation}
where $\chi$ denotes the unit-normalized ($\chi^\dagger\chi=1$)
nucleon Pauli spinor and $\nu=p\ndot{q}/M$ is the photon lab-frame
energy.  Simultaneously flipping the photon helicity and the nucleon
spin, which corresponds to the replacements
$\beps^*\leftrightarrow\beps$ and $\bsigma\to-\bsigma$, leaves the
amplitude \eq {T-dec} unaltered.  This reflects the conservation of
parity in strong and electromagnetic processes.  Note that in the lab
frame, the nucleon spin four-vector $s$ is given by $s^0=0$,
$\bs=\chi^\dagger\bsigma\chi$.

Decomposition \eq {T-dec} follows the notation of Drell and Hearn
\cite {Drell66}.\footnote {The notation of Gerasimov \cite
{Gerasimov65} is related to the one of Drell and Hearn by $G_1(\nu)=f_1(\nu)$,
$G_2(\nu)={\nu}f_2(\nu)$.} Observe that an explicit factor of $\nu$
renders the amplitude $f_2(\nu)$ even under crossing.

\subsubsection{Polarization}
If one lets the photon travel along the third coordinate axis,
$\bq=q^0\be_3$, then the left-circularly polarized state,
corresponding to photon helicity $+1$, is represented by
$\beps=(-\be_1-i\be_2)/\sqrt2$, which implies $\beps^*\ndot\beps=1$
and $\beps^*\ntimes\beps=i\be_3$.  Thus, $T(\nu) = f_1(\nu) - \nu
f_2(\nu)s^3$.  If nucleon and photon have antiparallel spin, then
$s^3=-1$ and the pertinent amplitude reads
\begin{subequations}
\begin{equation}  \label{T1/2}
 T_{1/2}(\nu)=f_1(\nu)+{\nu}f_2(\nu),
\end{equation}
where subscript 1/2 stands for ``total helicity 1/2''.  (Note,
however, that this is the difference -- not the sum -- of photon and
nucleon helicities in the center-of-mass frame, and that the nucleon
helicity in the lab frame is not even defined.)  Analogously, if
nucleon and photon spins are parallel, then $s^3=+1$ and
\begin{equation}  \label{T3/2}
 T_{3/2}(\nu)=f_1(\nu)-{\nu}f_2(\nu).
\end{equation} \label{T1/2-3/2}%
\end{subequations}

\subsubsection{Lowest-order electromagnetic coupling}
\label{lowest-order-story1}
As usual in the theory of electromagnetic interactions of hadrons, all
amplitudes are considered to lowest order in the electromagnetic
coupling, i.e., only the order-$\alpha$ pieces of the amplitudes
$f_{1,2}(\nu)$ are taken into account.  Pictorially, this means that
the blob in \Fig {forw} containes only strongly interacting
particles (quarks and gluons, or mesons and baryons, which do you
prefer?), and no photons and leptons.  As a consequence, all
quantities under consideration (cross sections, masses, magnetic
moments) are approximated by their lowest-order pieces.  This ought to
be kept in mind. I will further comment on it in the following
subsections.

\subsection{Low-energy theorem}
\label{Sect:LET}
In 1954, Low \cite {Low54} and Gell-Mann and Goldberger \cite
{Gell-Mann54} showed that to first order in the photon energy $\nu$,
the Compton amplitude $T(\nu)$ equals its Born contribution
\begin{equation} \label{T-Born}
 T^{\tBorn}(\nu) =
  \graph(3,2){
   \fmfleft{i1,i2} \fmfright{o1,o2}
   \fmf{fermion,width=thick}{i1,v1,v2,o1}
   \fmf{boson}{i2,v1}  \fmf{boson}{v2,o2}
   \fmfdot{v1,v2}
  } +
  \graph(3,2){
   \fmfleft{i1,i2} \fmfright{o1,o2}
   \fmf{fermion,width=thick}{i1,v1,v2,o1}
   \fmf{phantom}{i2,v1} \fmf{phantom}{v2,o2}
   \fmffreeze
   \fmf{boson}{i2,v2} \fmf{boson}{v1,o2}
   \fmfdot{v1,v2}
  }
\end{equation}
where the vertices incorporate the nucleon's charge $Ze$ and anomalous
magnetic moment $\mu_{\text{a}}=e\kappa/2M$,
\begin{equation}
 \graph(2,2){
  \fmftop{t} \fmfbottom{b1,b2}
  \fmf{boson}{t,v}
  \fmf{fermion,width=thick}{b1,v,b2}
  \fmfdot{v}
}
 = Ze\gamma^\mu + i \mu_{\text a} q_\rho \sigma^{\mu\rho}.
\end{equation}
The Born parts of amplitudes $f_{1,2}(\nu)$ read
\begin{subequations}
\begin{align}
 f_1^{\tBorn}(\nu) &\equiv -\frac{Z^2\alpha}{M}, \\
 f_2^{\tBorn}(\nu) &\equiv -\frac{\mu_{\text a}^2}{2\pi}.
\end{align}
\end{subequations}
Thus one has the zeroth-order low-energy theorem
\begin{subequations}
\begin{equation} \label{LET.Thirring}
 f_1(0) = -\frac{Z^2\alpha}{M},
\end{equation}
tracing back to Thirring \cite {Thirring50}, and on account of the explicit
factor of $\nu$ accompanying amplitude $f_2(\nu)$, the equality of the
first orders within $T(\nu)$ and $T^{\tBorn}(\nu)$ results into
the relation
\begin{equation} \label{LET.LGG}
 \boxed{f_2(0) = -\frac{\mu_{\text a}^2}{2\pi} = -\frac{\alpha \kappa^2}{2M^2}}
\end{equation} \label{LET}%
\end{subequations}
which was the novel finding of Low, Gell-Mann, and Goldberger.

Abarbanel and Goldberger \cite {Abarbanel68} re-derived the low-energy
theorem by means of dispersion relations.  The methods employed in
Refs.\ \plaincite {Low54,Gell-Mann54} and \plaincite {Abarbanel68}
have been summarized by Bardakci and Pagels \cite {Bardakci68}, who
also emphasized the now-following point in a footnote.\footnote
{\label{wisdom}A great deal of wisdom is frequently concealed in
footnotes.}

\subsubsection{Lowest-order electromagnetic coupling}
\label{lowest-order-story2}
It may be worth mentioning that the quantity $\mu_{\text{a}}$ in
\Eq {LET.LGG} is actually not the true, measureable anomalous magnetic
moment of the nucleon, but rather a fictitious lowest-order (in
$\alpha$) contribution to it.  This is due to the fact that Compton
amplitudes are considered only to lowest order in the electromagnetic
coupling.  Pictorially, the absence of virtual photons inside the blob
in \Fig {forw} is reflected by the absence of virtual photons from the
vertices of the Born contribution \eq {T-Born}.  We do not know this
ficticious lowest-order anomalous magnetic moment since we cannot
switch off electromagnetism inside hadrons.  But we believe that it is
a good approximation.  This belief is inspired by the smallness of the
Schwinger anomalous magnetic moment $\alpha/2\pi$ of the fundamental
charged leptons.  For hadrons, however, the approximate equality of
lowest-order and all-orders anomalous magnetic moments is by no means
necessary.  On the other hand, it is not clear whether the low-energy
theorem \eq {LET.LGG} would survive the inclusion of all orders of
$\alpha$.  Only for the next order, it has been proved by Cheng \cite
{Cheng68,Cheng69} and by Roy and Singh \cite {Roy68}, that
\Eq {LET.LGG} remains unchanged.  As far as the Thirring theorem \eq
{LET.Thirring} is concerned, the situation is completely different.
This low-energy theorem is exact to all orders of $\alpha$, and the
charge $Ze$ does, of course, not suffer radiative corrections, as does
the magnetic moment.

\subsection{Optical theorem}

\subsubsection{Total photoabsorption}
\begin{figure}[tb]
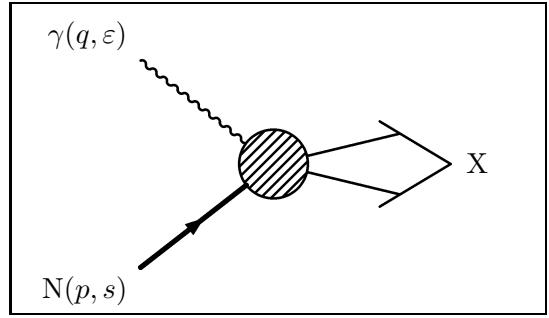

 \begin{displaymath} \boxedgraph(5,3){
  \fmfleft{i1,i2} \fmfright{o} \fmfbottom{b} \fmftop{t}
  \fmf{fermion,width=thick}{i1,v}
  \fmf{boson}{i2,v}
  \fmf{phantom,tension=1.5}{v,o}
  \fmfblob{1\ul}{v}
  \fmf{phantom}{t,a1} \fmf{plain,tension=5}{a1,a2} \fmf{plain,tension=2}{a2,o}
  \fmf{phantom}{b,a3} \fmf{plain,tension=5}{a3,a4} \fmf{plain,tension=2}{a4,o}
  \fmffreeze
  \fmf{plain}{a2,v,a4}
  \fmflabel{N$(p,s)$}{i1} \fmflabel{$\gamma(q,\varepsilon)$}{i2}
  \fmflabel{X}{o}
 } \end{displaymath}
 \caption[]{
  Polarized total photoabsorption on the nucleon,
  related to forward Compton scattering by means of the optical theorem.
  Cross sections for all final states X are summed over.
  \label{Fig:photoabs}}
\end{figure}

Generally, the optical theorem, which rests upon unitarity of the
$S$-matrix, relates the absorptive part of a forward amplitude to the
corresponding total absorption cross section.  In case of forward Compton
scattering off the nucleon, the optical theorem reads
\begin{subequations}
\begin{gather} \label{f1-opt}
 \im f_1(\nu) = \frac{\nu}{8\pi}
  \bigl(\sigma_{1/2}(\nu)+\sigma_{3/2}(\nu)\bigr), \\*
 \boxed{ \im f_2(\nu) = \frac{1}{8\pi}
  \bigl(\sigma_{1/2}(\nu)-\sigma_{3/2}(\nu)\bigr)} \label{f2-opt}
\end{gather} \label{opt-th}%
\end{subequations}
where $\sigma_{1/2}(\nu)$ and $\sigma_{3/2}(\nu)$ denote the
photoabsorption cross sections of the nucleon for total photon-nucleon
helicities 1/2 and 3/2, respectively.  The photoabsorption process is
depicted in \Fig {photoabs}.%

\subsubsection{Derivation}
The optical theorem \eq {opt-th} is formally
derived as follows.  The absorptive part of the forward Compton amplitude
\eq {forw-amp1} reads
\begin{equation} \label{abs-T}
 2\pi W^{\mu\nu} := \abs T^{\mu\nu} = \frac12\!\int\!\td^4x\, e^{iq\ndot x}
  \me {p,s} {[J^\mu(x),J^\nu(0)]} {p,s}.
\end{equation}
In this expression, the second part of the commutator,
$-J^\nu(0)J^\mu(x)$, can be omitted for positive photon energies
$q^0$, since no intermediate state coupling to $\gamma$N is lighter
than the nucleon itself:
\begin{align} \label{JnuJmu}
 \int\!\td^4x\, e^{iq\ndot x} & \me {p,s} {J^\nu(0)J^\mu(x)} {p,s} \notag\\*
 & = \sum_\tX \int\!\td^4x\, e^{iq\ndot x}
  \me {p,s} {J^\nu(0)} {\tX} \me {\tX} {J^\mu(x)} {p,s} \notag \\
 & = \sum_\tX \int\!\td^4x\, e^{i(p_\tX-p+q)\ndot x}
  \me {p,s} {J^\nu(0)} {\tX} \me {\tX} {J^\mu(0)} {p,s} \notag \\
 & = \sum_\tX (2\pi)^4 \dirac4(p_\tX-p+q)
  \me {p,s} {J^\nu(0)} {\tX} \me {\tX} {J^\mu(0)} {p,s}.
\end{align}
The delta function picks out intermediate states $|\tX\rangle$ with
four-momentum $p_\tX=p-q$, which lies below the nucleon-mass shell for
$q^0>0$.  Hence, expression \eq {JnuJmu} vanishes, as claimed above.

The absorptive part of the decomposition \eq {T-dec} is obtained by
simply replacing the amplitudes $f_{1,2}(\nu)$ by their imaginary
parts,
\begin{align}
 \abs T(\nu) &= \frac{e^2}{8\pi M}\,\eps_\mu^*\eps_\nu\abs T^{\mu\nu} \notag\\*
 &= \chi^\dagger\bigl[\im f_1(\nu) \,\beps^*\ndot\beps
  +i\nu\im f_2(\nu) \,\bsigma\ndot(\beps^*\ntimes\beps)\bigr]\chi.
\end{align}
The Coulomb-gauge photoabsorption cross section can be written
(see, e.g., Itzykson and Zuber \cite [App.~A] {Itzykson80})
\begin{equation} \label{photo}
 \sigma(\nu) = \frac{e^2}{4M\nu} \sum_{\text{X}}
  (2\pi)^4\dirac4(p+q-p_{\text{X}})\,
  \bigl|\langle\text{X}|\beps\ndot\bJ(0)|p,s\rangle\bigr|^2,
\end{equation}
where the sum runs over all possible final states
$\text{X}=\pi\tN,\pi\pi\tN,\ldots$, summing over spins and
integrating over phase space.  By virtue of the Fourier transform
of the unity, $\int\!\td^4x\,e^{ik\ndot{x}}=(2\pi)^4\dirac4(k)$, and
by employing the translational-invariance condition \eq {trans},
%\begin{equation}
% \me {p,s}{\bJ(x)}{\tX} = e^{i(p-p_\tX)\ndot x} \me {p,s}{\bJ(0)}{\tX},
%\end{equation}
the photoabsorption cross section can now be expressed in terms of the
absorptive amplitudes,
\begin{align}
 \sigma(\nu) &= \frac{e^2}{4M\nu} \int\!\td^4x\,e^{iq\ndot x}
  \sum_{\text{X}} \langle p,s|\beps^*\ndot\bJ(x)|\text{X}\rangle
  \langle\text{X}|\beps\ndot\bJ(0)|p,s\rangle \notag\\
 &= \frac{e^2}{2M\nu} \, \eps^*_\mu\eps_\nu \abs T^{\mu\nu}
  = \frac{4\pi}{\nu} \abs T(\nu).
\end{align}
Using relations \eq {T1/2-3/2}, one obtains
\begin{subequations}
\begin{equation}
 \sigma_{1/2}(\nu) = \frac{4\pi}{\nu}\bigl(\im f_1(\nu) + \nu\im f_2(\nu)\bigr)
\end{equation}
and
\begin{equation}
 \sigma_{3/2}(\nu) = \frac{4\pi}{\nu}\bigl(\im f_1(\nu) - \nu\im f_2(\nu)\bigr)
\end{equation}
\end{subequations}
for respective total helicities 1/2 and 3/2, from which \Eqs
{opt-th} follow immediately.

\subsubsection{Lowest-order electromagnetic coupling}
\label{lowest-order-story3}
As in the case of the low-energy theorem, a few words on the omission of
radiative corrections may be in order.  The optical theorem relates the
\emph {true} (all orders in $\alpha$) forward Compton amplitude to the \emph
{true} photoabsorption cross section.  What kind of cross section is the
\emph {lowest-order} amplitude (the one without virtual photons inside the
blob of \Fig {forw}) related to?  Again, this is a ficticious
cross section, depicted by \Fig {photoabs} if one removes all
photons and lepton pairs from the final state X \emph {and} from the
blob.  As far as the absence of photons and leptons from the final
state is concerned, this is actually desired, since experimentally,
a huge background of these particles is produced via Compton
scattering off electrons and Bethe-Heitler pair production on nuclei.
This background has to be disentangled from the signal anyway, which
is one of the main reasons for considering only the lowest order in
$\alpha$.  However, neglecting virtual photons inside the interaction
zone in \Fig {photoabs} is an approximation.  It means that
effects like electromagnetic final-state interactions are disregarded.
Again, it is believed that this approximation is a comparatively good one.
Finally, it should be noted that the inclusion of higher orders of $\alpha$
would lead to some other non-trivial complications in the mere \emph
{definition} of amplitudes and cross sections on account of the inevitable
radiation of soft photons in \emph {any} scattering process.

\subsection{Unsubtracted dispersion relation}
\label{Sect:UDR}

\subsubsection{Analytical continuation}
\begin{figure}[tb]
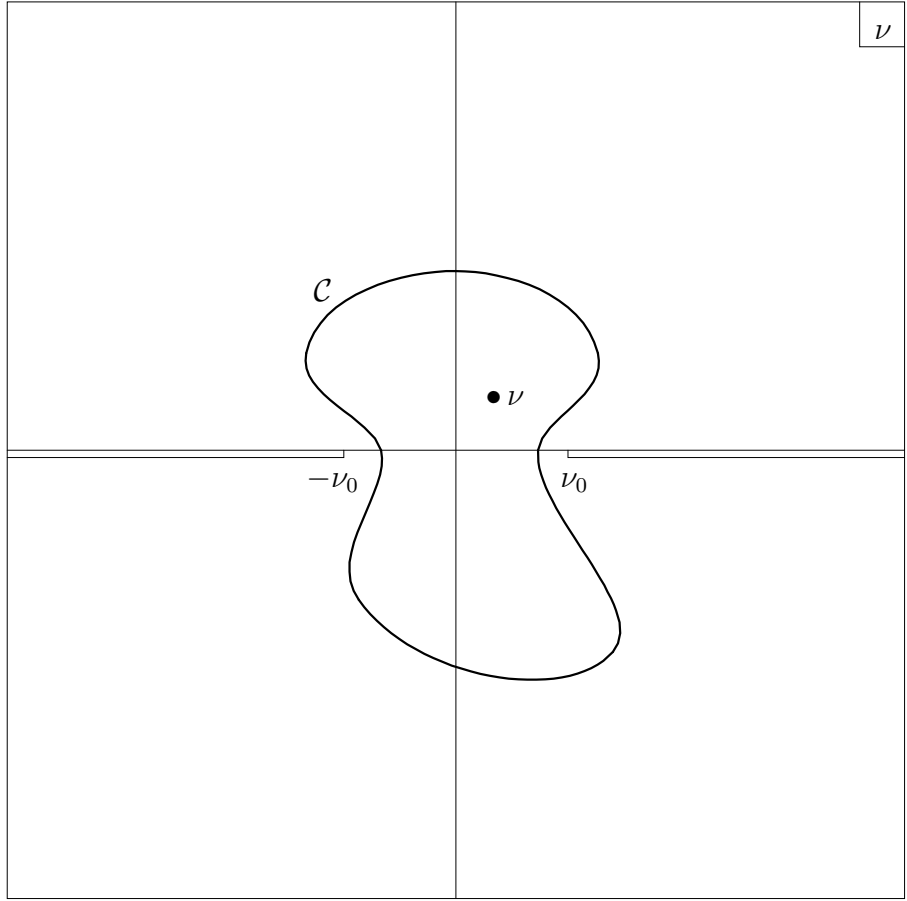

 \begin{center}
  \input cauchy1.eepic
 \end{center}
 \caption[]{
  Complex energy plane of the forward Compton scattering
  process: cut structure, finite Cauchy-integration contour
  \label{Fig:Cauchy1}}
\end{figure}

From causality, the forward Compton amplitudes $f_{1,2}(\nu)$ can be
analytically continued into the upper half of the complex energy plane.
The continuation into the lower half plane is done by demanding the
validity of the Schwarz reflection principle,
\begin{equation}  \label{Schwarz1}
 f_i(\nu^*) = f_i(\nu)^*,\quad i=1,2,\quad \nu\notin\Bbb{R}.
\end{equation}
Approaching the real axis from below, this reduces to
\begin{equation} \label{Schwarz2}
 f_i(\nu-i\eps)=f_i(\nu)^*,\quad \nu\in\Bbb{R}.
\end{equation}
Thus, the real parts of the amplitudes $f_i(\nu)$ are continuous,
while their imaginary parts change sign when crossing the real axis
and hence are discontinuous, leading to a \emph {cut} if $f_i(\nu)$ is
not real.  Since one works to lowest order in $\alpha$, the lowest
possible intermediate state in the Compton scattering process
illustrated by \Fig {forw} is the $\pi$N state with invariant mass
$(p+q)^2\ge(M+m_\pi)^2$.  In the lab frame, this inequality reads
$(M+\nu)^2-\nu^2\ge(M+m_\pi)^2$, or $\nu\ge\nu_0$ with
\begin{equation}
 \nu_0 := m_\pi+\frac{m_\pi^2}{2M},
\end{equation}
which is the inelastic \emph {threshold} of the photoproduction reaction.
The least-mass state in the $u$ channel of the same reaction is again
$\pi$N, opening at $(p-q)^2\ge(M+m_\pi)^2$, or $\nu\le-\nu_0$.  Thus
the above-mentioned cuts run from $-\infty$ to $-\nu_0$ and from
$\nu_0$ to $\infty$ along the real axis of the complex $\nu$ plane.
On the real axis itself, the values of functions $f_i(\nu)$ fit
to the upper half plane.  This structure of the complex $\nu$ plane is
indicated in \Fig {Cauchy1}.%
Observe that crossing symmetry reads $f_i(-\nu)=f_i(\nu)$ for complex
$\nu\notin\Bbb{R}$, so that for real energies,
$f_i(-\nu)=f_i(-\nu+i\eps)=f_i(\nu-i\eps)=f_i(\nu)^*$ due to the
reflection principle.  Hence the real parts of the amplitudes are even
functions of real $\nu$, while the imaginary parts are odd.

\subsubsection{Cauchy's theorem}
One may now apply Cauchy's theorem,
\begin{equation} \label{Cauchy}
 f_i(\nu) = \frac1{2\pi{i}} \oint_{\mathcal{C}}\!
  \frac{\td\nu'}{\nu'-\nu}\, f_i(\nu'),
\end{equation}
where the point $\nu$ is kept away from the cuts, see \Fig
{Cauchy1}.  Of course, the integration contour $\mathcal{C}$ must not
cross the cuts.  

\begin{figure}[tb]
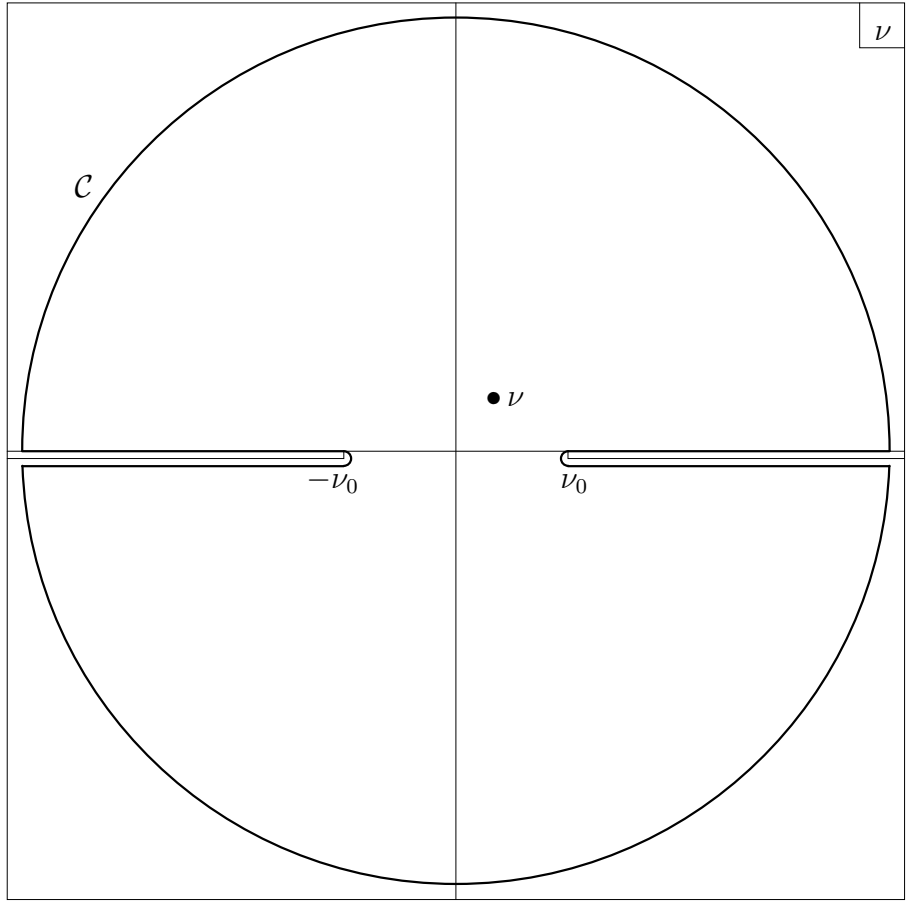

 \begin{center}
  \input cauchy2.eepic
 \end{center}
 \caption[]{
  Blown-up Cauchy-integration contour, the half circles running at infinite
  photon energies
  \label{Fig:Cauchy2}}
\end{figure}

Now the contour is enlarged to the infinite one depicted in \Fig {Cauchy2},
assuming for the moment that the integral stays finite on every
segment of the path.%
Later we will see whether this assumption can be justified from Regge
phenomenology.  The contribution from the piece on the right-hand cut
is just
\begin{subequations}
\begin{equation} \label{right.above}
% \frac1{2\pi i} \int_{\nu_0+i\eps}^{\infty+i\eps}\!
%  \frac{\td\nu'}{\nu'-\nu}\, f_i(\nu') =
 \frac1{2\pi i} \int_{\nu_0}^{\infty}\!
  \frac{\td\nu'}{\nu'-\nu}\, f_i(\nu').
\end{equation}
%since everything is continuous when approaching the real axis from above.
The contribution from the piece \emph {below} the right-hand cut reads
\begin{align} \label{right.below}
 \frac1{2\pi i} \int_{\infty-i\eps}^{\nu_0-i\eps}\!
  \frac{\td\nu'}{\nu'-\nu}\, f_i(\nu')
 &= -\frac1{2\pi i} \int_{\nu_0}^{\infty}\!
  \frac{\td\nu'}{\nu'-i\eps-\nu}\, f_i(\nu'-i\eps) \notag\\*
 &= -\frac1{2\pi i} \int_{\nu_0}^{\infty}\!
  \frac{\td\nu'}{\nu'-\nu}\, f_i(\nu')^*,
\end{align} \label{right}%
\end{subequations}
where the first equality is a simple substitution of the integration
variable and the second one is due to the continuity of the function
$\nu'-\nu$ and the fact that functions $f_i(\nu)$ get complex-conjugated
when crossing the real axis, \Eq {Schwarz2}.  Putting \Eqs
{right} together, one obtains a contribution
\begin{subequations}
\begin{equation} \label{cuts.right}
 \frac1\pi \int_{\nu_0}^{\infty}\!\frac{\td\nu'}{\nu'-\nu}\,
  \im f_i(\nu')
\end{equation}
from the right-hand cut.  Similarly, the pieces of the integration contour
enclosing the left-hand cut contribute
\begin{equation} \label{cuts.left}
 \frac1\pi \int^{-\nu_0}_{-\infty}\!\frac{\td\nu'}{\nu'-\nu}\,
  \im f_i(\nu') =
 \frac1\pi \int_{\nu_0}^{\infty}\!\frac{\td\nu'}{\nu'+\nu}\,
  \im f_i(\nu'),
\end{equation} \label{cuts}%
\end{subequations}
where the crossing properties of functions $f_i(\nu)$ have been employed.

\subsubsection{No-subtraction hypothesis}
Now one makes the crucial assumption that the half circles at infinite
complex photon energies, indicated in \Fig {Cauchy2}, do \emph {not}
contribute to the Cauchy integral \eq {Cauchy}.  This requirement is
fulfilled if functions $f_i(\nu)$ vanish asymptotically,
\begin{equation} \label{fito0}
 f_i(\nu) \xrightarrow[\nu\to\infty]{} 0.
\end{equation}
If, on the other hand, function $f_2(\nu)$ approaches a finite constant
$f_2(\infty)$ at $\nu\to\infty$, then the half circles contribute just
this constant.  I will revert to this possibility in \Sect
{fixed-pole}.  For the moment, let's stick to the assumption
$f_i(\infty) = 0$.  Then, the Cauchy integral \eq {Cauchy} is the
sum of expressions \eq {cuts},
\begin{equation}
 f_i(\nu) = \frac2\pi \int_{\nu_0}^{\infty}\!
  \frac{\td\nu'\,\nu'}{{\nu'}^2-\nu^2}\, \im f_i(\nu').
\end{equation}
Letting $\nu\to0$ on both sides of this \emph {unsubtracted dispersion
relation} gives
\begin{equation} \label{UDR}
 f_i(0) = \frac2\pi \int_{\nu_0}^{\infty}\!
  \frac{\td\nu'}{\nu'}\, \im f_i(\nu').
\end{equation}
Together with the low-energy theorems \eq {LET}
and the optical theorem \eq {opt-th} one arrives at
\begin{equation} \label{f1-dr}
 -\frac{4\pi^2Z^2\alpha}{M} =
  \int_{\nu_0}^\infty\! \td\nu
  \bigl(\sigma_{1/2}(\nu) + \sigma_{3/2}(\nu)\bigr)
\end{equation}
and
\begin{equation} \label{GDH2}
 \boxed{ -\frac{2\pi^2\alpha\kappa^2}{M^2} =
  \int_{\nu_0}^\infty\! \frac{\td\nu}\nu
  \bigl(\sigma_{1/2}(\nu) - \sigma_{3/2}(\nu)\bigr) }
\end{equation}

\subsubsection{Divergence of the unpolarized integral}
It can be seen at first sight that \Eq {f1-dr} is incorrect,
because the integrand is a \emph {positive} function of the photon
energy, while the constant on the left-hand side is \emph {negative}
for the proton and \emph {zero} for the neutron.  What went wrong?
Evidently, the Cauchy integration formula is valid for any \emph
{bounded} contour like the one in \Fig {Cauchy1}.  But in blowing
up the contour to the infinitely large shape depicted in \Fig
{Cauchy2}, the contribution from the pieces enclosing the cuts might
\emph {diverge}, say, to $+\infty$.  At the same time, the
contribution from the large half circles will go to $-\infty$, since
the sum of both is still the finite number $f(\nu)$.  In this
scenario, requirement \eq {fito0} will be violated.  In fact, as
far as the unpolarized cross section within the integral of \Eq
{f1-dr} is concerned, Regge theory tells us that its high-energy
behavior is at least that of a constant.  In case of the proton, for
which one has very-high-energy data (up to photon lab-frame energies
of roughly 20,000 GeV), there is even a significant \emph {rise} of
the total cross section \cite {Derrick92,Ahmed93} (see also the \emph
{Review of Particle Physics} \cite {PDG96}).  So this integral is
indeed badly divergent.

\subsubsection{Convergence of the GDH integral}
\Eq {GDH2} is the GDH sum rule.  Mueller and Trueman \cite
{Mueller67b} showed under the general assumptions of Regge theory,
that a ``worst-case'' high-energy behavior of
$\sigma_{1/2}-\sigma_{3/2}$ is caused by a cut in complex angular
momentum plane running from $-\infty$ to 1 (possibly due to
two-pomeron exchange \cite {Close88}), leading to
\begin{equation} \label{2Pom}
 \sigma_{1/2}(\nu) - \sigma_{3/2}(\nu) \xrightarrow[\nu\to\infty]{}
  \frac{\text{const}}{\ln^2\nu},
\end{equation}
which renders the integral finite,\footnote {Observe that an
asymptotically vanishing $\im{f_2(\nu)}$ alone does \emph {not} ensure
convergence.  For instance, a high-energy behavior of const$/\ln\nu$
would result in $$ \int\!\frac{\td\nu}{\nu}\,
\frac{\text{const}}{\ln\nu} = \text{const}
\!\int\!\frac{\td\ln\nu}{\ln\nu} \to \infty.$$}
\begin{equation}
 \int\!\frac{\td\nu}{\nu}\, \frac{\text{const}}{\ln^2\nu} =
 \text{const} \!\int\!\frac{\td\ln\nu}{\ln^2\nu} < \infty.
\end{equation}

\section{Equal-times current algebra}
\label{Sect:ETCA}

In this section, I present the current-algebra\footnote {For historical
reasons, ``current-algebra'' means ``equal-times current-algebra''
here and in the following.} derivation of the GDH sum rule, which is
based essentially on two premises.  Firstly, the operators of electric
charge densities are assumed to commute at equal times.  I will
demonstrate below how the validity of this assumption is \emph
{naively} obtained from canonical anticommutation relations among
fundamental Dirac fields, e.g., quarks.  \Sect {anomcomm}
is devoted to a more thorough investigation of this point.  Secondly,
one assumes that taking the infinite-momentum limit is legitimate.  At
the end of this section, the reader will hopefully have a solid notion
of what this really means -- both mathematically and physically.  In
the literature, one can find a couple of ans\"atze
\cite {Kawarabayashi66c,Khare75,Chang94a} that weaken the former
assumption (equal-times commuting charge-density operators), but
carelessly, the legitimacy of the infinite-momentum limit has never
been questioned seriously.  I will argue in \Sect {PRP} that in
view of possible modifications to the GDH sum rule, the validity of
the named assumptions is highly correlated.

I follow the idea of Hosoda and Yamamoto \cite{Hosoda66a}, but
I postpone the infinite-momentum limit to the very end of the calculation
in order to shed some light on its meaning.

\subsubsection{Timelike photons}
For reasons that will become transparent later, the forward Compton amplitude
\begin{equation}  \label{forw-amp2}
 T^{\mu\nu} = i\!\int\!\td^4x\, e^{iq\ndot x}
  \me {p,s} {\tT J^\mu(x)J^\nu(0)} {p,s}
\end{equation}
must be generalized to timelike photon virtualities $q^2>0$ in the
current-algebra approach to sum rules.  Since a virtual photon -- in
contrast to a real photon -- can have longitudinal polarization, the
number of invariant amplitudes increases from two (cf.\ \Eq {T-dec})
to four.  An appropriate decomposition of amplitude \eq{forw-amp2}
runs
\begin{align} \label{inv-dec}
 T^{\mu\nu} =
  &\left( -g^{\mu\nu} + \frac{q^\mu q^\nu}{q^2}\right) S_1(\nu,q^2) \notag\\*
  & + \frac1{M^2} \left( p^\mu - \frac{M\nu}{q^2} q^\mu \right)
    \left( p^\nu - \frac{M\nu}{q^2} q^\nu \right) S_2(\nu,q^2) \notag \\*
  & - \frac iM\, \eps^{\mu\nu\rho\sigma} q_\rho s_\sigma A_1(\nu,q^2)
    \notag\\*
  & - \frac i{M^3}\, \eps^{\mu\nu\rho\sigma} q_\rho
   \bigl( (M\nu) s_\sigma - (q\ndot s) p_\sigma \bigr) A_2(\nu,q^2),
\end{align}
where the multiplying powers of $M$ are such that all invariant amplitudes
$S_{1,2}(\nu,q^2)$ and $A_{1,2}(\nu,q^2)$ are dimensionless.
Observe that amplitudes $S_{1,2}(\nu,q^2)$ multiply tensors that are
\emph {symmetric} with respect to $\mu\leftrightarrow\nu$ and \emph
{independent} of the nucleon spin $s$, whereas amplitudes
$A_{1,2}(\nu,q^2)$ multiply tensors \emph {antisymmetric} w.r.t.\
$\mu\leftrightarrow\nu$ and \emph {linear} in $s$.  This coincidence reflects
the conservation of parity in strong and electromagnetic interactions,
analogously to the real-photon case (cf.\ the discussion following \Eq
{T-dec}).

At $q^2=0$, the above amplitudes are related to the amplitudes $f_{1,2}(\nu)$
of the previous section by
\begin{subequations}
\begin{equation}
 f_1(\nu) = \frac {\alpha}{2M}\, S_1(\nu,0)
\end{equation}
and \begin{equation}
 f_2(\nu) = \frac {\alpha}{2M^2}\, A_1(\nu,0).
\end{equation}
\end{subequations}

\subsubsection{Born contribution to forward virtual Compton amplitude}
For later use, I report the Born contibution to amplitude \eq
{inv-dec}, originating from the diagrams
\begin{equation}
 \graph(3,2){
  \fmfleft{i1,i2} \fmfright{o1,o2}
  \fmf{fermion,width=thick}{i1,v1,v2,o1}
  \fmf{boson}{i2,v1}  \fmf{boson}{v2,o2}
  \fmfblob{.6\ul}{v1,v2}
 } +
 \graph(3,2){
  \fmfleft{i1,i2} \fmfright{o1,o2}
  \fmf{fermion,width=thick}{i1,v1,v2,o1}
  \fmf{phantom}{i2,v1} \fmf{phantom}{v2,o2}
  \fmffreeze
  \fmf{boson}{i2,v2} \fmf{boson}{v1,o2}
  \fmfblob{.6\ul}{v1,v2}
 }
\end{equation}
The vertices incorporate the nucleon's Dirac and Pauli form factors,
normalized by $F_1(0)=Z$ and $F_2(0)=\kappa$,
\begin{equation} \label{NNgamma}
 \begin{gathered} \\
 \labelledgraph(2,2){
  \fmftop{t} \fmfbottom{b1,b2}
  \fmflabel{$\gamma(q)$}{t} \fmflabel{N$(k)$}{b1} \fmflabel{N$(p)$}{b2}
  \fmf{boson}{t,v}
  \fmf{fermion,width=thick}{b1,v,b2}
  \fmfblob{.6\ul}{v}
 } \\ \\
 \end{gathered}
 = eF_1(q^2)\,\gamma^\mu + \frac {ie}{2M}\,F_2(q^2)\, q_\rho \sigma^{\mu\rho},
\end{equation}
where $q=k-p$.  The Born contribution to the invariant amplitudes
$S_{1,2}(\nu,q^2)$ and $A_{1,2}(\nu,q^2)$ reads\footnote
{\Eqs {Born} remain unchanged if one supplements an on-shell
vanishing portion proportional to $q^\mu\qslash$ to the vertex \eq {NNgamma}.}
\begin{subequations}
\begin{align}
 S_1^\tBorn(\nu,q^2) & =
  -2F_1^2(q^2) - \frac {2(q^2)^2\,G_\tM^2(q^2)}{(2M\nu)^2-(q^2)^2}, \\
 S_2^\tBorn(\nu,q^2) & =
  2\, \frac {4M^2q^2\,F_1^2(q^2)-(q^2)^2\,F_2^2(q^2)}{(2M\nu)^2-(q^2)^2}, \\
 A_1^\tBorn(\nu,q^2) & =  \label{Born.A1}
  -F_2^2(q^2) + \frac {4M^2q^2\,F_1(q^2)G_\tM(q^2)}{(2M\nu)^2-(q^2)^2}, \\
 A_2^\tBorn(\nu,q^2) & =
  \frac {4M^3\nu\,F_2(q^2)G_\tM(q^2)}{(2M\nu)^2-(q^2)^2},
\end{align} \label{Born}%
\end{subequations}
with the magnetic Sachs form factor $G_\tM(q^2)=F_1(q^2)+F_2(q^2)$.
Observe that amplitudes $S_1(\nu,q^2)$ and $A_1(\nu,q^2)$ involve
non-pole parts $-2F_1^2(q^2)$ and $-F_2^2(q^2)$, which do not vanish
at $\nu\to\infty$.

In what follows, I will only need the ``polarized'' amplitudes
$A_{1,2}(\nu,q^2)$, in particular the linear combination
\begin{equation} \label{f2nuq2}
 f_2(\nu,q^2) := \frac\alpha{2M^2} \left(A_1(\nu,q^2) +
  \frac{q^2}{M\nu}A_2(\nu,q^2)\right),
\end{equation}
which reduces to the familiar amplitude $f_2(\nu)$ at $q^2=0$.
From \Eqs {Born}, its Born contribution is obtained as
\begin{equation} \label{f2nuq2-Born}
 f_2^\tBorn(\nu,q^2) = -\frac {\alpha\,F_2^2(q^2)}{2M^2}
 + \frac {2\alpha q^2\, G_\tM^2(q^2)}{(2M\nu)^2-(q^2)^2}.
\end{equation}
At non-zero $q^2$, the linear combination \eq {f2nuq2} represents
polarized forward scattering of a \emph {purely transverse} virtual
photon off a nucleon.  This will be demonstrated in \Ch {Q2},
where the notion of forward virtual Compton scattering recurs in the
context of inclusive electroproduction.  Observe, however, that presently
I deal with timelike photon virtualities, while the photon
being exchanged in electroproduction processes is always spacelike.

\subsection{Equal-times commutator of electric charge densities}
\label{Sect:ETCA.comm}

\subsubsection{Current density of a Dirac field}
The current density originating from a Dirac field $\psi(x)$ has the form
\begin{equation} \label{curr-def1}
 J^\mu(x) = \bar\psi(x) \gamma^\mu \psi(x)
 = \psi^\dagger(x) \gamma^0\gamma^\mu \psi(x).
\end{equation}
In particular, the charge density reads
\begin{equation}
 J^0(x) = \psi^\dagger(x) \psi(x).
\end{equation}
For the present, a single Dirac field is considered.  Multiple
flavors and colors will be included below.

\subsubsection{Naive current-density commutator}
In quantum field theory, the fields $\psi(x)$ and $\psi^\dagger(x)$
are taken to be operators.  Then the definition \eq {curr-def1} has
its deficiencies, owing to the equality of the spacetime arguments of
$\psi(x)$ and $\psi^\dagger(x)$.  I postpone the discussion of this
point to \Sect {anomcomm}.  For the moment, I want to stick to the
definition \eq {curr-def1} and work out the \emph {naive} equal-times
commutator of current densities, following from the canonical
anticommutation relations
\begin{subequations}
\begin{gather}
 \bigl\{\psi_\alpha(x),\psi^\dagger_\beta(y)\bigr\}_{\text{et}} =
  \delta_{\alpha\beta} \dirac3(\bx-\by), \\
  \bigl\{\psi^\dagger_\alpha(x),\psi^\dagger_\beta(y)\bigr\}_{\text{et}} =
  \bigl\{\psi_\alpha(x),\psi_\beta(y)\bigr\}_{\text{et}} = 0
\end{gather} \label{ACR}%
\end{subequations}
among Dirac fields.  Here, indices $\alpha$, $\beta$ denote spinor
components, and the subscript ``et'' stands for ``equal times'', i.e.,
$x^0=y^0$.  The anticommutation relations \eq {ACR} are employed
repeatedly in the following calculation,
\begin{align} \label{curr-comm}
 \etcomm {J^\mu(x)} {J^\nu(y)} &=
  (\gamma^0\gamma^\mu)_{\alpha\beta} (\gamma^0\gamma^\nu)_{\gamma\delta}
  \bigl[
  \psi^\dagger_\alpha(x) \psi_\beta(x) \psi^\dagger_\gamma(y) \psi_\delta(y)
  \notag \\*
 & \qquad\qquad\qquad\qquad {} -
  \psi^\dagger_\gamma(y) \psi_\delta(y) \psi^\dagger_\alpha(x) \psi_\beta(x)
  \bigr] \notag \\
 &= (\gamma^0\gamma^\mu)_{\alpha\beta} (\gamma^0\gamma^\nu)_{\gamma\delta}
  \bigl[
  \psi^\dagger_\alpha(x) \bigl( \delta_{\beta\gamma} \dirac3(\bx-\by) -
  \psi^\dagger_\gamma(y) \psi_\beta(x) \bigr) \psi_\delta(y)
  \notag \\*
 & \qquad\qquad\qquad\qquad {} -
  \psi^\dagger_\gamma(y) \bigl( \delta_{\delta\alpha} \dirac3(\bx-\by) -
  \psi^\dagger_\alpha(x) \psi_\delta(y) \bigr) \psi_\beta(x)
  \bigr] \notag\\
 &= \psi^\dagger(x) [\gamma^0\gamma^\mu,\gamma^0\gamma^\nu] \psi(x) \,
  \dirac3(\bx-\by).
\end{align}
Evidently, the commutator vanishes if $\mu=0$ or $\nu=0$, since
$(\gamma^0)^2=1$ commutes with any matrix.  This means that at equal
times, the charge density commutes with each component of the current
density,
\begin{equation} \label{naive-0mu}
 \etcomm {J^0(x)} {J^\mu(y)} = 0.
\end{equation}
For later use, I also work out the naive commutator of spatial
components.  One has
\begin{equation}
 \gamma^0[\gamma^0\gamma^i,\gamma^0\gamma^j] = -\gamma^0[\gamma^i,\gamma^j] =
  2i\,\eps^{ijk}\gamma^k\gamma_5.
\end{equation}
Thus,
\begin{equation}
 \etcomm {J^i(x)} {J^j(y)} = 2i\,\eps^{ijk}\,
  \bar\psi(x)\gamma^k\gamma_5\psi(x)\,\dirac3(\bx-\by),
\end{equation}
which can be merged with \Eq {naive-0mu}, giving
\begin{equation}
 \etcomm {J^\mu(x)} {J^\nu(y)} = 2i\,\eps^{0\mu\nu\lambda}\,
  \bar\psi(x)\gamma_\lambda\gamma_5\psi(x)\,\dirac3(\bx-\by).
\end{equation}

\subsubsection{Quarks}
I now want to introduce quarks of different flavors and colors.
This amounts to endowing the spinors with two additional indices and
supplementing appropriate Kronecker symbols to the canonical
anticommutation relations \eq {ACR}.  In matrix
notation, the electromagnetic current now reads
\begin{equation} \label{QCD-curr}
 J^\mu(x) = \sum_{q=u,d(,s)} Z_q\, \bar q(x) \gamma^\mu q(x)
 = \bar\psi(x) \gamma^\mu Z^{(\tq)} \psi(x),
\end{equation}
where $\psi(x)$ henceforth embodies diverse quark spinors and
$Z^{(\tq)}$ denotes the diagonal quark charge matrix:
\begin{subequations}
\begin{equation}
 \psi(x) = \begin{pmatrix} u(x)\\ d(x) \end{pmatrix} \qquad \text{and} \qquad
 Z^{(\tq)} = \begin{pmatrix} \frac23 & 0 \\ 0 & -\frac13 \end{pmatrix}
\end{equation}
for two flavors u and d, or
\begin{equation}
 \psi(x) = \begin{pmatrix} u(x)\\ d(x)\\s(x) \end{pmatrix}
  \qquad \text{and} \qquad
 Z^{(\tq)} = \begin{pmatrix}
   \frac23 & 0 & 0 \\ 0 & -\frac13 & 0 \\ 0 & 0 & -\frac13
  \end{pmatrix}
\end{equation}
\end{subequations}
for three flavors u, d, and s.  The notion $\gamma^\mu Z^{(\tq)}$ in
\Eq {QCD-curr} actually denotes the tensor product
$\gamma^\mu_{\text{Dirac}}\otimes
Z^{(\tq)}_{\text{flavor}}\otimes1_{\text{color}}$.  Re-doing the
above calculation of the naive current commutator, one arrives at
\begin{equation} \label{naive-curr-comm1}
 \etcomm {J^\mu(x)} {J^\nu(y)} = 2i\,\eps^{0\mu\nu\lambda}\,
  \bar\psi(x) \gamma_\lambda\gamma_5 \bigl(Z^{(\tq)}\bigr)^2
  \psi(x)\,\dirac3(\bx-\by).
\end{equation}
Again, the charge density commutes with each component of the current.
(This fact is independent of flavors and colors.)
In particular, the naive equal-times commutator of electric charge
densities, which is the one required by the current-algebra approach
to the GDH sum rule, reads
\begin{equation} \label{naive-comm}
 \boxed {\etcomm {J^0(x)} {J^0(y)} = 0}
\end{equation}

\subsubsection{Electric dipole-moment operator}
Most of the information contained in the charge-density commutator \eq
{naive-comm} is redundant with respect to the GDH sum rule. In fact,
only a certain \emph {moment} of it is needed, namely the commutator
of electric dipole moments, defined in the usual way as
\begin{equation} \label{D-def}
 \bD(x^0) = e \!\int\! \td^3x\,\bx J^0(x).
\end{equation}
Relation \eq {naive-comm} implies
\begin{equation} \label{naive-D-comm}
 \bigl[ D^i(0), D^j(0) \bigr] = 0
\end{equation}
for any pair of spatial indices $i,j$.  The dipole moments are
evaluated at time 0, but of course any other (equal) time would do.
Still, the derivation of the GDH sum rule requires by far not the full
information contained in \Eq {naive-D-comm}, but only the \emph
{nucleon matrix element} thereof.  I take the incoming nucleon to be
traveling along $\be_3$,
\begin{equation}
 p^\mu = \bigl( p^0, 0, 0, \sqrt{(p^0)^2 - M^2} \bigr),
\end{equation}
and define definite-chirality components of the dipole operator,
\begin{equation} \label{LR-def}
 D^{\tL,\tR}(0) = \frac1{\sqrt2}\, \bigl( D^1(0) \pm iD^2(0) \bigr),
\end{equation}
which correspond to circularly polarized photons.  Finally, both
nucleon states are taken to have positive helicity,
\begin{equation} \label{naive-comm-me}
 \me {p', \tfrac12} {[D^\tL(0), D^\tR(0)]} {p, \tfrac12} = 0.
\end{equation}

\subsubsection{Complete set of intermediate states}
As is standard in the current-algebra derivation of sum rules, one now
inserts a complete set of intermediate states and separates the
one-nucleon states from the continuum:
\begin{equation}  \label{1N+cont}
 1 = \sum_{\lambda=\pm\frac12}
  \!\int\!\frac{\td^3p''}{(2\pi)^3 2p^{\prime\prime0}}\,
  |p'',\lambda\rangle \langle p'',\lambda| +
  {\sum_\tX}' |\tX\rangle \langle\tX|.
\end{equation}
The primed sum runs over all hadronic intermediate states except the
one-nucleon states.  Non-hadronic particles are excluded from the sum,
corresponding to the fact that all quantities are consided to lowest
order in the electromagnetic coupling, as discussed in the previous section.

\subsubsection{One-nucleon contribution}
To work out the one-nucleon intermediate-state contribution to the
matrix element \eq {naive-comm-me},
\begin{align} \label{comm-1N-pre1}
 & \me {p',\tfrac12} {[D^\tL(0), D^\tR(0)]} {p, \tfrac12}_{\text{one-nucleon}}
   \notag \\*
 & \qquad = \sum_{\lambda=\pm\frac12}
  \!\int\!\frac{\td^3p''}{(2\pi)^3 2p^{\prime\prime0}}\,
  \me {p',\tfrac12} {D^\tL(0)} {p'',\lambda}
  \me {p'',\lambda} {D^\tR(0)} {p,\tfrac12} - \{\tL\leftrightarrow \tR\},
\end{align}
I need the nucleon matrix element of the electromagnetic current:
\begin{equation}
 \me {p_1} {J^\mu(0)} {p_2} = \bar{u}(p_1)\, \Gamma^\mu(p_1-p_2)\, u(p_2),
\end{equation}
where
\begin{subequations}
\begin{align} \label{Gamma-mu}
 \Gamma^\mu(q)
 & = \gamma^\mu F_1(q^2) + i\sigma^{\mu\nu} q_\nu \frac {F_2(q^2)} {2M}\\
 & = \gamma^\mu F_1(q^2)
   - \frac12\, [\gamma^\mu, \qslash]\, \frac {F_2(q^2)} {2M}.
\end{align}
\end{subequations}
In particular,
\begin{equation}
 \gamma^0\Gamma^0(q) = F_1(q^2) + \bgamma\ndot\bq \frac {F_2(q^2)} {2M}.
\end{equation}
Eventually, Dirac and Pauli form factors $F_{1,2}(q^2)$ will be
evaluated only at zero momentum transfer, where they are normalized to
$F_1(0)=Z$ and $F_2(0)=\kappa$.  The matrix element of the
dipole-moment operator \eq {D-def} can now be written\footnote {The
multiplication with coordinate $\bx$ in \Eq {D-def} corresponds to a
derivative in momentum space, while the $\bx$ integration gives rise
to a delta function.  This leads to the expression
$\nabla^i\dirac3(\bp_1-\bp_2)$ in \Eq {D-me}. Due to CP conservation,
the nucleon itself does of course \emph {not} possess an electric
dipole moment. If it did, expression \eq {Gamma-mu} would contain a
third term of the form $\gamma_5\sigma^{\mu\nu}q_\nu F_3(q^2)/2M$
(see, e.g., Itzykson and Zuber \cite [Eq.\ (3-203)] {Itzykson80}).}
\begin{equation} \label{D-me}
 \me {p_1} {D^i(0)} {p_2} = ie\,(2\pi)^3\, \nabla^i \dirac3(\bp_1-\bp_2)\,
 \bar{u}(p_1)\, \Gamma^0(p_1-p_2)\, u(p_2),
\end{equation}
and the  one-nucleon
intermediate-state contribution to the matrix element of the commutator
of dipole moments runs
\begin{align} \label{comm-1N-pre2}
 & \me {p',\tfrac12} {[D^i(0), D^j(0)]} {p, \tfrac12}_{\text{one-nucleon}}
   \notag \\*
 & \qquad = (2\pi)^3 \!\int\! \frac{\td^3p''}{2p^{\prime\prime0}}\,
   \nabla^i \dirac3(\bp'-\bp'')\, \nabla^j \dirac3(\bp-\bp'') \notag \\*
 & \qquad\quad \times \bar{u}(p',\tfrac12)\, \Gamma^0(p'-p'')\,
   (\pslash''+M)\, \Gamma^0(p''-p)\, u(p,\tfrac12) - \{i\leftrightarrow j\}.
\end{align}
Further manipulation of this expression is quite intricate.\footnote
{I proceeded by introducing a concrete representation for spinors
and gamma matrices in \Eq {comm-1N-pre2}, letting
computer-algebra system \textsl {Mathematica} handle the resulting large
expression.  Particular caution is advisable on account of the
momentum dependence of form factors and spinors.  I mention in passing
that if one forgets the latter, one arrives essentially at the
erroneous result of Pradhan and Khare \cite {Pradhan72}.} One has to
integrate by parts to get rid of the $\bp''$ integral together with
one of the delta functions.  Up to second derivatives of
$\dirac3(\bp'-\bp)$ will be left then.  The second derivative
$\nabla^i\nabla^j$ does not contribute, owing to the explicit
antisymmetrization in indices $i,j$ due to the commutator.
Subsequently, one has to repeatedly employ Leibnitz rule in the form
\begin{equation}
 \frac{\partial}{\partial\bp'}\dirac3(\bp'-\bp)\, \Phi(\bp,\bp')
 + \dirac3(\bp'-\bp)\, \frac\partial{\partial\bp'}\Phi(\bp,\bp') =
 \frac{\partial}{\partial\bp'}\bigl[ \dirac3(\bp'-\bp)\, \Phi(\bp,\bp') \bigr],
\end{equation}
which implies
\begin{equation}
 \bnabla\dirac3(\bp'-\bp)\, \Phi(\bp,\bp')
 = \bnabla\dirac3(\bp'-\bp)\, \Phi(\bp,\bp)
 - \dirac3(\bp'-\bp)\, \frac\partial{\partial\bp'}\Phi(\bp,\bp'),
\end{equation}
and it turns out that no derivative of $\dirac3(\bp'-\bp)$ remains at
all.  Introducing L and R components again, the result reads
\begin{align} \label{comm-1N}
 \me {p', \tfrac12} {[D^\tL(0), D^\tR(0)]} {p, \tfrac12}_{\text{one-nucleon}}
 = (2\pi)^3\, 2p^0\, \dirac3(\boldsymbol p' - \boldsymbol p)
     \left( \frac{2\pi\alpha\kappa^2}{M^2} -
     \frac{2\pi\alpha(Z+\kappa)^2}{(p^0)^2}\right).
\end{align}
Here I stress the presence of the second term, which vanishes in the
infinite-momentum limit $p^0\to\infty$.  Hitherto, its only appearance
in the literature was in \Ref {Pradhan72}, where it has an
incorrect form.

\subsubsection{Continuum contribution}
To obtain the continuum contribution to matrix element \eq
{naive-comm-me}, i.e., the sum over all intermediate states
$|\tX\rangle$ except the one-nucleon state,
\begin{align} \label{comm-cont-pre1}
 \me {p', \tfrac12} {[D^\tL(0), D^\tR(0)]} {p, \tfrac12}_\tcont
 = {\sum_{\tX}}' \me {p', \tfrac12} {D^\tL(0)} {\tX}
   \me {\tX} {D^\tR(0)} {p, \tfrac12} - \{\tL\leftrightarrow\tR\},
\end{align}
I employ current conservation $\partial_\mu J^\mu(x) = 0$ to get
\begin{equation}
 \dot D^i(x^0) = e \!\int\! \td^3x\, x^i \partial_0 J^0(x)
 = -e \!\int\! \td^3x\, x^i \bnabla\ndot\bJ(x)
 = e \!\int\! \td^3x\, J^i(x).
\end{equation}
Since the time derivative of any operator is related to the commutator
of that operator with the Hamiltonian (cf.\ \Eq {[A,p]}), one has
\begin{align} \label{D-R}
 \me {\tX} {D^\tR(0)} {p,\tfrac12}
% & = \frac1{p^0-p_\tX^0}\, \me {\tX} {[D^\tR(0),H]} {p,\tfrac12} \notag \\
 & = \frac{i}{p^0-p_\tX^0}\, \me {\tX} {\dot{D}^\tR(0)} {p,\tfrac12} \notag \\
 & = \frac{ie}{p^0-p_\tX^0} \!\int\! \td^3x\,
     \me {\tX} {J^\tR(\bx,0)} {p,\tfrac12} \notag \\
 & = \frac{ie}{p^0-p_\tX^0}\, (2\pi)^3 \dirac3(\bp_\tX-\bp)\,
     \me {\tX} {J^\tR(0)} {p,\tfrac12},
\end{align}
where definite-chirality components of the current are defined by
$J^{\tL,\tR}(x)=(J^1(x)\pm i J^2(x))/\sqrt2$.  The
translational-invariance condition \eq {trans} has been used to carry
out the spatial integrations.  Inserting this into \Eq
{comm-cont-pre1}, one obtains
\begin{align} \label{comm-cont-pre2}
 & \me {p', \tfrac12} {[D^\tL(0), D^\tR(0)]} {p, \tfrac12}_\tcont \notag \\*
 & \qquad = (2\pi)^3\, \dirac3(\boldsymbol p' - \boldsymbol p)\,
  {\sum_{\tX}}'
  (2\pi)^3\, \dirac3(\boldsymbol p_{\tX} - \boldsymbol p)\,
  \frac{4\pi\alpha
  |\langle p, {\tfrac12} | J^\tL(0) | \tX \rangle |^2}
  {(p^0-p_{\tX}^0)^2} - \{\tL\to\tR\}.
\end{align}
Owing to the factor of $\dirac3(\bp'-\bp)$ and since $J^\tL(x)$ is the
adjoint of $J^\tR(x)$, the right-hand side of \Eq {comm-cont-pre2}
could be written in terms of the square of a current matrix element
between a one-nucleon initial state and a hadronic final state being
summed over.  This looks very much like a photoabsorption cross
section, but there's some work left to do.

Introducing the timelike virtual photon momentum $q$ with $\bq = 0$,
I can substitute
\begin{equation} \label{q0}
 \dirac3(\bp_{\tX} - \bp) =
  \int_{q^0_{\tthr}}^\infty \!\td q^0\,\dirac4(p_{\tX} - p - q) =
  (2\pi)^{-4} \!\int_{q^0_{\tthr}}^\infty \!\td q^0
  \int\! \td^4x\, e^{-i(p_\tX-p-q)\ndot x},
\end{equation}
where the pion-production threshold $q^0_{\text{thr}} \equiv
M\nu_{\text{thr}}/p^0$ is determined by $(p+q_{\text{thr}})^2 =
(M+m_\pi)^2$, or explicitly,
\begin{equation}
 \nu_{\text{thr}} = \frac{p^0 q^0_{\text{thr}}}M =
  \left(m_\pi + \frac{m_\pi^2}{2M}\right)
  \left(\frac12 + \frac12\sqrt{1 + \frac{2Mm_\pi+m_\pi^2}{(p^0)^2}}
  \right)^{-1}.
\end{equation}
For $p^0\to\infty$, $\nu_{\text{thr}}$ approaches the
familiar pion-photoproduction threshold $\nu_0 = m_\pi + m_\pi^2/2M$.

Utilizing translational invariance once more, \Eq {comm-cont-pre2}
can be cast into the form
\begin{align} \label{comm-cont-pre3}
 & \me {p', \tfrac12} {[D^\tL(0), D^\tR(0)]} {p, \tfrac12}_\tcont
   = (2\pi)^3\, 2p^0\, \dirac3(\bp' - \bp)
  \int_{q^0_{\tthr}}^\infty \!\frac {\td q^0}{q^0}\, \frac\alpha{p^0q^0}
  \notag \\*
 & \qquad \times \!\int\! \td^4x\, e^{iq\ndot x} {\sum_{\tX}}'
  \me {p,\tfrac12} {J^\tL(x)} {\tX} \me {\tX} {J^\tR(0)} {p,\tfrac12}
  - \{\tL\to\tR\}.
\end{align}
Observe now that the prime at the sum over intermediate states can be
dropped, i.e., one-nucleon states can be added without altering the
result, as can most easily be seen by considering the four-dimensional
delta function within \Eq {q0}: For an on-shell nucleon and a
timelike (or even lightlike) photon with positive energy $q^0$, the
four-vector $p_\tX=p+q$ will always represent a state having a mass strictly
greater than that of the nucleon.\footnote {Note that this is not in
contradiction to the fact that the one-nucleon state yields a certain
finite contribution to the matrix element of the dipole-moment
commutator.  In fact, the exclusion of the one-nucleon state is
crucial for the manipulation presented in \Eq {D-R} not to produce
an artificial pole, and the exclusion of the point $q^0=0$ from the
$q^0$ integral of \Eq {comm-cont-pre3} enables one to
re-include the one-nucleon state.} In physical terms, this simply means
that a nucleon cannot absorb a photon without also emitting something.
Upon using the completeness relation $\sum_\tX|\tX\rangle\langle\tX|=1$,
one then has that
\begin{align} \label{comm-cont-pre4}
 & \me {p', \tfrac12} {[D^\tL(0), D^\tR(0)]} {p, \tfrac12}_\tcont
  = (2\pi)^3\, 2p^0\, \dirac3(\bp' - \bp)
  \int_{q^0_{\tthr}}^\infty \!\frac {\td q^0}{q^0}\, \frac{2\alpha}{M\nu}
  \notag \\*
 & \qquad\quad \times \frac12 \!\int\! \td^4x\, e^{iq\ndot x}
  \me {p,\tfrac12} {J^\tL(x)J^\tR(0)} {p,\tfrac12} - \{\tL\leftrightarrow\tR\},
\end{align}
where $\nu=p^0q^0/M=p\ndot q/M$ is the photon's lab-frame energy. The
last line is recognized as the antisymmetrized $\tL\tR$ component of
the absorptive part \eq {abs-T} of the forward virtual Compton
amplitude.  In view of the decomposition \eq {inv-dec} and using
$\bq=0$, $s^3=p^0/M$, and $\eps^{\tL\tR03}=-i$, one has
\begin{equation}
 \abs T^{\tL\tR} - \abs T^{\tR\tL} = \frac {\nu}{\pi M}\,
 \im \left( A_1(\nu,q^2) + \frac {q^2}{M\nu} A_2(\nu,q^2) \right),
\end{equation}
with $q^2=(q^0)^2$. Now the continuum contribution \eq
{comm-cont-pre4} can be written in terms of the imaginary part of the
forward virtual Compton amplitude $f_2(\nu,q^2)$, \Eq{f2nuq2},
\begin{equation} \label{comm-cont}
 \me {p', \tfrac12} {[D^\tL(0), D^\tR(0)]} {p, \tfrac12}_\tcont =
 (2\pi)^3\, 2p^0\, \dirac3(\bp' - \bp) \,
 8\!\!\int_{q^0_{\tthr}}^\infty\! \frac{\td q^0}{q^0}\, \im f_2(\nu,q^2).
\end{equation}

\subsubsection{The finite-momentum GDH sum rule}
\begin{figure}[tb]
 \begin{center}
  \input{q2nu}
 \end{center}
 \caption[]{
  Integration path of the finite-momentum GDH sum rule
  \eq{FMGDH} in the $(\nu,q^2)$ plane for nucleon energies $p^0=M$ and
  $p^0=2M$.  The heavy line represents the pion-production threshold.
  \label{Fig:q2nu}}
\end{figure}
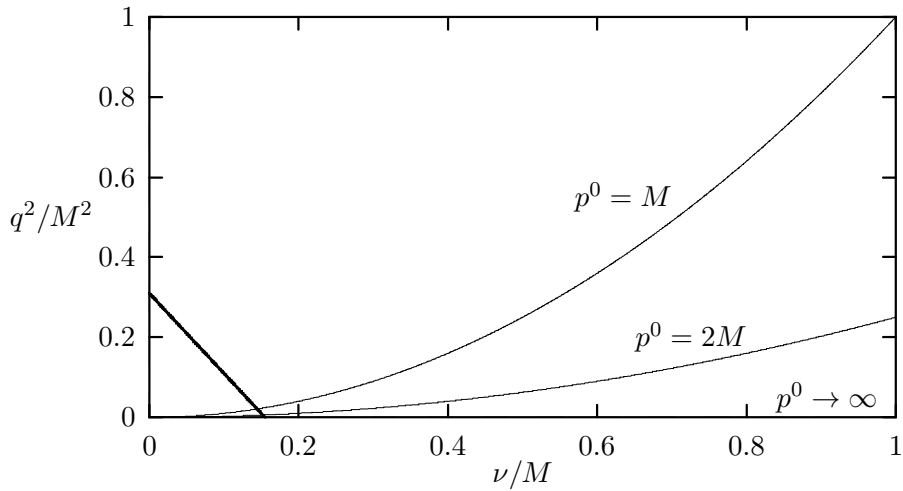

Since the one-nucleon part \eq {comm-1N} and the continuum part
\eq {comm-cont} sum up to give the commutator matrix element \eq
{naive-comm-me}, I conclude
\begin{equation} \label{FMGDH}
 \boxed{
 -\frac{2\pi^2\alpha\kappa^2}{M^2} + \frac{2\pi^2\alpha(Z+\kappa)^2}{(p^0)^2} =
  \int_{\nu_{\text{thr}}}^\infty\! \frac{\td \nu}\nu\,
  8\pi \im f_2\!\left(\nu, \frac{M^2\nu^2}{(p^0)^2}\right)}
\end{equation}
where I have performed a linear substitution $q^0\mapsto\nu=p^0q^0/M$
of the integration variable.
I call this equation the \emph {finite-momentum} GDH sum rule. It is
based solely on the naive charge-density commutator \eq
{naive-comm}, or on the weaker assumption presented by \Eq
{naive-comm-me}.  In particular, the integral on the right-hand side
of \Eq {FMGDH} \emph {converges}, since it relies only on the
validity of the completeness relation for the physical intermediate
states.  This is irrespective of the convergence of the genuine GDH
integral with its integrand $8\pi\im f_2(\nu,0)/\nu$.  The integration
path in the $(\nu,q^2)$ plane for various values of the energy $p^0$
is depicted in \Fig {q2nu}.  Note that for any finite value
of $p^0$, this path is a parabola that extends to arbitrarily high
timelike photon virtualities.%

\subsubsection{Correspondence between dispersion theory and current algebra}
At this point, a remark on the connection between dispersion-theoretic
and current-algebra derivation of the GDH sum rule is appropriate.
In a way, the one-nucleon intermediate-state contribution to the
commutator of electric charge densities corresponds to the Born
portion of the forward Compton amplitude, which in turn determines the
low-energy limit of the true amplitude due to the
Low-Gell-Mann-Goldberger theorem, as discussed in the previous
section.  Therefore it is not surprising that the left-hand side of
\Eq {FMGDH} has a part very similar to the constant on the
left-hand side of the GDH sum rule.  Furthermore, the continuum
contribution gives rise to an integral like the one on the right-hand
side of the GDH sum rule, whose integrand is ``almost'' fit to apply
the optical theorem.  By virtue of the Bjorken-Johnson-Low limit,
there is a third correspondence, namely the relation between the \emph
{high-energy} limit of the forward Compton amplitude and the current
commutator itself. This will be illustrated in the \Ch {mod}.  

The only thing that spoils exact one-to-one correspondence in all
three instances is the finiteness of the nucleon energy $p^0$,
resulting in a non-vanishing photon virtuality $q^2$.  Re-inspecting
the dipole-moment commutator matrix element \eq {naive-comm-me}, this
could actually have been evident from the very beginning of the
current-algebra derivation, because in momentum space, fixed time
corresponds to an \emph {integral over all photon energies} $q^0$,
while the space integral implicit in the definition of the dipole
moment corresponds to \emph {fixed photon three-momentum} $\bq=0$.
\label{q-story}
Therefore the photon virtuality will \emph {inevitably} run along with
the energy, and the derivation has to be supplemented by the
infinite-momentum limit, which will turn out to be most critical,
although least considered in the past.

\subsubsection{The finite-momentum GDH sum rule as a $q^0$ dispersion relation}
The finite-momentum GDH sum rule \eq {FMGDH} can be derived in a quite
different fashion, namely as an unsubtracted dispersion
relation in variable $q^0$ for the function $f_2(\nu,q^2)$ at $\bq=0$,
i.e., letting
\begin{equation} \label{nuq2}
 \nu=\frac{p^0q^0}M \quad \text{and} \quad q^2=(q^0)^2.
\end{equation}
Analogously to the fixed-$q^2$ dispersion relation \eq {UDR}, I write
\begin{equation} \label{UDR0}
 \lim_{q^0\to0} f_2(\nu,q^2) = \frac2\pi \int_{q^0_\tthr}^{\infty}\!
  \frac{\td q^0}{q^0}\, \im f_2(\nu,q^2).
\end{equation}
As illustrated in \Sect {LET}, the Low-Gell-Mann-Goldberger theorem
states that to first order in the photon energy $\nu$, the real
Compton amplitude equals its Born contribution.  It can be shown \cite
{Scherer96} that this also holds for the virtual Compton
amplitude.  Thus,
\begin{equation}
 \lim_{q^0\to0} f_2(\nu,q^2) = \lim_{q^0\to0} f_2^\tBorn(\nu,q^2),
\end{equation}
where the right-hand side can be calculated from \Eq {f2nuq2-Born}.
Note that it would be meaningless to simply write $f_2(0,0)$, because
function $f_2(\nu,q^2)$ is discontinuous at the point (0,0), which
lets the limit depend upon the way this point is approached.  For
instance, observe that
\begin{subequations}
\begin{equation}
 \lim_{\nu\to0} f_2^\tBorn(\nu,0) = f_2(0) = - \frac {\alpha\kappa^2}{2M^2},
\end{equation}
whereas
\begin{equation}
 \lim_{q^2\to0} f_2^\tBorn(0,q^2) = \infty.
\end{equation}
\end{subequations}
In terms of \Fig {q2nu}, the limit is finite as long as one runs
horizontally into the point (0,0), as does the parabola $q^0\to0$
taken in \Eq {UDR0}. Introducing the kinematics \eq
{nuq2} into \Eq {f2nuq2-Born} results into
\begin{equation} \label{q0to0}
 \lim_{q^0\to0} f_2(\nu,q^2) = - \frac {\alpha\kappa^2}{2M^2}
 + \frac {\alpha(Z+\kappa)^2}{2(p^0)^2}.
\end{equation}
The finite-momentum GDH sum rule \eq {FMGDH} is obtained by putting
together \Eqs {UDR0} and \eq {q0to0}.  I remark that this result
confirms the correctness of \Eq {comm-1N} as opposed to the result of
Pradhan and Khare \cite {Pradhan72}.

\subsection{Infinite-momentum limit}

Taking the infinite-momentum limit now constitutes the last
step of the derivation of the GDH sum rule.  I take the limit $p^0\to\infty$
on both sides of the finite-momentum GDH sum rule \eq {FMGDH}:
\begin{equation} \label{pre-GDH1}
 -\frac{2\pi^2\alpha\kappa^2}{M^2} =
  \lim_{p^0\to\infty} \int_{\nu_{\text{thr}}}^\infty\!
  \frac{\td \nu}\nu\, 8\pi \im f_2\!\left(\nu, \frac{M^2\nu^2}{(p^0)^2}
  \right).
\end{equation}
To get the accustomed form of the sum rule, one now has to interchange the
limit $p^0\to\infty$ with the $\nu$ integration, because then the integrand
is the imaginary part of forward Compton amplitude $f_2(\nu,q^2)$ at the
\emph {real} photon point $q^2=0$, where it is related to the observable
photoabsorption cross section via the optical theorem
\begin{equation} \label{opt-th2}
 8\pi \im f_2(\nu,0) = \sigma_{1/2}(\nu) - \sigma_{3/2}(\nu).
\end{equation}
As it stands, \Eq {pre-GDH1} is actually worthless, since no
observable is related to the timelike virtual Compton amplitude.
Pictorially, the infinite-momentum limit corresponds to stretching the
parabolae indicated in \Fig {q2nu} down to the straight line
running at $q^2=0$.

\subsubsection{Legitimacy of the infinite-momentum limit}
Will the properties of function $\im f_2(\nu,q^2)$ allow the limit
to be dragged into the integral? I stress that bare current algebra is
exhausted as regards this problem.  That is to say, no statement on
the permutation of $p^0\to\infty$ limit and $\nu$ integration can be
inferred from any kind of equal-times commutators.  In a way, current
algebra tells us something about the parabolae shown in \Fig {q2nu} and
nothing about the straight line $q^2=0$.  The legitimacy of the
infinite-momentum limit enters the current-algebra derivation of the
GDH sum rule as a mere conjecture!

\Sect {IML} of this thesis is devoted to a discussion of possible
modifications due to the infinite-momentum limit.

\section{Light-cone current algebra}
\label{Sect:LCCA}

In 1972, Dicus and Palmer \cite {Dicus72} presented a derivation of
the GDH sum rule from the algebra of currents on the light-cone, which
circumvents the infinite-momentum limit necessary in the usual
current-algebra approach.  In this section, I want to review the
basics of this technique.

\subsection{Light-cone coordinates}
For any four-vector $r$, one defines $\pm$ components by
\begin{equation}
 r^\pm = \frac1{\sqrt{2}}\, (r^0 \pm r^3).
\end{equation}
Remaining components are subsummed as
\begin{equation}
 \br_\perp = (r^1,r^2).
\end{equation}
The scalar product then reads
\begin{equation} \label{LC-SP}
 q\ndot x = q^+ x^- + q^- x^+ - \bq_\perp\ndot\bx_\perp.
\end{equation}

The derivation of the sum rule essentially takes the following course:
each instance of \emph {time} $x^0$ is replaced by \emph {light-cone
time} $x^+$; Ordinary space components $\bx$ are replaced by
$(x^-,\bx_\perp)$; \emph {light-cone energy} $q^-$ -- naturally
defined as the $q$ component accompanying variable $x^+$ in the scalar
product \eq {LC-SP} -- is substituted for ordinary energy $q^0$; etc.

Nevertheless, the reader shall beware of mistaking the light-cone
technique for a simple transform of the coordinate system.  The vital
foundation of the whole story is that the underlying field theory is
\emph {quantized on the light cone}.  For instance, the canonical
anticommutation relations of a fermionic theory are defined at equal
light-cone times\footnote {The fact that canonical (anti)commutation
relations are defined at equal $+$ components of spacetime is the
actual reason why these components are called light-cone \emph
{time}.} rather than equal times as in \Eqs {ACR}.

\subsection{Light-cone charge-density commutator}
Defining the fermionic current in the usual way, the naive current
commutator at equal light-cone times can be derived from the canonical
anticommutation relations analogously to the derivation of the naive
equal-times commutator presented on p.~\pageref {curr-comm} of this
thesis.  The result for the commutator of $+$ components of the
current (the light-cone charge density) reads
\begin{equation} \label{naive-LC-comm}
 \lccomm {J^+(x)} {J^+(y)} = 0,
\end{equation}
where subscript ``lc'' stands for $x^+=y^+$.\footnote {In Minkowski
space, condition $x^+=\tconst$ does actually not define a cone, but
rather a plane tangential to the light-cone, i.e., a
light-\emph{front}.  Thus, ``light-cone formalism'' should properly be
called ``light-front formalism'', and indeed a wee minority of theorists
does so.}

\subsubsection{Light-cone electric dipole moment}
One now defines the first moment of $J^+(x)$ \cite {Dicus72},
\begin{equation}
 \bD_\perp(x^+) = e \!\int\! \td x^-\, \td^2x_\perp\, \bx_\perp J^+(x),
\end{equation}
the light-cone analogue of the electric dipole moment \eq {D-def}.
In particular, one defines definite-chirality components $D^{\tL,\tR}(0)$ as
in \Eq {LR-def} and sandwiches the naive commutator
\begin{equation}
 \bigl[ D^\tL(0),D^\tR(0) \bigr] = 0,
\end{equation}
which follows from \Eq {naive-LC-comm}, between one-nucleon states
of positive helicity (taking the incoming nucleon to be propagating along
$\be_3$),
\begin{equation}
 \me {p', \tfrac12} {[D^\tL(0), D^\tR(0)]} {p, \tfrac12} = 0.
\end{equation}
Inserting a complete set of intermediate states and separating the one-nucleon
states from the continuum, one finds \cite {Dicus72}
\begin{equation} \label{LC-comm-1N}
 \me {p',\tfrac12} {[D^\tL(0),D^\tR(0)]} {p,\tfrac12}_{\text{one-nucleon}}
 = (2\pi)^3\, 2p^+\, \dirac{}(p^{\prime+} - p^+)\,
   \dirac2(\bp'_\perp)\, \frac{2\pi\alpha\kappa^2}{M^2}
\end{equation}
and
\begin{align} \label{LC-comm-cont}
 \me {p', \tfrac12} {[D^\tL(0), D^\tR(0)]} {p, \tfrac12}_\tcont
 & = (2\pi)^3\, 2p^+\, \dirac{}(p^{\prime+} - p^+)\,
   \dirac2(\bp'_\perp) \notag \\*
 & \quad \times 8\!\!\int_{q^-_\tthr}^\infty\!
   \frac{\td q^-}{q^-}\, \im f_2(\nu,0),
\end{align}
where $\nu = p\ndot q/M = p^+q^-/M$ and $q^-_\tthr=M\nu_0/p^+$.  The GDH sum
rule follows immediately upon application of the optical theorem \eq
{opt-th2}.

\subsection{Light cone vs.\ infinite momentum}
Most remarkably, this derivation works without employing the
infinite-momentum technique.  The one-nucleon contribution \eq
{LC-comm-1N} to the commutator matrix element is not contaminated by
an extra term like in \Eq {comm-1N}, and the continuum part \eq
{LC-comm-cont}, in contrast to \Eq {comm-cont}, exhibits just the
wanted integrand without further manipulation.  In fact, there is
actually not even the need for defining amplitude $f_2(\nu)$ off the
real-photon domain $q^2=0$.  Recalling the discussion on p.~\pageref
{q-story}, this is simply due to the fact that the photon virtuality
\begin{equation}
 q^2 = (q^0)^2 - \bq^2 = 2 q^+ q^- - \bq_\perp^2
\end{equation}
is \emph {linear} in the light-cone energy $q^-$ and vanishes
identically upon letting $q^+=0$ and $\bq_\perp=0$, whereas it is
quadratic in the ordinary energy $q^0$ and runs irresistibly with the
$q^0$ integral \eq {comm-cont}.  In terms of \Fig {q2nu}, we
have the contour of the $q^-$ integration of \Eq {LC-comm-cont}
right where we wanted it: on the abscissa $q^2=0$.  In a way, the
light-cone current-algebra method incorporates the infinite-momentum
limit from the very beginning.  This is a great advantage of the method.
A disadvantage is that little is known about possible non-naive
forms of the light-cone charge-density commutator \eq
{naive-LC-comm} -- even less than in the case of the equal-times
commutator \eq {naive-comm}.  I mention in advance that Dicus
and Palmer already suggest a non-naive form, namely the one
induced by a possible anomalous magnetic moment of quarks.  This
point will be presented in detail in \Sect {quark}.

\clearpage{\pagestyle{empty}\cleardoublepage}
\chapter{The GDH sum rule within perturbative models}
\label{Ch:pert}

In this chapter, I present some important tests of the GDH sum rule,
namely its investigation within perturbative models.  That is to
say, the polarized total photoabsorption cross section
$\sigma_{1/2}(\nu)-\sigma_{3/2}(\nu)$ (or, equivalently, the
imaginary part $\im f_2(\nu)$ of the polarized forward Compton
amplitude) of a given fermion, as well as its anomalous magnetic
moment $\kappa$, are calculated to lowest non-trivial order in a specific
perturbation theory, and it is checked whether both sides of the sum
rule coincide:
\begin{equation} \label{pertGDH}
 -\frac{2\pi^2\alpha\kappa^2}{M^2} \overset?=
  \int_{\nu_0}^\infty\! \frac{\td\nu}\nu
  \bigl(\sigma_{1/2}(\nu) - \sigma_{3/2}(\nu)\bigr) \qquad
  \text{(to specific order)}
\end{equation}
I mention in advance that all tests are concluded positively.  This
signals that any particular gauge invariant class of Feynman diagrams
obeys an unsubtracted dispersion relation,
\begin{equation}
 f_2(0) = \frac2\pi \int_{\nu_0}^{\infty}\!
  \frac{\td\nu'}{\nu'}\, \im f_2(\nu'),
\end{equation}
and that the low-energy theorem
\begin{equation} \label{pertLET}
 f_2(0) = -\frac{\alpha \kappa^2}{2M^2}
\end{equation}
holds order by order.  If you think about it, both of these points are
not \emph {too} surprizing, owing to the following reasons.  Firstly, once
you have a class of Feynman diagrams that gives rise to a \emph
{convergent} dispersion integral, it will generally not necessitate a
finite subtraction.  Secondly, the order-by-order validity of the
low-energy theorem gets plausible if you recall its derivation by
Gell-Mann and Goldberger \cite {Gell-Mann54}, which in fact \emph
{is} a perturbative one.  Nevertheless, all of the results presented
here are certainly non-trivial.

In \Sect {GM}, I present the investigation of the sum rule within a
model involving nucleons and pions, using pseudoscalar coupling \cite
{Gerasimov75}.  To lowest non-trivial order, the only final state of
the photoabsorption process is \piN.  This study is also quite useful
with respect to resonance saturation of the sum rule.  In \Sect {QED},
the fermion under consideration is taken to be an electron, utilizing
quantum electrodynamics to calculate cross sections \cite
{Altarelli72,Tsai75}.  The final state then is $\gamma\te^-$, i.e.,
Compton scattering.\footnote {Don't be puzzled by the double
appearance of the term Compton scattering.  The integrand of the GDH
integral is \emph {always} the imaginary part of the forward Compton
amplitude, being related, via the optical theorem, to the
photoabsorption cross section.  In terms of Feynman diagrams, the
optical theorem relates one-loop diagrams of the Compton process to
sqares of tree-order diagrams for photoabsorption, where intermediate
states of the former are final states of the latter.  These states
differ for different models, and in lowest-order QED they happen to
coincide with the initial state. (In the next order, one would have
$\gamma\gamma\te^-$ and $\te^+\te^-\te^-$.)  Of course,
``photoabsorption'' actually becomes a somewhat inappropriate notion
then.} The Weinberg-Salam model \cite {Altarelli72} (see also Brodsky
and Schmidt \cite {Brodsky95}), being most involved, is considered in
the last section of this chapter. In a sense, this model is an
extension of QED, since it incorporates weak gauge bosons Z$^0$ and
W$^\pm$, the Higgs boson H, and the neutrino $\nu_\te$ in addition to
$\gamma$ and e$^-$.  Final states are $\gamma\te^-$, Z$^0$e$^-$,
W$^-\nu_\te$, and H\,e$^-$.

\subsubsection{What does the inspection of perturbative models aim at?}
Perturbative models are not directly related to the experimentally
accessible GDH sum rule.  As far as the nucleon is concerned, this can
be attributed to the fact that effective perturbative descriptions
(like chiral perturbation theory) work at low energies only, while
perturbative QCD requires the photon to be highly virtual.  At $q^2=0$
and large $\nu$, the photon-nucleon interaction is conspicuously
non-perturbative.  The electron GDH sum rule, on the other hand, is
not measureable at all. Nevertheless, investigating a sum rule
perturbatively is surely very illustrative also in view of the ``real
world''.

However, the main purpose of presenting these considerations here is
that the effect of anomalous commutators and of the infinite-momentum
limit can be studied within the Weinberg-Salam model.  This will be
done in \Sect {PRP}, but I cannot resist anticipating what the
result will be: There is \emph {no} anomalous-commutator modification
of the GDH sum rule!

\section{Pseudoscalar pion-nucleon model}
\label{Sect:GM}
In 1975, Gerasimov and Moulin \cite {Gerasimov75} investigated the GDH
sum rule within the simplest model involving electromagnetic
interactions of pions and nucleons as well as strong interactions of
the named hadrons themselves.  Interaction vertices of this model are
shown in \Fig {GM-vert}.%
\begin{figure}[tb]
 \begin{displaymath} \boxed{ \begin{array}{cc} \\
  %
  % gamma-p coupling
  %
  {\rm (a)} &
  \graph(3,2){
   \fmfleft{i1,i2} \fmftop{t} \fmfright{o1,o2}
   \fmf{fermion}{i1,v,o1}
   \fmf{photon}{t,v}
  } \propto e \\[11ex]
  %
  % gamma-pi coupling
  %
  {\rm (b)} &
  \graph(3,2){
   \fmfleft{i1,i2} \fmftop{t} \fmfright{o1,o2}
   \fmf{dashes}{i1,v,o1}
   \fmf{photon}{t,v}
  } \propto e
  \qquad \qquad
  \graph(3,2){
   \fmfleft{i1,i2} \fmfright{o1,o2}
   \fmf{dashes}{i1,v,o1}
   \fmf{photon}{i2,v,o2}
  } \propto e^2 \\[11ex]
  %
  % pi-N coupling
  %
  {\rm (c)} &
  \graph(3,2){
   \fmfleft{i1,i2} \fmftop{t} \fmfright{o1,o2}
   \fmf{fermion}{i1,v,o1}
   \fmf{dashes}{t,v}
  } \propto g \\ \\
 \end{array} } \end{displaymath}
 \caption[]{
  Interaction vertices of the pseudoscalar pion-nucleon model.
  Solid, dashed, and wavy lines represent nucleons, pions, and photons,
  respectively. \\
  \hspace*{1em} (a) $\gamma$p coupling \\
  \hspace*{1em} (b) $\gamma\pi^\pm$ coupling \\
  \hspace*{1em} (c) pseudoscalar \piN\ coupling \\
  \label{Fig:GM-vert}}
\end{figure}
Photons are coupled to charged hadrons by means of minimal coupling.
(The $\gamma\gamma\pi\pi$ contact vertex of \Fig {GM-vert}(b) is
not needed in the present context.)  In particular, the nucleon has no
anomalous magnetic moment on the vertex level.  Interactions of pions with
nucleons are described by the Lagrangian
\begin{equation} \label{ps-int}
 \mathcal{L}_{\pi\tN\tN}(x) = g\, \bar\psi(x)\gamma_5\btau\ndot\bphi(x)\psi(x),
\end{equation}
where $\psi(x)$ denotes the isodoublet nucleon spinor and $\bphi(x)$
is the isotriplet of pion fields.  The interaction \eq {ps-int} is
called \emph {pseudoscalar} as opposed to \emph {pseudovector}
interaction, for which $\gamma_5\bphi$ is replaced by
$i\gamma_5\parslash\bphi$.  With pseudoscalar coupling, the coupling
constant $g$ is dimensionless.  In \Ref {Gerasimov75}, the
integrand $\sigma_{1/2}(\nu)-\sigma_{3/2}(\nu)$ of the GDH sum rule is
calculated within this model to lowest non-trivial order in
electromagnetic and strong coupling constants, i.e., to order $\alpha
g^2$.  The pertinent Feynman graphs are depicted in \Fig
{GM-graphs}.%
\begin{figure}[tb]
 \begin{displaymath} \begin{array}{|c@{\quad}c|}
  \hline
  & \\
  \textrm{(a)} &
  \begin{vmatrix} \\ \quad
  \begin{minipage}{3\unitlength} \begin{fmfgraph*}(3,2)
   \fmfleft{i1,i2} \fmfright{o1,o2}
   \fmflabel{p}{i1} \fmflabel{n}{o1}
   \fmflabel{$\gamma$}{i2} \fmflabel{$\pi^+$}{o2}
   \fmf{fermion}{i1,v1,o1}
   \fmf{dashes}{v1,v2,o2}
   \fmf{photon}{i2,v2}
  \end{fmfgraph*} \end{minipage}
  \quad+\quad
  \begin{minipage}{3\unitlength} \begin{fmfgraph*}(3,2)
   \fmfleft{i1,i2} \fmfright{o1,o2}
   \fmflabel{p}{i1} \fmflabel{n}{o1}
   \fmflabel{$\gamma$}{i2} \fmflabel{$\pi^+$}{o2}
   \fmf{fermion}{i1,v1,v2,o1}
   \fmf{boson}{i2,v1}  \fmf{dashes}{v2,o2}
  \end{fmfgraph*} \end{minipage} \quad \\ \\
  \end{vmatrix}^{\textrm{\normalsize2}} \\[11ex]
  & \qquad+\qquad
  \begin{vmatrix} \\
  \quad \begin{minipage}{3\unitlength} \begin{fmfgraph*}(3,2)
   \fmfleft{i1,i2} \fmfright{o1,o2}
   \fmflabel{p}{i1} \fmflabel{p}{o1}
   \fmflabel{$\gamma$}{i2} \fmflabel{$\pi^0$}{o2}
   \fmf{fermion}{i1,v1,v2,o1}
   \fmf{boson}{i2,v1}  \fmf{dashes}{v2,o2}
  \end{fmfgraph*} \end{minipage}
  \quad+\quad
  \begin{minipage}{3\unitlength} \begin{fmfgraph*}(3,2)
   \fmfleft{i1,i2} \fmfright{o1,o2}
   \fmflabel{p}{i1} \fmflabel{p}{o1}
   \fmflabel{$\gamma$}{i2} \fmflabel{$\pi^0$}{o2}
   \fmf{fermion}{i1,v1,v2,o1}
   \fmf{phantom}{i2,v1} \fmf{phantom}{v2,o2}
   \fmffreeze
   \fmf{boson}{i2,v2} \fmf{dashes}{v1,o2}
  \end{fmfgraph*} \end{minipage} \quad \\ \\
  \end{vmatrix}^{\textrm{\normalsize2}} \\[9ex]
  \hline
  & \\
  \textrm{(b)} &
  \begin{vmatrix} \\
  \quad \begin{minipage}{3\unitlength} \begin{fmfgraph*}(3,2)
   \fmfleft{i1,i2} \fmfright{o1,o2}
   \fmflabel{n}{i1} \fmflabel{p}{o1}
   \fmflabel{$\gamma$}{i2} \fmflabel{$\pi^-$}{o2}
   \fmf{fermion}{i1,v1,o1}
   \fmf{dashes}{v1,v2,o2}
   \fmf{photon}{i2,v2}
  \end{fmfgraph*} \end{minipage}
  \quad+\quad
  \begin{minipage}{3\unitlength} \begin{fmfgraph*}(3,2)
   \fmfleft{i1,i2} \fmfright{o1,o2}
   \fmflabel{n}{i1} \fmflabel{p}{o1}
   \fmflabel{$\gamma$}{i2} \fmflabel{$\pi^-$}{o2}
   \fmf{fermion}{i1,v1,v2,o1}
   \fmf{phantom}{i2,v1} \fmf{phantom}{v2,o2}
   \fmffreeze
   \fmf{boson}{i2,v2} \fmf{dashes}{v1,o2}
  \end{fmfgraph*} \end{minipage} \quad \\ \\
  \end{vmatrix}^{\textrm{\normalsize2}} \\[9ex]
  \hline
 \end{array} \end{displaymath}
 \caption[]{
  Feynman graphs determining the nucleon photoabsorption cross section to
  order $\alpha g^2$. \\
  \hspace*{1em} (a) proton \\
  \hspace*{1em} (b) neutron
  \label{Fig:GM-graphs}}
\end{figure}
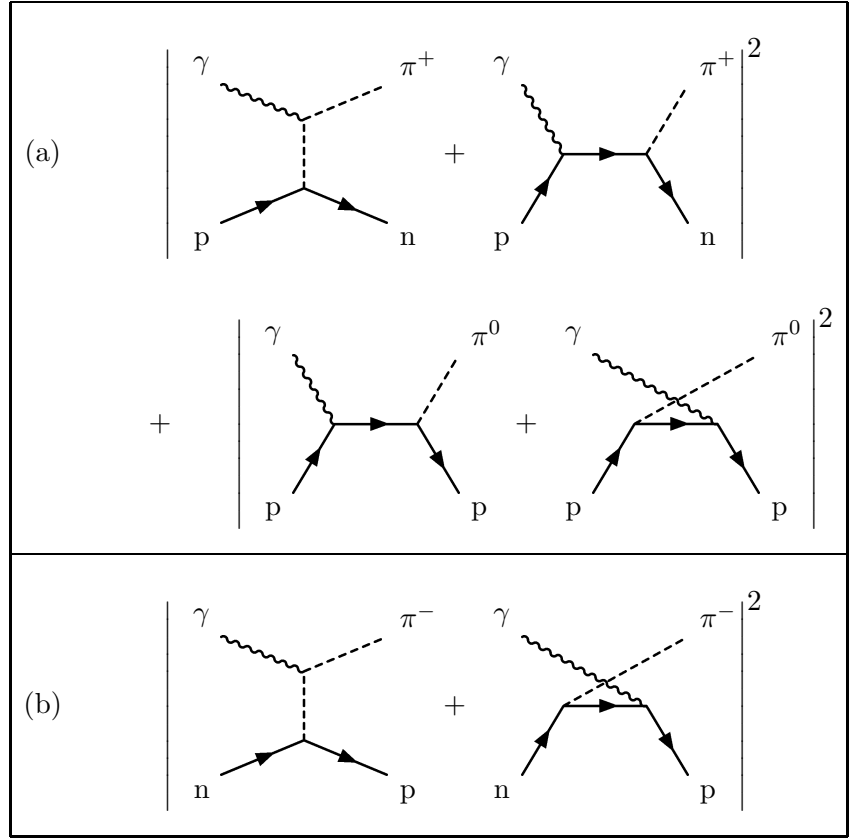
To this order, the only possible final state of the photoabsorption
reaction is \piN.  For the $\gamma$p initial state, there are
final-state isospin channels $\pi^+$n and $\pi^0$p, whereas for
$\gamma$n, there is only $\pi^-$p.  The cross section for process
$\gamma\tn\to\pi^0\tn$ vanishes identically to the order considered,
since the photon couples to neither of the neutral particles.

\subsubsection{Anomalous magnetic moment}
To lowest order, the anomalous magnetic moment $\kappa_\tp$ of the
proton is determined by the Feynman graphs
\begin{subequations}
\begin{equation} \label{kappa.p}
 \begin{minipage}{3\unitlength} \begin{fmfgraph*}(3,3)
  \fmfbottom{b1,b2} \fmftop{t}
  \fmf{fermion,label.side=left}{b1,v1}
  \fmf{fermion,label.side=left}{v3,b2}
  \fmf{fermion,label=p,label.side=left}{v1,v2,v3}
  \fmf{dashes,tension=0,label=$\pi^0$,label.side=right}{v1,v3}
  \fmf{photon,tension=2}{t,v2}
 \end{fmfgraph*} \end{minipage}
 \qquad+\qquad
 \begin{minipage}{3\unitlength} \begin{fmfgraph*}(3,3)
  \fmfbottom{b1,b2} \fmftop{t}
  \fmf{fermion,label.side=left}{b1,v1}
  \fmf{fermion,label.side=left}{v3,b2}
  \fmf{dashes,label=$\pi^+$,label.side=left}{v1,v2,v3}
  \fmf{fermion,tension=0,label=n,label.side=right}{v1,v3}
  \fmf{photon,tension=2}{t,v2}
 \end{fmfgraph*} \end{minipage}
\end{equation}
while $\kappa_\tn$ is determined by
\begin{equation} \label{kappa.n}
 \begin{minipage}{3\unitlength} \begin{fmfgraph*}(3,3)
  \fmfbottom{b1,b2} \fmftop{t}
  \fmf{fermion}{b1,v1}
  \fmf{fermion}{v3,b2}
  \fmf{fermion,label=p,label.side=left}{v1,v2,v3}
  \fmf{dashes,tension=0,label=$\pi^-$,label.side=right}{v1,v3}
  \fmf{photon,tension=2}{t,v2}
 \end{fmfgraph*} \end{minipage}
 \qquad+\qquad
 \begin{minipage}{3\unitlength} \begin{fmfgraph*}(3,3)
  \fmfbottom{b1,b2} \fmftop{t}
  \fmf{fermion}{b1,v1}
  \fmf{fermion}{v3,b2}
  \fmf{dashes,label=$\pi^-$,label.side=left}{v1,v2,v3}
  \fmf{fermion,tension=0,label=p,label.side=right}{v1,v3}
  \fmf{photon,tension=2}{t,v2}
 \end{fmfgraph*} \end{minipage}
\end{equation} \label{kappa}%
\end{subequations}
Thus, both quantities are of order $g^2$.  Consequently, the left-hand
side of the GDH sum rule \eq {pertGDH} is of order $\alpha g^4$, so
that to order $\alpha g^2$ it reads
\begin{equation} \label{GM-GDH}
 0 = \int_{\nu_0}^\infty\! \frac{\td\nu}\nu\,
  \bigl(\sigma_{1/2}(\nu) - \sigma_{3/2}(\nu)\bigr).
\end{equation}

\subsubsection{Polarized photoabsorption cross sections}
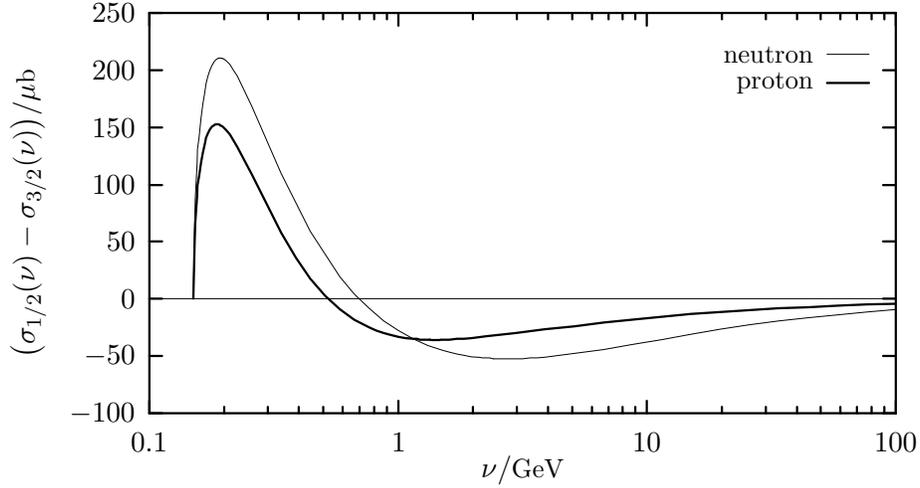
\begin{figure}[tb]
 \begin{center}
  \input{gm-xsect}
 \end{center}
 \caption[]{
  Polarized total photoabsorption cross section
  $\sigma_{1/2}(\nu) - \sigma_{3/2}(\nu)$ of proton (heavy line)
  and neutron (light line) to order $\alpha g^2$, calculated from
  the Feynman graphs of \Fig {GM-graphs}.
  Since $\text d\nu/\nu = \text d(\ln\nu)$, the vanishing of the integral
  \eq{GM-GDH} is reflected by the equality of the areas enclosed
  by the curves and the abscissa.  Cross sections vanish below pion-production
  threshold $\nu_0=m_\pi+m_\pi^2/2M=0.15$~GeV.
  \label{Fig:GM-xsect}}
\end{figure}

Gerasimov and Moulin \cite {Gerasimov75} calculated The integrand of
\Eq {GM-GDH} and showed that the integral indeed vanishes. Since
$\td\nu/\nu=\td(\ln\nu)$, this is reflected by the fact that the
curves in \Fig {GM-xsect} change sign and that the areas enclosed with
the abscissa to the left and to the right of the sign change are
equal.  To plot the cross sections, I have taken the experimental
values
\begin{equation} \label{g-sqared}
 \frac {g^2}{4\pi} = 14.6
\end{equation}
and $m_\pi=140$~MeV (the charged pion's mass, since $t$-channel pion
exchange is the dominant process).%

\label{g-story}
Of course, the value \eq {g-sqared} does not exactly suggest a
perturbative treatment.  The curves drawn in \Fig {GM-xsect} will
be far off the experimental photoabsorption cross sections, especially
at resonances and above two-pion threshold.  But numerical accuracy is
presently not on target.  Rather, inspection of perturbative models
aims at testing and illustrating simultaneously all assumptions on
which the derivation of the sum rule is based.

\subsubsection{Optical theorem}
At this point, a few words on the optical theorem may be appropriate.
As mentioned above, instead of calculating the cross section
difference $\sigma_{1/2}(\nu)-\sigma_{3/2}(\nu)$ by squaring the
tree-order amplitudes of, say, the $\gamma\tp\to\pi^0\tp$ process
depicted in the second line of \Fig {GM-graphs}(a), and by
subsequently integrating over final state configurations, i.e., pion
scattering angles, one can equally well calculate $8\pi\im f_2(\nu)$,
where forward Compton amplitude $f_2(\nu)$ is obtained from the
one-loop graphs
\begin{equation} \label{GM-imf2graphs}
 \graph(3,2){
  \fmfleft{i1,i2} \fmfright{o1,o2}
  \fmf{fermion}{i1,v1,v2,v3,v4,o1}
  \fmf{boson}{i2,v2} \fmf{boson}{v3,o2}
  \fmf{dashes,tension=0}{v1,v4}
 } +
 \graph(3,2){
  \fmfleft{i1,i2} \fmfright{o1,o2}
  \fmf{fermion}{i1,v1,v2,v3,v4}
  \fmf{fermion,tension=0.5}{v4,o1}
  \fmf{boson}{i2,v2} \fmf{boson}{v4,o2}
  \fmf{dashes,tension=0}{v3,v1}
 } +
 \graph(3,2){
  \fmfleft{i1,i2} \fmfright{o1,o2}
  \fmf{fermion}{v1,v2,v3,v4,o1}
  \fmf{fermion,tension=0.5}{i1,v1}
  \fmf{boson}{i2,v1} \fmf{boson}{v3,o2}
  \fmf{dashes,tension=0}{v2,v4}
 } +
 \graph(3,2){
  \fmfleft{i1,i2} \fmfright{o1,o2}
  \fmf{fermion,tension=0.5}{i1,v1}
  \fmf{fermion,tension=2}{v1,v2}
  \fmf{fermion,tension=1}{v2,v3}
  \fmf{fermion,tension=2}{v3,v4}
  \fmf{fermion,tension=0.5}{v4,o1}
  \fmf{boson}{i2,v1} \fmf{boson}{v4,o2}
  \fmf{dashes,right,tension=0}{v2,v3}
 }
\end{equation}
(plus the same graphs with their photon legs crossed).  In other
words: The optical theorem
\begin{equation}
 \im f_2(\nu) = \frac{1}{8\pi}
  \bigl[\sigma_{1/2}(\nu)-\sigma_{3/2}(\nu)\bigr]
\end{equation}
holds perturbatively.  Considering the graphs \eq {GM-imf2graphs} together with
the low-energy theorem \eq {pertLET} and the graphs \eq {kappa}
responsible for the anomalous magnetic moment, it is clear that the
left-hand side of the GDH sum rule will indeed vanish to the given order,
since a contribution to $f_2(0)$ proportional to $\kappa^2$ will not
arise until two-loop graphs like
\begin{equation}
 \graph(3,2){
  \fmfleft{i1,i2} \fmfright{o1,o2}
  \fmf{fermion}{i1,v1,v2,v3,v4,v5,v6,o1}
  \fmf{boson}{i2,v2} \fmf{boson}{v5,o2}
  \fmffreeze
  \fmf{dashes}{v1,v3} \fmf{dashes}{v4,v6}
 }
\end{equation}
are taken into account.

\subsubsection{Anomalous magnetic moment on the vertex level}
\label{kappa0-story}
\begin{figure}[tb]
 \begin{center}
  \input{piN-p} \\
  \input{piN-n}
 \end{center}
 \caption[]{
  Polarized total photoabsorption cross section
  $\sigma_{1/2}(\nu) - \sigma_{3/2}(\nu)$ of proton (upper plot)
  and neutron (lower plot) to order $\alpha g^2$,
  with (heavy lines) and without
  (light lines) anomalous magnetic moment $\kappa_0$ on the vertex level.
  As for the former case, where
  the physical values $\kappa_0^\tp=1.79$, $\kappa_0^\tn=-1.91$ have been
  adopted for definiteness, the cross section difference approaches a
  non-vanishing constant at $\nu\to\infty$.
  The $\kappa_0=0$ curves are the same as in \Fig {GM-xsect}.
  Cross sections vanish below pion-production
  threshold $\nu_0=m_\pi+m_\pi^2/2M=0.15$~GeV.
  \label{Fig:GM+kappa}}
\end{figure}
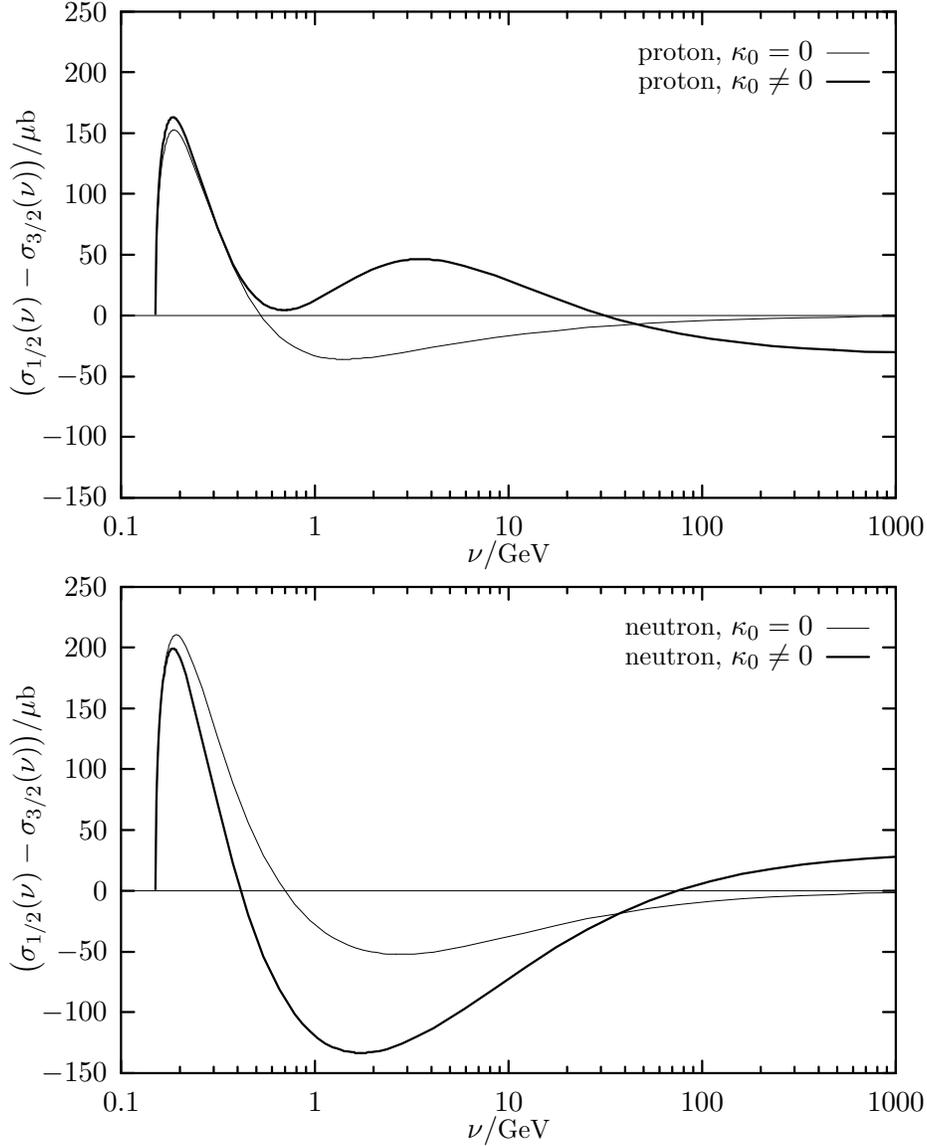
In \Fig {GM-vert}, the photon couples to the nucleon minimally, i.e.,
to its charge only.  I now investigate the case of non-minimal
coupling
\begin{equation}
 \graph(2,2){
  \fmftop{t} \fmfbottom{b1,b2}
  \fmf{boson}{t,v}
  \fmf{fermion}{b1,v,b2}
  \fmfdot{v}
}
 = Ze\gamma^\mu + \frac{ie\kappa_0}{2M}\, q_\rho \sigma^{\mu\rho}.
\end{equation}
That is to say, the nucleon adopts an anomalous magnetic moment
$\kappa_0$ on the vertex level.  In this scenario, the left-hand side
of the GDH sum rule is in fact divergent, due to the
(non-renormalizable) divergence of one-loop graphs of the form
\begin{equation}
 \graph(3,3){
  \fmfbottom{b1,b2} \fmftop{t}
  \fmf{fermion}{b1,v1}
  \fmf{fermion}{v3,b2}
  \fmf{fermion}{v1,v2,v3}
  \fmf{dashes,tension=0}{v1,v3}
  \fmf{photon,tension=2}{t,v2}
  \fmfdot{v2}
 }
\end{equation}

What happens to the right-hand side of the sum rule?  I calculated its
integrand $\sigma_{1/2}(\nu)-\sigma_{3/2}(\nu)$ by inserting the
pion-electroproduction magnetic Born terms of von Gehlen \cite
{vonGehlen69} into \Eq {GDH-CGLN} below, specializing to $Q^2=0$.
The result is depicted in \Fig {GM+kappa}.
Observe that the polarized cross section difference approaches
a non-zero constant as $\nu\to\infty$, indicating that the GDH integral
will diverge.  (This is of course consistent with the divergence of the
left-hand side of the sum rule.)  More precisely, one has that
\begin{subequations}
\begin{align}
 & \sigma^\tp_{1/2}(\nu) - \sigma^\tp_{3/2}(\nu) \xrightarrow[\nu\to\infty]{}
 \frac{\alpha g^2}{8M^2} \bigl( (\kappa_0^\tp)^2 - (\kappa_0^\tn)^2 \bigr)
 = -33~\mu\tb \\
\intertext{and}
 & \sigma^\tn_{1/2}(\nu) - \sigma^\tn_{3/2}(\nu) \xrightarrow[\nu\to\infty]{}
 \frac{\alpha g^2}{8M^2} \bigl( (\kappa_0^\tn)^2 - (\kappa_0^\tp)^2 \bigr)
 = 33~\mu\tb,
\end{align}
\end{subequations}
where the physical values $\kappa_0^\tp=1.79$, $\kappa_0^\tn=-1.91$
have been inserted.  Letting $\kappa_0^\tp=\kappa_0^\tn=0$
leads back to the Gerasimov-Moulin result depicted in \Fig {GM-xsect}.

\section{Quantum electrodynamics}
\label{Sect:QED}

For the remainder of this chapter, the fermion under consideration is
taken to be an electron rather than a nucleon.  In this case, quantum
electrodynamics (QED) as well as the Weinberg-Salam model of
electro-weak interactions suggest themselves as testing grounds for
the GDH sum rule.  QED -- certainly the most prominent perturbative
quantum field theory -- is actually \emph {part of} the Weinberg-Salam
model.  (As a matter of fact, the first study of the GDH sum rule
within QED, performed by Altarelli, Cabibbo, and Maiani \cite
{Altarelli72} in 1972, incorporated weak interactions, too.  Tsai,
DeRaad, and Milton \cite {Tsai75} investigated QED only, but were
additionally concerned with the Burkhardt-Cottingham sum rule \cite
{Burkhardt70}.)  Nevertheless, for pedagogical reasons I want to
present both models seperately.  This section is devoted to the less
involved QED case.

\subsubsection{Anomalous magnetic moment}
The procedure is analogous to the pion-nucleon model presented in the
preceding section, except that the pion is replaced by a photon.
To order $\alpha$, the anomalous magnetic moment of the electron is
determined by the graph
\begin{equation} \label{Schwinger-graph}
 \graph(2,2){
  \fmfbottom{b1,b2} \fmftop{t}
  \fmf{fermion}{b1,v1,v2,v3,b2} \fmf{photon,tension=2}{t,v2}
  \fmffreeze
  \fmf{photon}{v1,v3}
 }
\end{equation}
which gives rise to the famous Schwinger moment
\begin{equation} \label{Schwinger-moment}
 \kappa = \frac {\alpha}{2\pi} = 1.16\ntimes10^{-3}.
\end{equation}
(For the calculation, see, e.g., Itzykson and Zuber \cite
{Itzykson80}.)  Consequently, the left-hand side of the GDH sum rule
\eq {pertGDH} is of order $\alpha^3$, so that to order $\alpha^2$ it
reads
\begin{equation} \label{QED-GDH}
 0 = \int_{0}^\infty\! \frac{\td\nu}\nu\,
  \bigl(\sigma_{1/2}(\nu) - \sigma_{3/2}(\nu)\bigr).
\end{equation}

\subsubsection{Polarized photoabsorption cross sections}
Of course, the integrand on the right-hand side is of order
$\alpha^2$.  There is no order-$\alpha$ piece, since at least one
additional photon is needed to make up the final state. (There is no
such thing as a ``purely hadronic final state'' here.)  Observe that
the threshold $\nu_0$, i.e., the lower bound of the GDH integral \eq
{QED-GDH}, vanishes, since the final-state particles have the same
mass than the particles in the initial state.  This opens the question
for infrared convergence of integral \eq {QED-GDH}, in addition to the
usual question for ultraviolet convergence.  I mention in advance that
both of these questions will be affirmed.

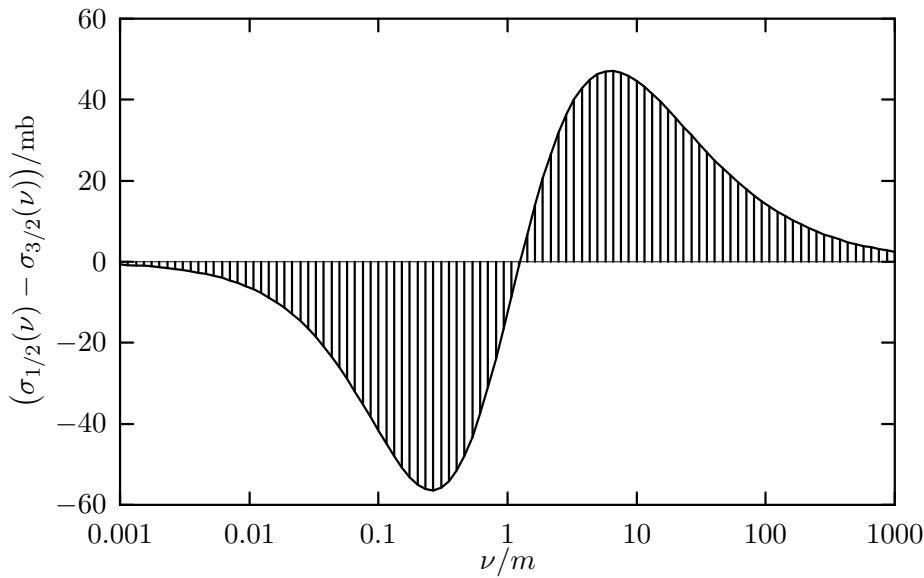
\begin{figure}[tb]
 \begin{center}
  \input{qed-xsect}
 \end{center}
 \caption[]{
  Polarized total photoabsorption cross section
  $\sigma_{1/2}(\nu) - \sigma_{3/2}(\nu)$ of the electron
  in order-$\alpha^2$ QED.
  Since $\text d\nu/\nu = \text d(\ln\nu)$, the vanishing of the integral
  \eq{QED-GDH} is reflected by the equality of the shaded areas.
  \label{Fig:QED-xsect}}
\end{figure}

Opposed to the graphs shown in \Fig {GM-graphs}, cross sections
$\sigma_{1/2}(\nu)$ and $\sigma_{3/2}(\nu)$ are now calculated from
\begin{equation}
 \left|
  \graph (3,2){
   \fmfleft{i1,i2} \fmfright{o1,o2}
   \fmf{fermion}{i1,v1,v2,o1}
   \fmf{boson}{i2,v1}  \fmf{boson}{v2,o2}
  } +
  \graph(3,2){
   \fmfleft{i1,i2} \fmfright{o1,o2}
   \fmf{fermion}{i1,v1,v2,o1}
   \fmf{phantom}{i2,v1} \fmf{phantom}{v2,o2}
   \fmffreeze
   \fmf{boson}{i2,v2} \fmf{boson}{v1,o2}
  }
 \right|^{\text{\normalsize2}}
\end{equation}
Again, the perturbative validity of the optical theorem can be checked
explicitly by computing $\im f_2(\nu)$ from the one-loop graphs
\begin{equation} \label{QED-imf2graphs}
 \graph(3,2){
  \fmfleft{i1,i2} \fmfright{o1,o2}
  \fmf{fermion}{i1,v1,v2,v3,v4,o1}
  \fmf{boson}{i2,v2} \fmf{boson}{v3,o2}
  \fmf{photon,tension=0}{v1,v4}
 } +
 \graph(3,2){
  \fmfleft{i1,i2} \fmfright{o1,o2}
  \fmf{fermion}{i1,v1,v2,v3,v4}
  \fmf{fermion,tension=0.5}{v4,o1}
  \fmf{boson}{i2,v2} \fmf{boson}{v4,o2}
  \fmf{photon,tension=0}{v3,v1}
 } +
 \graph(3,2){
  \fmfleft{i1,i2} \fmfright{o1,o2}
  \fmf{fermion}{v1,v2,v3,v4,o1}
  \fmf{fermion,tension=0.5}{i1,v1}
  \fmf{boson}{i2,v1} \fmf{boson}{v3,o2}
  \fmf{photon,tension=0}{v2,v4}
 } +
 \graph(3,2){
  \fmfleft{i1,i2} \fmfright{o1,o2}
  \fmf{fermion,tension=0.5}{i1,v1}
  \fmf{fermion,tension=2}{v1,v2}
  \fmf{fermion,tension=1}{v2,v3}
  \fmf{fermion,tension=2}{v3,v4}
  \fmf{fermion,tension=0.5}{v4,o1}
  \fmf{boson}{i2,v1} \fmf{boson}{v4,o2}
  \fmf{photon,right,tension=0}{v2,v3}
 }
\end{equation}
(plus the same graphs with their photon legs crossed).  The result
reads \cite {Altarelli72}
\begin{equation} \label{QED-xsect}
 \sigma_{1/2}(\nu)-\sigma_{3/2}(\nu) = \frac {2\pi\alpha^2}{m\nu}
 \left[ \left( 1 + \frac m\nu \right) \ln \left( 1 + \frac {2\nu}m \right)
 - 2 \left( 1 + \frac {\nu^2}{(m+2\nu)^2} \right) \right],
\end{equation}
which is depicted in \Fig {QED-xsect}.%

\subsubsection{Convergence of the GDH integral}
Expanding expression \eq {QED-xsect} about $\nu=0$ and $\nu\to\infty$,
one finds\footnote {Note that the \emph {average} of $\sigma_{1/2}(\nu)$
and $\sigma_{3/2}(\nu)$ does \emph {not} vanish as $\nu\to0$.  Rather,
it approaches the Thomson limit $8\pi\alpha^2/3m^2=665$~mb.}
\begin{align}
 & \sigma_{1/2}(\nu)-\sigma_{3/2}(\nu) \xrightarrow[\nu\to0]{}
 - \frac {8\pi\alpha^2\nu}{3m^3} \\
\intertext{and}
 & \sigma_{1/2}(\nu)-\sigma_{3/2}(\nu) \xrightarrow[\nu\to\infty]{}
 \frac {2\pi\alpha^2}{m\nu} \ln\frac{\nu}{m},
\end{align}
respectively, so that integral \eq {QED-GDH} converges.  An explicit analytical
integration\footnote {\textsl {Mathematica} copes with this task.} shows
that it indeed vanishes, as indicated by the shaded areas in \Fig {QED-xsect}.

\section{Weinberg-Salam model}
\label{Sect:WSM}

As noted above, the Weinberg-Salam model (for an introduction, I
recommend the textbook of Aitchison and Hey \cite {Aitchison89}) is an
extension of QED.  Neutrino, W$^\pm$, Z$^0$, and Higgs come into play
in addition to electron and photon.  Since these particles implicate
parity violation, some caution is in order concerning the definition
of cross sections $\sigma_{1/2}$ and $\sigma_{3/2}$.  For instance,
the cross section $\sigma_{+-}$ corresponding to photon helicity
$\lambda_\gamma=+1$ and electron helicity $\lambda_\te=-\frac12$ is no
longer equal to the cross section $\sigma_{-+}$ for
$\lambda_\gamma=-1$, $\lambda_\te=+\frac12$.  Both were formerly
called $\sigma_{3/2}$ (parallel helicities).  Now one must properly
define what is meant by parallel helicities.  An obvious choice is
$\sigma_{3/2}:=\frac12(\sigma_{+-}+\sigma_{-+})$ and likewise for
$\sigma_{1/2}$.  I note that Bac\'e and Hari Dass \cite {Bace75}
claimed that the parity violating portion of the Compton amplitude
vanishes in the one-loop order of the Weinberg-Salam model.

However, I intend to employ the model for learning something about
anomalous commutators and the infinite-momentum limit, and I am not
concerned with parity violation.  Therefore, I make life a bit easier
by letting the Weinberg angle $\theta_\tW$ adopt its ideal value,
i.e.,
\begin{equation}
 \sin\theta_\tW = \frac12,
\end{equation}
which, besides, is not far from what is observed experimentally, viz.\
$\sin\theta_\tW=0.48$ \cite {PDG96}.  Therewith, the Z$^0$ boson has a
purely axial-vector coupling to electrons, and graphs involving the
Z$^0$ boson are parity conserving.  Fixing the Weinberg angle implies
that the number of coupling constants is reduced from two
($g=e/\sin\theta_\tW$ and $g'=e/\cos\theta_\tW$) to one, which can be
chosen as the electric charge $e$.  Consequently, all quantities will
be expanded in the fine-structure constant $\alpha$, just as in QED.
The Fermi constant is then given by
\begin{equation}
 \frac {G_\tF}{\sqrt2} = \frac {2\pi\alpha}{M_\tW^2}
 = \frac {8\pi\alpha}{3M_\tZ^2}.
\end{equation}

For a compilation of Feynman rules, see App.\ F.3 of Aitchison and Hey \cite
{Aitchison89}.  I adopt the unitary gauge, which has the advantage
that the particle content of the theory is manifest.  That is to say,
Feynman graphs contain the above-mentioned physical particles only,
and no ficticious particles (so-called would-be-Goldstone bosons)
are needed.  In unitary gauge, the propagators of the massive gauge
bosons W$^\pm$, Z$^0$ read
\begin{equation}
 \labelledgraph(2,1){
  \fmfleft{l} \fmfright{r}
  \fmflabel{$\mu$}{l}  \fmflabel{$\nu$}{r}
  \fmf{boson,label.side=left,label=$k$}{l,r}
 } \qquad = \frac {i}{k^2-M_{\tW,\tZ}^2+i\eps}
 \left( -g_{\mu\nu} + \frac {k_\mu k_\nu}{M_{\tW,\tZ}^2} \right).
\end{equation}

\begin{figure}[tb]
 \begin{displaymath}  \begin{array}{|c@{\quad}c|}
  \hline
  & \\[-1ex]
  % gamma,Z intermediate states
  \text{(a)} &
   \labelledgraph(3,2){
    \fmfleft{i1,i2} \fmfright{o1,o2}
    \fmf{fermion}{i1,v1,v2,v3,v4,o1}
    \fmf{boson}{i2,v2} \fmf{boson}{v3,o2}
    \fmf{photon,tension=0,label.side=right,label=$\gamma$,,Z$^0$}{v1,v4}
   } +
   \labelledgraph(3,2){
    \fmfleft{i1,i2} \fmfright{o1,o2}
    \fmf{fermion}{i1,v1,v2,v3,v4}
    \fmf{fermion,tension=0.5}{v4,o1}
    \fmf{boson}{i2,v2} \fmf{boson}{v4,o2}
    \fmf{photon,tension=0,label.side=right,label=$\gamma$,,Z$^0$}{v1,v3}
   } +
   \labelledgraph(3,2){
    \fmfleft{i1,i2} \fmfright{o1,o2}
    \fmf{fermion}{v1,v2,v3,v4,o1}
    \fmf{fermion,tension=0.5}{i1,v1}
    \fmf{boson}{i2,v1} \fmf{boson}{v3,o2}
    \fmf{photon,tension=0,label.side=right,label=$\gamma$,,Z$^0$}{v2,v4}
   } +
   \labelledgraph(3,2){
    \fmfleft{i1,i2} \fmfright{o1,o2}
    \fmf{fermion,tension=0.5}{i1,v1}
    \fmf{fermion,tension=2}{v1,v2}
    \fmf{fermion,tension=1}{v2,v3}
    \fmf{fermion,tension=2}{v3,v4}
    \fmf{fermion,tension=0.5}{v4,o1}
    \fmf{boson}{i2,v1} \fmf{boson}{v4,o2}
    \fmf{photon,right,tension=0,label.side=right,label=$\gamma$,,Z$^0$}{v2,v3}
   } \\[5ex]
  \hline
  & \\[-1ex]
  % W-nu intermediate state
  \text{(b)} &
   \labelledgraph(3,2){
    \fmfleft{i1,i2} \fmfright{o1,o2}
    \fmf{fermion}{i1,v1} \fmf{fermion}{v4,o1}
    \fmf{boson}{v1,v2} \fmf{boson}{v3,v4}
    \fmf{boson,label.side=left,label=W$^-$}{v2,v3}
    \fmf{boson}{i2,v2} \fmf{boson}{v3,o2}
    \fmffreeze
    \fmf{phantom_arrow}{v1,v2,v3,v4}
    \fmf{fermion,label.side=right,label=$\nu_\te$}{v1,v4}
   } +
   \labelledgraph(3,2){
    \fmfleft{i1,i2} \fmfright{o1,o2}
    \fmf{fermion}{i1,v1} \fmf{fermion}{v3,v4} \fmf{fermion,tension=0.5}{v4,o1}
    \fmf{boson}{v1,v2,v3}
    \fmf{boson}{i2,v2} \fmf{boson}{v4,o2}
    \fmffreeze
    \fmf{phantom_arrow}{v1,v2,v3}
    \fmf{fermion,label.side=right,label=$\nu_\te$}{v1,v3}
   } +
   \labelledgraph(3,2){
    \fmfleft{i1,i2} \fmfright{o1,o2}
    \fmf{fermion}{v4,o1} \fmf{fermion}{v1,v2} \fmf{fermion,tension=0.5}{i1,v1}
    \fmf{boson}{v2,v3,v4}
    \fmf{boson}{i2,v1} \fmf{boson}{v3,o2}
    \fmffreeze
    \fmf{phantom_arrow}{v2,v3,v4}
    \fmf{fermion,label.side=right,label=$\nu_\te$}{v2,v4}
   } +
   \labelledgraph(3,2){
    \fmfleft{i1,i2} \fmfright{o1,o2}
    \fmf{fermion,tension=0.5}{i1,v1}
    \fmf{fermion,tension=2}{v1,v2}
    \fmf{fermion,tension=1,label.side=left,label=$\nu_\te$}{v2,v3}
    \fmf{fermion,tension=2}{v3,v4}
    \fmf{fermion,tension=0.5}{v4,o1}
    \fmf{boson}{i2,v1} \fmf{boson}{v4,o2}
    \fmffreeze
    \fmf{phantom_arrow,right}{v2,v3}
    \fmf{photon,right,label.side=right,label=W$^-$}{v2,v3}
   } \\[5ex]
  \hline
  & \\[-1ex]
  % H-e intermediate state
  \text{(c)} &
   \labelledgraph(3,2){
    \fmfleft{i1,i2} \fmfright{o1,o2}
    \fmf{fermion}{i1,v1,v2,v3,v4,o1}
    \fmf{boson}{i2,v2} \fmf{boson}{v3,o2}
    \fmf{dashes,tension=0,label.side=right,label=H}{v1,v4}
   } +
   \labelledgraph(3,2){
    \fmfleft{i1,i2} \fmfright{o1,o2}
    \fmf{fermion}{i1,v1,v2,v3,v4}
    \fmf{fermion,tension=0.5}{v4,o1}
    \fmf{boson}{i2,v2} \fmf{boson}{v4,o2}
    \fmf{dashes,tension=0,label.side=right,label=H}{v1,v3}
   } +
   \labelledgraph(3,2){
    \fmfleft{i1,i2} \fmfright{o1,o2}
    \fmf{fermion}{v1,v2,v3,v4,o1}
    \fmf{fermion,tension=0.5}{i1,v1}
    \fmf{boson}{i2,v1} \fmf{boson}{v3,o2}
    \fmf{dashes,tension=0,label.side=right,label=H}{v2,v4}
   } +
   \labelledgraph(3,2){
    \fmfleft{i1,i2} \fmfright{o1,o2}
    \fmf{fermion,tension=0.5}{i1,v1}
    \fmf{fermion,tension=2}{v1,v2}
    \fmf{fermion,tension=1}{v2,v3}
    \fmf{fermion,tension=2}{v3,v4}
    \fmf{fermion,tension=0.5}{v4,o1}
    \fmf{boson}{i2,v1} \fmf{boson}{v4,o2}
    \fmf{dashes,right,tension=0,label.side=right,label=H}{v2,v3}
   } \\[5ex]
  \hline
 \end{array} \end{displaymath}
 \caption[]{
  Feynman graphs determining the order-$\alpha^2$ forward Compton amplitude
  of the electron in the Weinberg-Salam model.  Addition of crossed graphs
  is understood. \\
  \hspace*{1em} (a) $\gamma\te^-$ and Z$^0\te^-$ intermediate states \\
  \hspace*{1em} (b) W$^-\nu_\te$ intermediate state \\
  \hspace*{1em} (c) H\,e$^-$ intermediate state \\
  External-line insertions like vacuum polarization are not shown in this
  figure.
  \label{Fig:WSM-graphs}
 }
\end{figure}

\subsubsection{Anomalous magnetic moment}
To order $\alpha$, the anomalous magnetic moment of the electron is
determined by the graphs
\begin{equation} \label{WSM-kappa-graphs}
 \labelledgraph(3,3){
  \fmfbottom{b1,b2} \fmftop{t}
  \fmf{fermion}{b1,v1} \fmf{fermion}{v3,b2}
  \fmf{fermion,label.side=left,label=e$^-$}{v1,v2,v3}
  \fmf{boson,tension=2}{t,v2}
  \fmf{boson,tension=0,label=$\gamma$,,Z$^0$}{v1,v3}
 } \quad+\quad
 \labelledgraph(3,3){
  \fmfbottom{b1,b2} \fmftop{t}
  \fmf{fermion}{b1,v1} \fmf{fermion}{v3,b2}
  \fmf{boson,label.side=left,label=W$^-$}{v1,v2,v3}
  \fmf{phantom_arrow,tension=0}{v1,v2,v3}
  \fmf{boson,tension=2}{t,v2} \fmf{fermion,tension=0,label=$\nu_\te$}{v1,v3}
 } \quad+\quad
 \labelledgraph(3,3){
  \fmfbottom{b1,b2} \fmftop{t}
  \fmf{fermion}{b1,v1} \fmf{fermion}{v3,b2}
  \fmf{fermion,label.side=left,label=e$^-$}{v1,v2,v3}
  \fmf{boson,tension=2}{t,v2} \fmf{dashes,tension=0,label=H}{v1,v3}
 }
\end{equation}
the first of which also appears in QED, \Eq {Schwinger-graph}.  The
graph involving the W boson has first been calculated by Brodsky and
Sullivan \cite {Brodsky67}. Its contribution to the electron's
anomalous magnetic moment reads\footnote {Note that the value reported
by Altarelli et al.\ \cite {Altarelli72} is a factor of eight too
small. Observe that, numerically, the value \eq {kappa-W} lies some
\emph {nine orders of magnitude} below the Schwinger moment \eq
{Schwinger-moment}, due to the smallness of the e-W mass ratio.  Yet,
it is taken to be of order $\alpha$, since one is not concerned with
numerics, but rather with the order-by-order consistency of the model,
which has no regard to the phenomenological masses of the particles
involved.  This state of affairs is similar to the case of the \piN\
coupling constant discussed on p.~\pageref {g-story} of this
thesis.\label {smallness}}
\begin{equation} \label{kappa-W}
 \Delta\kappa = \frac{10m^2}{3M_\tW^2\sin^2\theta_\tW}\,
\frac\alpha{2\pi} = 6.83\ntimes10^{-13}.
\end{equation}
Similar expressions are obtained for the Z$^0$- and Higgs-exchange graphs
of \Eq {WSM-kappa-graphs} \cite {Altarelli72}.

Thus, again, the left-hand side of the GDH sum rule \eq {pertGDH} is of order
$\alpha^3$, so that to order $\alpha^2$ it reads
\begin{equation} \label{WSM-GDH}
 0 = \int_{0}^\infty\! \frac{\td\nu}\nu\,
  \bigl(\sigma_{1/2}(\nu) - \sigma_{3/2}(\nu)\bigr).
\end{equation}

\subsubsection{Polarized photoabsorption cross sections}
The imaginary part of forward Compton amplitude $f_2(\nu)$, and
therewith, via the optical theorem, the cross section difference
$\sigma_{1/2}(\nu)-\sigma_{3/2}(\nu)$, is calculated from the Feynman
graphs of \Fig {WSM-graphs}.  It is found that the integral
converges\footnote {\label{kappaW}As a matter of fact, Altarelli et al.\ \cite
{Altarelli72} employed the GDH sum rule to fix the anomalous magnetic
moment $\kappa_\tW$ of the W boson.  They showed that the GDH integral
converges \emph {if and only if} $\kappa_\tW=1$.  This state of
affairs is similar to the case of the vertex-level anomalous magnetic
moment of the nucleon discussed on p.~\pageref {kappa0-story} of this
thesis.  The value $\kappa_\tW=1$ naturally emerges from what is
nowadays called the standard model, so no-one really doubts it
anymore.}, and \Eq {WSM-GDH} is explicitly affirmed
\cite {Altarelli72}.\footnote {For a plot of the W$^-\nu_\te$
contribution to $\sigma_{1/2}(\nu)-\sigma_{3/2}(\nu)$, determined by
the Feynman graphs of \Fig {WSM-graphs}(b), see Brodsky and Schmidt
\cite {Brodsky95}.  Numerically, this contribution does not exceed a
few pb, compared to tens of mb found as the QED contribution, \Fig
{QED-xsect}.  See also footnote \ref {smallness}.}

\clearpage{\pagestyle{empty}\cleardoublepage}
\chapter{Possible sources of modifications}
\label{Ch:mod}

In this chapter, I review the proposed sources of modifications of the
GDH sum rule. 

 \Sect {fixed-pole} is devoted to a discussion of a possible $J=1$
fixed pole in angular-momentum plane \cite {Abarbanel68}.  After a
short paragraph on the relevant ingredients of Regge theory, I
emphasize the crucial role that spin plays with respect to the
particular Regge singularity under consideration.  Next, I comment on
the question, which Compton-scattering observable the residue of the
fixed pole is related to.  Finally, I want to present a plausibility
argument \emph {against} the presence of a fixed pole.

In \Sect {anomcomm}, I discuss the claimed modification coming from an
anomalous charge-density algebra \cite {Chang94a}.  After an
introduction to anomalous commutators, sea\-gull amplitudes, and the
Bjorken-Johnson-Low technique \cite {Bjorken66,Johnson66}, I
comprehensively discuss this theory, focussing particularly on \Ref
{Chang94a}.  It is stressed that the commutator anomaly modifies the
finite-momentum GDH sum rule only, providing no information on the
legitimacy of the infinite-momentum limit.

\Sect {IML} aims at illustrating the significance of the infinite-momentum
limit itself, without regard to an anomalous charge-density algebra.
I give an example to shed some light on the way a modification of the
GDH sum rule might in principle be brought about by the
infinite-momentum limit.

In \Sect {PRP}, the Weinberg-Salam model is adopted to simultaneously
calculate the anomalous charge-density commutator \emph {and} the
effect of the infinite-momentum limit \cite {Pantfoerder98}.  It is
shown that both of these points yield a non-trivial contribution to
the GDH sum rule, in such a way that the individual modifications cancel
exactly.

 \Sect {a1} presents a short treatise on $t$-channel exchange of
axial-vector mesons, which in fact generally cannot contribute to the
GDH sum rule owing to fundamental symmetries.  This section shall
further illustrate what is actually done in anomalous-commutator
calculations. 

 \Sect {quark} is devoted to the discussion of a possible modification
of the GDH sum rule that arises as soon as one permits quarks to
possess anomalous magnetic moments.  The nature of the apparent
modification obtained in constituent-quark models is discussed.  In
\Sect {Ying}, I discuss the ``spontaneous breakdown of electromagnetic
gauge symmetry'' \cite {Ying96a}, while in \Sect {modLET}, I comment
on the question whether the Low-Gell-Mann-Goldberger low-energy
theorem \cite {Low54,Gell-Mann54} gets modified by a non-naive current
commutator.

\section{Fixed pole in angular momentum plane}
\label{Sect:fixed-pole}

In 1968, Abarbanel and Goldberger \cite {Abarbanel68} pointed out that
the GDH sum rule is modified if the nucleon Compton amplitude exhibits
a $J=1$ \emph {fixed pole} in complex angular-momentum plane in
the $t$ channel.  Within the scope of this thesis, I cannot go into
the details of Regge theory (for reviews, see, e.g., Collins and
Squires \cite {Collins68} or Drechsler \cite {Drechsler70}).
Nevertheless, I would like to emphasize a few points relevant to the
GDH sum rule that might be somewhat concealed in the existing
literature.

\subsubsection{Loss of bilinear unitarity in the $t$ channel}
Fixed poles cannot occur in purely hadronic processes owing to
bilinear unitarity in the $t$ channel \cite {Abarbanel67}.  However,
in electromagnetic interactions, bilinear unitarity is lost, because
the amplitude is considered to lowest order in the electromagnetic
coupling.  Therefore, fixed poles are not ruled out from first
principles.  On the other hand, there is also no a priori evidence
for their \emph {presence}, particularly as regards the $J=1$ case relevant
to the GDH sum rule.

\begin{figure}[tb]
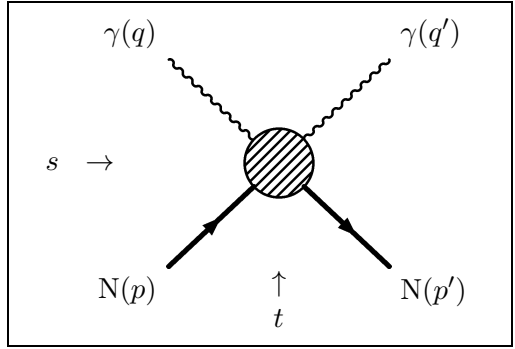

 \begin{displaymath}
  \boxed{ \quad \begin{matrix} \\[1ex]
   s & \to & \labelledgraph(4,3){
   \fmfleft{i1,i2} \fmfright{o1,o2}
   \fmflabel{N($p$)}{i1} \fmflabel{N($p'$)}{o1}
   \fmflabel{$\gamma(q)$}{i2} \fmflabel{$\gamma(q')$}{o2}
   \fmf{fermion,width=thick}{i1,v,o1}
   \fmf{boson}{i2,v,o2}
   \fmfblob{1\ul}{v}
   } \\
   & & \uparrow \\
   & & t
  \end{matrix} \qquad\quad }
 \end{displaymath}
 \caption[]{
  Nucleon Compton scattering in the $s$ channel corresponds to
  nucleon-antinucleon annihilation into two photons in the $t$ channel.
  Both processes are described by the same amplitudes.
  Regge theory relates the high-energy behavior of the
  $s$-channel amplitudes to the partial-wave expansions of the $t$-channel
  amplitudes, continued to complex angular momenta.
  The amplitudes relevant to the GDH sum rule correspond to 
  total helicity 2 in the $t$ channel.
  Hence their partial-wave expansions start at $J=2$, which
  renders possible a singularity at $J=1$.
 \label{Fig:Regge}}
\end{figure}

\subsubsection{Regge theory}
Reggeization of a scattering amplitude is constituted of three steps.
Firstly, the amplitude $A^\ts(s,\cos\theta_\ts)$ of the $s$-channel
process under consideration (in the present case: Compton scattering
$\gamma(q)\tN(p)\to\gamma(q')N(p')$, $s=(p+q)^2=(p'+q')^2$, cf.\ \Fig
{Regge}) is identified with the amplitude
$A^{\text{t}}(t,\cos\theta_{\text{t}})$ of the corresponding
$t$-channel process ($\tN(p)\bar\tN(-p')\to\gamma(-q)\gamma(q')$,
$t=(p-p')^2=(q'-q)^2$) by continuation to all values of $s$ and $t$.
(The scattering angles $\theta_\ts$ and $\theta_{\text{t}}$ are
specifically related to $s$ and $t$.)  If spin is involved, as in the
present case, then the relation between $s$-channel and $t$-channel
amplitudes might be intricate, but still it is trivial.  Secondly, the
$t$-channel amplitude is partial-wave expanded, thereby transforming
the dependence on the continuous variable $\cos\theta_{\text{t}}$ (or
$s$) into the discrete dependence on angular momentum $J$.  The
partial-wave amplitude $A^{\text{t}}(t,J)$ is analytically continued
to complex $J$, and the partial-wave series is written as a Cauchy
integral, whose contour must enclose all physical values of $J$, but
none of the poles and cuts that came in by analytic continuation
(Sommerfeld-Watson transform).  The integration contour is deformed to
a vertical line at $J=-\frac12$, closed by a half circle at infinity,
keeping trace of every pole and cut that is encountered.  At large
anergies, both the line at $J=-\frac12$ and the infinitely distant
half circle do not contribute \cite {Collins68}.  The scattering
amplitude is then expressed as the sum of the residues times some
explicit function of the position of all poles and cuts.  From this
expression, the large-$s$ behavior of the scattering amplitude is
inferred in the third step of the procedure.

\subsubsection{Spin}
Most surveys on Regge theory mainly treat the scattering of spin-0
particles to explain the basic concepts, because explicit formulae
become very lengthy if spin is involved.  However, in the present
context spin plays a vital role, since the sort of fixed poles
considered here (arising from continuation of Legendre functions)
always occurs at $J=0$ or at $J=1$ \cite {Abarbanel68}, while
generally, Regge singularities can only occur at angular momenta that
are not included in the partial-wave series.  The possibility of a
$J=1$ fixed pole is thus intimitely related to the occurrence of
$t$-channel total helicity \emph {two}.

In the present context, one wants to obtain a statement on the high-energy
behavior of the forward Compton amplitude with equal helicities in the initial
and final states (cf.\ \Sect {DT}), i.e., $s$-channel helicity amplitudes \cite
{Abarbanel68}
\begin{subequations}
\begin{equation}
 A^\ts_{\frac12 1, \frac12 1}(s,t) \quad \text{and} \quad
 A^\ts_{\frac12 -1, \frac12 -1}(s,t)
\end{equation}
at $t=0$.  These amplitudes correspond to $t$-channel helicity amplitudes \cite
{Abarbanel68}
\begin{equation}
 A^{\text{t}}_{\frac12 -\frac12, 1 -1}(t,s) \quad \text{and} \quad
 A^{\text{t}}_{-\frac12 \frac12, 1 -1}(t,s).
\end{equation}
\end{subequations}
Observe that the final-state photons have \emph {parallel} helicity.
Hence, the total helicity in the $t$ channel is \emph {two}.
Correspondingly, the partial-wave expansion starts at $J=2$, S and P
waves are absent.\footnote {This point is somewhat concealed in \Ref
{Abarbanel68}.  It is reflected in the appearance of the symbol
$e^{J\pm}_{\lambda\mu}(\cos\theta_{\text{t}})$ with indices
$\lambda=2,\mu=1$, which involves rotation matrix elements
$d^J_{\lambda\pm\mu}(\cos\theta_{\text{t}})$ that are defined for
$-J\le\lambda,\mu\le J$ only.} After the Sommerfeld-Watson transform,
the complex-$J$ integration must not enclose the entire
positive real axis, but only the values $J=2,3,\ldots$.  In deforming
the contour as described above, it is therefore possible to encounter
a pole at $J=1$.  Indeed, it is shown in \Ref {Abarbanel68} that such
a pole might occur due to the properties of the analytic continuation
of second-kind Legendre functions.  It has to be stressed, however,
that there is no model independent statement on the residue of that
pole.  Moreover, there is unfortunately not even a reasonable model
prediction on the residue.

The fixed pole resides at $J=1$ and contributes to an even-parity
combination of $t$-channel helicity amplitudes.  Hence it can be
assigned the quantum numbers $J^P=1^+$ of the a$_1$ meson (more
precicely: a$_1$ and f$_1$ mesons).  Observe that the pole
can\emph{not} be attributed to the $t$-channel exchange of a physical
a$_1$ meson.

\subsubsection{Effect on the GDH sum rule}
A $J=1$ fixed pole results \cite {Abarbanel68} in an asymptotically
non-vanishing real part of forward Compton amplitude $f_2(\nu)$,
\begin{subequations}
\begin{equation}
 \re f_2(\infty) \neq 0,
\end{equation}
while the imaginary part still vanishes:
\begin{equation}
 \im f_2(\infty) = 0.
\end{equation}
\end{subequations}
(This is consistent with the convergence of the GDH integral.)  The
effect of the non-vanishing real constant $f_2(\infty)$, which
essentially represents the residue\footnote {In the notation of \Ref
{Abarbanel68}, the residue of the fixed pole is given by
$$R_\tb(t{=}0)=\frac{4\pi}M\,f_2(\infty).$$} of the fixed pole, can be
seen in two equivalent ways.  Firstly, one may go back to the Cauchy
integration formula \eq {Cauchy} applied to function $f_2(\nu)$,
\begin{equation} \label{Cauchy-f2}
 f_2(\nu) = \frac1{2\pi{i}} \oint_{\mathcal{C}}\!
  \frac{\td\nu'}{\nu'-\nu}\, f_2(\nu'),
\end{equation}
where the contour $\mathcal{C}$ shall be the one depicted in \Fig
{Cauchy2}.  Now, since function $f_2(\nu)$ does not fall off at
$\nu\to\infty$, there is a contribution to integral \eq {Cauchy-f2}
coming from the half circles at infinite photon energies:
\begin{subequations}
\begin{equation}
 \frac1{2\pi{i}}
  \int_{\begin{subarray}{l}\text{circle}\\ \text{at }\infty\end{subarray}}\!
  \frac{\td\nu'}{\nu'-\nu}\, f_2(\nu')
 = \frac1{2\pi{i}} \oint\! \frac{\td\nu'}{\nu'-\nu}\, f_2(\infty)
 = f_2(\infty),
\end{equation}
where the last equality follows simply by application of Cauchy's
theorem to a constant function.
As demonstrated in \Sect {UDR}, the part of the integration contour that
encloses the cuts reads
\begin{equation}
 \frac1{2\pi{i}} \int_{\text{cuts}}\! \frac{\td\nu'}{\nu'-\nu}\, f_2(\nu')
 = \frac2\pi \int_{\nu_0}^{\infty}\!
  \frac{\td\nu'\,\nu'}{{\nu'}^2-\nu^2}\, \im f_2(\nu').
\end{equation}
\end{subequations}
Thus,  letting $\nu\to0$ and employing optical theorem \eq {f2-opt} and
low-energy theorem \eq {LET.LGG}, one has
\begin{equation} \label{GDH+f2(infty)}
 \boxed{ -\frac{2\pi^2\alpha\kappa^2}{M^2} =
  \int_{\nu_0}^\infty\! \frac{\td\nu}\nu
  \bigl(\sigma_{1/2}(\nu) - \sigma_{3/2}(\nu)\bigr) + 4\pi^2 f_2(\infty)}
\end{equation}

An equivalent way to see that this modification of the GDH sum rule is
brought about by a non-vanishing $f_2(\infty)$ is to write down a
\emph {subtracted} dispersion relation, i.e., one considers the
function
\begin{equation} \label{subtracted-f2}
 \frac {f_2(\nu)-f_2(0)}{\nu^2}
\end{equation}
instead of $f_2(\nu)$ itself.  Due to the denominator $\nu^2$,
function \eq {subtracted-f2} has an improved high-energy behavior. In
particular, it vanishes at $\nu\to\infty$.  (Observe that it does \emph
{not} exhibit an artificial pole at $\nu=0$, because $f_2(\nu)$ is
even under crossing.)  The dispersion relation for function \eq
{subtracted-f2} reads
\begin{equation} \label{SDR-f2}
 \frac {f_2(\nu)-f_2(0)}{\nu^2} = \frac2\pi
 \!\int_{\nu_0}^\infty\! \frac {\td\nu'\,\nu'}{\nu^{\prime2}-\nu^2}\,
 \frac{\im f_2(\nu')-\im f_2(0)}{\nu^{\prime2}},
\end{equation}
where, as before, the point $\nu$ must not lie on one of the cuts.  Taking into
account that $\im f_2(0)=0$, this becomes
\begin{equation} \label{SDR}
 f_2(0) = \frac2\pi
 \!\int_{\nu_0}^\infty\! \frac {\td\nu'(-\nu^2)}{\nu'(\nu^{\prime2}-\nu^2)}\,
 \im f_2(\nu') + f_2(\nu).
\end{equation}
Letting $\nu$ approach \emph {infinity} in \Eq {SDR}, one gets
\begin{equation} \label{SDR-infty}
 f_2(0) = \frac2\pi
 \!\int_{\nu_0}^\infty\! \frac {\td\nu'}{\nu'}\, \im f_2(\nu') + f_2(\infty),
\end{equation}
which gives \Eq {GDH+f2(infty)} upon application of optical theorem
and low-energy theorem.  Note that the subtraction here was \emph
{not} enforced by a divergent integral, in which case it would have
been impossible to drag the limit $\nu\to\infty$ inside the $\nu'$
integral of \Eq {SDR}.  This type of subtraction is conveniently called
a ``subtraction at infinity''.

\subsubsection{High-energy polarized forward Compton scattering}
Considering the modified GDH sum rule \eq {GDH+f2(infty)}, the
question might arise, which Compton-scattering observable the magnitude
of quantity $f_2(\infty)$ is related to.  In view of \Sect {forwComp},
it is clear that the answer will be some polarized forward
differential Compton cross section at large photon energy.
Nevertheless, one has to go a little further than \Sect {forwComp},
where only helicity eigenstates of photon and nucleon were considered,
and initial and final particles were taken to be in the same spin
state.  Analogously to \Eq {T-dec}, I define the Coulomb-gauge forward
Compton amplitude
\begin{equation}  \label{T-dec'}
 T(\nu) = \chi^{\prime\dagger}\bigl[f_1(\nu) \,\beps^{\prime*}\ndot\beps
  +i\nu f_2(\nu) \,\bsigma\ndot(\beps^{\prime*}\ntimes\beps)\bigr]\chi,
\end{equation}
where the polarization vectors $\beps,\beps'$ of incident and scattered
photon, as well as Pauli spinors $\chi,\chi'$ of initial and final
nucleon are allowed to differ.  As in \Sect {forwComp}, I take the
photon to be travelling along $\be_3$.  I use the term ``nucleon
helicity'' synonymously with center-of-mass nucleon helicity, although
all quantities are considered in the lab frame.  Positive nucleon
helicity corresponds to polarization towards the incident photon beam.  Of
course, if all particles are in a helicity eigenstate, then owing to
the forward direction, the amplitude \eq {T-dec'} vanishes unless
$\beps'=\beps$ and $\chi'=\chi$.  According to \Eqs {T1/2-3/2}, one
obtaines the two amplitudes $T_{1/2,3/2}(\nu)=f_1(\nu)\pm\nu f_2(\nu)$
for antiparallel and parallel helicity, respectively.  To project out
$f_2(\nu)$ alone, one has to either polarize the nucleon transversely
(instead of longitudinally) or polarize the photon linearly (instead
of circularly).  Adopting the latter alternative, I assume the
incident photon to be linearly polarized in the direction of the first
coordinate axis, i.e., $\beps=\be_1$.  The nucleon shall have helicity
$+\frac12$.  Then, \Eq {T-dec'} yields
\begin{equation}
 T_{\text{cons}}(\nu) = f_1(\nu)
\end{equation}
if the photon polarization is conserved ($\beps'=\be_1$), and
\begin{equation}
 T_{\text{flip}}(\nu) = i\nu f_2(\nu)
\end{equation}
if it is flipped by 90 degrees ($\beps'=\be_2$).  (It vanishes if the
final-state nucleon has negative helicity.)  Falling back on \Eq
{sigmaComp}, I conclude that at large photon energies, the forward
differential cross section for 90-degrees spin-flip scattering of
a linearly polarized photon off a longitudinally polarized nucleon,
\begin{equation} \label{sigma-flip}
 \left.\frac{\td\sigma_{\text{flip}}}{\td\Omega^\tlab}\right|_{\text{forward}}
 = \nu^2 |f_2(\nu)|^2,
\end{equation}
grows quadratically with energy if $f_2(\infty)\neq0$.  Practically,
however, this is certainly not measureable directly.  Moreover, it is perhaps
worth noting that no lower bound on the \emph {total} Compton cross
section can be read off from \Eq {sigma-flip}, because there is no
statement on the scattering-angle dependence.

Very recently, L'vov, Scopetta, Drechsel, and Scherer \cite {Lvov98}
claimed that the forward Compton amplitudes $f_{1,2}(\nu)$ can be
determined in a certain energy range by measuring small-angle
photoproduction of electron-positron pairs.

\subsubsection{Discussion}
As noted before, in case of the nucleon no sensible model prediction
for the magnitude of $f_2(\infty)$, i.e., the residue of the $J=1$
fixed pole, can be found in the literature.  This is the state of the
art three decades after Abarbanel and Goldberger suggested the fixed
pole.  It can be traced back to the fact that all models of hadrons
fail in the particular kinematical domain under consideration,
namely $Q^2=0$ and $\nu\to\infty$ (high-energy real Compton
scattering).  Perturbative QCD, i.e., the parton model and its
refinements, require high spacelike photon virtualities $Q^2$, while
effective models work at low or intermediate energies only.

Nevertheless, as illustrated in the previous chapter, the GDH sum rule
can be studied within perturbative models, if the nucleon is replaced
by a fundamental fermion like the electron (\Sects {QED} and \sect
{WSM}), or if one employs a toy model of the nucleon interacting
perturbatively with some fundamental pseudoscalar meson, called
a pion for simplicity  (\Sect {GM}).  It has to be stressed that
all of these models yield the unmodified GDH sum rule, i.e.,
$f_2(\infty)=0$.

I emphasize that in principle, simple perturbative models are by all
means capable of yielding fixed poles in electro-weak amplitudes.
Consider, for instance, the amplitude for the scattering of isovector
photons off pions.  Naive current algebra gives rise to the
Fubini\footnote {In \nocite {Fubini66b}\Refs {Bronzan67a} and
\plaincite {Bronzan67b}, the Fubini sum rule is called
Fubini-Dashen-Gell-Mann sum rule.  Occasionally, it is also called
Fubini-Furlan-Rossetti sum rule.} sum rule \cite {Fubini66b}, which in
fact \emph {requires} a certain non-vanishing residue of a fixed pole
at $J=1$, proportional to the pion form factor [\plaincite
{Bronzan67a}\nocite {Bronzan67b}--\plaincite {Singh67}].  (This is
analogous to the fact that, as regards Compton scattering off
nucleons, naive current algebra gives rise to the unmodified GDH sum
rule, which implies a \emph {vanishing} fixed-pole residue.)  Indeed,
the correct residue is obtained within a sigma model \cite
{Bronzan67a,Bronzan67b}.

Of all arguments against the presence of the fixed pole that I can
offer you, here is my personal favourite (although I have to admit
that there is some handwaving in it): Inspired by the successful
tests of the GDH sum rule within QED and the Weinberg-Salam model,
which are \emph {per se} of higher order in $\alpha$, it is likely
that the nucleon's \emph {full} forward Compton amplitude $\bar f_2(\nu)$,
involving all orders of $\alpha$,
\begin{equation} \label{bar-f2}
 \bar f_2(\nu) = \sum_{n=1}^\infty \alpha^n f_2^{(n)}(\nu),
\end{equation}
is well-defined, which I want to assume in the following.  Observe
that the first term in \Eq {bar-f2}, $\alpha f_2^{(1)}(\nu)$, is the
quantity otherwise called $f_2(\nu)$.  Since all orders of $\alpha$
are considered, bilinear unitarity is restored.  As noted above,
this rules out the presence of a fixed pole.  Hence, function \eq
{bar-f2} obeys an unsubtracted dispersion relation
\begin{equation} \label{bar-DR}
 \bar f_2(0) = \frac 2\pi \!\int_0^\infty\! \frac {\td\nu}\nu\,
 \im \bar f_2(\nu),
\end{equation}
i.e., $\bar f_2(\infty)=0$.  Now, insert expansion \eq {bar-f2}
into \Eq {bar-DR}:
\begin{equation} \label{exp-DR}
 \sum_{n=1}^\infty \alpha^n f_2^{(n)}(0)
 = \frac 2\pi \sum_{n=1}^\infty \alpha^n \!\int_0^\infty\! \frac {\td\nu}\nu\,
 \im f_2^{(n)}(\nu).
\end{equation}
Since, by definition, all of the functions $f_2^{(n)}(\nu)$ are independent
of $\alpha$, I can seperately equate the coefficients on both sides of \Eq
{exp-DR}.  In particular,
\begin{equation}
 f_2^{(1)}(0)
 = \frac 2\pi \!\int_0^\infty\! \frac {\td\nu}\nu\, \im f_2^{(1)}(\nu),
\end{equation}
and that's it!  

Yet, I would like to re-formulate this point somewhat differently.
Assume that \emph {there are} orders $n$ that need a subtraction at
infinity:
\begin{equation}
 f_2^{(n)}(0) = \frac 2\pi \!\int_0^\infty\! \frac {\td\nu}\nu\,
 \im f_2^{(n)}(\nu) + f_2^{(n)}(\infty).
\end{equation}
Introducing these relations into \Eq {exp-DR}, one obtains
\begin{equation} \label{sum=0}
 \sum_{n=1}^\infty \alpha^n f_2^{(n)}(\infty) = 0.
\end{equation}
If not all coefficients $f_2^{(n)}(\infty)$ were zero, then this
relation (in fact, there is one such relation for each nucleon) would
implicitly determine one of the fundamental parameters of the standard
model (the unit charge) from some of the others, since functions
$f_2^{(n)}(\nu)$ embody ``pure QCD''.  This, however, would be crazy.

\section{Anomalous charge-density commutator}
\label{Sect:anomcomm}

In view of the current-algebra derivation of the GDH sum rule
presented in \Sect {ETCA}, a modification may arise if the charge-density
commutator does not have the naive form
\begin{equation} \label{naive-comm2}
 \etcomm {J^0(x)} {J^0(y)} = 0.
\end{equation}
Any deviation from this form I call ``non-naive'' commutator.
However, in this thesis I pay particular attention to modifications of
\eq {naive-comm2} stemming from the chiral anomaly, which I call
``anomalous commutators''.

This subject already has a quite long history: Seminal work was done
in 1969 by Jackiw and Johnson \cite {Jackiw69} (see also Jackiw \cite
{Jackiw85}) and by Adler and Boulware \cite {Adler69b}, who were
concerned with spinor electrodynamics and the decay of the neutral
pion, and who considered the commutator of a vector current with an
axial-vector current.  Charge-density commutators within the
non-linear sigma model were considered by Kramer, Palmer, and Meetz
\cite {Kramer86a,Kramer86b}.  Fadeev \cite {Fadeev84} investigated a
chiral gauge theory, i.e., a theory of chiral fermions interacting by
means of quantized gauge fields with quantum numbers that couple to
both the fermionic current \emph {and} its commutator.  These studies
were elaborated by Jo [\plaincite {Jo85a}\nocite {Jo85b}--\plaincite
{Jo87}], who employed a Bjorken-Johnson-Low (BJL) technique \cite
{Bjorken66,Johnson66}, and by Ghosh and Banerjee \cite
{Ghosh88,Ghosh89} using a point-splitting method.

Note that in the present instance, we are concerned with the
electromagnetic current, whose commutator -- as can be read off, e.g.,
from \Eq {naive-curr-comm1} -- involves the quantum numbers of an
\emph {axial}-vector current, which is \emph {not} coupled to any
gauge field, at least in the standard model.  (Weak bosons are out of
the question with regard to the hadronic GDH sum rule.)  Moreover,
since all amplitudes are considered to lowest non-trivial order in
$\alpha$, the photon is presently regarded as an external field, i.e.,
it is not quantized.  Hence the above-mentioned methods do not seem to
be applicable to the problem under consideration.  To overcome this
deficiency, Chang and Liang \cite {Chang91,Chang92} proposed an
anomalous charge-density algebra that had no apparent regard to any
gauge field.  (As you will see below, this is not exactly true,
because a field newly introduced ad hoc in \Refs {Chang91} and
\plaincite{Chang92} turns out to be just the gauge field of chiral
transformations of quarks.)  The proposed charge-density algebra was
applied to the GDH sum rule by Chang, Liang, and Workman \cite
{Chang94a}.  This work has recently caused some attention, since the
result sounds quite promising.
\Ref {Chang94a} \emph {appeared} to provide a theoretical solution
to the puzzle initiated by the multipole analyses mentioned in the
introduction.  However, I am able to show \cite {Pantfoerder98} that
the proposed modification is actually a delusion!  This is because
anomalous commutators modify the \emph {finite-momentum}
GDH sum rule only, while the infinite-momentum limit \emph {takes back}
the modification!

Explaining this fact is the purpose of the present and the following
sections.  First, I try to shed some light on questions that might
arise about non-naive charge-density commutators.  What renders them
possible and what are they good for?  Then, I want to motivate the use
of anomalous commutators by presenting their ``classical''
application, namely $\pi^0$ decay.  In the two following short
subsections, I explain the difference between physical scattering
amplitudes and T-product amplitudes, as well as the BJL limit.  Next,
I illustrate the work of Chang and Liang \cite {Chang91,Chang92} and
show what modification of the finite-momentum GDH sum rule arises.
\Sect {IML} will be devoted to a detailed discussion of the
infinite-momentum limit.  Finally, in \Sect {PRP}, I compute the
anomalous commutator within the Weinberg-Salam model, where the
legitimacy of the infinite-momentum limit can explicitly be checked.

\subsection{Why non-naive charge-density commutators?}
\label{Sect:why}

For the moment, let us consider a single charged Dirac field $\psi(x)$
with current-density four-vector
\begin{equation} \label{curr-def2}
 J^\mu(x) = \bar\psi(x)\gamma^\mu\psi(x),
\end{equation}
interacting with a gauge field $A^\mu(x)$.

\subsubsection{Gauge invariance and commutators}
The local gauge transformation of fields $\psi(x)$ and $A^\mu(x)$
reads
\begin{equation} \begin{split} \label{gauge-trans}
 eA^\mu(x) &\mapsto e{A'}^\mu(x) = eA^\mu(x) + \partial^\mu\chi(x), \\
 \psi(x) &\mapsto \psi'(x) = e^{-i\chi(x)}\psi(x),
\end{split} \end{equation}
where $\chi(x)$ is an arbitrary function.  The part of the
transformation acting on the matter fields $\psi(x)$, $\bar\psi(x)$ is
generated by the charge-density operator
$J^0(x)=\bar\psi(x)\gamma^0\psi(x)=\psi^\dagger(x)\psi(x)$.  This
means that the action of an infinitesimal transformation
$\delta\psi(x)=-i\delta\chi(x)\psi(x)$ on any local operator
$O_\tm(x)$ \emph {containing only matter fields} is given by
\begin{equation} \label{deltaOm}
 \delta O_\tm(x)
 = -i \int\!\td^3y \etcomm {J^0(y)} {O_\tm(x)} \delta\chi(y),
\end{equation}
where subscript ``et'' denotes equal times, i.e., $x^0=y^0$.
Local gauge invariance of $O_\tm(x)$ means that this vanishes
for every function $\delta\chi(x)$.  Consequently,
\begin{equation}
 [J^0(y),O_\tm(x)]_{\text{et}} = 0.
\end{equation}

Now, as a matter of course, the current density itself is a gauge invariant
observable.  Thus, naively,
\begin{equation}  \label{naivej0jmu}
 \etcomm {J^0(x)} {J^\mu(y)} = 0.
\end{equation}

\subsubsection{Conventional c-number Schwinger term}
On the other hand, as Julian Schwinger showed \cite {Schwinger59} in
the 50's, for the spatial current density ($\mu=1,2,3$) the above
equal-times commutator is not only \emph {allowed} to be
non-vanishing, it in fact
\emph {has} to be non-vanishing.  This can be understood (see also Ref.\
\plaincite [Sect.\ 11-3-1] {Itzykson80}) by considering the
divergence of the space part of the naive commutator \eq {naivej0jmu},
\begin{equation}
 \etcomm {J^0(x)} {\bnabla\ndot\bJ(y)} = 0,
\end{equation}
employing current conservation,
\begin{equation}
 \etcomm {J^0(x)} {\partial_0 J^0(y)} = 0,
\end{equation}
taking the vacuum expectation value, and inserting a sum over a complete set
of energy eigenstates, $1=\sum_n|n\rangle\langle n|$, which gives in
the limit $\by\to\bx$
\begin{equation}
 \sum_n E_n \bigl|\langle n|J^0(x)|0\rangle\bigr|^2 = 0.
\end{equation}
From the positivity of energy this would imply that $\langle
n|J^0(x)|0\rangle=0$ for all excited states $|n\rangle$.  Thus, the
vacuum would be an eigenvector of the charge density, which is
``excluded physically for the example of the charge flux vector''
\cite {Schwinger59}. The simplest non-naive form of the commutator reads
\begin{equation}
 \etcomm {J^0(x)} {\bJ(y)}
  = i K \bnabla\dirac3(\bx - \by),
\end{equation}
with some (c-number) constant $K$.  A specfic instance of a c-number
Schwinger term, motivated by vector-meson dominance, will be presented in \Sect
{CLW} of this thesis.

So, what about gauge invariance then?

\subsubsection{Point splitting}
As Schwinger pointed out \cite {Schwinger59}, the current \eq
{curr-def2}, given as a fermion bilinear with \emph {equal} spacetime
coordinates within $\bar\psi$ and $\psi$, is in fact ill-defined.  To
make it a well-defined quantity, the spacetime points have to be \emph
{separated} by a small spacelike distance $\eps$, letting $\eps\to0$
eventually.  Now, the quantity
$\bar\psi(x-\eps/2)\gamma^\mu\psi(x+\eps/2)$ is indeed no longer
locally gauge invariant on account of the different action of
transformation \eq {gauge-trans} at points $x+\eps/2$ and $x-\eps/2$.
To restore gauge invariance of the so-defined current, a factor
involving the gauge field $A^\mu(x)$ has to be supplemented,
\begin{equation} \label{point-splitting}
 J^\mu(x;\eps) := \bar\psi(x-\eps/2)\gamma^\mu\psi(x+\eps/2)
  (1 + ie\, \beps\ndot\bA(x)).
\end{equation}
The crucial thing to be noted is that by means of this so-called
point-splitting method, the current becomes gauge-field dependent!
Then, however, \Eq {deltaOm} must be generalized.
Considering a local operator $O(x)$ that containes both matter and
gauge fields,
the action of an infinitesimal gauge transformation
$\delta A^\mu(x)=\partial^\mu\delta\chi(x)$,
$\delta\psi(x)=-i\delta\chi(x)\psi(x)$ reads
\begin{equation} \label{deltaO}
 \delta O(x)
 = -i \int\!\td^3y \etcomm {G(y)} {O(x)} \delta\chi(y),
\end{equation}
where
\begin{equation}
 G(x) = J^0(x) + \frac ie \nabla^k\frac\delta{\delta A^k(x)}
\end{equation}
represents the \emph {full} generator of gauge transformations, whose
second component acts on the gauge field.

Consequently, gauge invariance manifests itself as
\begin{equation}
 \etcomm {G(x)} {J^\mu(y)} = 0,
\end{equation}
while a non-vanishing commutator of $J^\mu(y)$ with the charge
density, which acts on the matter field only, is by no means
prohibited.

\subsubsection{Caution!}
As a conclusion, one ought to keep in mind that regarding the matter current
density as a purely matter-field dependent quantity, which would have
to commute with the charge-density operator (being the matter part of
the generator of the gauge transformation), is too naive an idea
to yield the correct commutation relations.  Some caution is in order!

\subsection{Decay of the neutral pion}

To motivate the use of anomalous commutators in particle physics, I
now sketch the current-algebra calculation of the
$\pi^0\to\gamma\gamma$ decay width as given by Jackiw and Johnson
in 1969 \cite {Jackiw69} (see also Jackiw \cite {Jackiw85}).

The neutral pion is observed to decay into two photons with a width of
roughly 8~eV (branching ratio 98.8\%).  As you will see below, this
rather large value is naively unexpected.  Indeed, among the few
experimentally observed manifestations of the chiral anomaly \cite
{Kramer84}, $\pi^0$ decay is the clearest one.\footnote {The so-called
proton spin crisis is nowadays widely accepted to be a manifestation
of the U(1)$_\tA$ anomaly, which has much in common with the chiral
anomaly but is not the same!} Its width measures the $S$-matrix
element
\begin{equation} \label{pi0-amp}
 S_\tfi = \me {\gamma(k_1,\eps_1)\, \gamma(k_2,\eps_2)} S {\pi^0(p)},
\end{equation}
where $p=k_1+k_2$.  I study the limit $p\to0$, which is not too far
from the pion mass shell due to the smallness of $m_\pi^2$.  (This is
a rather lax phrase for an actually much deeper meaning, which has to
do with the Goldstone-boson character of the pion.)  Applying the
Lehmann-Symanzik-Zimmermann reduction formulae (see, e.g., Bjorken and
Drell \cite {Bjorken65}) to the pion and one of the photons, one gets
\begin{equation} \label{pi0-amp1}
 S_\tfi \propto m_\pi^2\, \eps_{1\mu}
 \!\int\!\td^4x\, e^{ik_1\ndot x} \!\int\!\td^4y\,
 \me {0}{\tT\, J^\mu(x)\phi(y)}{\gamma(k_2,\eps_2)},
\end{equation}
where $\phi(x)$ is the $\pi^0$ field operator and T, as usual,
represents time ordering.  I omit any constant factor except the
pion mass, which I keep trace of in order to show that the final
result will not exhibit a zero or a pole at $m_\pi^2=0$.  The pion
field can be expressed in terms of an axial-vector current by virtue of the
PCAC (``Partially Conserved Axial-vector Current'') relation
[\plaincite {Nambu60}\nocite {Gell-Mann60}--\plaincite{Chou61}]
\begin{equation} \label{PCAC}
 \partial_\mu J_5^\mu(x) = F_{\pi} m_\pi^2 \phi(x),
\end{equation}
where $F_\pi=93$~MeV is the pion decay constant.  The axial-vector current
$J_5^\mu(x)$ has the same quantum numbers than the pion.  In QCD it reads
\begin{equation}
 J_5^\mu(x) = \tfrac12 \bigl( \bar u(x)\, \gamma^\mu\gamma_5\, u(x)
 - \bar d(x)\, \gamma^\mu\gamma_5\, d(x) \bigr).
\end{equation}
By means of the step function, the time-ordered product can be written
\begin{align}
 \tT\, J^\mu(x)\, \partial_\nu J_5^\nu(y)
 & = \theta(x^0-y^0)\, J^\mu(x)\, \partial_\nu J_5^\nu(y)
   + \theta(y^0-x^0)\, \partial_\nu J_5^\nu(y)\, J^\mu(x) \notag \\*
 & = \theta(x^0-y^0)\, \bigl[ J^\mu(x), \partial_\nu J_5^\nu(y) \bigr]
   + \partial_\nu J_5^\nu(y)\, J^\mu(x).
\end{align}
The second term does not contribute to the amplitude \eq {pi0-amp1},
since it is a total divergence with respect to $y$.  As for the first
term, integrate partially and observe that
\begin{subequations}
\begin{align}
 \frac \partial{\partial y^0}\, \theta(x^0-y^0) & = -\dirac{}(x^0-y^0), \\*
 \frac \partial{\partial y^k}\, \theta(x^0-y^0) & = 0.
\end{align}
\end{subequations}
Therewith, amplitude \eq {pi0-amp1} runs
\begin{equation} \label{pi0-amp2}
 S_\tfi \propto \eps_{1\mu}
 \!\int\!\td^4x\, e^{ik_1\ndot x} \!\int\!\td^3y\,
 \me {0}{\etcomm{J^\mu(x)}{J_5^0(y)}}{\gamma(k_2,\eps_2)}.
\end{equation}
Observe how the time-ordered product in \Eq {pi0-amp1} translates
into the equal-times commutator in \Eq {pi0-amp2} due to the
divergence appearing in the PCAC relation \eq {PCAC} and owing to
the fact that the derivative of the step function is the delta
function.  It is always for this little trick that commutators emerge in
current-algebra treatment of soft-pion processes.

\subsubsection{Sutherland-Veltman theorem}
Analogously to the naive commutator \eq {naive-0mu} of a \emph
{vector} charge density with a vector current, the commutator
$[J^\mu(x),J_5^0(y)]_\tet$ involving an ``axial charge density''
vanishes on employing canonical anticommutation relations among quark
fields.  Thus, naively the decay amplitude \eq {pi0-amp} vanishes
at $p^2=0$.  This misbelief is called \emph {Sutherland-Veltman theorem} \cite
{Sutherland67,Veltman67}.  If the decay amplitude is not dramatically
varying from $p^2=0$ to $p^2=m_\pi^2$, then, according to the
Sutherland-Veltman theorem, the decay width of the neutral pion ought
to be much smaller than 8 eV.

After Sutherland and Veltman pointed out this ``experimental failure
of PCAC'' \cite {Jackiw85}, the most widely accepted explanation was
that the decay amplitude \emph {was} in fact rapidly varying with
$p^2$, for unknown reasons.  Although this is principally not
impossible, no reason could be found for the putative rapid variation,
and it became clear that either PCAC or the naive commutator
$[J^\mu(x),J_5^0(y)]_\tet=0$ has to be sacrificed.  As it turns out,
one can indeed \emph {choose} which of these sacrifices to make \cite
{Jackiw85}.  Since I am concerned with anomalous commutators, I of
course focus on the latter one.

\subsubsection{Anomalous commutator}
Anomalous commutators like the one under consideration are
conveniently calculated by virtue of the Bjorken-Johnson-Low (BJL)
technique \cite {Bjorken66,Johnson66}.  More precicely, this technique
fetches \emph {matrix elements} -- not commutators themselves, but
after all, we are only \emph {interested} in matrix elements.  The BJL
method will be explained in \Sect {BJL}.  It reduces the commutator
problem to the calculation of a certain high-energy limit of the
according amplitude, which is obtained from the Feynman graphs
\begin{equation} \label{triangle}
 \begin{matrix}
 \begin{minipage}{3\unitlength} \begin{fmfgraph*}(3,3)
  \fmfbottom{b} \fmftop{t1,t2}
  \fmflabel{$J_5^\mu(x)$}{b} \fmflabel{$\gamma$}{t1} \fmflabel{$\gamma$}{t2}
  \fmf{boson}{t1,v1}
  \fmf{boson}{t2,v2}
  \fmf{fermion,tension=0}{v1,v2}
  \fmf{fermion}{v2,v3,v1}
  \fmf{boson,tension=2}{b,v3}
 \end{fmfgraph*} \end{minipage}
 \qquad+\qquad
 \begin{minipage}{3\unitlength} \begin{fmfgraph*}(3,3)
  \fmfbottom{b} \fmftop{t1,t2}
  \fmflabel{$J_5^\mu(x)$}{b} \fmflabel{$\gamma$}{t1} \fmflabel{$\gamma$}{t2}
  \fmf{phantom}{t1,v1}
  \fmf{phantom}{t2,v2}
  \fmf{fermion,tension=0}{v1,v2}
  \fmf{fermion}{v2,v3,v1}
  \fmf{boson,tension=2}{b,v3}
  \fmffreeze
  \fmf{boson}{t1,v2}
  \fmf{boson}{t2,v1}
 \end{fmfgraph*} \end{minipage} \\ \big.
 \end{matrix}
\end{equation}
Such fermion triangle (or higher polygon) graphs with an odd number of
axial-vector currents attached to the vertices always signalize
appearance of the anomaly.  I stress that the result does not depend
on the kind of fermions that make up the triangle.  Since the current
$J_5^\mu(x)$ couples to the third component of ``axial isospin'', the
graphs \eq {triangle} are proportional to
\begin{subequations}
\begin{equation}
 3\bigl[ \bigl(\tfrac23\bigr)^2\ntimes\tfrac12
 + \bigl(-\tfrac13\bigr)^2\ntimes\bigl(-\tfrac12\bigr) \bigr]
 = \tfrac12
\end{equation}
upon letting u and d quarks -- one of each color --
circulate within the loop, and
\begin{equation}
 \bigl(1\bigr)^2\ntimes\tfrac12 + \bigl(0\bigr)^2\ntimes\bigl(-\tfrac12\bigr)
 = \tfrac12
\end{equation} \label{loops}%
\end{subequations}
if one takes ``bare'' nucleons with axial charge $g_\tA=1$,
which has been the historical version.
The equality of expressions \eq {loops} implies independence on the kind
of fermions used.

I will comprehensively illustrate the method in \Sect {PRP}, where I
am concerned with the anomalous \emph {vector}-charge-density
commutator -- the one underlying the GDH sum rule.  Here I only cite
the result of Jackiw and Johnson \cite {Jackiw69}:
\begin{subequations}
\begin{align}
 \etcomm {J^0(x)}{J_5^0(y)} \label{pi0-comm.0}
 & = \frac {ie}{2\pi^2}\, \tilde F^{0j}(y)\, \nabla^j\dirac3(\bx-\by), \\*
 \etcomm {J^i(x)}{J_5^0(y)} \label{pi0-comm.i}
 & = \frac {ie}{4\pi^2}\, \tilde F^{ij}(x)\, \nabla^j\dirac3(\bx-\by),
\end{align} \label{pi0-comm}%
\end{subequations}
where $\tilde F^{\mu\nu}(x)$ denotes the dual electromagnetic tensor,
\begin{equation} \label{dual}
 \tilde F^{\mu\nu} = \tfrac12\, \eps^{\mu\nu\alpha\beta}
 (\partial_\alpha A_\beta - \partial_\beta A_\alpha).
\end{equation}

For later use, I re-write the charge-charge commutator \eq {pi0-comm.0}
in terms of the electromagnetic field $A^\mu(x)$, employing \Eq {dual}:
\begin{equation} \label{pi0-comm0}
 \etcomm {J^0(x)}{J_5^0(y)} = - \frac {ie}{2\pi^2}\,
 \bigl( \bnabla\ntimes\bA(x) \bigr) \ndot \bnabla \dirac3(\bx-\by).
\end{equation}

Since $A^\mu(x)$ is the field operator of the photon, it has a non-vanishing
matrix element between vacuum and the one-photon state.  Therefore,
commutators \eq {pi0-comm} lead to a non-zero decay amplitude \eq
{pi0-amp2}.  Performing the calculation thoroughly, keeping all factors,
one obtaines a $\pi^0$ decay width that is indeed amazingly close to
the experimental value.  Taking fractional-charge quarks to make
up the triangles in \Eq {triangle}, this result lends support to
the idea that quarks carry color degrees of freedom.

\subsection{Time ordering and seagulls}
\label{Sect:T+seagulls}

Up to now, I throughout treated physical amplitudes of electromagnetic
processes as matrix elements of \emph {time-ordered products} of
electromagnetic currents.  As I will demonstrate shortly, this is not
necessarily correct.  The difference between a physical amplitude and its
associated T-product amplitude, the so-called sea\-gull amplitude, did
not play any role in this thesis so far, but it is relevant to the
discussion of anomalous charge-density commutators, particularly as
regards my own investigation within the Weinberg-Salam model, which I
will present in \Sect {PRP}.  Therefore, I now give an introduction
into the origin and properties of sea\-gulls.  (In \Sect
{PRP.additionalgraphs}, you will also learn where their funny name comes
from.)

By means of the BJL technique, the matrix element of the
charge-density commutator is given by a certain high-energy limit of
the (generally non-forward) Compton amplitude of the fermion under
consideration.  Therefore I am concerned with the $S$-matrix element
\begin{equation} \label{S-Compton}
 S_\tfi = \me {q',\eps';p',\lambda'} S {q,\eps;p,\lambda}
\end{equation}
for the scattering of a photon with momentum $q$ and polarization
$\eps$ off a fermion (either nucleon or electron) with momentum $p$
and helicity $\lambda$.  In terms of the $S$-matrix element \eq
{S-Compton}, the (physical) Compton amplitude $T^{\mu\nu}$ is defined
by the relation
\begin{equation} \label{S-T}
 S_\tfi = -ie^2\, (2\pi)^4\dirac4(p+q-p'-q')\, \eps_\mu^{\prime*}\, \eps_\nu
 T^{\mu\nu}.
\end{equation}
For definiteness, the T-product Compton amplitude is now endowed with
a subscript 0:
\begin{equation}
 T_0^{\mu\nu} = i\!\int\!\td^4x\, e^{iq'\ndot x}
  \me {p',\lambda'} {\tT J^\mu(x)J^\nu(0)} {p,\lambda},
\end{equation}
To find the difference $T^{\mu\nu}-T_0^{\mu\nu}$, one has to recall
the way relation \eq {S-T} emerges by virtue of the
Lehmann-Symanzik-Zimmermann reduction technique (see, e.g., Bjorken
and Drell \cite {Bjorken65}). The crucial point is that the reduction
formula is applied \emph {twice}, since there are two external
photons.

Applying the reduction formula to the incident photon yields
\begin{equation}
 S_\tfi = i\, \eps_\nu \!\int\! \td^4y\, e^{-iq\ndot y}\,
 \me {p',\lambda';q',\eps'} {\square A^\nu(y)} {p,\lambda},
\end{equation}
where $A^\mu(x)$ is the operator of the electromagnetic field, and the
d'Alembert differential operator is
$\square=\partial_\mu\partial^\mu=\partial_0^2-\bnabla^2$.  Employing
Maxwell's equations
\begin{equation}
 \square A^\mu(x) = J^\mu(x),
\end{equation}
as well as translational invariance \eq {trans}, one has
\begin{equation}
 S_\tfi = ie\, \eps_\nu \!\int\! \td^4y\, e^{-i(p+q-p'-q')\ndot y}\,
 \me {p',\lambda';q',\eps'} {J^\nu(0)} {p,\lambda}.
\end{equation}
The $y$ integration gives rise to a four-dimensional delta function:
\begin{equation} \label{S-J}
 S_\tfi = ie\, (2\pi)^4\dirac4(p+q-p'-q')\, \eps_\nu\,
 \me {p',\lambda';q',\eps'} {J^\nu(0)} {p,\lambda}.
\end{equation}
Now the final-state photon is reduced out:
\begin{equation}
 \me {p',\lambda';q',\eps'} {J^\nu(0)} {p,\lambda} = -i\, \eps_\mu^{\prime*}
 \!\int\! \td^4x\, e^{iq'\ndot x}\,
 \me {p',\lambda'} {\square\bigl(\tT A^\mu(x) J^\nu(0)\bigr)} {p,\lambda}.
\end{equation}
By comparison with \Eqs {S-T} and \eq {S-J}, one identifies\footnote
{Occasionally, the expression $\square(\tT A^\mu(x)J^\nu(0))$ is
abbreviated as $e\,\tT^*J^\mu(x)J^\nu(0)$.  Note, however, that this
is only a notation, because strictly speaking, there is no operator T$^*$
acting  on the currents which yields such an expression.}
\begin{equation} \label{T-quabla}
 e\, T^{\mu\nu} = i \!\int\! \td^4x\, e^{iq'\ndot x}\,
 \me {p',\lambda'} {\square\bigl(\tT A^\mu(x)J^\nu(0)\bigr)} {p,\lambda}.
\end{equation}
Observe that the d'Alembert operator acts on the full time-ordered
product and not directly on the field $A^\mu(x)$.  Writing out the T
product,
\begin{subequations}
\begin{align}
 \tT A^\mu(x) J^\nu(0)
 & = \theta(x^0) A^\mu(x) J^\nu(0) + \theta(-x^0) J^\nu(0) A^\mu(x)
   \label{T-prod.1} \\
 & = \theta(x^0) \bigl[ A^\mu(x), J^\nu(0) \bigr] + J^\nu(0) A^\mu(x),
\end{align} \label {T-prod}%
\end{subequations}
one realizes that the time-derivative part of
$\square=\partial_0^2-\bnabla^2$ is caught on the step function,
\begin{align}
 \partial_0^2 \bigl( \tT A^\mu(x) J^\nu(0) \bigr)
 & = \partial_0 \Bigl( \dirac{}(x^0) \bigl[ A^\mu(x), J^\nu(0) \bigr]
     + \theta(x^0) \bigl[ \partial_0 A^\mu(x), J^\nu(0) \bigr] \Bigr)
     + J^\nu(0) \partial_0^2 A^\mu(x) \notag\\
 & = \partial_0 \Bigl( \dirac{}(x^0) \bigl[ A^\mu(x), J^\nu(0) \bigr] \Bigr)
     + \dirac{}(x^0) \bigl[ \partial_0 A^\mu(x), J^\nu(0) \bigr]
     + \tT\, \partial_0^2 A^\mu(x) J^\nu(0),
\end{align}
whereas the spatial derivatives go through,
\begin{equation}
 \bnabla^2\bigl( \tT A^\mu(x) J^\nu(0) \bigr)
 = \tT\, \bnabla^2 A^\mu(x) J^\nu(0).
\end{equation}
Hence, the d'Alembert operator does generally not commute with
the time ordering:
\begin{align} \label{quablaT}
 \square \bigl( \tT A^\mu(x) J^\nu(0) \bigr) = \tT\, \square A^\mu(x) J^\nu(0)
 + \partial_0 \Bigl( \dirac{}(x^0) \bigl[A^\mu(x),J^\nu(0)\bigr] \Bigr)
 + \dirac{}(x^0) \bigl[ \partial_0 A^\mu(x), J^\nu(0) \bigr].
\end{align}

Inserting \Eq {quablaT} into \Eq {T-quabla} and employing Maxwell's
equations again, one arrives at
\begin{equation}
 \boxed{ T^{\mu\nu} = T_0^{\mu\nu} + S^{\mu\nu} }
\end{equation}
where the sea\-gull amplitude $S^{\mu\nu}$ is given by
\begin{align}
 e\, S^{\mu\nu} = i \!\int\! \td^4x\, e^{iq'\ndot x}\, \Bigl[
 & \partial_0 \Bigl( \dirac{}(x^0)
   \me {p',\lambda'}{[A^\mu(x),J^\nu(0)]}{p,\lambda} \Bigr) \notag\\*
 & + \dirac{}(x^0) \me {p',\lambda'}{[\partial_0A^\mu(x),J^\nu(0)]}{p,\lambda}
 \Bigr].
\end{align}
After a partial integration, one can immediately carry out the $x^0$
integration:
\begin{align} \label{S-explicit}
 e\, S^{\mu\nu}
 & = q^{\prime0} \!\int\! \td^3x\, e^{-i\bq'\ndot\bx}\,
   \me {p',\lambda'}{[A^\mu(x),J^\nu(0)]_\tet}{p,\lambda} \notag\\*
 & \quad + i \!\int\! \td^3x\, e^{-i\bq'\ndot\bx}\,
   \me {p',\lambda'}{[\partial_0A^\mu(x),J^\nu(0)]_\tet}{p,\lambda}.
\end{align}

Observe that the sea\-gull \eq {S-explicit} is a first-order\footnote
{More generally, it can be shown \cite {Jackiw85} that the sea\-gull is
always a (possibly higher-order) polynomial, even if one deals with an
amplitude that does not involve conserved currents coupled to a gauge
field as in the present context.} polynomial in the energy
$q^{\prime0}$.  In all cases I am concerned with, it even turns out to
be independent of $q^{\prime0}$, i.e., the linear term in
\Eq {S-explicit} vanishes.\footnote {This circumstance gets plausible
by considering the point-splitting definition \eq {point-splitting} of
the electromagnetic current.  As illustrated in \Sect {why}, one
can generally work with canonical commutators among fundamental fields
as long as one recalls that the current must not be defined as a fermion
bilinear at coincident spacetime points.  The explicit dependence of
the point-splitting current \eq {point-splitting} on the gauge field
$A^\mu(x)$ leaves its equal-times commutator with the gauge field
itself canonically vanishing, whereas it will generally not commute
with the canonically conjugated field, i.e., the electric field, which
occurs in the constant portion of expression \eq {S-explicit}.}
Nevertheless, presently the only important thing to be kept in mind is
the fact that $S^{\mu\nu}$ is polynomial in $q^{\prime0}$.

%Observe, further,  that the constant term in \Eq {S-explicit} involves
%the operator
%\begin{equation}
% i\,\partial_0 A^\mu(x) = \bigl[ A^\mu(x), H(x^0) \bigr],
%\end{equation}
%where $H$ denotes the Hamilton operator.  Thus, albeit at equal times,
%the commutator appearing involves the full dynamics of the system under
%consideration.

\subsection{Bjorken-Johnson-Low limit}
\label{Sect:BJL}

In 1966, Bjorken \cite {Bjorken66} and Johnson and Low \cite
{Johnson66} proposed a method for computing an equal-times commutator
from the pertinent T-product amplitude.  I now want to illustrate this
techique, following Bjorken's original presentation \cite {Bjorken66}
(see also Jackiw \cite {Jackiw85}).

Let $A(x)$ and $B(x)$ be any two local operators (eventually, I will
of course take currents), and sandwich the time-ordered product
between any two hadronic or leptonic states $|\tX\rangle$ and $|\tY\rangle$:
\begin{equation} \label{T-AB}
 T := i \!\int\! \td^4x\, e^{iq'\ndot x}\,
 \me {\tY}{\tT A(x) B(0)}{\tX}.
\end{equation}
Writing out the T product occurring in \Eq {T-AB} as in \Eq {T-prod.1}
and introducing the Fourier transform \eq {theta-Fourier} of the step
function, one has
\begin{align} \label{T-AB2}
 T & = -\frac 1{2\pi} \!\int\! \td^4x\, e^{iq'\ndot x}
   \!\int_{-\infty}^{\infty}\! \frac {\td E}{E+i\eps}
   \Bigl( \me {\tY}{A(x)B(0)}{\tX}\, e^{-iEx^0}
   + \me {\tY}{B(0)A(x)}{\tX}\, e^{iEx^0} \Bigr) \notag\\
 & = -\!\int_{-\infty}^{\infty}\! \frac {\td E}{E+i\eps} \bigl(
   \rho(q^{\prime0}-E) + \bar\rho(q^{\prime0}+E) \bigr),
\end{align}
where functions $\rho(q^{\prime0})$ and $\bar\rho(q^{\prime0})$ are defined by
\begin{subequations}
\begin{equation}
 \rho(q^{\prime0}) = \frac 1{2\pi} \!\int\! \td^4x\, e^{iq'\ndot x}\,
 \me {\tY}{A(x)B(0)}{\tX}
\end{equation}
and
\begin{equation}
 \bar\rho(q^{\prime0}) = \frac 1{2\pi} \!\int\! \td^4x\, e^{iq'\ndot x}\,
 \me {\tY}{B(0)A(x)}{\tX}
\end{equation}
\end{subequations}
respectively.  Linear substitutions of the integration variable transform \Eq
{T-AB2} into
\begin{equation}
 T = -\!\int_{-\infty}^{\infty}\! \td E' \left(
 \frac {\rho(E')}{q^{\prime0}-E'+i\eps}
 + \frac {\bar\rho(E')}{-q^{\prime0}+E'+i\eps} \right).
\end{equation}
Multiplying with $q^{\prime0}$ and letting $q^{\prime0}$ approach infinity,
this reduces to
\begin{align}
 \lim_{q^{\prime0}\to\infty} q^{\prime0}\, T
 & = - \!\int\! \td E' \bigl( \rho(E') - \bar\rho(E') \bigr) \notag\\
 & = - \frac 1{2\pi} \!\int\! \td E' \!\int\! \td^4x\,
   e^{i(E'x^0-\bq'\ndot\bx)}\, \me {\tY} {[A(x),B(0)]} {\tX}.
\end{align}
The $E'$ and $x^0$ integrations are performed straightforwardly:
\begin{align} \label{BJL-AB}
 \lim_{q^{\prime0}\to\infty} q^{\prime0}\, T
 & = - \!\int\! \td^4x\, e^{-i\bq'\ndot\bx}\, \dirac{}(x^0)\,
   \me {\tY} {[A(x),B(0)]} {\tX} \notag \\*
 & = - \!\int\! \td^3x\, e^{-i\bq'\ndot\bx}\,
   \me {\tY} {[A(x),B(0)]_\tet} {\tX}.
\end{align}
This is the BJL formula for the matrix element of an equal-times commutator.
Re-introducing current operators and one-fermion states, I conclude
\begin{equation}
 \boxed{ \int\! \td^3x\, e^{-i\bq'\ndot \bx}\,
  \me {p',\lambda'}{[J^\mu(x),J^\nu(0)]_\tet}{p,\lambda}
  = - \lim_{q^{\prime0}\to\infty} q^{\prime0}\, T_0^{\mu\nu} }
\end{equation}

\subsubsection{Identifying seagulls}
In \Sect {PRP}, I aim at a perturbative calculation of the anomalous
charge-density commutator within the Weinberg-Salam model, i.e., from
(one-loop) Feynman graphs, which do not yield the T-product amplitude
$T_0^{\mu\nu}$, but rather the physical amplitude $T^{\mu\nu}$.
However, it is clear from the above considerations that the difference
$S^{\mu\nu}=T^{\mu\nu}-T_0^{\mu\nu}$, i.e., the sea\-gull amplitude, is
readily identified by its polynomial behavior in energy
$q^{\prime0}$.  Thus, the procedure simply amounts to calculating
$T^{\mu\nu}$ and subsequently dropping all polynomials in
$q^{\prime0}$.  In other words, the above limit picks out the linear
term of the Laurent series of amplitude $T^{\mu\nu}$ in the variable
$1/q^{\prime0}$.

\subsection{Anomalous commutator of Chang and Liang}
\label{Sect:CLW}

To illustrate the work of Chang and Liang \cite {Chang91,Chang92}, I
need some preliminaries.  The electromagnetic current
\eq {QCD-curr} has to be generalized to a U(3) nonet of currents
\begin{subequations}
\begin{equation} \label{J-def.V}
 J_a^\mu(x) = \bar\psi(x)\,\gamma^\mu\frac{\lambda_a}2\,\psi(x)
\end{equation}
and a nonet of axial-vector currents
\begin{equation} \label{J-def.A}
 J_{5a}^\mu(x) = \bar\psi(x)\,\gamma^\mu\gamma_5\frac{\lambda_a}2\,\psi(x),
\end{equation} \label{J-def}%
\end{subequations}
where, as usual,
\begin{equation}
 \psi(x) = \begin{pmatrix} u(x)\\ d(x)\\s(x) \end{pmatrix},
\end{equation}
and $\lambda_{0\ldots8}$ denote Gell-Mann matrices, which are normalized by
\begin{equation} \label{lambda-norm}
 \tr(\lambda_a\lambda_b) = 2\delta_{ab}.
\end{equation}
Of particular interest are the three diagonal matrices
\begin{equation}
 \lambda_0 = \sqrt{\frac23}
 \begin{pmatrix} 1&0&0 \\ 0&1&0 \\ 0&0&1 \end{pmatrix}, \qquad
 \lambda_3 =
 \begin{pmatrix} 1&0&0 \\ 0&-1&0 \\ 0&0&0 \end{pmatrix}, \qquad
 \lambda_8 = \frac1{\sqrt3}
 \begin{pmatrix} 1&0&0 \\ 0&1&0 \\ 0&0&-2 \end{pmatrix}.
\end{equation}
You have met a certain combination of $\lambda_3$ and $\lambda_8$
before:
\begin{equation}
 \frac {\lambda_3}2 + \frac 1{\sqrt3} \frac {\lambda_8}2 =
 \begin{pmatrix} \frac23&0&0 \\ 0&-\frac13&0 \\ 0&0&-\frac13 \end{pmatrix} =
 Z^{(\tq)},
\end{equation}
the quark charge matrix.  So, the familiar electromagnetic current is
given by
\begin{equation}
 J^\mu(x) = J_3^\mu(x) + \frac1{\sqrt3}\, J_8^\mu(x).
\end{equation}
Gell-Mann matrices $\lambda_{0\ldots8}$ are generators of the group
U(3) in the sense that any unitary $3\ntimes3$ matrix $U$ can be
written $U=\exp(i\alpha_a\lambda_a/2)$.  Matrix $\lambda_0$ is special
in that it has a non-vanishing trace, i.e., it generates the
baryon-number U(1) subgroup of flavor U(3).  Matrices
$\lambda_{1\ldots8}$ accordingly generate the conventional flavor
SU(3) subgroup.  Since quarks carry baryon number $1/3$, the
baryon-number density reads
\begin{equation}
 \frac13\, \psi^\dagger(x)\,\psi(x) = \sqrt{\frac23}\, J_0^0(x).
\end{equation}
Commutators and anticommutators of Gell-Mann matrices satisfy
\begin{equation}
 \left[ \frac{\lambda_a}2, \frac{\lambda_b}2 \right]
 = i f_{abc} \frac{\lambda_c}2, \qquad
 \left\{ \frac{\lambda_a}2, \frac{\lambda_b}2 \right\}
 = d_{abc} \frac{\lambda_c}2,
\end{equation}
where $f_{abc}$ is totally antisymmetric and $d_{abc}$ is totally
symmetric.  Both, $f$ and $d$ symbols are real numbers.  Since the
matrix $\lambda_0$ commutes with all others, $f_{abc}$ vanishes upon
letting any of its indices equal zero, while the $d$ symbol obeys
$d_{ab0}=\sqrt{2/3}\,\delta_{ab}$.

\subsubsection{Naive current commutators}
From definitions \eq {J-def} of vector and axial-vector currents,
naive commutation relations can be derived utilizing canonical
anticommutation relations among quark fields as in \Sect
{ETCA.comm}.  The result reflects the fact that the charge densities
$J_a^0(x)$ and $J_{5a}^0(x)$ generate local U(3)$\ntimes$U(3)
transforms:
\begin{equation} \begin{split} \label{naive-VA-comm}
 \etcomm {J_a^0(x)}{J_b^\mu(y)} = \etcomm {J_{5a}^0(x)}{J_{5b}^\mu(y)}
 & = i f_{abc}\, J_c^\mu(x)\, \dirac3(\bx-\by), \\[.5ex]
 \etcomm {J_a^0(x)}{J_{5b}^\mu(y)} = \etcomm {J_{5a}^0(x)}{J_{b}^\mu(y)}
 & = i f_{abc}\, J_{5c}^\mu(x)\, \dirac3(\bx-\by).
\end{split} \end{equation}
Observe that these relations incorporate the algebra of charge
densities ($\mu=0$) as well as charge-current commutators
($\mu=1,2,3$).  Both are free of anomalies and Schwinger terms.

\subsubsection{Vector-meson dominance}
For later use, I give the form of the charge-current algebra involving
finite c-number Schwinger terms as inspired by vector-meson dominance
(VMD, occasionally also called ``current-field identity'') \cite
{Sakurai69,DeAlfaro73}:
\begin{subequations}
\begin{align}
 \etcomm {J_{a}^0(x)}{\bJ_{b}(y)} = \etcomm {J_{5a}^0(x)}{\bJ_{5b}(y)}
 & = i f_{abc}\, \bJ_{c}\, \dirac3(\bx-\by)
   + 2iF_\pi^2\, \delta_{ab}\, \bnabla\dirac3(\bx-\by), \label{VMD-comm.ST} \\*
 \etcomm {J_{a}^0(x)}{\bJ_{5b}(y)} = \etcomm {J_{5a}^0(x)}{\bJ_{b}(y)}
 & = i f_{abc}\, \bJ_{5c}\, \dirac3(\bx-\by).
\end{align} \label{VMD-comm}%
\end{subequations}
The idea of VMD is easy to explain.  Since the currents $J_a^\mu(x)$
and $J_{5a}^\mu(x)$ have the same quantum numbers as vector and
axial-vector mesons, respectively, they can couple directly to
these mesons.  As far as the $\rho$ meson and the isovector part of
the photon are concerned, hadronic form factors are observed to be
\emph {dominated} by this process at low momentum transfer.  If vector
mesons are regarded as fundamental particles with field operators
$v_a^\mu(x)$, then this dominance can be expressed as \cite
{Gell-Mann61,Kroll67}
\begin{equation} \label{CFI}
 J_a^\mu(x) = \frac {m_\tV^2}{f_\tV}\, v_a^\mu(x)
\end{equation}
where $m_\tV$ and $f_\tV$ denote universal mass and coupling constant
of fields $v_a^\mu(x)$.  On the other hand, these fields obey standard
canonical commutation relations of massive spin-1 fields \cite
{Sakurai69,DeAlfaro73}:
\begin{equation}
 m_\tV^2 \etcomm {v_a^0(x)}{\bv_b(y)}
 = if_\tV f_{abc}\, \bv_c(x)\, \dirac3(\bx-\by)
   + i\delta_{ab}\, \bnabla\dirac3(\bx-\by),
\end{equation}
and likewise for the axial-vector mesons.  If one substitutes currents
for the meson fields by virtue of the VMD relation \eq {CFI},
\Eqs {VMD-comm} follow upon using the
Kawarabayashi-Suzuki-Riazuddin-Fayyazuddin (KSRF) relation \cite
{Kawarabayashi66d,Riazuddin66}
\begin{equation} \label{KSRF}
 \frac {m_\tV^2}{f_\tV^2} = 2F_\pi^2.
\end{equation}
By the way: The equality of vector and axial-vector Schwinger terms in
\Eq {VMD-comm.ST} can actually be proved from much weaker
assumptions than those demanded by VMD.  The proof, given by
Weinberg \cite {Weinberg67}, is recapitulated nicely and comprehensibly in
a footnote\footnote {See footnote \ref {wisdom} on p.~\pageref
{wisdom} of this thesis.} in the textbook of de Alfaro, Fubini, Furlan,
and Rossetti \cite [p.~365]{DeAlfaro73}.

\subsubsection{Triple-commutator vacuum expectation value}
With these formulae at hand, I can write down the quantity that has
actually been calculated by Chang and Liang, namely the vacuum
expectation value of a triple commutator of four charge densities,
incorporating an odd number of axial charge densities, e.g. \cite
{Chang91,Chang92}\footnote {I remark that the pertinent Eq.\ (11) in
\Ref {Chang92} has an evident misprint since its right-hand side
is real while the left-hand side is purely imaginary.  Eq.\ (23) of
\Ref {Chang91} does not suffer from this kind of inconsistency,
but its sign is opposite to the one in my \Eq {3commVEV}, which I
obtained by re-doing the calculation of \Ref {Chang92}.  So, the
anomalous charge-density algebra I gained has opposite sign compared
to \Refs {Chang94a}\nocite {Chang91}--\plaincite {Chang92}, but there
is a compensating second sign change in the inferred modification of
the finite-momentum GDH sum rule (not the \emph {genuine} sum rule, of
course).}
\begin{align} \label{3commVEV}
 & \me 0 {\etcomm{J_{5a}^0(x)}{\comm{J_b^0(y)}{\comm{J_c^0(w)}{J_d^0(z)}}}} 0
   \notag \\*
 & \qquad = \frac {iN_\tc}{24\pi^2}\, f_{abe}\, d_{cde}\,
   \bigl( \bnabla\dirac3(\bx-\by)\ntimes\bnabla\dirac3(\by-\bz) \bigr)
   \ndot \bnabla\dirac3(\bz-\bw),
\end{align}
where $N_\tc=3$ is the number of colors. All time components are
equal: $x^0=y^0=z^0=w^0$. Expression \eq {3commVEV} is the same if
the axial charge density is at any other position or if there are
three axial charge densities.  It vanishes if their number is zero,
two, or four.  Observe that the symmetry of the $d$ symbol does not
contradict the antisymmetry of the innermost commutator, since
$\bnabla\dirac3(\bz-\bw)$ changes sign upon interchanging $\bw$ and
$\bz$.

The anomalous charge-density commutator is then \emph {proposed} as
the simplest one reproducing the triple-commutator \eq {3commVEV}
upon iterative application, together with the standard conventional
Schwinger term \eq{VMD-comm} inferred from VMD.  I postpone the
presentation of these ideas and first comment on the calculation of \Eq
{3commVEV}.

Two different methods have been employed to yield the
triple-commutator vacuum expectation value \eq {3commVEV}: firstly
an anomaly-functional method \cite {Chang91}, which utilizes a triple
Bjorken-Johnson-Low limit to calculate the triple commutator from a
four-point function; secondly a rather direct computation \cite
{Chang92}, performing all possible Wick contractions among properly
normal-ordered products of currents.  The anomaly is caused
by a subtle type of divergence, which can be remedied by a shift of
the integration variable.  Both calculations start from a
theory of massless, non-interacting quarks.  The vanishing of the
quark mass has the effect that right- and left-handed quark fields
$\tfrac12(1\pm\gamma_5)\psi(x)$ are entirely independent, so that
right- and left-handed charge densities $\tfrac12(J^0(x)\pm J_5^0(x))$
commute.  The fact that quarks are taken to be non-interacting has the
effect that the lowest-order result is exact.  Here, ``lowest order''
means the following: In a way, either of these approaches is
equivalent to calculating box diagrams
\begin{equation} \label{box}
 \fmfstraight
 \graph(3,3){
  \fmfleft{l1,l2} \fmfright{r1,r2}
  \fmf{photon}{l1,v1}  \fmf{photon}{l2,v2}
  \fmf{photon}{r1,v4}  \fmf{photon}{r2,v3}
  \fmf{fermion,tension=.5}{v1,v2,v3,v4,v1}
 }
 \qquad + \text{permutations},
 \fmfcurved
\end{equation}
with an odd number of axial-vector currents attached.  If quarks
interact, say, via gluon exchange, then there is an infinite number of
gluon-radiative corrections to these diagrams.

Without any mass scale in the theory, it is clear from dimensional
considerations why a product of a minimum number of \emph {four}
charge densities is needed to gain a non-vanishing vacuum expectation
value \eq {3commVEV}:  In 3+1-dimensional spacetime, charge density
has dimension three, i.e., GeV$^3$.  Thus, the innermost commutator
in \Eq {3commVEV} has dimension six, and every further charge
density adds three.  On the other hand, there is clearly one delta
function of dimension three per commutator, so three gradients
altogether have to be supplied to balance dimensions.  For a single or
double commutator, no scalar expression incorporating three gradients
can be found, but for a triple commutator, there is the exterior
product displayed in
\Eq {3commVEV}.  Moreover, this expression is unique up to a
constant factor.  Note that in 1+1-dimensional spacetime, matters
stand differently: Since currents and delta functions have
dimension one, a \emph {single} charge-density commutator can exhibit
a non-vanishing vacuum expectation value without introducing any mass
scale, namely a multiple of the derivative of $\delta(x^1-y^1)$.  This
commutator is known as \emph {Kac-Moody algebra}.

\subsubsection{The number of colors}
Considering the box diagrams \eq {box}, it is also clear why the number
of colors occurs in the result \eq{3commVEV}, since of course one
quark of each color can circulate in the box.

\subsubsection{Anomaly cancellation}
In \Refs {Chang91} and \plaincite {Chang92}, quarks are the only
fermionic degrees of freedom.  If leptons were added consistently,
i.e., if the currents \eq {J-def} were complemented for leptonic pieces
so that leptons circulated in the box diagrams \eq {box}, too, then the
famous mechanism of anomaly cancellation would cause the triple commutator
\eq {3commVEV} to vanish.  As you shall see below, hadrons (more
precicely: spin-1 mesons) will be coupled to the external currents
when the whole thing is applied to the GDH sum rule.  This is the only
argument for the omission of leptons.  That is to say, in view of
anomaly cancellation it is clear from the beginning that additional
particles other than quarks will have to be introduced and that these
particles will have to be hadrons.

\subsubsection{A first-guess charge-density algebra}
Now comes the most critical step of the procedure: An algebra of
charge densities is \emph {proposed} which reproduces the
triple-commutator vacuum expectation value \eq {3commVEV} in an
iterative computation.  That is to say: one simply \emph {guesses} a
charge-density commutator that -- upon commuting with two further
charge densities and taking the vacuum expectation value -- fetches
expression \eq {3commVEV}.  To this end, two new nonets of fields
$b_{\tV a}^\mu(x)$, $b_{\tA a}^\mu(x)$ have to be introduced ad hoc.
The simplest such guess reads [\plaincite {Chang94a}\nocite
{Chang91}--\plaincite{Chang92}]\footnote {Note that I have corrected
for sign errors, and that I have taken a factor of $N_\tc$ out of the
definition of the $b$ fields, so that the Schwinger term in \Eq
{Jb-comm.ST} has the simplest possible form.  This normalization
appeares most natural to me.}
\begin{subequations}
\begin{align}
 & \etcomm {J_a^0(x)}{J_b^0(y)} = \etcomm {J_{5a}^0(x)}{J_{5b}^0(y)} =
   \notag \\*
 & \qquad\quad i f_{abc}\, J_c^0(x)\, \dirac3(\bx-\by) -
   \frac{iN_\tc}{48\pi^2}\, d_{abc} \bigl( \bnabla\ntimes\bb_{\tA c}(x) \bigr)
   \ndot \bnabla\dirac3(\bx-\by), \label{Jb-comm.JJ} \\
 & \etcomm {J_a^0(x)}{J_{5b}^0(y)} = \etcomm {J_{a}^0(x)}{J_{5b}^0(y)} =
   \notag \\*
 & \qquad\quad i f_{abc}\, J_{5c}^0(x)\, \dirac3(\bx-\by) -
   \frac{iN_\tc}{48\pi^2}\, d_{abc} \bigl( \bnabla\ntimes\bb_{\tV c}(x) \bigr)
     \ndot \bnabla\dirac3(\bx-\by), \label{Jb-comm.JJ5}
\end{align}
Furthermore, commutators between charge densities and the newly
introduced $b$ fields have to be supplied:
\begin{align}
 \etcomm {J_{a}^0(x)}{\bb_{\tV b}(y)} = \etcomm {J_{5a}^0(x)}{\bb_{\tA b}(y)}
 & = i f_{abc}\, \bb_{\tV c}\, \dirac3(\bx-\by)
   + 2i\, \delta_{ab}\, \bnabla\dirac3(\bx-\by), \label{Jb-comm.ST} \\*
 \etcomm {J_{a}^0(x)}{\bb_{\tA b}(y)} = \etcomm {J_{5a}^0(x)}{\bb_{\tV b}(y)}
 & = i f_{abc}\, \bb_{\tA c}\, \dirac3(\bx-\by).
 \label{Jb-comm.can}
\end{align} \label{Jb-comm}%
\end{subequations}
\Eq {3commVEV} can indeed be derived from \Eqs {Jb-comm}:
one needs \eq {Jb-comm.JJ} for the innermost commutator, next the
canonical form \eq {Jb-comm.can}, and finally the c-number
Schwinger term \eq {Jb-comm.ST} -- plus a bit of vector algebra, of course.

Considering that algebra \eq {Jb-comm} is claimed to be the
simplest possible one, the reader may find that it looks pretty
complex.  So, let me try to motivate its form.  From causality, each
of the given commutators must vanish unless $\bx=\by$, i.e., it must
be a finite sum of derivatives of $\dirac3(\bx-\by)$. (In momentum
space, this corresponds to a polynomial in the according
three-momentum.)  A \emph {single} derivative is of course the
simplest ansatz.  Thus, the additional term in the charge-density
commutator $[J_a^0(x),J_b^0(y)]_\tet$ is proposed as
$i\bB_{ab}(x)\ndot\bnabla\dirac3(\bx-\by)$.  Due to antisymmetry of
the commutator, this expression has to change sign upon simultaneously
interchanging $a$ with $b$ and $\bx$ with $\by$, i.e.,
$\bB_{ab}(x)\ndot\bnabla\dirac3(\bx-\by)\overset{!}=
-\bB_{ba}(y)\ndot\bnabla\dirac3(\by-\bx)
=\bB_{ba}(x)\ndot\bnabla\dirac3(\bx-\by)
+\bnabla\ndot\bB_{ba}(x)\,\dirac3(\bx-\by)$, which is most easily
achieved by choosing $\bB_{ab}(x)$ symmetric in its indices and source
free.  Recalling that we intend to reproduce the triple-commutator
\eq {3commVEV}, the $d$ symbol suggests itself:
$\bB_{ab}(x)=d_{abc}\bB_c(x)$, where the source free field $\bB_c(x)$
can be expressed as the curl of another field.  Finally, parity
considerations reveal that the new field is an axial vector in \Eq
{Jb-comm.JJ} and a vector in \Eq {Jb-comm.JJ5}.  

As for relations \eq {Jb-comm.ST} and \eq {Jb-comm.can}, these are
just in the style of the VMD-inspired charge-current commutators \eq
{VMD-comm}.

So far, no qualitative statement has been given concerning matrix
elements of the fields $b_{\tV a}^\mu(x)$ and $b_{\tA a}^\mu(x)$,
which is of course our desire in view of the GDH sum rule.  To proceed
further, yet another guess is made in \Ref {Chang94a}, namely that
the fields are \emph {proportional} to the currents,
$b_{{\tV}a}^\mu(x)\propto J_a^\mu(x)$, $b_{{\tA}a}^\mu(x)\propto
J_{5a}^\mu(x)$, at least as far as low-lying matrix elements are
concerned.  If one now confronts \Eq {Jb-comm.ST} with \Eq
{VMD-comm.ST}, it is clear that
\begin{equation} \label{b-CLW}
 b_{\tV a}^\mu(x) = \frac 1{F_\pi^2}\, J_a^\mu(x)
 \qquad \text{and} \qquad
 b_{\tA a}^\mu(x) = \frac 1{F_\pi^2}\, J_{5a}^\mu(x),
\end{equation}
so that the anomalous charge-density commutator \eq {Jb-comm.JJ},
for electric charge-densities $J^0=J_3^0+J_8^0/\sqrt3$, becomes
\begin{equation} \label{a-comm}
 \boxed{ \etcomm {J^0(x)} {J^0(y)} = i\,
 \bigl( \bnabla \ntimes \ba(x) \bigr) \ndot \bnabla \dirac3(\bx-\by)}
\end{equation}
with\footnote {I note in passing that the combination
of axial-vector currents appearing in \Eq {a-CLW} is the same as
that occurring in the naive commutator \eq {naive-curr-comm1} of
spatial electromagnetic current densities.}
\begin{align} \label{a-CLW}
 a^\mu(x) & = - \frac {1}{8\pi^2F_\pi^2}\, 
 \bar\psi(x)\, \gamma^\mu\gamma_5\, \bigl(Z^{(\tq)}\bigr)^2\, \psi(x) \notag\\*
 & = - \frac {1}{24\pi^2F_\pi^2}
 \left( \sqrt{\tfrac83}\, J_{5\,0}^\mu(x) + J_{5\,3}^\mu(x)
     + \tfrac 1{\sqrt3}\, J_{5\,8}^\mu(x) \right).
\end{align}
Axial-vector currents $J_{5a}^\mu(x)$ are defined in \Eq {J-def.A}.

The anomalous commutator \eq {a-comm} implies a modification of the
\emph {finite-momentum} GDH sum rule \eq {FMGDH}.  If the
infinite-momentum limit were legitimate, then the (genuine) GDH sum rule
would be modified, too.  This legitimacy is \emph {assumed} in \Ref
{Chang94a}.  (More precisely, the possibility of its failure is not
taken into account.)  As explained in \Sect {ETCA}, the problem of the
infinite-momentum limit cannot be addressed within the present
context, since the charge-density commutator alone does not tell us
anything about the validity of the infinite-momentum limit.  This can
only be achieved by considering a specific field theoretic model and
calculating the forward virtual Compton amplitude in the whole
$(\nu,q^2)$ plane \cite {Pantfoerder98}.  This check is presented in
\Sect {PRP}, and it reveales that there is \emph {no modification of
the GDH sum rule due to anomalous commutators!}

\subsubsection{The Emperor's New Clothes}
\label{emperor-story}
The ``revocation'' of the modification by virtue of the
infinite-momentum limit is of course the central point of my criticism
of \Ref {Chang94a}, because it obviously invalidates its result.
For completeness, I would nonetheless like to remark on the nature of
the somewhat obscure fields $b_{\tV a}^\mu(x)$ and $b_{\tA a}^\mu(x)$.
\Eqs {b-CLW} may be misleading in that they suggest that these
fields are directly related to pions, which is not true.  It is
instructive to recall that the factors $1/F_\pi^2$ in \Eqs {b-CLW}
originated from the KSRF relation \eq {KSRF}.  Taking back the
manipulation performed by applying KSRF, and in view of the VMD
relations \eq {CFI}, \Eqs {b-CLW} run
\begin{equation} \label{b=2v}
 b_{\tV a}^\mu(x) = 2f_\tV\, v_a^\mu(x)
 \qquad \text{and} \qquad
 b_{\tA a}^\mu(x) = 2f_\tA\, a_a^\mu(x).
\end{equation}
Hence, up to a factor of 2, fields $b_{\tV a}^\mu(x)$ and $b_{\tA
a}^\mu(x)$ are simply the respective field operators $v_a^\mu(x)$ and
$a_a^\mu(x)$ of vector and axial-vector mesons (with the coupling
constants absorbed), and what was done by Chang and Liang was
effectively the calculation of the commutator anomaly of quark
currents in the presence of U(3)$\ntimes$U(3) gauge fields represented
by spin-1 mesons.  (In view of this new insight, also observe the
similarity of \Eq {Jb-comm.JJ5} with the Jackiw-Johnson commutator
\eq {pi0-comm0}.)  Thus, contrary to their claim, Chang and Liang did
not really do something new.  Their result reproduces a lowest-order
approximation to, e.g., the results of Jo \cite {Jo85a}.\footnote
{\label{CLWvsJo}More precisely: If one considers only the left-handed
current commutators of \Refs {Chang94a}\nocite {Chang91}--\plaincite
{Chang92}, or if one suitably generalizes Jo's result \cite [Eq.\
(2.20)] {Jo85a} to vector currents, then the latter turns out to be a
factor of three larger.  This inconsistency may be related to the
problem of inferring a single commutator from the triple-commutator
vacuum expectation value (see discussion ``Iterative vs.\ simultaneous
triple commutator'' below).  Also note the domain dependence of
regularized linearly divergent integrals, which is characteristic for
the anomaly.  Chang and Liang integrated a linearly divergent
three-dimensional integral \cite [Eq.\ (5)] {Chang92} over a solid
ball of radius $R$, subsequently letting $R\to\infty$.  If instead,
the domain of integration is taken to be the space between two
parallel planes with distance $H\to\infty$, then one obtains
a supplementary factor of three.}

\subsubsection{Spin-1 mesons as gauge fields}
\label{spin-1-story}
The idea that chiral transforms are gauged -- with spin-1 mesons
emerging as gauge fields and a mass term that breaks the gauge
invariance locally but preserves it globally -- is essentially the
idea of VMD and current algebra \cite {Sakurai69,DeAlfaro73}, and it
has a couple of deficiencies, at least in its primitive form as
employed in \Ref {Chang94a}.  For instance, it demands that all spin-1
mesons couple to all other hadrons with one universal coupling
constant $f_\tA=f_\tV$, and that the masses within the vector-meson
nonet ($\rho$, $\omega$, $\phi$, K$^*$) are degenerate.  The masses
within the axial-vector-meson nonet (a$_1$, f$_1$(1285), f$_1$(1420),
K$_1$) must also be degenerate, and mutually, the nonets are related
by $m_\tA^2=2m_\tV^2$ \cite {Weinberg67}.  Experimentally, all this is
quite far from truth.\footnote {In historical applications of this
idea, neither strange nor flavor-singlet mesons were involved, but
only the isovector fields $\rho$ and a$_1$, i.e., SU(2)$\ntimes$SU(2)
symmetry was considered instead of U(3)$\ntimes$U(3).  The Weinberg
relation $m_\tA^2=2m_\tV^2$ implied a mass of 1090~MeV for the a$_1$,
and indeed there was a bump in the $\rho$-$\pi$ spectrum at 1080~MeV.
Nowadays, however, the mass of the a$_1$ meson is known to be
$1230\pm40$ MeV \cite {PDG96}.} Moreover, the chiral anomaly is known 
to yield much worse numerical results for spin-1 mesons than for
the pion \cite {Kramer84}.

\subsubsection{Iterative vs.\ simultaneous triple commutator}
Finally, I want to draw your attention to a discussion on the risks of
iterative computation of multiple commutators.  As has been pointed out
in Refs.\ \plaincite {Takahashi81}\nocite {Banerjee90}--\plaincite
{Banerjee91}, the iterative procedure need not necessarily yield the
same result as a simultaneous calculation like the one underlying
\Eq {3commVEV}.  Moreover, the Jacobi identity might fail. \Eq
{3commVEV}, however, can be shown to be consistent with the Jacobi
identity.

\subsubsection{Nonetheless...}
In spite of all that, one may adopt the following point of view.
Anomalous charge-density commutators are not prohibited.  And just in
case someone calculated the commutator in a more reasonable model
than the one just described, it is by all means desireable to learn
more about the possibility of a modification of the GDH sum rule
and about the role of the infinite-momentum limit.

\subsection{Modification of the finite-momentum GDH sum rule}
\label{Sect:mod-FMGDH}

Consider the general non-naive charge-density commutator \eq
{a-comm}. (In view of other applications, I temporarily do not assume a
specific form of the axial-vector field $a^\mu(x)$.)  With this
commutator at hand, I re-inspect the current-algebra derivation of the GDH sum
rule as presented in \Sect {ETCA}.

Instead of the naive dipole-moment commutator \eq {naive-D-comm},
one now has
\begin{align} \label{pre-anom-D}
 \comm {D^i(0)}{D^j(0)}
 & = e^2 \!\int\! \td^3x\, x^i \!\int\! \td^3y\, y^j\,
  \etcomm {J^0(x)} {J^0(y)} \notag \\
 & = ie^2 \!\int\! \td^3x\, x^i \!\int\! \td^3y\, y^j\,
  \bigl( \bnabla \ntimes \ba(x) \bigr) \ndot \bnabla \dirac3(\bx-\by).
\end{align}
(Time $x^0$ is understood to be 0 throughout.)  Observe that
\begin{equation}
 \int\! \td^3y\, y^j\, \bnabla \dirac3(\bx-\by) = \bnabla x^j = \be_j,
\end{equation}
leaving
\begin{equation}
 \comm {D^i(0)}{D^j(0)}
  = ie^2\, \eps^{jlk} \!\int\! \td^3x\, x^i\, \nabla^l a^k(x).
\end{equation}
Partial integration yields
\begin{equation}
 \comm {D^i(0)}{D^j(0)}
  = ie^2\, \eps^{ijk} \!\int\! \td^3x\, a^k(x).
\end{equation}
The integral in this expression is simply the total ``current'' of the
axial-vector field $a^\mu(x)$.  Proceeding along the same lines as in
\Sect {ETCA}, I define left- and right-handed components
$D^{\tL,\tR}(0)$ as in \Eq {LR-def}, and sandwich the resulting
operator equation
\begin{equation}
 \comm {D^\tL(0)}{D^\tR(0)}
  = e^2 \!\int\! \td^3x\, a^3(x)
\end{equation}
between one-nucleon states of positive helicity.
Utilizing translational invariance \eq {trans}, this leads to
\begin{align}
 \me {p',\tfrac12} {[D^\tL(0),D^\tR(0)]} {p,\tfrac12}
 & = e^2 \!\int\! \td^3x\, e^{-i(\bp'-\bp)\ndot x}\,
  \me {p',\tfrac12} {a^3(0)} {p,\tfrac12} \notag \\*
 & = (2\pi)^3 \dirac3(\bp'-\bp)\, e^2\,
  \me {p,\tfrac12} {a^3(0)} {p,\tfrac12}.
\end{align}
By Lorentz covariance and the quantum numbers of field $a^\mu(x)$, the
matrix element emerging here is fixed up to a constant $\tilde g_\tA$,
which I define by
\begin{equation} \label{gA-tilde}
 \boxed{
 \tilde g_\tA\, \bar u(p,\tfrac12)\, \gamma^\mu\gamma_5\, u(p,\tfrac12) =
 \me {p,\tfrac12} {a^\mu(0)} {p,\tfrac12} }
\end{equation}
Since $\bp\propto\be_3$, the third component of the spinor expression
obeys
\begin{equation}
 \bar u(p,\tfrac12)\, \gamma^3\gamma_5\, u(p,\tfrac12) = 2Ms^3 = 2p^0,
\end{equation}
so that finally
\begin{equation} \label{anom-comm-me}
 \me {p',\tfrac12} {[D^\tL(0),D^\tR(0)]} {p,\tfrac12}
 = (2\pi)^3\, 2p^0\, \dirac3(\bp'-\bp)\, 4\pi\alpha\, \tilde g_\tA.
\end{equation}
On the other hand, one directly evaluates the commutator matrix element
by inserting a complete set of intermediate states and separating the
one-nucleon states from the continuum, which has been done in \Sect
{ETCA}:
\begin{align} \label{comm-1N+cont}
 & \me {p', \tfrac12} {[D^\tL(0), D^\tR(0)]} {p, \tfrac12} \notag \\*
 & \qquad = \me {p', \tfrac12} {[D^\tL(0), D^\tR(0)]} {p, \tfrac12}_\toneN
 + \me {p', \tfrac12} {[D^\tL(0), D^\tR(0)]} {p, \tfrac12}_\tcont
   \notag \\*
 & \qquad = (2\pi)^3\, 2p^0\, \dirac3(\boldsymbol p' - \boldsymbol p)
   \left( \frac{2\pi\alpha\kappa^2}{M^2} -
   \frac{2\pi\alpha(Z+\kappa)^2}{(p^0)^2}
   + 8\!\!\int_{q^0_{\tthr}}^\infty\! \frac{\td q^0}{q^0}\, \im f_2(\nu,q^2)
   \right).
\end{align}
Equating \eq {anom-comm-me} with \eq {comm-1N+cont} gives
\begin{equation} \label{mod-FMGDH}
 -4\pi^2\alpha \left( \frac{\kappa^2}{2M^2}
  - \frac{(Z+\kappa)^2}{2(p^0)^2} - \tilde g_\tA \right) =
  \int_{\nu_{\text{thr}}}^\infty\! \frac{\td \nu}\nu\,
  8\pi \im f_2\!\left(\nu, \frac{M^2\nu^2}{(p^0)^2}\right).
\end{equation}

\subsubsection{Infinite-momentum limit}
Like the unmodified finite-momentum GDH sum rule \eq {FMGDH}, the
modified version \eq {mod-FMGDH}, as it stands, is useless, because
for any finite value of the nucleon energy $p^0$, the integration on the
right-hand side runs over arbitrarily high timelike photon
virtualities (see \Fig {q2nu}), where the Compton amplitude is not
related to observables.  Upon taking the limit $p^0\to\infty$ on both
sides of \Eq {mod-FMGDH},
\begin{equation}
 -4\pi^2\alpha \left( \frac{\kappa^2}{2M^2} - \tilde g_\tA \right) =
  \lim_{p^0\to\infty} \!\int_{\nu_{\text{thr}}}^\infty\! \frac{\td\nu}\nu\,
  8\pi \im f_2\!\left(\nu, \frac{M^2\nu^2}{(p^0)^2}\right),
\end{equation}
one arrives at a sensible relation only if one knows (or assumes) what
modification is brought about by dragging the limit into the integral.
If there was \emph {no} additional modification (as conjectured in
\Ref {Chang94a}), then one would arrive at
\begin{equation} \label{mod-GDH}
 \boxed{ -4\pi^2\alpha \left( \frac{\kappa^2}{2M^2} - \tilde g_\tA \right) =
  \int_{\nu_0}^\infty\! \frac{\td\nu}\nu
  \bigl(\sigma_{1/2}(\nu) - \sigma_{3/2}(\nu)\bigr) }
\end{equation}
upon using the optical theorem \eq{f2-opt}.

I now report the explicit numbers that have been found by Chang,
Liang, and Workman \cite {Chang94a}.  \Eqs {a-CLW} and \eq {gA-tilde}
determine the constant $\tilde g_\tA$ in terms of axial-vector
currents $J_{5a}^\mu(x)$:
\begin{equation} \label{gAtilde-J5}
 \tilde g_\tA\, \bar u(p,\tfrac12)\, \gamma^\mu\gamma_5\, u(p,\tfrac12) =
 - \frac {1}{24\pi^2F_\pi^2}\,
   \me {p,\tfrac12} {\Bigl( \sqrt{\tfrac83}\, J_{5\,0}^\mu(0) + J_{5\,3}^\mu(0)
     + \tfrac 1{\sqrt3}\, J_{5\,8}^\mu(0) \Bigr)} {p,\tfrac12}.
\end{equation}
On the other hand, the forward matrix elements emerging in
\Eq{gAtilde-J5} define axial charges $g_{\tA a}$ of the nucleon by
means of
\begin{equation}
 g_{\tA a}\, \bar u(p,\tfrac12)\, \gamma^\mu\gamma_5\, u(p,\tfrac12) =
   \me {p,\tfrac12} {J_{5a}^\mu(0)} {p,\tfrac12},
\end{equation}
which implies
\begin{equation}
 4\pi^2\alpha\, \tilde g_\tA = - \frac {\alpha}{6F_\pi^2}\,
 \left( \sqrt{\tfrac83}\, g_{\tA0} + g_{\tA3}
     + \tfrac 1{\sqrt3}\, g_{\tA8} \right).
\end{equation}
Assuming isospin symmetry, the proton and neutron isosinglet axial
charges are degenerate, 
\begin{subequations}
\begin{equation}
 g_{\tA0}^\tp = g_{\tA0}^\tn \quad\text{and}\quad g_{\tA8}^\tp = g_{\tA8}^\tn,
\end{equation}
whereas the isotriplet axial charges have opposite sign and equal
magnitude:
\begin{equation} \label{iso-sym.3}
 g_{\tA3}^\tp = - g_{\tA3}^\tn = g_\tA = 1.26
\end{equation} \label{iso-sym}%
\end{subequations}
(the neutron-beta-decay constant).  Values for $g_{\tA0}$ and
$g_{\tA8}$ are difficult to fit experimentally.  They can be estimated
from hyperon beta decay under the (severely restrictive) additional
assumption of SU(3) symmetry \cite {Chang94a}.  The proton-neutron
difference, however, involves $g_\tA$ only.  In this channel, the
modification constant occurring in \Eq {mod-GDH} is given by
\begin{equation} \label{CLW-p-n}
 4\pi^2\alpha\, (\tilde g_\tA^\tp - \tilde g_\tA^\tn)
 = - \frac {\alpha\,g_\tA}{3F_\pi^2} = - 137~\mu\tb,
\end{equation}
where $F_\pi=93$~MeV has been used.  Accidentally, the value \eq
{CLW-p-n} is relatively close to the discrepancy that was estimated
from pion-photoproduction multipole analyses.  For instance, the
latest such estimate, given by Sandorfi, Whisnant, and Khandaker, is
$-158~\mu$b \cite {Sandorfi94}, while Karliner \cite {Karliner73}
found $-107~\mu$b.

\section{The infinite-momentum limit}
\label{Sect:IML}

In \Sect {ETCA}, it was shown that, provided merely that the operators of
electric charge densities commute at equal times, one arrives at the following
preliminary form of the GDH sum rule:
\begin{subequations}
\begin{equation} \label{pre-GDH2.unmod}
 -\frac{2\pi^2\alpha\kappa^2}{M^2} =
  \lim_{p^0\to\infty} \int_{\nu_{\text{thr}}}^\infty\!
  \frac{\td \nu}\nu\, 8\pi \im f_2\!\left(\nu, \frac{M^2\nu^2}{(p^0)^2}
  \right).
\end{equation}
I stressed that current algebra alone cannot tell whether the
$p^0\to\infty$ limit can be dragged into the $\nu$ integral without
bringing about a modification.  Starting from the anomalous
charge-density commutator \eq {a-comm}, a modified form of \Eq
{pre-GDH2} occurred in the previous section, viz.
\begin{equation} \label{pre-GDH2.mod}
 -4\pi^2\alpha \left( \frac{\kappa^2}{2M^2} - \tilde g_\tA \right) =
  \lim_{p^0\to\infty} \!\int_{\nu_{\text{thr}}}^\infty\! \frac{\td\nu}\nu\,
  8\pi \im f_2\!\left(\nu, \frac{M^2\nu^2}{(p^0)^2}\right).
\end{equation} \label{pre-GDH2}%
\end{subequations}
Again, consideration of current commutators will not help us getting ahead.

\subsubsection{Why some caution is advisable}
Physicists commonly don't make a fuss about permuting limits or integrations.
They simply do it and see whether the result makes any sense.

\begin{figure}[tb]
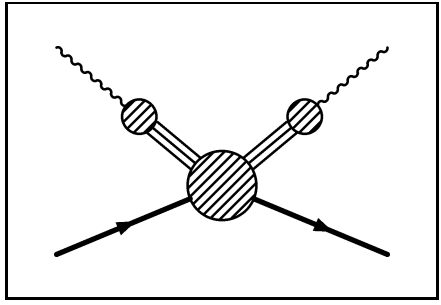

 \begin{displaymath}
  \boxed{ \begin{matrix} \\ \graph(6,3){
   \fmfleft{i1,i2} \fmfright{o1,o2}
   \fmf{fermion,width=thick}{i1,v1,o1}
   \fmf{boson}{i2,v2}
   \fmf{boson}{v3,o2}
   \fmf{plain}{v2,v1,v3}
   \fmfblob{1\ul}{v1}
   \fmfblob{.5\ul}{v2,v3}
   \fmffreeze
   \fmfi{plain}{vpath (__v2,__v1) shifted (thick*(1,1))}
   \fmfi{plain}{vpath (__v2,__v1) shifted (thick*(-1,-1))}
   \fmfi{plain}{vpath (__v1,__v3) shifted (thick*(-1,1))}
   \fmfi{plain}{vpath (__v1,__v3) shifted (thick*(1,-1))}
  } \\ \\ \end{matrix} }
 \end{displaymath}
 \caption[]{
  Intermediate hadron states in virtual Compton scattering,
  leading to singularities in the photon mass $q^2$
  \label{Fig:inthad}
 }
\end{figure}

Nevertheless, one can easily see the possible origin of difficulties
on quite general grounds.  Recall that the timelike virtual Compton
amplitude meets singularities in the photon mass $q^2$ due to
intermediate hadron states, as indicated in \Fig {inthad}.
Thus one expects
for the amplitude $f_2(\nu,q^2)$ a spectral representation of the form
\begin{equation}  \label{spec-rep}
 \im f_2(\nu,q^2) = \frac1\pi \int_{q_\tmin^2}^\infty\!
  \frac{\td q^{\prime2}}{q^2 - q^{\prime2}}\, \rho(\nu,q^{\prime2}),
\end{equation}
where $q_\tmin^2$ is the mass of the lowest-lying state that couples to
the photon.  Inserting \Eq{spec-rep} into one of \Eqs {pre-GDH2}, it is
evident that for finite $p^0$, the $\nu$ integration also meets the
$q^2$ singularities, since $q^2=M^2\nu^2/(p^0)^2$.  Only if one can
drag the limit $p^0\to\infty$ inside the integral in \Eq {pre-GDH2},
these singularities play no explicit role.

\subsubsection{An example}
To further illustrate the possibility of a modification due to
non-interchangeability of limit and integration in \Eqs {pre-GDH2},
suppose that function $f_2(\nu,q^2)$ had a piece
$\tilde{f}_2(\nu,q^2)$ independent of photon lab-frame
energy $\nu$, i.e.,
\begin{equation}
 \tilde{f}_2(\nu,q^2) = \tilde{f}(q^2),
\end{equation}
where the imaginary part of function $\tilde{f}(q^2)$ is assumed to be
logarithmically integrable and, for simplicity, vanishing at
$q^2<\nu_0^2$.  (As a matter of fact, exactly such an expression will
emerge in \Sect {PRP}.)  The simplest non-trivial function
incorporating these properties is a zero-width pole
$\tilde{f}(q^2)\propto(q^2-\mu^2-i\eps)^{-1}$ with imaginary part
$\im\tilde{f}(q^2)\propto\dirac{}(q^2-\mu^2)$.  One has
\begin{equation}
 \int\! \frac {\td\nu}{\nu}\, \im \tilde{f}_2
  \left(\nu,\frac{M^2\nu^2}{(p^0)^2}\right) =
 \frac12 \!\int\! \frac {\td(q^2)}{q^2}\, \im \tilde{f}(q^2) =: C,
\end{equation}
which might be any real constant you like.  Consequently, the $p^0\to\infty$
limit of above integral equals just that constant $C$.  In contrast,
\begin{equation}
 \int\! \frac {\td\nu}{\nu}\, \lim_{p^0\to\infty} \im \tilde{f}_2
  \left(\nu,\frac{M^2\nu^2}{(p^0)^2}\right) =
 \int\! \frac {\td\nu}{\nu}\, \im \tilde{f} (0) = 0,
\end{equation}
since I assumed $\im\tilde{f}(0)=0$.  This scenario exemplifies the
possibility of a \emph {finite} modification brought about by the
infinite-momentum limit.  In principle, there might equally well be an
\emph {infinite} modification.

\subsubsection{The state of the art}
To conclude this section, I want to stress that during several years
of research, I found no instance of the \emph {unmodified}
``preliminary'' GDH sum rule \eq {pre-GDH2.unmod} that would have been
modified by virtue of the infinite-momentum limit.  Yet, there is an
instance of the \emph {modified} form \eq {pre-GDH2.mod} that is
affected by interchanging the $p^0\to\infty$ limit with the $\nu$
integration, in such a way that the modification disappears!  This
very strong evidence for the validity of the unmodified GDH sum rule
is presented in the now-following section.

\section{Anomalous commutator within Weinberg-Salam model}
\label{Sect:PRP}

In this section, I present the calculation of the anomalous
charge-density commutator by means of the Bjorken-Johnson-Low (BJL)
technique \cite {Bjorken66,Johnson66} within the order-$\alpha^2$
Weinberg-Salam model, which I introduced in \Sect {WSM}.  More
precisely, the one-electron matrix element of that commutator is
calculated.  In fact, since the BJL method starts from an
amplitude, i.e., a matrix element, it is only capable to yield matrix
elements of commutators, not commutators themselves.  But after all,
the one-electron matrix element is exactly what is needed for the
discussion of the GDH sum rule.

The contents of my surveys on T-product amplitudes and sea\-gulls (\Sect
{T+seagulls}), as well as on the BJL technique (\Sect {BJL}) apply to
the electron in just the same manner as to the nucleon.

In addition to the Feynman graphs discussed in \Sect {WSM} (\Fig
{WSM-graphs}), a few more one-loop graphs have to be considered, since
one has to go off the real photon point $q^2=0$.  

I employ the BJL technique to calculate the matrix element of the
charge-density commutator
\begin{equation} \label{WSM-BJL}
 \int\! \td^3x\, e^{-i\bq'\ndot \bx}\,
  \me {p',\tfrac12}{[J^0(x),J^0(0)]_\tet}{p,\tfrac12}
  = - \lim_{q^{\prime0}\to\infty} q^{\prime0}\, T^{00}
\end{equation}
within the Weinberg-Salam model to order $\alpha^2$, i.e., to one
loop.  The Weinberg angle $\theta_\tW$ is assumed to adopt its ideal
value, i.e.,
\begin{equation}
 \sin\theta_\tW = \frac12.
\end{equation}
The Fermi constant is then given by
\begin{equation} \label{Fermi}
 \frac {G_\tF}{\sqrt2} = \frac {2\pi\alpha}{M_\tW^2}
 = \frac {8\pi\alpha}{3M_\tZ^2}.
\end{equation}

The external particles in \Eq {WSM-BJL} are electrons with positive
helicity.  Although the result will exhibit a factor of
$\dirac3(\bp'-\bp)$, it is essential to go (slightly) off the forward
direction during the calculation.  (One can show that this is owing to
the fact that the information contained in \emph {all} components of
the \emph {forward} Compton amplitude $T^{\mu\nu}$ can be translated
into the \emph {time-time} component $T^{00}$ of the \emph
{off-forward} amplitude.)  As explained in \Sect {BJL}, the sea\-gull
portion of amplitude $T^{\mu\nu}$ has to be dropped after it is
identified on account of its polynomial dependence on $q^{\prime0}$.

This investigation has been published in \Ref {Pantfoerder98},
but it is presented here in much more detail.

\subsection{Additional Feynman graphs}
\label{Sect:PRP.additionalgraphs}

Since the energy $q^{\prime0}$ of the scattered photon  (and with it the
energy $q^0=q^{\prime0}+p^{\prime0}-p^0$ of the incident photon) runs
while its three-momentum $\bq'$ is kept fixed, the photon necessarily
gets virtual.  Therefore one has to consider a couple of Feynman graphs in
addition to the ones depicted in \Fig {WSM-graphs}, \Sect {WSM}.
the newly emerging graphs are shown in \Fig {additionalgraphs}.%
\begin{figure}[tb]
 \begin{displaymath} \begin{array}{|c@{\qquad}c|}
  \hline
  & \\[-1ex]
  % WW-gamma-gamma contact graph
  \text{(a)} &
   \labelledgraph(2.5,2){
    \fmfleft{i1,i2} \fmfright{o1,o2}
    \fmf{fermion,tension=2}{i1,v2}
    \fmf{fermion,label=$\nu_{\rm e}$}{v2,v3}
    \fmf{fermion,tension=2}{v3,o1}
    \fmf{boson,label=W$^-\!\!$,label.side=left}{v2,v1}
    \fmf{boson,label=W$^-$,label.side=left}{v1,v3}
    \fmf{phantom_arrow,tension=0}{v2,v1,v3}
    \fmf{boson,tension=2}{i2,v1,o2}
   } \\[5ex]
  \hline
  & \\[-1ex]
  % Higgs exchange via fermion loop
  \text{(b)} &
   \labelledgraph(2.5,2.5){
    \fmfleft{i1,i2} \fmfright{o1,o2}
    \fmf{fermion}{i1,v1,o1}
    \fmf{dashes,tension=2,label=H,label.side=left}{v1,v2}
    \fmf{fermion,tension=0}{v3,v4}
    \fmf{fermion}{v4,v2,v3}
    \fmf{boson}{i2,v3}
    \fmf{boson}{v4,o2}
   }
   +
   \labelledgraph(2.5,2.5){
    \fmfleft{i1,i2} \fmfright{o1,o2}
    \fmf{fermion}{i1,v1,o1}
    \fmf{dashes,tension=2,label=H,label.side=left}{v1,v2}
    \fmf{fermion,tension=0}{v3,v4}
    \fmf{fermion}{v4,v2,v3}
    \fmf{phantom}{i2,v3}
    \fmf{phantom}{v4,o2}
    \fmffreeze
    \fmf{boson}{i2,v4}
    \fmf{boson}{v3,o2}
   } \\[7ex]
  \hline
  & \\[-1ex]
  % Higgs exchange via W loop
  \text{(c)} &
   \labelledgraph(2.5,2.5){
    \fmfleft{i1,i2} \fmfright{o1,o2}
    \fmf{fermion}{i1,v1,o1}
    \fmf{dashes,tension=2,label=H,label.side=left}{v1,v2}
    \fmf{boson,tension=0}{v3,v4}
    \fmf{boson,label=W$^-$,label.side=left}{v4,v2}
    \fmf{boson}{v2,v3}
    \fmf{phantom_arrow,tension=0}{v2,v3,v4,v2}
    \fmf{boson}{i2,v3}
    \fmf{boson}{v4,o2}
   }
   +
   \labelledgraph(2.5,2.5){
    \fmfleft{i1,i2} \fmfright{o1,o2}
    \fmf{fermion}{i1,v1,o1}
    \fmf{dashes,tension=2,label=H,label.side=left}{v1,v2}
    \fmf{boson,tension=0}{v3,v4}
    \fmf{boson,label=W$^-$,label.side=left}{v4,v2}
    \fmf{boson}{v2,v3}
    \fmf{phantom_arrow,tension=0}{v2,v3,v4,v2}
    \fmf{phantom}{i2,v3}
    \fmf{phantom}{v4,o2}
    \fmffreeze
    \fmf{boson}{i2,v4}
    \fmf{boson}{v3,o2}
   }
   +
   \labelledgraph(2.5,2.5){
    \fmfleft{i1,i2} \fmfright{o1,o2}
    \fmf{fermion}{i1,v1,o1}
    \fmf{dashes,tension=2,label=H,label.side=left}{v1,v2}
    \fmf{boson,left}{v2,v3}
    \fmf{boson,left,label=W$^-$}{v3,v2}
    \fmf{phantom_arrow,left,tension=0}{v2,v3,v2}
    \fmf{boson}{i2,v3,o2}
   } \\[7ex]
  \hline
  & \\[-1ex]
  % Z exchange
  \text{(d)} &
   \labelledgraph(2.5,2.5){
    \fmfleft{i1,i2} \fmfright{o1,o2}
    \fmf{fermion}{i1,v1,o1}
    \fmf{boson,tension=2,label=Z$^0$,label.side=left}{v1,v2}
    \fmf{fermion,tension=0}{v3,v4}
    \fmf{fermion}{v4,v2,v3}
    \fmf{boson}{i2,v3}
    \fmf{boson}{v4,o2}
   }
   +
   \labelledgraph(2.5,2.5){
    \fmfleft{i1,i2} \fmfright{o1,o2}
    \fmf{fermion}{i1,v1,o1}
    \fmf{boson,tension=2,label=Z$^0$,label.side=left}{v1,v2}
    \fmf{fermion,tension=0}{v3,v4}
    \fmf{fermion}{v4,v2,v3}
    \fmf{phantom}{i2,v3}
    \fmf{phantom}{v4,o2}
    \fmffreeze
    \fmf{boson}{i2,v4}
    \fmf{boson}{v3,o2}
   } \\[7ex]
  \hline
 \end{array} \end{displaymath}
 \caption[]{
  One-loop Feynman graphs contributing to the virtual
  Compton amplitude of the electron in the Weinberg-Salam model,
  in addition to those depicted if \Fig {WSM-graphs}.  \\
  \hspace*{1em} (a) WW$\gamma\gamma$ contact graph \\
  \hspace*{1em} (b) Higgs exchange via fermion loop \\
  \hspace*{1em} (c) Higgs exchange via W-boson loop \\
  \hspace*{1em} (d) Z$^0$ exchange \\
  \label{Fig:additionalgraphs}
 }
\end{figure}
Their distinguishing mark is the fact that they cannot be cut into two
parts such that both pieces represent the \emph {absorption} of a
(real) photon.  The crucial graph is the Z$^0$-exchange graph, because
the coupling of the Z$^0$ boson to the photons by virtue of an
electron-triangle subgraph gives rise to the chiral anomaly.

Yet a few more one-loop diagrams can be drawn but are individually vanishing:
\begin{subequations}
\begin{equation}
 \begin{gathered} \\
 \labelledgraph(2.5,2.5){
  \fmfbottom{b} \fmftop{t1,t2}
  \fmflabel{$\gamma,\tZ^0$}{b} \fmflabel{$\gamma$}{t1} \fmflabel{$\gamma$}{t2}
  \fmf{photon}{t1,v1,t2}
  \fmf{photon,tension=2}{b,v2}
  \fmf{boson,left}{v2,v1}
  \fmf{boson,left,label=W$^-$}{v1,v2}
  \fmf{phantom_arrow,left,tension=0}{v2,v1,v2}
 } \quad = \quad
 \labelledgraph(2.5,2.5){
  \fmfbottom{b} \fmftop{t1,t2}
  \fmflabel{$\gamma,\tZ^0$}{b} \fmflabel{$\gamma$}{t1} \fmflabel{$\gamma$}{t2}
  \fmf{photon}{b,v1,t2}
  \fmf{photon,tension=2}{t1,v2}
  \fmf{boson,left}{v2,v1}
  \fmf{boson,left,label=W$^-$}{v1,v2}
  \fmf{phantom_arrow,left,tension=0}{v2,v1,v2}
 } \quad = \quad
 \labelledgraph(2.5,2.5){
  \fmfbottom{b} \fmftop{t1,t2}
  \fmflabel{$\gamma,\tZ^0$}{b} \fmflabel{$\gamma$}{t1} \fmflabel{$\gamma$}{t2}
  \fmf{photon}{b,v1,t1}
  \fmf{photon,tension=2}{t2,v2}
  \fmf{boson,left}{v1,v2}
  \fmf{boson,left,label=W$^-$}{v2,v1}
  \fmf{phantom_arrow,left,tension=0}{v2,v1,v2}
 } \quad = 0 \\ \\
 \end{gathered}
\end{equation}
or cancel each other:
\begin{equation}
 \begin{matrix} \\
 \labelledgraph(2.5,2.5){
  \fmfbottom{b} \fmftop{t1,t2}
  \fmflabel{$\gamma,\tZ^0$}{b} \fmflabel{$\gamma$}{t1} \fmflabel{$\gamma$}{t2}
  \fmf{boson}{t1,v1}
  \fmf{boson}{t2,v2}
  \fmf{boson,label.side=left,label=W$^-$}{v2,v3}
  \fmf{boson}{v3,v1}
  \fmf{boson,tension=2}{b,v3}
  \fmffreeze
  \fmf{boson}{v1,v2}
  \fmf{phantom_arrow}{v1,v2,v3,v1}
 } \quad+\quad
 \labelledgraph(2.5,2.5){
  \fmfbottom{b} \fmftop{t1,t2}
  \fmflabel{$\gamma,\tZ^0$}{b} \fmflabel{$\gamma$}{t1} \fmflabel{$\gamma$}{t2}
  \fmf{phantom}{t1,v1}
  \fmf{phantom}{t2,v2}
  \fmf{boson,label.side=left,label=W$^-$}{v2,v3}
  \fmf{boson}{v3,v1}
  \fmf{boson,tension=2}{b,v3}
  \fmffreeze
  \fmf{boson}{v1,v2}
  \fmf{boson}{t1,v2}
  \fmf{boson}{t2,v1}
  \fmf{phantom_arrow}{v1,v2,v3,v1}
 } \quad = 0, \\ \\
 \end{matrix}
\end{equation}
\begin{equation}
 \begin{matrix} \\
 \labelledgraph(2.5,2.5){
  \fmfbottom{b} \fmftop{t1,t2}
  \fmflabel{$\gamma$}{b} \fmflabel{$\gamma$}{t1} \fmflabel{$\gamma$}{t2}
  \fmf{boson}{t1,v1}
  \fmf{boson}{t2,v2}
  \fmf{fermion}{v2,v3,v1}
  \fmf{boson,tension=2}{b,v3}
  \fmffreeze
  \fmf{fermion}{v1,v2}
 }
 \quad+\quad
 \labelledgraph(2.5,2.5){
  \fmfbottom{b} \fmftop{t1,t2}
  \fmflabel{$\gamma$}{b} \fmflabel{$\gamma$}{t1} \fmflabel{$\gamma$}{t2}
  \fmf{phantom}{t1,v1}
  \fmf{phantom}{t2,v2}
  \fmf{fermion}{v2,v3,v1}
  \fmf{boson,tension=2}{b,v3}
  \fmffreeze
  \fmf{fermion}{v1,v2}
  \fmf{boson}{t1,v2}
  \fmf{boson}{t2,v1}
 } \quad = 0. \\ \\
 \end{matrix}
\end{equation}
\end{subequations}
As far as the three-photon coupling is concerned, the vanishing of the
sum of these diagrams represents the one-loop version of Furry's
theorem.  Then, naturally, in case of the W-boson loops, one of the photons
can be replaced by a Z$^0$ boson, since up to a factor of $\tan\theta_\tW$,
the WW$\gamma$ coupling equals the WWZ coupling.

\subsubsection{WW$\gamma\gamma$ contact graph}
Let us first have a look at the WW$\gamma\gamma$ contact graph of \Fig
{additionalgraphs}(a).  The Feynman rule for the quadrilinear coupling
reads \cite {Aitchison89}
\begin{equation}
 \begin{gathered}\\ \labelledgraph(2,2){
  \fmfleft{i1,i2} \fmfright{o1,o2}
  \fmflabel{$\alpha$}{i1} \fmflabel{$\mu$}{i2}
  \fmflabel{$\beta$}{o1} \fmflabel{$\nu$}{o2}
  \fmf{boson}{i1,v,o1}
  \fmf{boson}{i2,v,o2}
  \fmffreeze
  \fmf{phantom_arrow}{i1,v,o1}
 }\\ \\ \end{gathered} \quad = -ie^2 (2g^{\mu\nu}g^{\alpha\beta}
  - g^{\mu\alpha}g^{\nu\beta} - g^{\mu\beta}g^{\nu\alpha}).
\end{equation}
Without having to perform the loop integration, one realizes that this
graph is \emph {independent of the photon energy} $q^{\prime0}$.
(This is a general feature of contact graphs.)  Hence, according to
\Sect {BJL}, it contributes solely to the sea\-gull amplitude and must be
dismissed in the calculation of the BJL limit \eq {WSM-BJL}.  By the
way: what animal\footnote {\label {animals}Theorists love animals
(sea\-gulls, penguins, even tadpoles), parts of animals (cat ears) and
parts of ex-animals (handbags).} (if any) does \Fig
{additionalgraphs}(a) look like?  You guess it!  This is the origin of
the name ``sea\-gull''.

\subsubsection{Higgs exchange}
Next, consider the Higgs-exchange graphs depicted in \Fig
{additionalgraphs}(b,c).  Since the Higgs boson has spin 0, it does not
couple to the electron's spin.  Hence, a contribution to the GDH sum rule
is not expected.  While this is evident in the dispersion-theoretic
approach, the spin argument is somewhat concealed as regards the anomalous
charge-density commutator.  Nevertheless, one can show that the Higgs-exchange
graphs do not contribute to the matrix element of the dipole-moment
commutator in the BJL limit, independently of how the Higgs boson is
coupled to the photons.  I want to sketch the proof in the following.

Upon crossing symmetry and gauge invariance, the H$\gamma\gamma$ vertex
can be written
\begin{equation} \label{Hgg}
 \begin{gathered} \\
 \labelledgraph(2,2){
  \fmfbottom{b} \fmftop{t1,t2}
  \fmflabel{H}{b} \fmflabel{$\gamma(q)$}{t1} \fmflabel{$\gamma(q')$}{t2}
  \fmf{photon}{t1,v,t2}
  \fmf{dashes,tension=1.5}{b,v}
  \fmfblob{.8\ul}{v}
 } \\ \\
 \end{gathered} \quad =
 \begin{aligned}[t]
  & \left( g^{\mu\nu} - \frac {q^\mu q^{\prime\nu}}{q\ndot q'} \right)
    H_1(q^2,q^{\prime2},q\ndot q') \\
  & + \left( q^{\prime\mu} - \frac {q^{\prime2}}{q\ndot q'}\, q^{\mu} \right)
    \left( q^{\nu} - \frac {q^{2}}{q\ndot q'}\, q^{\prime\nu} \right)
    H_2(q^2,q^{\prime2},q\ndot q') \\
  & + \eps^{\mu\nu\alpha\beta}\, q^{}_{\alpha}q'_{\beta}\,
    H_3(q^2,q^{\prime2},q\ndot q'), \Big.
 \end{aligned}
\end{equation}
where Lorentz index $\mu$ ($\nu$) corresponds to the scattered
(incident) photon.  (The parity violating piece proportional to
invariant function $H_3$ emerges if graphs incorporating the W boson,
like the ones depicted in \Fig {additionalgraphs}(c), are taken into
account.)  Crossing symmetry imposes the additional constraints
\begin{equation}
 H_i(q^2,q^{\prime2},q\ndot q') =  H_i(q^{\prime2},q^2,q\ndot q'), \quad
 i=1,2,3.
\end{equation}
Letting $\mu=\nu=0$ in \Eq {Hgg} and taking the BJL limit
$q^{\prime0}\to\infty$ gives rise to terms proportional to
$\bq^{\prime2}$ and $\bq\ndot\bq'$.  The four-momentum $q$ of the
incident photon is fixed by energy- and momentum conservation,
$p+q=p'+q'$.  In other words, it is accounted a shorthand of
expression $q'+p'-p$ when the BJL limit is taken.  Hence one has terms
proportional to $\bq^{\prime2}$ and $(\bp'-\bp)\ndot\bq'$.  On account
of \Eq {WSM-BJL}, this has to be Fourier transformed to coordinate
space, yielding $\bnabla^2\dirac3(\bx)$ and
$(\bp'-\bp)\ndot\bnabla\dirac3(\bx)$.  The crucial point now is that
only a certain moment of the charge-density is considered, namely the
dipole moment $D^i(x^0)=e\int\td^3x\,x^iJ^0(x)$.  Thus the second
derivative $\bnabla^2\dirac3(\bx)$ gives no contribution, while the
term involving the first derivative of $\dirac3(\bx)$ gives something
proportional to $(p'-p)^i$.  As explained in detail below, taking the
dipole moment $D^j$ of the other charge-density involved in the
commutator simply gives a factor of $\nabla^j\dirac3(\bp'-\bp)$.
Observe now that
\begin{equation} \label{pp-p}
 (p'-p)^i\, \nabla^j\dirac3(\bp'-\bp) = g^{ij}\, \dirac3(\bp'-\bp),
\end{equation}
which is multiplied by a factor of
\begin{equation}
 \frac {\bar u(p',\tfrac12)\, \gamma_5\, u(p,\tfrac12)}{(p'-p)^2-M_\tH^2}
\end{equation}
coming from the propagator of the Higgs boson and its coupling to the
electron line.  Finally, observe that \Eq {pp-p} picks out the forward
direction, and that
\begin{equation}
 \bar u(p,\tfrac12)\, \gamma_5\, u(p,\tfrac12) = 0.
\end{equation}
So, indeed, Higgs-exchange graphs do not contribute to the commutator
matrix element under consideration.

I note in passing that I have explicitly calculated the electron-loop
graphs of \Fig {additionalgraphs}(b), adopting a Pauli-Villars
regularization scheme following Steinberger \cite [Sect.\ III.A.1]
{Steinberger49}, with the result that the H$\gamma\gamma$ coupling \eq
{Hgg} by virtue of the electron triangle is of order
$(q^{\prime0})^{-2}$, thus vanishing in the BJL limit, regardless of
the lower parts of the Higgs-exchange diagrams, i.e., the coupling of
the Higgs to the electron line.

\subsubsection{Z$^0$ exchange}
The contribution of the Z$^0$-exchange graphs of \Fig {additionalgraphs}(d)
to the Compton amplitude $T^{\mu\nu}$ reads
\begin{equation} \label{TZ}
 T^{\mu\nu}_{(\text Z)} = - \frac{M_{\text Z}^2G_{\text F}}{2\sqrt2}\,
  Z^{\mu\nu\rho}(q,q')\,
  \frac{-g_{\rho\sigma}+(p'-p)_\rho(p'-p)_\sigma/M_{\text Z}^2}
   {(p'-p)^2-M_{\text Z}^2}\,
  \bar u(p',s') \gamma^\sigma\gamma_5 u(p,s),
\end{equation}
where the integral
\begin{align} \label{Z-def}
 Z^{\mu\nu\rho}
 & = \int\!\frac{\text d^4k}{(2\pi)^4}
  \tr\left(\gamma^\mu\frac i{\kslash-m+i\eps}
  \gamma^\nu\frac i{\kslash-\qslash-m+i\eps}
  \gamma^\rho\gamma_5\frac i{\kslash-\qslash'-m+i\eps} \right) \notag\\*
 & \quad + \left\{ \begin{matrix} \mu\leftrightarrow\nu \\ q\leftrightarrow-q'
  \end{matrix} \right\}
\end{align}
embodies the electron-loops, and the constant
$M_\tZ^2G_\tF/2\sqrt2=e^2/3$ is due to the coupling of the Z$^0$ boson to
the electron lines:
\begin{equation} \label{Zee}
 \graph(2,2){
  \fmfbottom{b1,b2} \fmftop{t}
  \fmf{fermion}{b1,v,b2} \fmf{boson}{t,v}
 } = - \frac {ie}{\sqrt{3}}\, \gamma^\rho\gamma_5.
\end{equation}
the Z$^0$ propagator in \Eq {TZ},
\begin{equation} \label{Z0-prop}
 i\, \frac{-g_{\rho\sigma}+(p'-p)_\rho(p'-p)_\sigma/M_{\text Z}^2}
  {(p'-p)^2-M_{\text Z}^2},
\end{equation}
is written in unitary gauge.  However, all results will be explicitly
independent of the second term in its numerator, so that any other
gauge would yield the same result.  (The $i\eps$ prescription is
omitted, since the momentum of the Z$^0$ boson is not integrated over,
and since we are far off its mass shell.)  Finally, \Eq {TZ} exhibits
a factor of $-1$ due to the closed fermion loop.

For the moment, I consider arbitrary external-electron spins $s,s'$
and all Lorentz indices $\mu,\nu$ of the amplitude.  For the
calculation of the charge-density commutator, only $\mu=\nu=0$ and
positive helicities will be needed.  Later on, however, I want to
study the infinite-momentum limit, for which I need all Lorentz
indices and spins, but only the forward direction $p=p',s=s'$.

The triangle loop integral \eqref{Z-def} can be cast into the form
\begin{align} \label{Z-dec}
 Z^{\mu\nu\rho} & =
  \eps^{\nu\rho\alpha\beta}q'_\alpha q^{}_\beta
  (q^\mu Z_1 + q^{\prime\mu} Z_2)
  + \eps^{\mu\nu\rho\alpha} q_\alpha Z_3
  \notag\\
 &\quad
  + \eps^{\mu\rho\alpha\beta}q'_\alpha q^{}_\beta
  (q^{\prime\nu} Z'_1 + q^\nu Z'_2)
  + \eps^{\mu\nu\rho\alpha} q'_\alpha Z'_3,
\end{align}
where $Z_{1,2,3}$ are functions of the Lorentz invariants $q^2$,
$q^{\prime2}$, and $q\ndot{q'}$.  From crossing symmetry, the primed
functions are given by $Z'_{1,2,3}(q^2,q^{\prime2},q\ndot{q'}) =
Z_{1,2,3}(q^{\prime2},q^2,q\ndot{q'})$.  Gauge invariance
$q^{}_{\nu}T_{(\text Z)}^{\mu\nu}=0$, $q'_{\mu}T_{(\text Z)}^{\mu\nu}=0$,
imposes the constraint
\begin{equation}  \label{Z-gaugeinv}
 q\ndot{q'}Z_1 + q^{\prime2}Z_2 + Z_3 = q\ndot{q'}Z'_1 + q^2Z'_2 + Z'_3 = 0.
\end{equation}

Evaluation of the integral \eq {Z-def} proceeds as follows.  By
definition of the fermion propagators, the Dirac trace becomes
\begin{align}
 & \tr \left( \gamma^\mu\frac i{\kslash-m+i\eps}
  \gamma^\nu\frac i{\kslash-\qslash-m+i\eps}
  \gamma^\rho\gamma_5\frac i{\kslash-\qslash'-m+i\eps} \right) = \notag \\*
 & \qquad -i\, \frac
  {\tr \left( \gamma^\mu (\kslash+m) \gamma^\nu (\kslash-\qslash+m)
   \gamma^\rho\gamma_5 (\kslash-\qslash'+m) \right)}
  {\bigl( k^2-m^2+i\eps \bigr) \bigl( (k-q)^2-m^2+i\eps \bigr)
   \bigl( (k-q')^2-m^2+i\eps \bigr)},
\end{align}
where the numerator is evaluated using trace theorems, while the denominator
is made capable by means of the Feynman formula
\begin{equation}
 \frac 1{ABC} = 2 \!\int_0^1\! \td x \!\int_0^{1-x}\! \td y\,
  \frac 1{\bigl( (A-C)x + (B-C)y + C \bigr)^3},
\end{equation}
valid for arbitrary non-zero constants $A,B,C$.  The result, being
linearly divergent (as, besides, is already apparent from the defining
equation \eq {Z-def}), is Pauli-Villars regularized by subtracting
from the integrand the same expression with a large cut-off $\Lambda$
substituted for the electron mass $m$, letting $\Lambda\to\infty$
after the loop integration is performed.  Functions
$Z_{1,2}(q^2,q^{\prime2},q\ndot q')$ turn out to be finite in the
limit $\Lambda\to\infty$, and are given by
\begin{subequations}
\begin{equation} \label{Z.1}
 Z_1 = \frac{i}{\pi^2}\!\int_0^1\!\text dx \!\int_0^{1-x}\!\text dy\,
 \frac{xy}
  {x(1-x)\,q^2 + y(1-y)\,q^{\prime2} - 2xy\,q\ndot q' - m^2 + i\eps}
\end{equation}
and
\begin{equation} \label{Z.2}
 Z_2 = \frac{i}{\pi^2}\!\int_0^1\!\text dx \!\int_0^{1-x}\!\text dy\,
 \frac{-x(1-x)}
  {x(1-x)\,q^2 + y(1-y)\,q^{\prime2} - 2xy\,q\ndot q' - m^2 + i\eps}.
\end{equation}
On the other hand, function $Z_3(q^2,q^{\prime2},q\ndot q')$ is
formally divergent, but can be fixed by means of the gauge-invariance
condition \eq {Z-gaugeinv}, giving
\begin{equation} \label{Z.3}
 Z_3 = \frac{i}{\pi^2}\!\int_0^1\!\text dx \!\int_0^{1-x}\!\text dy\,
 \frac{x(1-x)\,q^{\prime2}-xy\,q\ndot q'}
  {x(1-x)\,q^2 + y(1-y)\,q^{\prime2} - 2xy\,q\ndot q' - m^2 + i\eps},
\end{equation}
This fact was first observed by Rosenberg \cite {Rosenberg63}, who was
concerned with the same Feynman integral applied to a different
process (see also Adler \cite {Adler69a}).  As noted above, the primed
invariant functions $Z'_{1,2,3}$ are obtained from the unprimed
functions by interchanging $q^2$ and $q^{\prime2}$.  Subsequently
re-naming the Feynman parameters, $x\leftrightarrow y$, one arrives at
\begin{equation} \label{Z.1p}
 Z'_1 = \frac{i}{\pi^2}\!\int_0^1\!\text dx \!\int_0^{1-x}\!\text dy\,
 \frac{xy}
  {x(1-x)\,q^2 + y(1-y)\,q^{\prime2} - 2xy\,q\ndot q' - m^2 + i\eps},
\end{equation}
\begin{equation} \label{Z.2p}
 Z'_2 = \frac{i}{\pi^2}\!\int_0^1\!\text dx \!\int_0^{1-x}\!\text dy\,
 \frac{-y(1-y)}
  {x(1-x)\,q^2 + y(1-y)\,q^{\prime2} - 2xy\,q\ndot q' - m^2 + i\eps},
\end{equation}
and
\begin{equation} \label{Z.3p}
 Z'_3 = \frac{i}{\pi^2}\!\int_0^1\!\text dx \!\int_0^{1-x}\!\text dy\,
 \frac{y(1-y)\,q^{2}-xy\,q\ndot q'}
  {x(1-x)\,q^2 + y(1-y)\,q^{\prime2} - 2xy\,q\ndot q' - m^2 + i\eps}.
\end{equation} \label{Z}%
\end{subequations}

\subsection{Anomalous charge-density commutator}
With the result \eq {Z} at hand, I am now in a position to
calculate the anomalous charge-density commutator due to  Z$^0$
exchange by virtue of the BJL limit \eq {WSM-BJL}:
\begin{align} \label{Z0-comm1}
 & \int\! \td^3x\, e^{-i\bq'\ndot \bx}\,
  \me {p',\tfrac12}{[J^0(x),J^0(0)]_\tet}{p,\tfrac12}
  = - \lim_{q^{\prime0}\to\infty} q^{\prime0}\, T_{(\tZ)}^{00} =
  \notag\\*
 & \qquad \frac{M_{\text Z}^2G_{\text F}}{2\sqrt2}
  \lim_{q^{\prime0}\to\infty} q^{\prime0}Z^{00\rho}(q,q')\,
  \frac{-g_{\rho\sigma}+(p'-p)_\rho(p'-p)_\sigma/M_{\text Z}^2}
   {(p'-p)^2-M_{\text Z}^2}\,
  \bar u(p',\tfrac12) \gamma^\sigma\gamma_5 u(p,\tfrac12),
\end{align}
from \Eq {TZ}.  Inserting the decompostion \eq {Z-dec}, one has
\begin{equation} \label {Z00}
 Z^{00\rho} = \eps^{0\rho\alpha\beta} q'_\alpha q^{}_\beta
 (q^0 Z_1 + q^{\prime0} Z_2 + q^{\prime0} Z_1' + q^0 Z_2').
\end{equation}
The invariant functions $Z_3^{(\prime)}$ do not occur.  Recall that
$q^0=q^{\prime0}+p^{\prime0}-p^0$, so that $q^0$ is of the same order
than $q^{\prime0}$ in the limit $q^{\prime0}\to\infty$.  Observe that
from \Eqs {Z}, functions $Z_{1,2}^{(\prime)}$ are of order
$(q^{\prime0})^{-2}$.  Hence, expression $q^{\prime0}Z^{00\rho}$ is
finite.  For the specific combination of functions
$Z_{1,2}^{(\prime)}$ appearing in \Eq {Z00}, the Feynman-parameter
integration gets particularly simple in the limit $q^{\prime0}\to\infty$:
\begin{align} \label{Z1212}
 & (q^{\prime0})^2 (Z_1+Z_2+Z_1'+Z_2') \notag\\*
 & \qquad = \frac i{\pi^2} (q^{\prime0})^2
   \!\int_0^1\! \td x \!\int_0^{1-x}\! \td y\,
   \frac {2xy - x(1-x) - y(1-y)}
    {x(1-x)\,q^2 + y(1-y)\,q^{\prime2} - 2xy\,q\ndot q' - m^2 + i\eps}
   \notag\\
 & \qquad \xrightarrow[q^{\prime0}\to\infty]{}
   - \frac i{\pi^2} \!\int_0^1\! \td x \!\int_0^{1-x}\! \td y
   = - \frac i{2\pi^2}.
\end{align}
Note that the dependence on the mass $m$ of the particle making up the
triangle loop dropped out.  This is characteristic for the anomaly.
By means of \Eq {Z1212}, the expression occurring in the charge-density
commutator \eq {Z0-comm1} can be written
\begin{equation}
 \lim_{q^{\prime0}\to\infty} q^{\prime0}Z^{00\rho}
 = - \frac i{2\pi^2}\, \eps^{0\rho\alpha\beta} q'_\alpha q^{}_\beta
 = - \frac i{2\pi^2}\, \eps^{0\rho\alpha\beta} q'_\alpha (p'-p)_\beta,
\end{equation}
where in the last expression, $q$ has been replaced by $q'+p'-p$, and
the antisymmetry of the epsilon tensor has been employed.  Observe
that due to the term $\eps^{0\rho\alpha\beta}$, the indices
$\alpha,\beta$ are spatial only.  Inserting this into \Eq {Z0-comm1},
it is clear that the (gauge dependent) second part of the propagator
gives no contribution, since
$\eps^{0\rho\alpha\beta}(p'-p)_\beta(p'-p)_\rho=0$, while the $g_{\rho\sigma}$
part gives
\begin{align} \label{Z0-comm2}
 & \int\! \td^3x\, e^{-i\bq'\ndot \bx}\,
  \me {p',\tfrac12}{[J^0(x),J^0(0)]_\tet}{p,\tfrac12} \notag\\*
 & \qquad = i\, \frac{M_{\text Z}^2G_{\text F}}{4\pi^2\sqrt2}\,
  \eps^{0\rho\alpha\beta}\, q'_\alpha (p'-p)_\beta\,
  \frac{\bar u(p',\tfrac12) \gamma_\rho\gamma_5 u(p,\tfrac12)}
   {(p'-p)^2-M_{\text Z}^2} \notag\\
 & \qquad = -i\, \frac{M_{\text Z}^2G_{\text F}}{4\pi^2\sqrt2}\,
  \eps^{kij}\, q^{\prime i} (p'-p)^j\,
  \frac{\bar u(p',\tfrac12) \gamma^k\gamma_5 u(p,\tfrac12)}
   {(p'-p)^2-M_{\text Z}^2}.
\end{align}
The $\bq'$ dependence lies solely in the factor $q^{\prime i}$,
whose Fourier transform is $-i\nabla^i\dirac3(\bx)$.
Hence, it is a trivial matter to transform into coordinate space:
\begin{align} \label{Z0-comm3}
 \me {p',\tfrac12}{[J^0(x),J^0(0)]_\tet}{p,\tfrac12} =
  - \frac{M_{\text Z}^2G_{\text F}}{4\pi^2\sqrt2}\,
  \eps^{kij}\, (p'-p)^j\,
  \frac{\bar u(p',\tfrac12) \gamma^k\gamma_5 u(p,\tfrac12)}
   {(p'-p)^2-M_{\text Z}^2}\,
  \nabla^i\dirac3(\bx).
\end{align}

\subsubsection{Anomalous commutator in terms of gauge field}
I point out that the same result is obtained upon first considering
the vacuum-to-Z$^0$ matrix element\footnote {\Eq {Z-comm} is
consistent with the lowest-order approximation to the result obtained
by Jo \cite [Eq.\ (2.20)] {Jo85a}, suitably generalized from
left-handed currents and left-handed gauge fields to vector currents and
axial-vector gauge fields.} of the commutator, which yields in terms
of the Z$^0$-boson field operator $Z^\mu(x)$:
\begin{equation} \label{Z-comm}
 \etcomm {J^0(x)} {J^0(y)} = \frac {-ie}{2\pi^2\sqrt3}\,
 \bigl( \bnabla \ntimes \bZ(x) \bigr) \ndot \bnabla \dirac3(\bx-\by),
\end{equation}
subsequently sandwiching this relation between one-electron states,
using the coupling \eq {Zee} and the equation of motion of the Z$^0$
boson.  This procedure amounts to seperately considering the upper and
lower halfs of the Z$^0$-exchange diagrams depicted in \Fig
{additionalgraphs}(d).  Observe that expression \eq {Z-comm} has the
form of the general anomalous charge-density commutator \eq {a-comm}
discussed in \Sect {anomcomm}.

\subsubsection{Electric dipole moment}
The reader may now recall that we are interested in the dipole-moment
commutator.  I proceed from charge-density $J^0(x)$ to the dipole
moment $D^i(0)$ by computing the first moment of \Eq {Z0-comm3}:
\begin{align}
 \me {p',\tfrac12}{[D^i(0),J^0(0)]}{p,\tfrac12} & =
   e\!\int\!\td^3x\, x^i\,
   \me {p',\tfrac12}{[J^0(x),J^0(0)]_\tet}{p,\tfrac12} \notag\\*
 & = \frac{eM_{\text Z}^2G_{\text F}}{4\pi^2\sqrt2}\,
  \eps^{kij}\, (p'-p)^j\,
  \frac{\bar u(p',\tfrac12) \gamma^k\gamma_5 u(p,\tfrac12)}
   {(p'-p)^2-M_{\text Z}^2}.
\end{align}
Translational invariance \eq {trans} gives rise to a phase factor,
\begin{align}
 \me {p',\tfrac12}{[D^i(0),J^0(y)]}{p,\tfrac12} =
 \me {p',\tfrac12}{[D^i(0),J^0(0)]}{p,\tfrac12}\, e^{-i(\bp'-\bp)\ndot\by},
\end{align}
which in turn fetches a gradient of $\dirac3(\bp'-\bp)$ when the other
dipole-moment operator is introduced:
\begin{equation}
 \int\! \td^3y\, y^j\, e^{-i(\bp'-\bp)\ndot\by} =
 i (2\pi)^3\nabla^j\dirac3(\bp'-\bp).
\end{equation}
Finally, the identity
\begin{equation}
 (p'-p)^{j'} \nabla^j\dirac3(\bp'-\bp) = -\delta^{j'j}\,\dirac3(\bp'-\bp)
\end{equation}
leads to
\begin{align}
 & \me {p',\tfrac12}{[D^i(0),D^j(0)]}{p,\tfrac12} =
  -i\, \frac{e^2M_{\text Z}^2G_{\text F}}{4\pi^2\sqrt2}\,
  \eps^{kij}\, (2\pi)^3\dirac3(\bp'-\bp)\,
  \frac{\bar u(p,\tfrac12) \gamma^k\gamma_5 u(p,\tfrac12)}{-M_{\text Z}^2}.
\end{align}
As usual, I take the electron to be travelling along $\be_3$ and
introduce circularly polarized components
\begin{equation}
 D^{\tL,\tR}(0) = \frac1{\sqrt2}\, \bigl( D^1(0) \pm iD^2(0) \bigr),
\end{equation}
which gives rise to the spinor expression
\begin{equation}
 \bar u(p,\tfrac12)\,\gamma^3\gamma_5\, u(p,\tfrac12) = 2ms^3 = 2p^0.
\end{equation}
Consequently,
\begin{equation} \label{WSM-D-comm}
 \boxed{ \me{p',\tfrac12}{[D^\tL(0),D^\tR(0)]}{p,\tfrac12} =
  (2\pi)^3\, 2p^0\, \dirac3(\bp'-\bp)\, \frac\alpha\pi\, \frac{G_\tF}{\sqrt2} }
\end{equation}
instead of the naive, i.e., vanishing, commutator, on which the conventional
current-algebra derivation of the GDH sum rule is based.

\subsubsection{Modified finite-momentum GDH sum rule}
Observe that \Eq {WSM-D-comm} has the same form as \Eq {anom-comm-me}.
Thus, analogously to the procedure presented in \Sect {mod-FMGDH},
the finite-momentum GDH sum rule obtains a modification:
\begin{equation} \label{WSM-mod-FMGDH}
 \boxed{ \alpha\, \frac {G_\tF}{\sqrt2} =
  \lim_{p^0\to\infty} \!\int_{0}^\infty\! \frac{\td\nu}\nu\,
  8\pi \im f_2\!\left(\nu, \frac{m^2\nu^2}{(p^0)^2}\right) }
\end{equation}
where the polarized forward virtual Compton amplitude $f_2(\nu,q^2)$ is
defined as in \Sect {ETCA}:
\begin{equation} \label{WSM-f2nuq2}
 f_2(\nu,q^2) := \frac\alpha{2m^2} \left(A_1(\nu,q^2) +
  \frac{q^2}{m\nu}A_2(\nu,q^2)\right).
\end{equation}

\subsubsection{Infinite-momentum limit}
\begin{figure}[tb]
 \begin{center}
  \input{wsm-q2nu}
 \end{center}
 \caption[]{
  The integration path of the finite-momentum GDH sum rule
  \eq{WSM-mod-FMGDH} in the $(\nu,q^2)$ plane for electron energies $p^0=m$,
  $3m$, and $6m$.  For any finite value of $p^0$, the integration
  passes the two-electron threshold $q^2=4m^2$, picking up the constant on
  the left-hand side of Eq.\ \eqref{WSM-mod-FMGDH}.
  The genuine GDH sum rule integrates along the straight line at $q^2=0$.
  \label{Fig:WSM-q2nu}}
\end{figure}
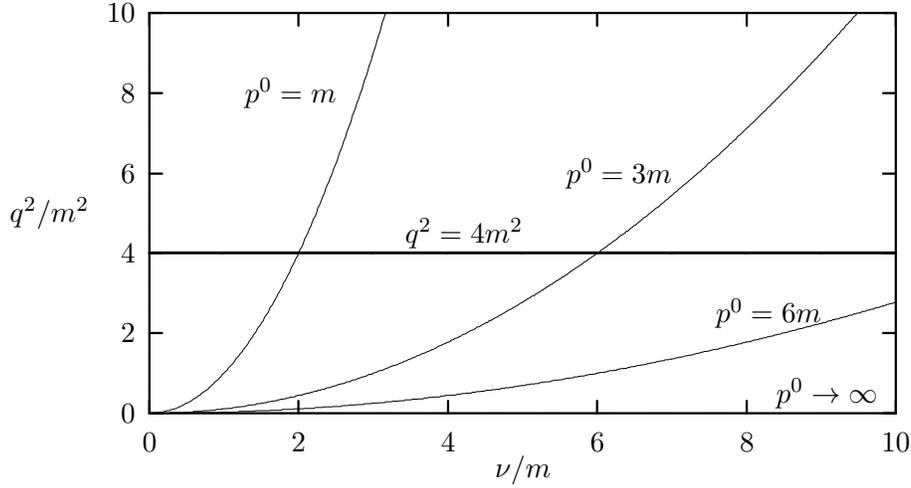

For any finite value of the electron energy $p^0$, the integral on the
right-hand side of \Eq {WSM-mod-FMGDH} runs along a parabola in
$(\nu,q^2)$ plane, as depicted in \Fig {WSM-q2nu}.%
What if the $p^0\to\infty$ limit and the $\nu$ integration were
interchangeable in \Eq {WSM-mod-FMGDH}?  Upon application of the
optical theorem, one would then of course have the familiar GDH integral
\begin{equation} \label{GDHint}
 \int_0^\infty\! \frac{\td\nu}{\nu}\,
 \bigl( \sigma_{1/2}(\nu) - \sigma_{3/2}(\nu) \bigr)
\end{equation}
on the right-hand side, while the left-hand side is a specific
order-$\alpha^2$ constant. (Recall that the Fermi constant \eq {Fermi}
is of order $\alpha$).  However, I illustrated in \Sect {WSM} that the
left-hand side of the sum rule is of order $\alpha^3$ in the
Weinberg-Salam model, and that Altarelli, Cabibbo, and Maiani \cite
{Altarelli72} proved that the integral on the right-hand side confirms
this, i.e., it vanishes to order $\alpha^2$.  Hence, at this point it
actually has to be strongly \emph {expected} that the limit cannot be
dragged inside the integral in \Eq {WSM-mod-FMGDH} without further
modification.

Of course, Altarelli et al.\ \cite {Altarelli72} considered only the
kinematical domain of the genuine GDH sum rule, i.e., $q^2=0$ (the
abscissa in \Fig {WSM-q2nu}).  As a consequence, the Z$^0$-exchange
graph did simply not emerge.  On the other hand, I inspect the sum
rule at finite momenta and take the $p^0\to\infty$ limit eventually,
which is certainly \emph {the long way} to the sum rule.  Let me
remind you that I do this in order to \emph {test} for the legitimacy
of the infinite-momentum limit, particularly if there is an
anomalous-commutator modification.  I work in a perturbative model,
because this enables me to explicitly calculate the ``limit of an
integral'' \eq {WSM-mod-FMGDH} \emph {and} the ``integral of a limit''
\eq {GDHint}, in order to confront both quantities.

To make a long story short: In the following I will show that
permuting limit and integration in \Eq {WSM-mod-FMGDH} gives rise to a
second modification that exactly cancels the previous one.  This modification
is due to the same Feynman graphs that give rise to the commutator anomaly,
namely the Z$^0$-exchange diagrams of \Fig {additionalgraphs}(d).
More specifically,  the corresponding contribution
to Compton amplitude $f_2(\nu,q^2)$ picks up an imaginary part if the
photon mass $q^2$ exceeds the two-electron threshold $4m^2$, which of course
it does for every finite energy $p^0$.  At $q^2=0$, these graphs do not
contribute to $f_2(\nu,q^2)$.

\subsection{Infinite-momentum limit}
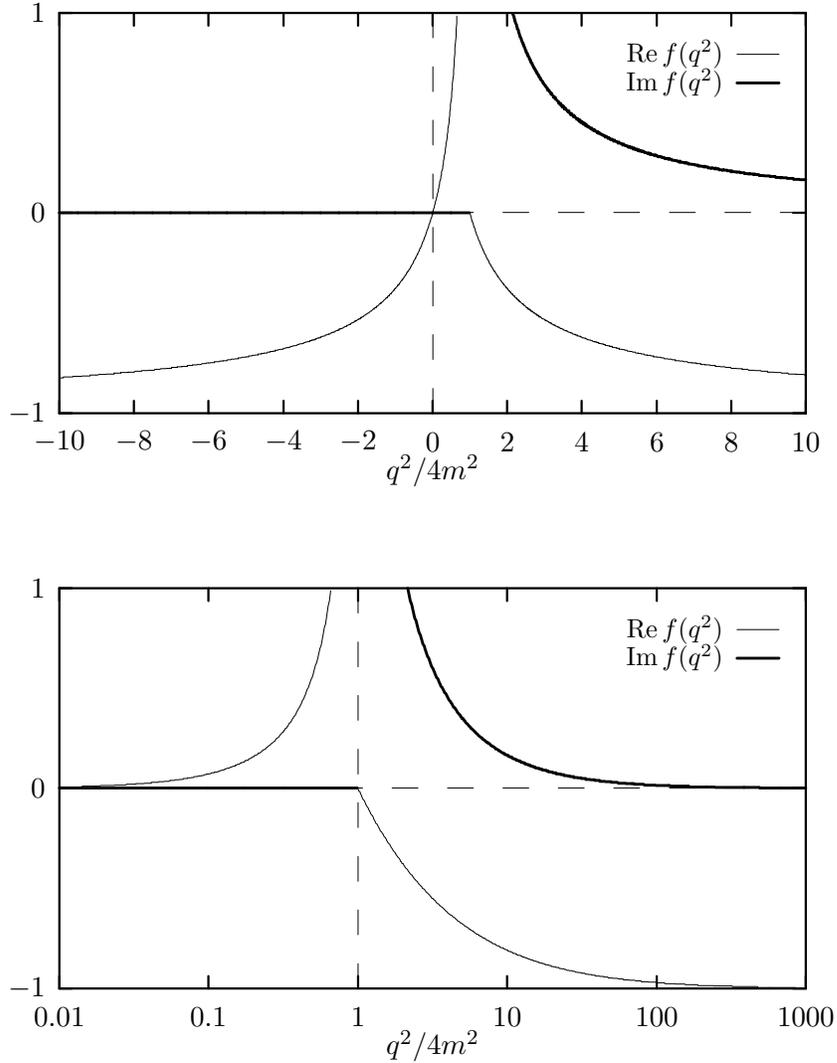
\begin{figure}[tb]
 \begin{center}
  \input{fq2-lin} \\
  \input{fq2-log}
 \end{center}
 \caption[]{
  Real part (light line) and imaginary part (heavy line)
  of the function $f(q^2)$ occurring in the Z$^0$-exchange contribution
  $f_2^{(\text Z)}(\nu,q^2) \propto f(q^2)$
  to the forward virtual Compton amplitude of the
  order-$\alpha^2$ Weinberg-Salam model.
  Upper plot: linear $q^2$ scale;
  lower plot: logarithmic $q^2$ scale, only $q^2>0$.
  A branch-point singularity occurs at the two-electron threshold $q^2=4m^2$.
  \label{Fig:fq2}}
\end{figure}

I want to calculate the contribution of the Z$^0$-exchange diagrams to
the forward virtual Compton amplitude $f_2(\nu,q^2)$.  That is to say,
\Eqs {TZ} and \eq {Z-dec} are specialized to the drastically simplifying
case $p=p'$, $q=q'$,
and $s=s'$:
\begin{equation} \label{TZ-forw1}
 T^{\mu\nu}_{\text{(Z)}} = - \frac{M_{\text Z}^2G_{\text F}}{2\sqrt2}\,
  Z^{\mu\nu\rho}(q)\,
  \frac{-g_{\rho\sigma}}{-M_{\text Z}^2}\,
  \bar u(p,s)\, \gamma^\sigma\gamma_5\, u(p,s)
\end{equation}
with
\begin{equation}
 Z^{\mu\nu\rho}(q) = \eps^{\mu\nu\rho\alpha} q_\alpha\,
 \bigl( Z_3(q^2)+Z_3'(q^2) \bigr).
\end{equation}
By virtue of the identity
\begin{equation}
  \bar u(p,s)\, \gamma^\sigma\gamma_5\, u(p,s) = 2ms^\sigma,
\end{equation}
one obtains
\begin{equation} \label{TZ-forw2}
 T^{\mu\nu}_{\text{(Z)}} = m\, \frac{G_{\text F}}{\sqrt2}\,
 \eps^{\mu\nu\rho\sigma} q_\rho s_\sigma \bigl( Z_3(q^2)+Z_3'(q^2) \bigr).
\end{equation}
From \Eqs {Z.3} and \eq {Z.3p}, I read off
\begin{align}
 Z_3+Z_3' & = \frac{i}{\pi^2}\!\int_0^1\!\text dx \!\int_0^{1-x}\!\text dy\,
 \frac{x(1-x)\,q^2 + y(1-y)\,q^{2} - 2xy\,q^2}
  {x(1-x)\,q^2 + y(1-y)\,q^{2} - 2xy\,q^2 - m^2 + i\eps} \notag\\
 & = \frac{i}{\pi^2}\!\int_0^1\!\text dx \!\int_0^{1-x}\!\text dy\,
 \frac{(x-y)(1-x-y)\,q^2}
  {(x+y)(1-x-y)\,q^2 - m^2 + i\eps},
\end{align}
which involves the sum $x+y$ only.  Hence, substituting $x\mapsto z=x+y$
renders the y integration trivial:
\begin{equation} \label{Z3}
 Z_3+Z_3' = \frac{i}{\pi^2}\!\int_0^1\!\text dz \!\int_0^{z}\!\text dy\,
 \frac{z(1-z)\,q^2}{z(1-z)\,q^2 - m^2 + i\eps}
 = - \frac i{2\pi^2}\, f(q^2),
\end{equation}
where
\begin{equation} \label{f-def}
 f(q^2) := -2 \!\int_0^1\!\text dz\,
 \frac{z^2(1-z)\,q^2}{z(1-z)\,q^2 - m^2 + i\eps}.
\end{equation}
The integration in \Eq {f-def} can be performed explicitly.  The
result
\begin{equation} \label{f-res}
 f(q^2) =
 \begin{cases}
  -1 + \frac{4m^2}{\sqrt{(q^2-4m^2)q^2}} \arcoth{\sqrt{1-\frac{4m^2}{q^2}}}
  & \text{if } q^2<0 \\
  -1 + \frac{4m^2}{\sqrt{(4m^2-q^2)q^2}} \arccot{\sqrt{\frac{4m^2}{q^2}-1}}
  & \text{if } 0<q^2<4m^2 \\
  -1 + \frac{4m^2}{\sqrt{(q^2-4m^2)q^2}}
  \left( \artanh{\sqrt{1-\frac{4m^2}{q^2}}} + \frac {i\pi}2 \right)
  & \text{if } q^2>4m^2
 \end{cases}
\end{equation}
is depicted in \Fig {fq2}.%
Note that function $f(q^2)$ is continuous at $q^2=0$.

\subsubsection{Z$^0$-exchange contribution to forward Compton amplitude}
Inserting \Eq {Z3} into \Eq {TZ-forw2}, I conclude that
\begin{equation} \label{TZ-forw3}
 T^{\mu\nu}_{\text{(Z)}} = -\frac{im}{2\pi^2}\, \frac{G_\tF}{\sqrt2}\,
 \eps^{\mu\nu\rho\sigma} q_\rho s_\sigma\, f(q^2).
\end{equation}
On comparison with the invariant decomposition of the polarized,
i.e., antisymmetric part of the forward virtual Compton amplitude,
\begin{align}
 \tfrac12 \bigl(T^{\mu\nu}-T^{\nu\mu}\bigr) =
  & - \frac im\, \eps^{\mu\nu\rho\sigma} q_\rho s_\sigma\, A_1(\nu,q^2)
    \notag\\*
  & - \frac i{m^3}\, \eps^{\mu\nu\rho\sigma} q_\rho
   \bigl( (M\nu) s_\sigma - (q\ndot s) p_\sigma \bigr) A_2(\nu,q^2),
\end{align}
one realizes that Z$^0$ exchange contributes to amplitude $A_1(\nu,q^2)$ only:
\begin{align}
 A_1^{(\tZ)}(\nu,q^2) & = \frac {m^2}{2\pi^2}\, \frac {G_\tF}{\sqrt2}\, f(q^2),
 \notag\\*
 A_2^{(\tZ)}(\nu,q^2) & = 0.
\end{align}
By definition \eq {WSM-f2nuq2}, this implies
\begin{equation} \label{WSM-f2}
 \boxed{
 f_2^{(\tZ)}(\nu,q^2) = \frac{\alpha}{4\pi^2}\, \frac{G_\tF}{\sqrt2}\, f(q^2)}
\end{equation}

\subsubsection{Interchanging limit and integration}
Let us now have a closer look at the result \eq {WSM-f2} together with \Eq
{f-def}.  The first and most essential thing to be noted is that
function $f_2^{(\tZ)}(\nu,q^2)$ depends solely on the photon's virtuality
$q^2$.  At $q^2=0$, it vanishes:
\begin{equation}
 f_2^{(\tZ)}(\nu,0) = 0,
\end{equation}
according to the fact that Z$^0$ exchange does not contribute to the real
Compton process.  Hence, of course,
\begin{equation} \label{int-f2Z0}
 \int_0^\infty\! \frac{\td\nu}\nu \im f_2^{(\tZ)}(\nu,0) = 0,
\end{equation}
Now, what happens if $q^2$ rises?  Since the expression $z(1-z)$
attaines a maximum of $\frac14$ at $z=\frac12$, the integrand in \Eq
{f-def} exhibits poles in the vicinity of the real axis if $q^2>4m^2$,
thereupon giving function $f(q^2)$ an imaginary part.  Considering
\Fig {additionalgraphs}(d), this branch point singularity is owing
to the fact that at $q^2>4m^2$, the two electrons that couple to the
photon can be put on their mass shells.  So, what about the
Z$^0$-exchange contribution to the \emph {finite-momentum} GDH
integral \eq {WSM-mod-FMGDH}?  Using \Eqs {f-def} and \eq {WSM-f2}, I
infer
\begin{align} \label{pre-f2Z}
 \int_0^\infty\! \frac{\td\nu}{\nu}\, 8\pi
 \im f_2^{(\tZ)} \left(\nu,\frac{m^2\nu^2}{(p^0)^2}\right)
 & = \frac {2\alpha}\pi\, \frac{G_\tF}{\sqrt2} \!\int_0^\infty\!
 \frac{\td\nu}{\nu} \im f \left(\frac{m^2\nu^2}{(p^0)^2}\right)  \notag\\
 & = \frac\alpha\pi\, \frac{G_\tF}{\sqrt2} \!\int_{4m^2}^\infty\!
 \frac{\td(q^2)}{q^2} \im f(q^2).
\end{align}
Observe that owing to the $\nu$ independence of function
$f_2^{(\tZ)}(\nu,q^2)$, the result is independent of the electron's energy
$p^0$.  This means that on each parabola in the $(\nu,q^2)$ plane of
\Fig {WSM-q2nu}, the GDH integral picks up the \emph {same} constant from
the part of the parabola that lies above the two-electron threshold,
while \Eq {int-f2Z0} reflects the fact that there is no such
contribution if the integration path lies entirely below the
two-electron threshold.  The integral \eq {pre-f2Z} can be calculated
either directly from the imaginary part of \Eq {f-res}, or by
recalling that it represents a $q^2$ dispersion integral, subtracted at
infinity on account of the limit $f(\infty)=-2\int_0^1\text dz\,z=-1$.
The result reads
\begin{equation} \label{int-f2Z}
 \int_0^\infty\! \frac{\td\nu}{\nu}\, 8\pi
 \im f_2^{(\tZ)} \left(\nu,\frac{m^2\nu^2}{(p^0)^2}\right)
 = \alpha\, \frac{G_\tF}{\sqrt2}.
\end{equation}
Note that, again, the $m$ dependence dropped out.  Confrontation of \Eqs
{int-f2Z0} and \eq {int-f2Z} reveals that interchanging $p^0\to\infty$
limit and $\nu$ integration is non-trivial owing to the Z$^0$-exchange
graphs:
\begin{equation}
 \left(\lim_{p^0\to\infty}\int_0^\infty\!\frac{\text d\nu}\nu -
  \int_0^\infty\!\frac{\text d\nu}\nu\lim_{p^0\to\infty}\right)
  8\pi\im f_2\!\left(\nu,\frac{m^2\nu^2}{(p^0)^2}\right) =
  \alpha\,\frac{G_{\text F}}{\sqrt2}.
\end{equation}
Thus, an additive constant occurs in the relation of the GDH
integral \eq {GDHint} to the finite-momentum GDH integral on the right-hand
side of \Eq {WSM-mod-FMGDH}:
\begin{equation} \label{WSM-IML}
 \boxed{ \int_0^\infty\! \frac{\td\nu}{\nu}\,
 \bigl( \sigma_{1/2}(\nu) - \sigma_{3/2}(\nu) \bigr) =
  \lim_{p^0\to\infty} \!\int_{0}^\infty\! \frac{\td\nu}\nu\,
  8\pi \im f_2\!\left(\nu, \frac{m^2\nu^2}{(p^0)^2}\right)
 - \alpha\,\frac{G_{\text F}}{\sqrt2} }
\end{equation}
Inserting the finite-momentum GDH sum rule \eq {WSM-mod-FMGDH} leads back
to the undisturbed sum rule \eq {WSM-GDH}.

\subsection{Summary and discussion}
\label{Sect:PRP.summary}
The Weinberg-Salam model serves as an ideal testing ground for the
legitimacy of the infinite-momentum limit, particularly in the presence
of an anomalous-commutator modification of the finite-momentum GDH sum
rule.  I have shown that via the BJL limit, the Z$^0$-exchange Feynman
graphs of \Fig {additionalgraphs}(d) lead to an anomalous charge-density
commutator \eq {Z0-comm3} (or dipole-moment commutator \eq
{WSM-D-comm}), which modifies the finite-momentum GDH sum rule \eq
{WSM-mod-FMGDH}.  By explicit calculation of the pertinent Compton
amplitude at all values of $\nu$ and $q^2$, I proved that the naive
infinite-momentum limit fails in the presence of this modification.
Rather, letting $p^0\to\infty$ gives rise to a second modification \eq
{WSM-IML} that cancels the previous one.  In other words, \emph {the
anomalous-commutator modification does not survive the infinite-momentum
limit!} I stress that this result is consistent with the fact that no
order-$\alpha^2$ modification of the GDH sum rule within the
Weinberg-Salam model is observed by direct evaluation at $q^2=0$ \cite
{Altarelli72}.

\subsubsection{The role of gauge invariance}
It is perhaps worth noting that the exact cancellation of the two
modifications of the GDH sum rule can be traced back to the
gauge-invariance condition \eqref {Z-gaugeinv}, which relates the
function $Z_3$ occurring in \Eq {TZ-forw2} to functions $Z_{1,2}$ of
\Eq {Z1212}.

\subsubsection{BJL limit and Regge limit}
\label{BJL+Regge}
Observe that the Z$^0$-exchange contribution \eq {WSM-f2} to the
forward virtual Compton amplitude presents an explicit instance of the
non-coincidence of BJL limit ($q^0\to\infty$, $\bq$ fixed) and Regge
limit ($\nu\to\infty$, $q^2$ fixed):
\begin{subequations}
\begin{equation}
 \lim_{\substack{q^0\to\infty\\ \bq~\text{fixed}}} f_2^{(\tZ)}(\nu,q^2)
 = \frac{\alpha}{4\pi^2}\, \frac{G_\tF}{\sqrt2}\, f(\infty),
\end{equation}
whereas
\begin{equation} \label{lim-f2Z.Regge}
 \lim_{\substack{\nu\to\infty\\ q^2~\text{fixed}}} f_2^{(\tZ)}(\nu,q^2)
 = \frac{\alpha}{4\pi^2}\, \frac{G_\tF}{\sqrt2}\, f(q^2).
\end{equation} \label{lim-f2Z}%
\end{subequations}
Function $f(q^2)$ (see \Eq {f-res}, \Fig {fq2}) obeys $f(0)=0$ and
$f(\pm\infty)=-1$.  Thus, the limits \eq {lim-f2Z} coincide at large
$q^2$.  This also holds for large \emph {spacelike} $q^2$, i.e.,
$Q^2:=-q^2\to\infty$, indicating the equivalence of BJL and Regge
limits in deep-inelastic scattering.

Observe, further, that \Eq {lim-f2Z.Regge} presents a $J=1$ fixed pole
in complex angular-momentum plane (cf.\ \Sect {fixed-pole}), whose
residue is non-polynomial in the photon virtuality $q^2$, contrary to
the common sense of early studies on fixed poles (see, e.g., Cheng and
Tung \cite {Cheng70}).  At $q^2=0$, the residue vanishes.

\subsubsection{Anomaly cancellation}
Finally, I remark that if quarks were included into the model, i.e., if
not only electrons, but also u and d quarks were taken to circulate in
the loop of \Fig {additionalgraphs}(d), then the customary effect of
anomaly cancellation would remove the modification of the charge-density
commutator as well as the one due to the infinite-momentum limit.
This can be traced back to the $m$ independence of \Eqs {Z1212} and
\eq {int-f2Z}.

\section{$t$-channel a$_1$ exchange}
\label{Sect:a1}

The possible fixed pole in angular-momentum plane, discussed in \Sect
{fixed-pole}, possesses the quantum numbers $J^P=1^+$ of the a$_1$
meson (more precisely: a$_1$ and f$_1$ mesons).  This fact is
misunderstood once in a while. I stress that fixed poles of given spin
$J$ can generally not be attributed to $t$-channel exchange of a
physical spin-$J$ particle.  To shed some light on this point, the
present section is devoted to a discussion of the process represented
by the following Feynman graph:
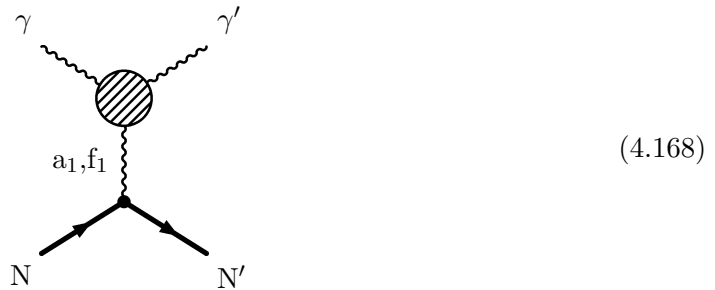
\begin{equation} \label{a1ex-graph}
 \begin{gathered} \\
 \labelledgraph(3,3){
  \fmfleft{i1,i2} \fmfright{o1,o2}
  \fmflabel{N}{i1} \fmflabel{N$'$}{o1}
  \fmflabel{$\gamma$}{i2} \fmflabel{$\gamma'$}{o2}
  \fmf{fermion,width=thick}{i1,v1,o1} \fmf{photon}{i2,v2,o2}
  \fmf{boson,label=a$_1$,,f$_1$,label.side=left}{v1,v2}
  \fmfblob{.8\ul}{v2}
  \fmfdot{v1}
 } \\ \\
 \end{gathered}
\end{equation}

Apart from doing away with the above-mentioned misbelief, I aim at a
further discussion of the alleged anomalous-commutator modification
of the GDH sum rule, claimed by Chang, Liang, and Workman \cite{Chang94a}.

\subsubsection{Landau-Yang theorem}
Half a century ago, Landau \cite {Landau48} and Yang \cite {Yang50}
inferred from symmetry considerations, that two real photons never are
in a state of total angular momentum one.\footnote {More precisely, a
two-photon state cannot have total angular momentum $J=1$ or $J=3,5,\ldots$
and odd parity \cite {Landau48,Yang50}.}
Hence, one expects the upper vertex
\begin{equation} \label{AVV-graph}
 V^{\mu\nu\rho}(q,q') = \quad
 \begin{gathered} \\
 \labelledgraph(2,2){
  \fmfbottom{b} \fmftop{t1,t2}
  \fmflabel{a$_1(q'-q)$}{b}
  \fmflabel{$\gamma(q)$}{t1} \fmflabel{$\gamma(q')$}{t2}
  \fmf{photon}{t1,v,t2}
  \fmf{boson,tension=1.5}{b,v}
  \fmfblob{.8\ul}{v}
 } \\ \\
 \end{gathered}
\end{equation}
of diagram \eq {a1ex-graph} to vanish identically if real photons are
attached.  In \Eq {AVV-graph}, Lorentz indices $\mu$, $\nu$, and
$\rho$ represent scattered and incident photon and the meson,
respectively.  I let $q^2=q^{\prime2}=0$.  By parity conservation and
crossing symmetry, the general invariant decomposition of vertex \eq
{AVV-graph} runs
\begin{align} \label{AVV-dec}
 V^{\mu\nu\rho}
 & = \eps^{\mu\nu\alpha\beta}(q^{\prime}-q)^\rho q'_\alpha q^{}_\beta\,
   V_1(q\ndot q') \notag \\*
 & \quad + (\eps^{\nu\rho\alpha\beta}q^{\prime\mu}
   + \eps^{\mu\rho\alpha\beta}q^{\nu}) q'_\alpha q^{}_\beta\, V_2(q\ndot q')
   \notag \\*
 & \quad + \eps^{\mu\nu\rho\alpha} (q+q')_\alpha\, V_3(q\ndot q').
\end{align}
(A fourth tensor $(\eps^{\nu\rho\alpha\beta}q^\mu +
\eps^{\mu\rho\alpha\beta}q^{\prime\nu}) q'_\alpha q^{}_\beta$ is proven
to be linearly dependent on the others by virtue of identity \eq
{anti5}.)  Gauge invariance, $q'_\mu V^{\mu\nu\rho}=0$, implies
\begin{equation}
 V_3(q\ndot q') = 0,
\end{equation}
while due to the condition $\eps\ndot q=\eps'\ndot q'=0$, function
$V_2(q\ndot q')$ does not contribute when the vertex function \eq
{AVV-dec} is contracted with real photon polarization vectors:
\begin{equation} \label{epsepsV}
 \eps'_\mu \eps_\nu V^{\mu\nu\rho} = \eps^{\mu\nu\alpha\beta}
 \eps'_\mu \eps^{}_\nu q'_\alpha q^{}_\beta (q'-q)^\rho\, V_1(q\ndot q').
\end{equation}
In accordance with the Landau-Yang theorem, this expression vanishes
upon contraction with a third polarization vector $\eps''_{\rho}$ representing
the axial-vector meson and satisfying $\eps''\ndot(q'-q)=0$.  Contracted
with a propagator, expression \eq {epsepsV} does not necessarily
vanish, due to the fact that an off-shell spin-1 propagator has a
spin-0 part, unless it is written in Landau gauge.  Nevertheless,
taking $q=q'$ in \Eq {epsepsV} explicitly reveals that the graph \eq
{a1ex-graph} does not contribute to real forward Compton scattering.
Hence it does not affect the GDH sum rule.  However, it \emph {does}
affect the \emph {finite-momentum} GDH sum rule, as shown in the following.

\subsubsection{Anomalous charge-density commutator}
A striking analogy can be revealed between a$_1$ exchange and Z$^0$
exchange as discussed in \Sect {PRP}, if the axial-vector meson
is coupled to the photons by means of quark triangle-loop diagrams:
\begin{equation} \label{a1q-graphs}
 \begin{minipage}{3\unitlength}\begin{fmfgraph*}(3,3)
  \fmfleft{i1,i2} \fmfright{o1,o2}
  \fmf{fermion,width=thick}{i1,v1,o1}
  \fmf{boson,tension=2,label=a$_1$,label.side=left}{v1,v2}
  \fmf{fermion,tension=0}{v3,v4}
  \fmf{fermion,label=u,,d,label.side=left}{v4,v2}
  \fmf{fermion}{v2,v3}
  \fmf{boson}{i2,v3}
  \fmf{boson}{v4,o2}
  \fmfdot{v1}
 \end{fmfgraph*}\end{minipage}
 \quad + \quad
 \begin{minipage}{3\unitlength}\begin{fmfgraph*}(3,3)
  \fmfleft{i1,i2} \fmfright{o1,o2}
  \fmf{fermion,width=thick}{i1,v1,o1}
  \fmf{boson,tension=2,label=a$_1$,label.side=left}{v1,v2}
  \fmf{fermion,tension=0}{v3,v4}
  \fmf{fermion,label=u,,d,label.side=left}{v4,v2}
  \fmf{fermion}{v2,v3}
  \fmf{phantom}{i2,v3}
  \fmf{phantom}{v4,o2}
  \fmffreeze
  \fmf{boson}{i2,v4}
  \fmf{boson}{v3,o2}
  \fmfdot{v1}
 \end{fmfgraph*}\end{minipage}
\end{equation}
This means that spin-1 mesons are regarded as chiral gauge fields with
pointlike coupling to quarks.  (For a discussion, I refer the reader to
p.~\pageref {spin-1-story} of this thesis.)  For simplicity,
I consider the proton-neutron difference in order to project out isovector
exchange, i.e., a$_1$ exchange only, rather than a$_1$ and f$_1$ exchange.
Falling back on \Eqs {TZ}--\eq {Z0-prop}, I can re-use the results of \Sect
{PRP} by translating the lower Z$^0\te^-\te^-$ vertex into the a$_1$NN
vertex of graphs \eq {a1q-graphs} according to
\begin{subequations}
\begin{equation} \label{subst.lower}
 - \frac {ie}{\sqrt3}\, \gamma^\rho\gamma_5 \mapsto
 ig_\tA f_\tA\, \gamma^\rho\gamma_5,
\end{equation}
and the upper Z$^0\te^-\te^-$ vertex into the a$_1$qq vertex according
to
\begin{equation} \label{subst.upper}
 - \frac {ie}{\sqrt3}\, \gamma^\sigma\gamma_5 \mapsto
 \frac i2\, f_\tA\, \gamma^\sigma\gamma_5,
\end{equation}
where the constant $f_\tA$ couples the physical meson to the
axial-vector current (cf.\ \Eq {b=2v}), $g_\tA=1.26$ is the neutron
beta-decay constant (the nucleon matrix element of the axial-vector
current), and a factor of two in \Eq {subst.lower} accounts for taking
the proton-neutron difference.  The triangle loops couple to the axial
charge times the square of the electric charge.  Moreover, there are
$N_\tc=3$ quarks of each flavor.  Thus, one has to substitute\footnote
{Readers familiar with the chiral anomaly may recognize \Eq {subst.anom-canc}
as being the reason for the  anomaly-cancellation mechanism.}
\begin{equation} \label{subst.anom-canc}
 -1 = -\bigl(-1\bigr)^2 \mapsto N_\tc
 \bigl[ \bigl(\tfrac23\bigr)^2 - \bigl(-\tfrac13\bigr) \bigr] = 1.
\end{equation} \label{subst}%
\end{subequations}
Combining \Eqs {subst}, one encounters the substitution
\begin{equation}
 \frac {e^2}3 = \frac {M_\tZ^2G_\tF}{2\sqrt2} \mapsto \frac12\, g_\tA f_\tA^2.
\end{equation}
The Z$^0$ propagator \eq {Z0-prop} has to be replaced by the a$_1$-meson
propagator
\begin{equation} \label{a1-prop}
 i\, \frac{-g_{\rho\sigma}+(p'-p)_\rho(p'-p)_\sigma/m_\tA^2}{(p'-p)^2-m_\tA^2}.
\end{equation}
Putting these pieces together, the calculation of anomalous charge-density
and dipole-moment commutators proceeds in perfect analogy to \Sect {PRP}.
One obtains
\begin{align}
 & \me {\tp(p',\tfrac12)}{[J^0(x),J^0(0)]_\tet}{\tp(p,\tfrac12)} -
   \me {\tn(p',\tfrac12)}{[J^0(x),J^0(0)]_\tet}{\tn(p,\tfrac12)} \notag \\*
 & \qquad = \frac{g_\tA f_\tA^2}{4\pi^2}\, \eps^{kij}\, (p'-p)^j\,
  \frac{\bar u(p',\tfrac12) \gamma^k\gamma_5 u(p,\tfrac12)}{(p'-p)^2-m_\tA^2}\,
  \nabla^i\dirac3(\bx)
\end{align}
in place of \Eq {Z0-comm3}, and
\begin{align}
 & \me{\tp(p',\tfrac12)}{[D^\tL(0),D^\tR(0)]}{\tp(p,\tfrac12)} -
   \me{\tn(p',\tfrac12)}{[D^\tL(0),D^\tR(0)]}{\tn(p,\tfrac12)} \notag \\*
 & \qquad = - (2\pi)^3\, 2p^0\, \dirac3(\bp'-\bp)\,
 \frac\alpha\pi\, \frac{g_\tA f_\tA^2}{m_\tA^2}
\end{align}
in place of \Eq {WSM-D-comm}.  Proceeding as in \Ref {Chang94a}, I
employ the KSRF relation \eq {KSRF} together with Weinberg's identity
$m_\tA^2=2m_\tV^2$ \cite {Weinberg67}, as well as the fact that within the
present context, vector- and axial-vector-meson couplings necessarily
are degenerate, $f_\tV=f_\tA$ \cite {Sakurai69,DeAlfaro73}.  I arrive at
\begin{align} \label{a1ex-comm}
 & \me{\tp(p',\tfrac12)}{[D^\tL(0),D^\tR(0)]}{\tp(p,\tfrac12)} -
   \me{\tn(p',\tfrac12)}{[D^\tL(0),D^\tR(0)]}{\tn(p,\tfrac12)} \notag \\*
 & \qquad = - (2\pi)^3\, 2p^0\, \dirac3(\bp'-\bp)\,
 \frac{\alpha g_\tA}{\pi F_\pi^2}.
\end{align}

Up to a factor of three,\footnote {See footnote \ref {CLWvsJo} on page
\pageref {CLWvsJo}.} \Eq {a1ex-comm} reproduces the result of Chang,
Liang, and Workman \cite {Chang94a} as summarized in \Eqs
{anom-comm-me} and \eq {CLW-p-n} of this thesis.  In other words: If
you look hard at the investigation presented in \Refs
{Chang94a}\nocite {Chang91}--\plaincite {Chang92}, you realize that
actually the $t$-channel exchange of axial-vector mesons is
studied.\footnote {Cf.\ discussion ``The Emperor's New Clothes'' on
page \pageref {emperor-story} of this thesis.} Hence, it is equivalent
to the investigation of the anomalous commutator within the
Weinberg-Salam model, as long as all particles and couplings are
substituted appropriately.  This again shows that indeed the
commutator anomaly  modifies merely the \emph {finite-momentum} GDH sum
rule, being remedied upon taking the infinite-momentum limit.

\section{Anomalous magnetic moments of quarks}
\label{Sect:quark}

This section is devoted to the discussion of possible modifications of
the GDH sum rule that arise in a non-naive, yet canonical manner if
one assumes quarks to possess anomalous magnetic moments $\kappa_q$
\cite {Kawarabayashi66c,Khare75}.  The reader may wonder why this is a
field of investigation, because in the standard model quarks are
pointlike, thus having $\kappa_q=0$.  (As a matter of fact, some of
the most vital foundations of the standard model -- like
renormalizability -- depend upon the exact vanishing of $\kappa_q$.)
But ``quarks'' can also mean ``constituent quarks'' -- those nice
objects having masses on hadronic scales and quantum numbers of bare
quarks.  One can fancy a constituent quark as a bare quark being
surrounded by a cloud of mesons or quark-antiquark pairs and gluons.
Such an object clearly has an internal structure.  Moreover,
constituent-quark models are non-perturbative models, so there is no
reason -- apart from aesthetics, if you want -- for the anomalous
magnetic moment of a constituent quark to vanish.  And as the GDH sum
rule is occasionally investigated in the context of constituent-quark
models (see, e.g., \Refs {Drechsel93}\nocite
{DeSanctis94,DeSanctis96,Li93}--\plaincite {Cardarelli97}), it is by
all means worth while considering the case $\kappa_q\neq0$.

\subsubsection{Modified charge-density algebra}
To derive the GDH sum rule within the current-algebra approach, one
needs the equal-times commutator $[J^0(x),J^0(y)]_\tet$ of electric
charge densities.  In \Sect {ETCA.comm}, I presented a derivation
of this commutator from canonical anticommutation relations \eq {ACR}
among quark fields, assuming the electromagnetic current to have the
form
\begin{equation} \label{QCD-curr2}
 J^\mu(x) = \sum_{q=u,d,s} Z_q\, \bar q(x)\, \gamma^\mu\, q(x)
 = \bar\psi(x)\, \gamma^\mu Z^{(\tq)} \, \psi(x),
\end{equation}
where
\begin{equation}
 \psi(x) = \begin{pmatrix} u(x)\\ d(x)\\s(x) \end{pmatrix}
 \qquad \text{and} \qquad Z^{(\tq)} =
 \begin{pmatrix}
  \frac23 & 0 & 0 \\
  0 & -\frac13 & 0 \\
  0 & 0 & -\frac13
 \end{pmatrix}
\end{equation}
for three flavors u, d, and s.\footnote {In this section, I only consider the
case of three flavors, but all formulae can be applied to the two-flavor
case by letting $Z_\ts=\kappa_\ts=0$ throughout.}  The charge-density algebra
then reads
\begin{equation} \label{naive-comm3}
 \etcomm {J^0(x)} {J^0(y)} = 0.
\end{equation}

However, if the current does not have the minimal form \eq {QCD-curr2}
-- that is to say, if it does not arise from a Lagrangian density
$\bar\psi(i\parslash-m)\psi$ by virtue of minimal coupling -- then the
algebra \eq {naive-comm3} may be modified.  A particularly interesting
example is the addition of a Pauli term to the Lagrangian, which adds
to the current a derivative of a tensor current \cite
{Kawarabayashi66c,Khare75}:
\begin{equation} \label{mod-curr}
 J^\mu(x) = \bar\psi(x)\, \gamma^\mu Z^{(\tq)}\, \psi(x)
 + \frac{\partial}{\partial x^\nu}
 \left( \bar\psi(x)\, \sigma^{\mu\nu} \frac{\kappa^{(\tq)}}{2m^{(\tq)}}\,
 \psi(x) \right).
\end{equation}
Here, $\kappa^{(\tq)}$ and $m^{(\tq)}$ are diagonal $3\ntimes3$
matrices of quark anomalous magnetic moments and quark masses,
respectively:
\begin{equation}
 \kappa^{(\tq)} =
 \begin{pmatrix}
  \kappa_\tu & 0 & 0 \\
  0 & \kappa_\td & 0 \\
  0 & 0 & \kappa_\ts
 \end{pmatrix}, \qquad m^{(\tq)} =
 \begin{pmatrix}
  m_\tu & 0 & 0 \\
  0 & m_\td & 0 \\
  0 & 0 & m_\ts
 \end{pmatrix}.
\end{equation}
Thus,
\begin{equation}
 \frac{\kappa^{(\tq)}}{2m^{(\tq)}} =
 \tfrac12\, \kappa^{(\tq)}\, \bigl( m^{(\tq)} \bigr)^{-1} =
 \begin{pmatrix}
  \frac{\kappa_\tu}{2m_\tu} & 0 & 0 \\
  0 & \frac{\kappa_\td}{2m_\td} & 0 \\
  0 & 0 & \frac{\kappa_\ts}{2m_\ts}
 \end{pmatrix}.
\end{equation}
From $\gamma^0\sigma^{0k}=i\gamma^k$, and taking $\mu=0$ in \Eq {mod-curr},
the charge density is obtained as
\begin{equation} \label{mod-charge}
 J^0(x) = \psi^\dagger(x)\, Z^{(\tq)} \, \psi(x)
 + i\bnabla\ndot \left( \psi^\dagger(x)\, \bgamma
 \frac{\kappa^{(\tq)}}{2m^{(\tq)}}\, \psi(x) \right).
\end{equation}
The equal-times commutator $[J^0(x),J^0(y)]_\tet$ is worked out by
repeatedly employing the canonical anticommutation relations \eq
{ACR} analogously to the procedure presented in \Sect {ETCA.comm},
p.~\pageref {curr-comm} of this thesis.  Most generally, the commutator of two
fermion bilinears is obtained as
\begin{align} \label{pre-gen-comm}
 & \etcomm {\psi^\dagger(x)\, \Gamma\Lambda\, \psi(x)}
  {\psi^\dagger(y)\, \Gamma'\Lambda'\, \psi(y)} \notag \\*
 & \qquad = \frac12\, \psi^\dagger(x)\,
 \bigl( [\Gamma,\Gamma'] \{\Lambda,\Lambda'\}
  + \{\Gamma,\Gamma'\} [\Lambda,\Lambda'] \bigr)\, \psi(x)\, \dirac3(\bx-\by),
\end{align}
where $\Gamma,\Gamma'$ act in Dirac space and $\Lambda,\Lambda'$ act
in flavor space.  That is to say, $\Gamma,\Gamma'$ are any linear
combinations of the matrices 1, $\gamma^\mu$, $\sigma^{\mu\nu}$,
$\gamma^\mu\gamma_5$, and $\gamma_5$, while $\Lambda,\Lambda'$ are
linear combinations of Gell-Mann matrices $\lambda_{0\ldots8}$.
Presently, I consider only the diagonal flavor matrices $Z^{(\tq)}$
and $\kappa^{(\tq)}/2m^{(\tq)}$, which mutually commute.  Thus, the
second term of the sum in \Eq {pre-gen-comm} can be omitted,
leaving
\begin{equation} \label{general-comm}
 \etcomm {\psi^\dagger(x)\, \Gamma\Lambda\, \psi(x)}
  {\psi^\dagger(y)\, \Gamma'\Lambda'\, \psi(y)} = \psi^\dagger(x)\,
 [\Gamma,\Gamma'] \Lambda\Lambda' \, \psi(x)\, \dirac3(\bx-\by).
\end{equation}
From this relation it can be seen immediately that the canonical piece
$\psi^\dagger(x)\,Z^{(\tq)}\,\psi(x)$ of the charge density \eq
{mod-charge} commutes not only with itself, but also with the additional
piece.  Thus one is left with
\begin{align}
 \etcomm {J^0(x)} {J^0(y)} & = - \frac{\partial^2}{\partial x^i\partial y^j}\,
  \etcomm {\psi^\dagger(x)\,\gamma^i\tfrac{\kappa^{(\tq)}}{2m^{(\tq)}}\,\psi(x)}
   {\psi^\dagger(y)\,\gamma^j\tfrac{\kappa^{(\tq)}}{2m^{(\tq)}}\,\psi(y)}
  \notag \\
 & = - \frac{\partial^2}{\partial x^i\partial y^j}\,
  \Bigl( \psi^\dagger(x)\, [\gamma^i,\gamma^j]
  \bigl(\tfrac{\kappa^{(\tq)}}{2m^{(\tq)}}\bigr)^2\, \psi(x)\, \dirac3(\bx-\by)
  \Bigr).
\end{align}
Using the relation
\begin{equation}
 [\gamma^i,\gamma^j] = - 2i\,\eps^{ijk}\gamma^0\gamma^k\gamma_5,
\end{equation}
one arrives at
\begin{equation} \label{kappa-comm}
 \boxed{ \etcomm {J^0(x)} {J^0(y)}
  = 2i\, \Bigl[ \bnabla \ntimes \Bigl(\bar\psi(x)\, \bgamma\gamma_5
  \bigl( \tfrac{\kappa^{(\tq)}}{2m^{(\tq)}} \bigr)^2\, \psi(x) \Bigr)\Bigr]\,
  \ndot \bnabla \dirac3(\bx-\by) }
\end{equation}

\subsubsection{Modified finite-momentum GDH sum rule}
Observe that the charge-density algebra \eq {kappa-comm} has
exactly the form \eq {a-comm} with the axial-vector field $a^\mu(x)$
given by
\begin{align} \label{a-kappa}
 a^\mu(x) & = 2\, \bar\psi(x)\, \gamma^\mu\gamma_5
  \bigl( \tfrac{\kappa^{(\tq)}}{2m^{(\tq)}} \bigr)^2\, \psi(x) =
 \sum_{q=u,d,s} \frac{\kappa_q^2}{2m_q^2}\,
 \bar q(x)\, \gamma^\mu\gamma_5\, q(x) \notag \\*
 & = \sqrt{\frac23} \left(
   \frac{\kappa_\tu^2}{2m_\tu^2} + \frac{\kappa_\td^2}{2m_\td^2} +
   \frac{\kappa_\ts^2}{2m_\ts^2} \right) J_{5\,0}^\mu(x) + 
   \left( \frac{\kappa_\tu^2}{2m_\tu^2} - \frac{\kappa_\td^2}{2m_\td^2} \right)
   J_{5\,3}^\mu(x) \notag \\*
 & \quad + \frac1{\sqrt3} \left(
   \frac{\kappa_\tu^2}{2m_\tu^2} + \frac{\kappa_\td^2}{2m_\td^2} -
   \frac{\kappa_\ts^2}{m_\ts^2} \right) J_{5\,8}^\mu(x).
\end{align}
Axial-vector currents $J_{5a}^\mu(x)$ are defined in \Eq {J-def.A}.
Their coefficients in \Eq {a-kappa} are easily obtained upon using
orthogonality of Gell-Mann matrices, \Eq {lambda-norm}.  According to
\Sect {mod-FMGDH}, \Eq {a-kappa} leads to the following modification of
the finite-momentum GDH sum rule:
\begin{equation} \label{kappa-FMGDH}
 -4\pi^2\alpha \left( \frac{\kappa^2}{2M^2}
  - \frac{(Z+\kappa)^2}{2(p^0)^2} - \tilde g_\tA \right) =
  \int_{\nu_{\text{thr}}}^\infty\! \frac{\td \nu}\nu\,
  8\pi \im f_2\!\left(\nu, \frac{M^2\nu^2}{(p^0)^2}\right),
\end{equation}
where the constant $\tilde g_\tA$ is determined by
\begin{equation} \label{gA(a)}
 \tilde g_\tA\, \bar u(p,\tfrac12)\, \gamma^\mu\gamma_5\, u(p,\tfrac12) =
 \me {p,\tfrac12} {a^\mu(0)} {p,\tfrac12}.
\end{equation}
Introducing \Eq {a-kappa}, as well as axial charges $g_{\tA a}$ of the
nucleon, defined by means of 
\begin{equation}
 g_{\tA a}\, \bar u(p,\tfrac12)\, \gamma^\mu\gamma_5\, u(p,\tfrac12) =
   \me {p,\tfrac12} {J_{5a}^\mu(0)} {p,\tfrac12},
\end{equation}
one obtaines the modification constant
\begin{subequations}
\begin{align}
 \tilde g_\tA & = \sqrt{\frac23} \left(
   \frac{\kappa_\tu^2}{2m_\tu^2} + \frac{\kappa_\td^2}{2m_\td^2} +
   \frac{\kappa_\ts^2}{2m_\ts^2} \right) g_{\tA0} + 
   \left( \frac{\kappa_\tu^2}{2m_\tu^2} - \frac{\kappa_\td^2}{2m_\td^2} \right)
   g_{\tA3} \notag \\*
 & \quad + \frac1{\sqrt3} \left(
   \frac{\kappa_\tu^2}{2m_\tu^2} + \frac{\kappa_\td^2}{2m_\td^2} -
   \frac{\kappa_\ts^2}{m_\ts^2} \right) g_{\tA8}.
\end{align}
Assuming isospin symmetry $g_{\tA0}^\tp=g_{\tA0}^\tn$,
$g_{\tA8}^\tp=g_{\tA8}^\tn$, and $g_{\tA3}^\tp=-g_{\tA3}^\tn=g_\tA$
(cf.\ \Eqs {iso-sym}), the proton-neutron difference is obtained as
\begin{equation} \label{kappa-gA.p-n}
 \tilde g_\tA^\tp - \tilde g_\tA^\tn = 2 g_\tA
 \left( \frac{\kappa_\tu^2}{2m_\tu^2} - \frac{\kappa_\td^2}{2m_\td^2} \right),
\end{equation} \label{kappa-gA}%
\end{subequations}
where $g_\tA=1.26$ is the familiar neutron-beta-decay constant.

\subsubsection{Infinite-momentum limit}
To gain something useful from the (as-such needless) finite-momentum
GDH sum rule \eq {kappa-FMGDH}, one has to assume the legitimacy of
the infinite-momentum limit, i.e., function $f_2(\nu,q^2)$ is
conjectured to allow for the $p^0\to\infty$ limit to be dragged into
the $\nu$ integral.  Then one can write
\begin{equation} \label{kappa-GDH}
 \boxed{ -4\pi^2\alpha \left( \frac{\kappa^2}{2M^2} - \tilde g_\tA \right) =
  \int_{\nu_0}^\infty\! \frac{\td\nu}\nu
  \bigl(\sigma_{1/2}(\nu) - \sigma_{3/2}(\nu)\bigr) }
\end{equation}
where the modification constant $\tilde g_\tA$ is given by \Eqs {kappa-gA}.

What evidence is there for assuming the legitimacy of the
infinite-momentum limit?  As pointed out in \Sect {ETCA}, from the
commutator \eq {kappa-comm} nothing can be inferred concerning this
problem.  Still, the modification, if present, must be expected to
survive the infinite-momentum limit.  This is because, as claimed by
Dicus and Palmer \cite {Dicus72}, anomalous magnetic moments of quarks
also affect current commutators on the light-cone, and as I explained
in \Sect {LCCA}, the light-cone method circumvents the
infinite-momentum limit.  As a rule of thumb, one may say: ``If you
have a non-naive commutator of whatever origin, leading to a
modification of the finite-momentum GDH sum rule, and if you'd like to
know whether this modification will survive the infinite-momentum
limit, try and check whether the same kind of commutator emerges
within light-cone current-algebra!''\footnote {Fortunately, there is
no need to examine the effect of the \emph {chiral anomaly} on
light-cone current commutators, since the investigation presented in
\Sect {PRP} clearly shows that the anomaly does not modify the GDH sum
rule.}

\subsubsection{Discussion}
It was Kawarabayashi and Suzuki \cite {Kawarabayashi66c}, who first
observed that if the current-algebra derivation is applied, anomalous
magnetic moments of quarks bring about a modification of the GDH sum
rule (Provided, of course, that this modification survives the
infinite-momentum limit).  Khare \cite {Khare75} gives a more
comprehensive treatment, discussing other sum rules, too.  Note,
however, that the extended electromagnetic current \eq {mod-curr}
involves the piece
\begin{equation} \label {kappa-tr}
 \frac13 \left( \frac {\kappa_\tu}{2m_\tu} + \frac {\kappa_\td}{2m_\td} +
 \frac {\kappa_\ts}{2m_\ts} \right)
 \frac {\partial}{\partial x^\nu}
 \bigl( \bar\psi(x) \sigma^{\mu\nu} \psi(x) \bigr),
\end{equation}
i.e., the derivative of the \emph {flavor-singlet} tensor current,
which is not included in \Refs {Kawarabayashi66c} and \plaincite
{Khare75}.  This is a severe deficiency of \Refs {Kawarabayashi66c}
and \plaincite {Khare75}, since it restricts their results to the case
that the sum of anomalous magnetic moments occurring in \Eq {kappa-tr}
vanishes.

It may be instructive to confront my results with the ones of de
Sanctis, Drechsel, and Giannini \cite {DeSanctis94}, who effectively
perform the equal-times current-algebra derivation of the GDH sum
rule, starting from a constituent-quark current that includes
relativistic corrections.  In Eq.\ (15) of \Ref {DeSanctis94}, a
modification of the sum rule is suggested, which has the form \eq
{kappa-GDH} with
\begin{equation} \label{DeSanctis}
 \tilde g_\tA = \sum_{i=1}^3 \frac {\kappa_i^2}{2m_i^2}\,
 \langle \sigma_{\tz i} \rangle,
\end{equation}
where $\langle\sigma_{\tz i}\rangle$ is the spin of the $i$th
constituent, projected onto the nucleon's spin.  (Note that the upper
limit 3 of the sum in \Eq {DeSanctis} is the number of baryon
constituents, i.e., $N_\tc$, \emph {not} the number of flavors
considered.)  \Eq {DeSanctis} is in qualitative agreement with my \Eqs
{a-kappa} and \eq {gA(a)}, since the operator $\bar
q\gamma^\mu\gamma_5q$ tests each constituent of flavor $q$ for its
spin.

As I pointed out at the beginning of this section, QCD does not predict
a modification of the GDH sum rule due to quark anomalous magnetic moments,
since current quarks are pointlike.  On the other hand, constituent-quark
models apparently predict a modification, as soon as constituent quarks
are allowed to have $\kappa_q\neq0$.  What is the solution of this
paradox?

In other words: Does \Eq {kappa-GDH} really tell us that the GDH
experiment will measure the anomalous magnetic moments of constituent
quarks?  My answer is: definitely no!  Rather, the modification
constant in \Eq {kappa-GDH} should better be written on the other
side of the equation, because it gives the ``amount of principle
incorrectness'' of the integral due to neglecting all final states
other than three-quark bound states.  That is to say, while the QCD
current operator may transform a one-nucleon state into any QCD state
having the quantum numbers of $\gamma$N, the constituent-quark-model
current only transforms the nucleon into another three-quark bound
state subsequently decaying into $\pi$N or other hadronic states,
i.e., a \emph {resonance}.  So, the full non-resonant background is
missing from the integral of \Eq {kappa-GDH}.  I call this lack a
``principle incorrectness'' since I do \emph {not} mean numerical
uncertainties coming from such things as narrow-width approximation,
simplicity of the nucleon wave function or whatsoever.  My persuasion
that the missing non-resonant part of the GDH integral equals just the
modification constant on the left-hand side of \Eq {kappa-GDH}
feeds on the following consideration: As mentioned before, anomalous
magnetic moments of quarks signal some internal structure.  For
instance, if you recall the pion-cloud picture of the
constituent quark, it is natural that the photon can strike out a pion
from the quark.  This, however, is a non-resonant process!  In the
non-relativistic limit, the equality of modification constant on the
one hand and non-resonant part of the GDH integral on the other hand,
both being generated by anomalous magnetic moments of quarks, was
recently proved by Cardarelli, Pasquini, and Simula \cite
{Cardarelli97}.  Note, however, that if one introduces into \Eq
{kappa-gA} the value $g_\tA=5/3$, which is standard in nonrelativistic
constituent-quark models, then Eq.\ (12) of \Ref {Cardarelli97} is
only half of my result (or the one of \Refs
{Kawarabayashi66c,Khare75}).  Honestly, I do not know whether this is
due to an error or has a deeper meaning.

\subsubsection{Conclusion}
I conclude that a modification of the GDH sum rule must not be
expected from anomalous magnetic moments of quarks, neither in QCD
(because current quarks do not possess anomalous magnetic moments) nor
in constituent-quark models (because the apparent modification merely
represents a portion missing from the modelled GDH integral). After
all, this conclusion restores consistency between constituent-quark
models and QCD.

%\markboth {Chapter \thechapter.~~Possible sources of modifications}
%{\ref{Sect:Ying}.~~``Spontaneous breakdown of electromagnetic gauge symmetry''}

\section[``Localized spontaneous breakdown of electromagnetic gauge symmetry'']
 {``Localized spontaneous breakdown of electromagnetic\\ gauge symmetry''}
%\markright{\thesection.~~``Spontaneous breakdown of electromagnetic gauge symmetry''}
\label{Sect:Ying}

In this section, I comment on a proposed modification of the GDH sum
rule stated by Ying \cite {Ying96a,Ying96b} in 1996.  Ying claims
that this modification is possible even if both the naive current
commutation relation is valid and the infinite momentum limit is
legitimate.  It is due to ``the presence of a localized region inside
the nucleon, in which the electromagnetic gauge symmetry is
spontaneously broken down'' \cite {Ying96a}.  I mention in advance
that I do not agree with this point of view.

\subsubsection{Erratum}
To begin with, I want to point out two mistakes in the article under
consideration.  Firstly, it has to be noted that there are too many
invariant amplitudes in Ying's decomposition of the forward virtual
Compton amplitude, Ref.\ \plaincite {Ying96a}, Eq.\ (4).  The correct
amplitude without requiring gauge invariance reads
\begin{align}  \label{TmunuYing}
 T^{\mu\nu}(p,q) = \bar u(p,s) \bigl[
 & F_1g^{\mu\nu} - F_2q^\mu q^\nu + F_3p^\mu p^\nu
   - F_4(p^\mu q^\nu+p^\nu q^\mu) + iF_5\sigma^{\mu\nu} \notag\\*
 & + iF_6(p^\mu\sigma^{\nu\alpha}q_\alpha-p^\nu\sigma^{\mu\alpha}q_\alpha)
   + iF_7(q^\mu\sigma^{\nu\alpha}q_\alpha-q^\nu\sigma^{\mu\alpha}q_\alpha)
   \bigr] u(p,s),
\end{align}
where I omitted a factor of $1/2M$ on account of differing normalizations of
one-nucleon states.  (The amplitude \eq {TmunuYing} is
dimensionless, while Ying's conventions are such that $T^{\mu\nu}$ has
dimension GeV$^{-1}$.)  Ying's eighth tensor
$i\eps^{\mu\nu\alpha\beta}q_{\alpha}p_{\beta}\qslash\gamma_5$ is in
fact linearly dependent on the others, owing to the relation
\begin{align}
 \bar u(p,s) \bigl[ & m(\nu^2-q^2)\sigma^{\mu\nu}
  + \nu(p^\mu\sigma^{\nu\alpha}q_\alpha-p^\nu\sigma^{\mu\alpha}q_\alpha)
  \notag\\*
 & - m(q^\mu\sigma^{\nu\alpha}q_\alpha-q^\nu\sigma^{\mu\alpha}q_\alpha)
  + \eps^{\mu\nu\alpha\beta}q_{\alpha}p_{\beta}\qslash\gamma_5 \bigr] u(p,s)
  = 0,
\end{align}
which is most easily verified by writing
$\bar{u}\Gamma{u}=\tr(\Gamma{u}\bar{u})$, where $\Gamma$ is any
linear combination of the matrices 1, $\gamma^\mu$, $\sigma^{\mu\nu}$,
$\gamma^\mu\gamma_5$, and $\gamma_5$, and by using the relation
$u(p,s)\bar{u}(p,s)=\frac12(\pslash+M)(1+\gamma_5\sslash)$, trace
theorems, and identity \eq {anti5}.  Sometimes it is
quite tricky to realize the linear dependence of tensors in the
$\bar{u}\Gamma{u}$ notation, and eliminating the spinors in the manner
just described often helps.  Note that because gauge invariance imposes
\emph {three} constraints (see \Eqs {gauge-inv} below), and since, of
course, the usual number of \emph {four} invariant amplitudes must
remain in the ultimate (gauge invariant) forward virtual Compton
amplitude, the total number of $3+4=7$ (not 8) non-gauge-invariant
amplitudes must indeed be expected.

As a matter of fact, Ying's amplitude $F_8(\nu,q^2)$ can consistently
be dropped throughout the entire paper without affecting the main
message.

Secondly, as I demonstrated in \Sect {ETCA.comm}, the naive
equal-times commutator of electric current densities reads
\begin{equation} \label{naive-curr-comm2}
 \etcomm {J^\mu(x)} {J^\nu(y)} = 2i\,\eps^{0\mu\nu\lambda}\,
  \bar\psi(x)\, \gamma_\lambda\gamma_5\bigl(Z^{(\tq)}\bigr)^2\psi(x)\,
  \dirac3(\bx-\by),
\end{equation}
where $Z^{(\tq)}$ is the quark charge matrix.  The right-hand side of
\Eq {naive-curr-comm2} involves an axial-vector current, in contrast to
Eq.\ (17) of Ref. \plaincite {Ying96a}, which involves a tensor current.
In particular, since $\eps^{00\nu\lambda}=0$, the charge density
$J^0(x)$ naively commutes with \emph {each} component of the current.
This is vital since naively, the charge density generates the local gauge
transform, and the current is a gauge invariant quantity.

Replacing Ying's erroneous commutator by the correct one, \Eq
{naive-curr-comm2}, amounts to substituting the axial charge of the
nucleon for its tensor charge in the entire paper.  Again, this will
not affect the message of the paper, since the actual values of these
charges are never used.

\subsubsection{Imposing gauge invariance}
The gauge invariance condition, $q_\mu T^{\mu\nu} = 0$, imposed on the
invariant decomposition \eq {TmunuYing}, gives rise to the
constraints
\begin{subequations}
\begin{align}
 F_1(\nu,q^2) &= q^2F_2(\nu,q^2) + M\nu F_4(\nu,q^2), \\*
 M\nu F_3(\nu,q^2) &= q^2 F_4(\nu,q^2), \\
 F_5(\nu,q^2) &= M\nu F_6(\nu,q^2) + q^2 F_7(\nu,q^2), \label{F5}
\end{align} \label{gauge-inv}%
\end{subequations}
which are Eqs.\ (5)--(7) of Ying \cite {Ying96a}.

Since I am concerned with spin physics, I focus my attention on the
antisymmetric part of amplitude \eq {TmunuYing}, i.e., the part
involving the invariant amplitudes $F_{5,6,7}$.  Equating
decompositions \eq {inv-dec} and \eq {TmunuYing}, one finds the
following relation between the familiar spin-dependent forward Compton
amplitudes\footnote {Observe that these are not the functions
$A_{1,2}(\nu,q^2)$ of Ying.  The relation is
$A_1=2M^2A_1^{\text{Ying}}$, $A_2=2M^4A_2^{\text{Ying}}$.}
$A_{1,2}(\nu,q^2)$ and Ying's amplitudes $F_{6,7}(\nu,q^2)$,
\begin{subequations}
\begin{align}
 A_1 &= 2M^3F_6,\\*
 A_2 &= 2M^3F_7.
\end{align}
\end{subequations}
%Notwithstanding all the talking about when the
%constraint of gauge invariance has to be imposed, the quantity $F_5$
%can simply be regarded as a shorthand of the specific combination
%given by \Eq {F5}.
The quantity $F_5$ ought to be regarded as a mere shorthand of the
specific combination given by \Eq {F5}.  In view of
\Eq {f2nuq2}, one has the simple relation $\nu f_2(\nu,q^2) =
\alpha F_5(\nu,q^2)$.

\subsubsection{High-energy behavior from naive current commutator?}
Ying employs the Bjorken-Johnson-Low (BJL) technique \cite{Bjorken66,Johnson66}
to derive the high-energy behavior\footnote {As mentioned
above, the constant in this relation is not a tensor charge, as in
Eq.\ (20) of Ref.\ \cite {Ying96a}, but an axial charge, or more
precisely: a definite linear combination of isovector axial charge and
flavor-singlet and flavor-octet isoscalar axial charges.}
\begin{equation} \label{F5-high}
 F_5(\nu,q^2) \xrightarrow[\nu\to\infty]{} \frac{\text{const}}{\nu}
\end{equation}
from the naive current commutator \eq {naive-curr-comm2}.
It should be pointed out, however, that using the naive commutator in this
context is unjustified, as Johnson and Low already demonstrated in their
original paper \cite {Johnson66} (see also \cite {Jackiw85}).  In fact,
the BJL technique may only be employed to derive a (non-naive) commutator
from the high-energy behavior of an amplitude, i.e., just the other
way round.

\subsubsection{BJL limit and Regge limit}
Ying claims that one can proceed from the BJL limit $q^0\to\infty$,
$\bq$ fixed, to the desired Regge limit $\nu\to\infty$, $q^2$ fixed,
by using an infinite-momentum frame.  However, as I  demonstrated
in \Sects {IML} and \sect {PRP}, the legitimacy of the infinite-momentum
technique is a highly fragile assumption.  (See \Sect {PRP.summary} for
an explicit instance of non-coincidence of BJL limit and Regge limit.)

\subsubsection{Invalidation of the Burkhardt-Cottingham sum rule}
One of the more serious problems with the results of Ref.\ \plaincite
{Ying96a} is that they obviously invalidate the Burkhardt-Cottingham
sum rule \cite {Burkhardt70}.  This sum rule is a
superconvergence relation for the amplitude $A_2(\nu,q^2)$, which requires
the high-energy behavior
\begin{equation}
 \lim_{\nu\to\infty} \nu A_2(\nu,q^2) = 0,
\end{equation}
while Ying concludes $F_7\propto\nu$, and hence $\nu A_2\propto\nu^2$.
Although it is indeed not entirely certain whether the
Burkhardt-Cottingham integral converges at finite values of $Q^2=-q^2$
(see Jaffe \cite {Jaffe90} for an excellent review), such a dramatic
behavior vehemently contradicts widely accepted Regge
asymptotics of amplitude $A_2$.  According to Ioffe, Khoze, and
Lipatov \cite {Ioffe84}, the worst-case singularity in complex
angular momentum plane that contributes to $A_2$ is a three- or
more-pomeron cut, giving rise to the high-energy behavior
$A_2\propto1/\ln^5\nu$.

\subsubsection{Discontinuous $q^2$ evolution}
Ying's argumentation in favor of a modification of the GDH sum rule proceeds
as follows.  Define the function
\begin{equation}
 \rho(q^2) := -\lim_{\nu\to\infty} \frac{q^2F_7(\nu,q^2)}{\nu}.
\end{equation}
Since amplitude $F_7(\nu,q^2)$ lacks massless poles, the numerator
$q^2F_7$ vanishes at $q^2=0$, and thus $\rho(0)=0$.  On the other hand,
the limit
\begin{equation}
 \rho_\infty := \lim_{\substack{q^2\to0\\ q^2\neq0}} \rho(q^2)
\end{equation}
is claimed to be non-vanishing due to ``spontaneous breakdown of
electromagnetic gauge symmetry'' \cite {Ying96a}.  Hence the function
$q^2F_7(\nu,q^2)$ is asymptotically linear in $\nu$ for any finite
$q^2$.  Moreover, as noted above, the sum $F_5 = M\nu F_6 + q^2 F_7$
is pretended to be asymptotically proportional to the inverse of
$\nu$.  Consequently, amplitude $F_6(\nu,q^2)$ must approach a
constant that cancels the high-energy behavior of $F_7$,
\begin{equation}
 F_6(\infty,q^2) = -\lim_{\nu\to\infty} \frac{q^2F_7(\nu,q^2)}{M\nu} =
  \frac{\rho(q^2)}M,
\end{equation}
or
\begin{equation}
 A_1(\infty,q^2) = 2M^2\rho(q^2),
\end{equation}
which enforces a subtraction at infinity in the dispersion relation
\begin{equation} \label{I1Ying}
 \frac14 \bigl( \bar A_1(0,q^2) - A_1(\infty,q^2) \bigr) =
  \!\int_{\nu_0}^\infty\! \frac {\td\nu}{\nu}\, G_1(\nu,q^2) =: I_1(q^2).
\end{equation}
Here, $2\pi\,G_1=\im A_1$ denotes the familiar spin dependent nucleon
structure function (the absorptive part of the forward virtual Compton
amplitude), and
\begin{equation}
 \bar A_1(\nu,q^2)
 := A_1(\nu,q^2) - \frac {4M^2q^2\,F_1(q^2)G_\tM(q^2)}{(2M\nu)^2-(q^2)^2}
\end{equation}
is the amplitude with its Born pole subtracted (cf.\ \Eq {Born.A1}),
obeying the low-energy theorem $\bar A_1(0,0)=-\kappa^2$, where
$\kappa$ is the anomalous magnetic moment of the nucleon.

Now, if all this was true, then the integral \eq {I1Ying} would be \emph
{discontinuous} as a function of $q^2$,
\begin{equation} \label{I10Ying}
 I_1(0) = -\frac{\kappa^2}4,
\end{equation}
while
\begin{equation}
 I_1(0\pm) := \lim_{\substack{q^2\to0\\ q^2\neq0}}I_1(q^2) =
  -\frac{\kappa^2}4 - \frac{m^2\rho_{\infty}}2,
\end{equation}
with
$\rho_{\infty}^\tp\approx-\rho_{\infty}^\tn\approx1\text{~GeV}^{-2}$
\cite {Ying96a}.  However, as long as there are no massless particles
degenerate with the photon in all quantum numbers, the forward Compton
amplitude -- and with it the integral of its absorptive part -- should
be \emph {analytic} in $q^2$ with all cuts and poles well separated
from the point $q^2=0$.  But even if it \emph {was} discontinuous,
then the genuine GDH integral, which simply has nothing to do with
virtual photons, would certainly not acquire the value $I_1(0\pm)$,
but rather $I_1(0)$, \Eq {I10Ying}, which is left unmodified according
to Ying.  So what?

\subsubsection{Conclusion}
To conclude this section, I stress that the mechanism proposed in
\Refs {Ying96a} and \plaincite {Ying96b} -- albeit possibly relevant
with respect to other fields of elementary particle physics --
certainly does \emph {not} affect the GDH sum rule, mainly because the
author misunderstood the BJL technique, which cannot be utilized to
derive the high-energy behavior of an amplitude from a commutator,
and secondly because he defined the GDH integral by an inappropriate
$q^2$ limiting procedure.

\section{Modified low-energy theorem?}
\label{Sect:modLET}

A number of people (including myself) has been fooled by the fact that
Low's derivation \cite {Low54} of the low-energy theorem (LET) \eq
{LET.LGG} \emph {looks like} it would be modified by a non-naive
current commutator.  In a few instances this delusion even came to
publication \cite {Kawarabayashi66c,LEGS94}.  However, as already
noted by Khare \cite {Khare75}, one has to recall that the LET holds
for the \emph {physical} Compton amplitude, while the current
commutator, in the low-energy limit, has impact on the \emph
{T-product} amplitude only.  For any specific form of the commutator,
the sea\-gull, i.e., the difference of physical and T-product amplitudes,
is uniquely determined by the requirement of gauge invariance of the
physical amplitude \cite {Jackiw85}.  Taking into account the
additional contribution of the sea\-gull to the LET, one soon realizes
that it cancels the contribution of the commutator in any reasonable
case, albeit in this way the derivation gets unnecessarily awkward.  As
a matter of fact, the easiest way to proceed is to forget about Low's
derivation of the LET (just keeping in mind that it does by no means enforce
the current commutator to have its canonical form), focussing instead
on the derivation by means of Feynman diagrams given by Gell-Mann and
Goldberger \cite {Gell-Mann54}, or the one based on dispersion
relations for helicity amplitudes, as presented 14 years later by
Abarbanel and Goldberger \cite {Abarbanel68}.  After all, these
approaches pleasantly remind one of the fact that LETs follow from
symmetries of nature, which are much more fundamental than any kind of
anomalous charge-density algebra.

\clearpage{\pagestyle{empty}\cleardoublepage}
\chapter{$Q^2$ evolution of the GDH sum rule}
\label{Ch:Q2}

\begin{figure}[tb]
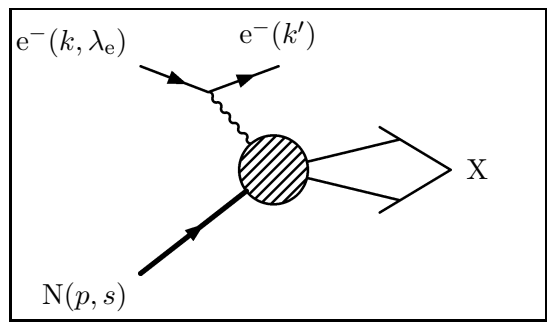

 \begin{displaymath}
  \boxedgraph(5,3){
   \fmfleft{i1,i2} \fmfright{o} \fmfbottom{b} \fmftop{t}
   \fmf{fermion,width=thick}{i1,v}
   \fmf{phantom}{i2,v}
   \fmf{phantom,tension=1.5}{v,o}
   \fmfblob{1\ul}{v}
   \fmf{phantom}{t,a1} \fmf{plain,tension=5}{a1,a2} \fmf{plain,tension=2}{a2,o}
   \fmf{phantom}{b,a3} \fmf{plain,tension=5}{a3,a4} \fmf{plain,tension=2}{a4,o}
   \fmffreeze
   \fmf{fermion,tension=2}{i2,f}
   \fmf{fermion}{f,t}
   \fmf{photon}{f,v}
   \fmf{plain}{a2,v,a4}
   \fmflabel{N$(p,s)$}{i1} \fmflabel{X}{o}
   \fmflabel{e$^-(k,\lambda_{\rm{e}})$}{i2} \fmflabel{e$^-(k')$}{t}
  }
 \end{displaymath}
 \caption[]{
  Polarized inclusive electron-nucleon scattering
  in the one-photon-exchange approximation.
  Cross sections for all final states X are summed over.
  \label{Fig:incl}}
\end{figure}

In this chapter, I investigate the $Q^2$ evolution of the GDH sum rule,
i.e., the transition from the photoabsorption process depicted in
\Fig {photoabs} (p.~\pageref {Fig:photoabs}) to polarized inclusive
electron-nucleon scattering in the one-photon-exchange approximation
as shown in \Fig {incl}.  Experimentally, the charged lepton
might as well be a positron or a muon, but this makes no difference
here, so I would like to stick to the name ``electron'' for
simplicity.%
Data on polarized inclusive electroproduction have been acquired at
SLAC [\plaincite {Baum83}\nocite {Anthony93}--\plaincite {Abe97b,Abe97a}]
and at CERN \cite {Ashman89,Adeva93}.

\begin{figure}[tb]
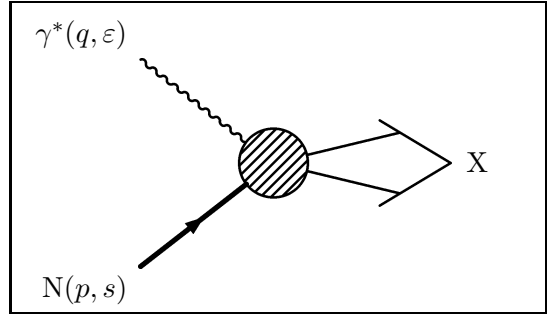

 \begin{displaymath}
  \boxedgraph(5,3){
   \fmfleft{i1,i2} \fmfright{o} \fmfbottom{b} \fmftop{t}
   \fmf{fermion,width=thick}{i1,v}
   \fmf{boson}{i2,v}
   \fmf{phantom,tension=1.5}{v,o}
   \fmfblob{1\ul}{v}
   \fmf{phantom}{t,a1} \fmf{plain,tension=5}{a1,a2} \fmf{plain,tension=2}{a2,o}
   \fmf{phantom}{b,a3} \fmf{plain,tension=5}{a3,a4} \fmf{plain,tension=2}{a4,o}
   \fmffreeze
   \fmf{plain}{a2,v,a4}
   \fmflabel{N$(p,s)$}{i1} \fmflabel{$\gamma^*(q,\varepsilon)$}{i2}
   \fmflabel{X}{o}
  }
 \end{displaymath}
 \caption[]{
  ``Virtual photoabsorption'' on the nucleon.
  Cross sections for all final states X are summed over.
  Conventions for the virtual-photon polarization vector $\eps$ are
  specified in \Sect {kin+X.virt}.
  \label{Fig:virt}}
\end{figure}

The inclusive electroproduction cross section is differential in the
scattering (solid) angle and in the energy of the scattered electron.
Owing to the one-photon-exchange approximation, it is possible to
express this double differential cross section in terms of a
ficticious absorption cross section of a polarized virtual photon on a
polarized nucleon.  This reduction not only simplifies notations, it
also projects out the physically interesting polarization degrees of
freedom, namely transverse and longitudinal polarization of the
virtual photon.  The ``virtual photoabsorption'' process is depicted
in \Fig {virt}.%

This chapter is organized as follows.  In \Sect {kin+X}, kinematical
variables are defined, and polarized inclusive electroproduction cross
sections are presented and expressed in terms of nucleon structure
functions $G_{1,2}(\nu,Q^2)$. Moreover, the relation to virtual
photoabsorption cross sections is given. In \Sect {gen}, I present
different generalizations of the GDH integral to non-zero photon
virtuality $Q^2$, focussing particularly on the physical polarization
degrees of freedom mentioned above.  I argue that at intermediate
$Q^2$, large deviations among different generalizations of the GDH
integral have to be expected.  \Sect {piN} gives the contribution of
pion electroproduction to these integrals, as well as expansions of
this contribution in terms of lowest multipoles.  This last section
aims at providing some important formulae necessary for performing the
saturation of different $Q^2$ evolutions of the GDH integral.  The
saturation will become feasible as soon as experimental data on
polarized pion electroproduction at low $Q^2$ allow for reliably
extracting the multipoles.  Pertinent experiments are under preparation
at Jefferson Lab. 

\section{Kinematics and cross sections}
\label{Sect:kin+X}
The four-momenta of incident and scattered electron are denoted by $k$ and
$k'$, respectively.  The four-momentum of the virtual photon is
$q=k-k'$.  Being the difference of two timelike momenta, $q$ is
always spacelike, i.e., $Q^2:=-q^2>0$.  As usual, the electron's mass is
neglected in explicit calculations.  Lab-frame energies of incident and
scattered electron are denoted by $E=p\ndot{k}/M$ and $E'=p\ndot{k'}/M$,
respectively.  The photon's energy $\nu$ and virtuality $-Q^2$ are
related to $E$, $E'$, and the lab-frame scattering angle $\theta_\te$
via
\begin{equation}
 \begin{split}
  \nu &= E - E', \\
  Q^2 &= 2EE'(1-\cos\theta_\te) = 4EE'\sin^2\frac{\theta_\te}{2}.
 \end{split}
\end{equation}

\begin{figure}[tb]
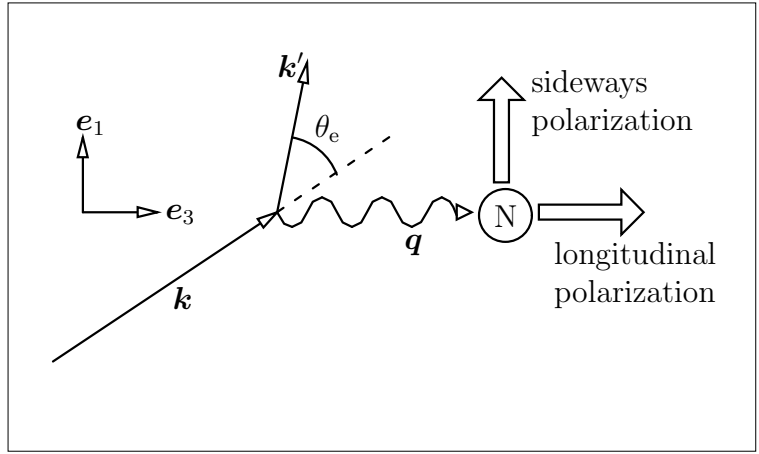

 \begin{center}
  \input spin.eepic
 \end{center}
 \caption[]{
  Kinematics of polarized inclusive electron-nucleon scattering in the
  lab frame.  The incident electron is taken to have definite helicity,
  while the spin vector $\bs$ of the struck nucleon equals $\be_1$ or $\be_3$
  for respective sideways or longitudinal polarization.  Final-state hadrons
  are not shown.
  \label{Fig:spin}}
\end{figure}

As indicated in \Fig {incl}, the incoming electron and
nucleon are polarized, and all final-state polarizations are summed
over.  The incident electron's helicity is denoted by
$\lambda_{\te}=\pm\frac12$. I discriminate nucleon polarizations with
respect to the direction given by the lab-frame three-momentum $\bq$
of the virtual photon.  Adopting the nomenclature of Raskin and
Donnelly \cite {Raskin89}, I distinguish between \emph {longitudinal}
and \emph {sideways} nucleon polarization.  In the former case, the
nucleon's spin $\bs=\bq/|\bq|$, while in the latter case, $\bs$ is
perpendicular to $\bq$ but coplanar with $\bk$ and $\bk'$, i.e., it
lies within the scattering plane.  \Fig {spin} shall illustrate
this state of affairs.%

I now give an explicit coordinate decomposition of the four-vectors
$k$, $k'$, $q$, $p$, and $s$.  Since I work in the lab frame, the
nucleon is at rest,
\begin{subequations}
\begin{equation}
 p^\mu = (M,0,0,0),
\end{equation}
while the photon is taken to be travelling along the positive $\be_3$ axis,
\begin{equation}
 q^\mu = k^\mu - k^{\prime\mu} = (\nu,0,0,\sqrt{\nu^2+Q^2}). \label{kin.q}
\end{equation}
Longitudinal polarization means
\begin{equation} \label{kin.long}
 s^\mu = (0,0,0,1),
\end{equation}
while sideways polarization is given by
\begin{equation} \label{kin.side}
 s^\mu = (0,1,0,0).
\end{equation}
The electron momenta are a little more intricate to work out.  They read
\begin{equation}
 k^\mu = \left( E,\, \frac {EE'\sin\theta_\te}{\sqrt{\nu^2+Q^2}},\, 0,\,
  \frac {E(E-E'\cos\theta_\te)}{\sqrt{\nu^2+Q^2}} \right)
\end{equation}
and
\begin{equation}
 k^{\prime\mu} = \left( E',\,
 \frac {EE'\sin\theta_\te}{\sqrt{\nu^2+Q^2}},\, 0,\,
 \frac {E'(E\cos\theta_\te-E')}{\sqrt{\nu^2+Q^2}} \right).
\end{equation} \label{kin}%
\end{subequations}
The reader may check that $q=k-k'$, $k^2=k^{\prime2}=0$ (electron mass
neglected!), and $\bk\ndot\bk'=|\bk|\,|\bk'|\cos\theta_\te$.

\subsection{Polarized inclusive electroproduction}
In inclusive electroproduction experiments, energy $E'$ and scattering
angles $\theta_{\text{e}}^{\tlab}$, $\phi_{\text{e}}^{\tlab}$ of the
outgoing electron are detected, i.e., one is interested in the
double differential cross section
$\td\sigma_{\lambda_{\te}}/\td{E'}\td\Omega_{\te}^{\tlab}$.  From now
on, I will omit the superscript ``lab'', since all quantities are
taken to be in the lab frame.  To project out polarization
observables, cross sections for negative and postive electron
helicities are subtracted,
\begin{equation} \label{el-Xsect}
  \frac {\td\sigma_{-1/2}} {\td E'\,\td\Omega_{\te}} -
  \frac {\td\sigma_{+1/2}} {\td E'\,\td\Omega_{\te}}.
\end{equation}

\subsubsection{Leptonic and hadronic tensors}
I intend to express these cross sections in terms of polarized nucleon
structure functions $G_{1,2}(\nu,Q^2)$.  To this end, I introduce the
leptonic and hadronic tensors defined by
\begin{equation} \label{Lmunu}
 L_{\mu\nu}(\lambda_{\te}) = \sum_{\lambda_{\te}'=\pm\frac12}
  \bar u(k,\lambda_{\te}) \gamma_\mu u(k',\lambda_{\te}') \,
  \bar u(k',\lambda_{\te}') \gamma_\nu u(k,\lambda_{\te})
\end{equation}
and
\begin{align} \label{Wmunu}
 W^{\mu\nu} &= \frac1{4\pi}
  \sum_{\text{X}} (2\pi)^4\dirac4(p+q-p_{\text{X}})\,
  \langle p,s|J^\mu(0)|\tX\rangle\,
  \langle \tX|J^\nu(0)|p,s\rangle \notag\\
 &= \frac1{4\pi}\!\int\!\td^4x\, e^{iq\ndot x}
  \me {p,s} {[J^\mu(x),J^\nu(0)]} {p,s},
\end{align}
where the latter equality is obtained by writing
\begin{equation}
 (2\pi)^4\dirac4(q+p-p_\tX) = \!\int\!\td^4x\, e^{i(q+p-p_\tX)\ndot x}
\end{equation}
and utilizing translational invariance \eq {trans},
%\begin{equation}
% \me {p,s}{J^\mu(0)}{\tX}\, e^{i(p-p_\tX)\ndot x} = \me {p,s}{J^\mu(x)}{\tX},
%\end{equation}
as well as the completeness relation $\sum_\tX|\tX\rangle\langle\tX|=1$.
The reader may recall that ${2\pi}W_{\mu\nu}$
is the absorptive part of the forward virtual Compton amplitude
\begin{equation}  \label{forw-amp3}
 T^{\mu\nu} = i\!\int\!\td^4x\, e^{iq\ndot x}
  \me {p,s} {\tT J^\mu(x)J^\nu(0)} {p,s}.
\end{equation}
The antisymmetric parts of $W^{\mu\nu}$ and $T^{\mu\nu}$ have invariant
decompositions
\begin{equation} \label{W-dec}
 W^{\mu\nu}_\tA =
  -\frac iM\, \eps^{\mu\nu\rho\sigma} q_\rho s_\sigma G_1(\nu,Q^2)
   -\frac i{M^3}\, \eps^{\mu\nu\rho\sigma} q_\rho
   \bigl( (M\nu) s_\sigma - (q\ndot s) p_\sigma \bigr) G_2(\nu,Q^2)
\end{equation}
and
\begin{equation} \label{TA-dec}
 T^{\mu\nu}_\tA =
  -\frac iM\, \eps^{\mu\nu\rho\sigma} q_\rho s_\sigma A_1(\nu,Q^2)
   -\frac i{M^3}\, \eps^{\mu\nu\rho\sigma} q_\rho
   \bigl( (M\nu) s_\sigma - (q\ndot s) p_\sigma \bigr) A_2(\nu,Q^2),
\end{equation}
respectively, where the forward amplitudes $A_{1,2}(\nu,Q^2)$
and structure functions $G_{1,2}(\nu,Q^2)$ are simply related by
\begin{equation}
 \im A_{1,2}(\nu,Q^2) = 2\pi G_{1,2}(\nu,Q^2).
\end{equation}

Let us now have a closer look at the leptonic tensor \eq {Lmunu}.
For an arbitrary spin four-vector $s_\te$ one has
\begin{equation}
 u(k,s_\te) \, \bar u(k,s_\te) =
  \frac{1+\gamma_5\seslash}2 (\kslash+m_\te).
\end{equation}
In the limit $m_\te\to0$, individual components of $s_\te$ go to infinity
in such a way that $m_\te s_\te$ stays finite.  For definite helicity
$\lambda_\te$ one has
\begin{equation}
 m_\te s_\te \xrightarrow[m_\te\to0]{} \pm k \quad \text{for} \quad
  \lambda_\te=\pm\frac12,
\end{equation}
so that above projection matrix reads
\begin{equation}
 u(k,\pm\tfrac12) \, \bar u(k,\pm\tfrac12) =
  \frac{1\pm\gamma_5}2 \kslash,
\end{equation}
and I infer
\begin{equation} \label{L-dec}
 L_{\mu\nu}(\pm\tfrac12) =
  \tr \left(\gamma_\mu \kslash' \gamma_\nu \frac{1\pm\gamma_5}2 \kslash
  \right) =
  2(k_\mu k_\nu' + k_\nu k_\mu' - \tfrac12Q^2g_{\mu\nu})
  \pm 2i\,\eps_{\mu\nu\alpha\beta} k^\alpha q^\beta.
\end{equation}

\subsubsection{Inclusive electroproduction cross sections}
Returning now to the differential cross sections \eq {el-Xsect}, I
apply standard formulae of scattering theory to get \cite {Itzykson80}
\begin{equation}
 \td\sigma_{\lambda_\te} = \frac1{4ME}\, \frac{\td^3k'}{(2\pi)^3\,2k^{\prime0}}
 \sum_{\lambda_\te'=\pm\frac12} \sum_\tX
 \bigl|T_{\text{eN$\to$eX}}(\lambda_\te,\lambda_\te')\bigr|^2\,
 (2\pi)^4 \dirac4(q+p-p_\tX),
\end{equation}
where all hadronic final states X are summed over.  The invariant
scattering amplitude is given by
\begin{equation}
 T_{\text{eN$\to$eX}} = \bar{u}(k',\lambda_\te') \gamma_\nu u(k,\lambda_\te)\,
 \frac{e^2}{q^2}\, \me{\tX}{J^\nu(0)}{p,s}.
\end{equation}
In terms of lab-frame energy and scattering angle, the Lorentz
invariant momentum-space element reads
\begin{equation}
 \frac{\td^3k'}{(2\pi)^3\,2k^{\prime0}} = \frac{E'}{16\pi^3}\,
 \td E'\, \td\Omega_\te.
\end{equation}
Thus one has
\begin{equation}
 \frac{\td\sigma_{\lambda_\te}}{\td E'\,\td\Omega_\te} =
 \frac{\alpha^2}{MQ^4}\, \frac{E'}E\, L_{\mu\nu}(\lambda_\te) W^{\mu\nu},
\end{equation}
where definitions \eq {Lmunu} and \eq {Wmunu} have been employed.
Subtracting electron helicities $\lambda_\te=\pm\frac12$, I project out
the antisymmetric part of the leptonic tensor \eq {L-dec}, which in
turn vanishes when contracted with the symmetric part of the hadronic tensor,
so that
\begin{equation}
 \frac {\td\sigma_{-1/2}} {\td E'\,\td\Omega_{\te}} -
 \frac {\td\sigma_{+1/2}} {\td E'\,\td\Omega_{\te}} =
 \frac{2\alpha^2}{MQ^4}\, \frac{E'}E\, L^\tA_{\mu\nu} W_\tA^{\mu\nu},
\end{equation}
where
\begin{equation}
 L^\tA_{\mu\nu} =
 \frac12 \bigl( L_{\mu\nu}(-\tfrac12) - L_{\mu\nu}(+\tfrac12) \bigr) =
 -2i\,\eps_{\mu\nu\alpha\beta}k^\alpha q^\beta,
\end{equation}
and $W_\tA^{\mu\nu}$ is given in terms of structure functions
$G_{1,2}(\nu,Q^2)$ by \Eq {W-dec}.  The contraction is worked out by
means of the relation
\begin{equation}
 \eps_{\mu\nu\alpha\beta}\eps^{\mu\nu\rho\sigma} k^\alpha q^\beta =
 -2 k^\rho q^\sigma + 2 k^\sigma q^\rho
\end{equation}
and leads to
\begin{align} \label{diff-X}
 \frac {\td\sigma_{-1/2}} {\td E'\,\td\Omega_{\te}} -
  \frac {\td\sigma_{+1/2}} {\td E'\,\td\Omega_{\te}}
  = \frac{4\alpha^2}{M^3Q^2}\, \frac{E'}E\, \bigl[
 & -M\bigl((k+k')\ndot s\bigr) G_1(\nu,Q^2) \notag\\*
 & +2\bigl(E'(k\ndot s)-E(k'\ndot s)\bigr) G_2(\nu,Q^2) \bigr].
\end{align}
Inserting the explicit kinematics \eq {kin}, one obtains
\begin{subequations}
\begin{equation} \label{incl1.long}
 \boxed{
  \left[
   \frac {\td\sigma_{-1/2}} {\td E'\,\td\Omega_{\te}} -
   \frac {\td\sigma_{+1/2}} {\td E'\,\td\Omega_{\te}}
  \right]_{\text{long}}
  = \frac{4\alpha^2}{M^3Q^2}\, \frac{E'}{E}\, \frac{E+E'}{\sqrt{\nu^2+Q^2}}
    \bigl[ M\nu G_1(\nu,Q^2) - Q^2 G_2(\nu,Q^2) \bigr]
 }
\end{equation}
and
\begin{equation} \label{incl1.side}
 \boxed{
  \left[
   \frac {\td\sigma_{-1/2}} {\td E'\,\td\Omega_{\te}} -
   \frac {\td\sigma_{+1/2}} {\td E'\,\td\Omega_{\te}}
  \right]_{\text{side}}
  = \frac{8\alpha^2}{M^3Q^2}\, \frac{E'}{E}\,
    \frac{EE'\sin\theta_\te}{\sqrt{\nu^2+Q^2}}
    \bigl[ M G_1(\nu,Q^2) + \nu G_2(\nu,Q^2) \bigr]
 }
\end{equation} \label{incl1}%
\end{subequations}

\subsection{Virtual photoabsorption}
\label{Sect:kin+X.virt}
One commonly expresses inclusive electroproduction cross sections in
terms of ``virtual photoabsorption cross sections'' defined by
\begin{equation} \label{virt-def}
 \sigma(\nu,Q^2) = \frac{e^2}{4MK}\, \eps_\mu^{\prime*}\eps_\nu\,
  \sum_{\text{X}} (2\pi)^4\dirac4(p+q-p_{\text{X}})\,
  \langle p,s|J^\mu(0)|\tX\rangle\,
  \langle \tX|J^\nu(0)|p,s\rangle,
\end{equation}
with appropriate choices for the ``virtual-photon polarization
vectors'' $\eps$ and $\eps'$.  This definition generalizes, in a
minimal way, the photoabsorption cross section \eq {photo} to
non-zero $Q^2$.  Since the flux of a virtual particle is not a firmly
defined quantity, there is some freedom in choosing the factor $K$
appearing in \Eq {virt-def}.  The convention adopted here,
\begin{equation}
 K = \frac1M \sqrt{(p\ndot q)^2-p^2q^2} = \sqrt{\nu^2+Q^2},
\end{equation}
follows Ioffe, Khoze, and Lipatov \cite {Ioffe84}.  Different choices of the
virtual-photon flux can be found in the literature, all of which
reduce to the (well-defined) photoabsorption cross section at photon
virtuality $Q^2=0$.  Using \Eq {Wmunu}, one has
\begin{equation}
 \sigma(\nu,Q^2) = \frac {4\pi^2\alpha}{M\sqrt{\nu^2+Q^2}}\,
 \eps_\mu^{\prime*}\eps_\nu W^{\mu\nu}.
\end{equation}

\subsubsection{Virtual-photon polarization vector}
Taking the photon to be propagating in the direction of the positive
$\be_3$ axis, diverse polarizations are represented as follows.
Transverse polarizations are given in the same manner as for real
photons, namely
\begin{subequations}
\begin{equation} \label{pol.+}
 \eps^\mu(+1) = \frac1{\sqrt2} (0, -1, -i, 0)
\end{equation}
for left-circular polarization (helicity $+1$), and
\begin{equation} \label{pol.-}
 \eps^\mu(-1) = \frac1{\sqrt2} (0, 1, -i, 0)
\end{equation}
for right-circular polarization (helicity $-1$).  Linear polarization
is specified by
\begin{equation}\label{pol.perp}
 \eps^\mu(\perp)
 = \frac i{\sqrt2}\, \bigl[ \eps^\mu(-1) + \eps^\mu(+1) \bigr] = (0,0,1,0)
\end{equation}
(normal to the scattering plane) and
\begin{equation}
 \eps^\mu(\parallel) =
 \frac 1{\sqrt2}\, \bigl[ \eps^\mu(-1) - \eps^\mu(+1) \bigr] = (0,1,0,0)
\end{equation}
(tangential).  Longitudinal\footnote {The reader should not be confused
by the different meanings of ``longitudinal polarization'' for nucleon
and photon.  A longitudinally polarized nucleon has its \emph {spin}
parallel to the momentum, whereas in the case of a longitudinal
photon, \emph {electric field} and momentum are parallel, as can be
inferred from \Eqs {kin.q} and \eq {pol.0} by considering the
electromagnetic field tensor
$F^{\mu\nu}=\eps^{\mu}q^{\nu}-\eps^{\nu}q^{\mu}$.  Therefore, if a
nucleon and a photon are in helicity eigenstates with both helicities
non-zero, then the nucleon is called longitudinal while the photon is
called transverse.} polarization is represented by
\begin{equation} \label{pol.0}
 \eps^\mu(0) =
 \left( \sqrt{\frac{\nu^2}{Q^2}+1},\, 0,\, 0,\, \frac{\nu}{Q} \right).
\end{equation} \label{pol}%
\end{subequations}
Note that the normalization $\eps(0)^2=1$ is a matter of convention.

\subsubsection{Virtual-photoabsorption cross sections}
I consider two scenarios in more detail.  Firstly, I take the nucleon to have
longitudinal polarization, \Eq {kin.long}, while the photon is
assumed to be polarized transversely, $\eps=\eps'=\eps(\pm1)$. I subtract
left-circular from right-circular polarization and define
\begin{subequations}
\begin{align} \label{sigma-W.GDH}
 \sigma_{1/2}(\nu,Q^2) - \sigma_{3/2}(\nu,Q^2)
  = \frac {4\pi^2\alpha} {M\sqrt{\nu^2+Q^2}}\, \bigl[
  \eps_\mu^*(-1)\eps_\nu(-1) - \eps_\mu^*(+1)\eps_\nu(+1)
 \bigr]\, W^{\mu\nu}_\tlong
\end{align}
Secondly, the nucleon is taken to have sideways polarization, and the
\emph {interference} between longitudinal and transverse (linear)
photon polarization is considered,
\begin{equation} \label{sigma-W.LT}
 \sigma_{\tLT}(\nu,Q^2) = \frac {4\pi^2\alpha} {M\sqrt{\nu^2+Q^2}}
  \im\left( \eps_\mu^*(0)\, \eps_\nu(\perp)\, W_\tside^{\mu\nu} \right).
\end{equation} \label{sigma-W}%
\end{subequations}
Using decomposition \eq {W-dec} together with the explicit
kinematics \eq {kin} yields
\begin{subequations}
\begin{gather} \label{sigma-G.GDH}
 \boxed{ \sigma_{1/2}(\nu,Q^2) - \sigma_{3/2}(\nu,Q^2) =
  \frac {8\pi^2\alpha} {M^2}\, \frac {\nu} {\sqrt{\nu^2+Q^2}}
  \left[ G_1(\nu,Q^2) - \frac{Q^2}{M\nu} G_2(\nu,Q^2) \right] } \\*
 \boxed{ \sigma_{\tLT}(\nu,Q^2) = \label{sigma-G.LT}
  \frac {4\pi^2\alpha} {M^2}\, \frac {Q} {\sqrt{\nu^2+Q^2}}
  \left[ G_1(\nu,Q^2) + \frac \nu M G_2(\nu,Q^2) \right] }
\end{gather} \label{sigma-G}%
\end{subequations}

\subsubsection{Relation to electroproduction cross sections}
By means of \Eqs {sigma-G}, I can cast the inclusive electroproduction
cross sections \eq {incl1} into the form
\begin{subequations}
\begin{gather} \label{incl2.long}
 \boxed{ \left[
  \frac {\td\sigma_{-1/2}} {\td E'\,\td\Omega_{\te}} -
  \frac {\td\sigma_{+1/2}} {\td E'\,\td\Omega_{\te}}
 \right]_{\text{long}} =
 \Gamma \sqrt{1-\eps^2}\,
  \bigl(\sigma_{1/2}(\nu,Q^2) - \sigma_{3/2}(\nu,Q^2)\bigr) } \\
 \boxed{ \left[ \label{incl2.side}
  \frac {\td\sigma_{-1/2}} {\td E'\,\td\Omega_{\te}} -
  \frac {\td\sigma_{+1/2}} {\td E'\,\td\Omega_{\te}}
 \right]_{\text{side}} =
 2\Gamma \sqrt{2\eps(1-\eps)}\, \sigma_{\tLT}(\nu,Q^2) }
\end{gather} \label{incl2}%
\end{subequations}
where
\begin{equation}
 \eps = \left[ 1 + 2 \left( 1+\frac{\nu^2}{Q^2} \right)
  \tan^2\frac {\theta_{\text{e}}} {2} \right]^{-1}
\end{equation}
is the ``degree of polarization of the virtual photon'', and
\begin{equation}
 \mathit\Gamma = \frac {\alpha} {2\pi^2}\, \frac {E'} {E}\,
  \frac {\sqrt{\nu^2+Q^2}} {Q^2}\, \frac 1 {1-\eps}
\end{equation}
is the ``virtual-photon flux per electron'' (nomenclature of Raskin
and Donnelly \cite {Raskin89}).  The reader may verify the relations
\begin{equation}
 \frac {1+\eps}{1-\eps} = \frac {(E+E')^2}{\nu^2+Q^2}
\end{equation}
and
\begin{equation}
 \frac {2\eps}{1-\eps} = \frac {4EE'-Q^2}{\nu^2+Q^2}.
\end{equation}

\subsubsection
 {The ``most physical'' linear combinations of structure functions}
As mentioned above, \Eqs {incl2} reduce the
differential electroproduction cross sections to ``virtual
photoabsorption cross sections'' depending solely on lab-frame energy
$\nu$ and virtuality $-Q^2$ of the photon.
The indices 1/2 and 3/2 appearing on the left-hand side of
\Eq{sigma-W.GDH} denote -- as in the case of real photoabsorption
considered so far -- antiparallel and parallel photon-nucleon
helicity, respectively.  However, the virtual photon can also be
polarized longitudinally, i.e., have zero helicity.  And since the
cross section is the absolute square of the amplitude depicted by
\Fig {incl}, which is a sum over all three photon
helicities, there is also an interference between longitudinal and
transverse polarization, showing up in \Eq {sigma-W.LT}.  The
particular combinations of structure functions $G_{1,2}(\nu,Q^2)$
occurring in \Eqs{sigma-G} are the ``most physical'' ones owing to
their direct relation to the polarization of the exchanged photon.  At
$Q^2=0$, the combination \eq {sigma-G.GDH} reduces to the familiar
combination of real-photoabsorption cross sections
\begin{equation} \label{G1(0)}
 \sigma_{1/2}(\nu,0) - \sigma_{3/2}(\nu,0) =
  \frac {8\pi^2\alpha} {M^2}\, G_1(\nu,0) =
  \sigma_{1/2}(\nu) - \sigma_{3/2}(\nu),
\end{equation}
whereas, of course, the LT-interference cross section \eq {sigma-G.LT}
vanishes for real photons,
\begin{equation}
 \sigma_{\tLT}(\nu,0) = 0.
\end{equation}

\section{Generalizations of the GDH integral}
\label{Sect:gen}

This section is devoted to different generalizations of the GDH
integral to non-zero photon virtuality.  To contact to
deep-inelastic-scattering (DIS) sum rules like the Bjorken sum rule
\cite {Bjorken66}, I additionally give all integrals in terms of the
functions
\begin{equation}
 \boxed{ g_1(x,Q^2) = \frac \nu M G_1(\nu,Q^2) \quad \text {and} \quad
  g_2(x,Q^2) = \frac {\nu^2} {M^2} G_2(\nu,Q^2) }
\end{equation}
where $x=Q^2/2M\nu$ is the Bjorken scaling variable.  These structure
functions have the simple scaling behavior (modulo logarithms)
\begin{equation} \label{scaling}
 g_i(x,Q^2) \xrightarrow[Q^2\to\infty]{} g_i(x), \quad i=1,2.
\end{equation}

Adopting the notation of Soffer and Teryaev \cite {Soffer93,Soffer95}
and translating the photon-energy threshold
\begin{equation}
 \nu_0 = m_\pi + \frac {m_\pi^2+Q^2} {2M}
\end{equation}
into a Bjorken-$x$ threshold\footnote {If $Q^2$ exceeds a few GeV$^2$,
then $x_0$ is very near to unity.  Therefore, the deviation from unity
is usually ignored in deep-inelastic scattering, and Bjorken-$x$ runs
from zero to \emph {one}.  This is why the upper bound of the integrals
\eq {I} may look a little unfamiliar at first sight.  Above $x_0$,
structure functions $g_{1,2}(x,Q^2)$ vanish except for a delta
function at $x=1$ coming from the elastic contribution to
inclusive electroproduction, i.e., a single nucleon as
hadronic final state X in \Fig {incl}.  The elastic contribution is
insignificant at high $Q^2$.  For a discussion, see Ji \cite {Ji93a}.}
\begin{equation}
 x_0 = \frac {Q^2}{2M\nu_0} = \frac {Q^2}{2Mm_\pi+m_\pi^2+Q^2},
\end{equation}
I define
\begin{subequations}
\begin{equation} \label{I.1}
 I_1(Q^2) := \!\int_{\nu_0}^\infty\! \frac {\td\nu}{\nu}\, G_1(\nu,Q^2)
  = \frac {2M^2}{Q^2} \!\int_0^{x_0}\! \td x\, g_1(x,Q^2),
\end{equation}
\begin{equation} \label{I.2}
 I_2(Q^2) := \frac1M \!\int_{\nu_0}^\infty\! \td\nu\, G_2(\nu,Q^2)
  = \frac {2M^2}{Q^2} \!\int_0^{x_0}\! \td x\, g_2(x,Q^2),
\end{equation}
and
\begin{align} \label{I.1+2}
 I_{1+2}(Q^2) := I_1(Q^2) + I_2(Q^2)
  & = \!\int_{\nu_0}^\infty\! \frac {\td\nu}{\nu}\,
   \left( G_1(\nu,Q^2) + \frac\nu M G_2(\nu,Q^2) \right) \notag \\
  & = \frac {2M^2}{Q^2} \!\int_0^{x_0}\! \td x\,
   \bigl( g_1(x,Q^2) + g_2(x,Q^2) \bigr).
\end{align} \label{I}%
\end{subequations}
Up to a factor of $2M^2/Q^2$, the integrals $I_{1,2}(Q^2)$ are the zeroth
moments\footnote {The moments $\int\td x\,g_{1,2}(x,Q^2)$ are commonly denoted
as $\Gamma_{1,2}(Q^2)$.} of structure functions $g_{1,2}(x,Q^2)$.
Integral $I_{1+2}(Q^2)$ has a more physical meaning, since it involves
the linear combination $G_1+\frac{\nu}{M}G_2$ of structure functions,
which is related to the LT-interference cross section defined above.
In fact, on account of \Eq {sigma-G.LT} one may write
\begin{equation}
 I_{1+2}(Q^2) = \frac {M^2}{4\pi^2\alpha} \!\int_{\nu_0}^{\infty}\!
 \frac{\td\nu}\nu\, \frac{\sqrt{\nu^2+Q^2}}Q\, \sigma_{\tLT}(\nu,Q^2).
\end{equation}
Analogously, I define in view of \Eq {sigma-G.GDH} 
\begin{subequations}
\begin{align} \label{I-GDH.plain}
 \tilde I_{\tGDH}(Q^2) & := \frac {M^2}{8\pi^2\alpha} \!\int_{\nu_0}^{\infty}\!
 \frac{\td\nu}\nu\, \frac{\sqrt{\nu^2+Q^2}}\nu\,
 \bigl( \sigma_{1/2}(\nu,Q^2) - \sigma_{3/2}(\nu,Q^2) \bigr) \notag \\*
 & = \int_{\nu_0}^\infty\! \frac {\td\nu}{\nu}\,
  \left( G_1(\nu,Q^2) - \frac{Q^2}{M\nu} G_2(\nu,Q^2) \right) \notag \\*
 & = \frac {2M^2}{Q^2} \!\int_0^{x_0}\! \td x\,
  \left( g_1(x,Q^2) - \frac{4M^2x^2}{Q^2} g_2(x,Q^2) \right).
\end{align}
and
\begin{equation} \label{I-GDH.flux}
 \boxed{ \begin{split}
  I_\tGDH(Q^2)
  & := \frac {M^2}{8\pi^2\alpha} \!\int_{\nu_0}^{\infty}\! \frac{\td\nu}\nu\,
   \bigl( \sigma_{1/2}(\nu,Q^2) - \sigma_{3/2}(\nu,Q^2) \bigr) \\
  & = \int_{\nu_0}^{\infty}\! \frac{\td\nu}{\sqrt{\nu^2+Q^2}}
   \left( G_1(\nu,Q^2) - \frac{Q^2}{M\nu} G_2(\nu,Q^2) \right) \\
  & = \frac{2M^2}{Q^2} \!\int_0^{x_0}\!
   \frac {\td x}{\sqrt{1+\frac{4M^2x^2}{Q^2}}}
   \left( g_1(x,Q^2) - \frac{4M^2x^2}{Q^2} g_2(x,Q^2) \right)
 \end{split} }
\end{equation} \label{I-GDH}%
\end{subequations}
where the last definition is most directly related to the (transverse)
virtual photoabsorption cross sections.

\subsubsection{GDH, Bjorken, and Burkhardt-Cottingham sum rules}
In terms of the integrals \eq {I}, the GDH sum rule reads (provided it holds,
of course)
\begin{equation} \label{I1(0)=GDH}
 I_1(0) = \!\int_{\nu_0}^\infty\! \frac {\td\nu}{\nu}\, G_1(\nu,0)
  = \frac{M^2}{8\pi^2\alpha} \!\int_{\nu_0}^\infty\! \frac {\td\nu}{\nu}\,
  \bigl( \sigma_{1/2}(\nu) - \sigma_{3/2}(\nu) \bigr)
  = -\frac{\kappa^2}{4},
\end{equation}
while the Bjorken sum rule \cite {Bjorken66} (see also \Ref
{Bjorken70}) is given by
\begin{equation} \label{Bjorken-SR}
 I_1^\tp(Q^2) - I_1^\tn(Q^2)
 \xrightarrow[Q^2\to\infty]{} \frac{2M^2}{Q^2} \!\int_0^1\! \td x\,
 \bigl( g_1^\tp(x) - g_1^\tn(x) \bigr)
 = \frac{2M^2}{Q^2}\, \frac{g_\tA}6,
\end{equation}
where $g_\tA=1.26$ is the phenomenological neutron beta-decay constant. \Eq
{Bjorken-SR} is well established experimentally \cite {Abe97b}.

In contrast to GDH and Bjorken sum rules, the Burkhardt-Cottingham sum
rule \cite {Burkhardt70} (see also Jaffe \cite {Jaffe90}) applies to all
values of $Q^2$:
\begin{equation} \label{BC}
 I_2(Q^2) = \frac 14\, F_2(Q^2) \bigl( F_1(Q^2) + F_2(Q^2) \bigr) = \frac14\,
 \frac {G_\tM(Q^2) - G_\tE(Q^2)}{1 + Q^2/4M^2}\, G_\tM(Q^2).
\end{equation}
Dirac and Pauli form factors $F_{1,2}(Q^2)$ are normalized as
$F_1(0)=Z$ and $F_2(0)=\kappa$, respectively.  Sachs form factors
$G_{\tE,\tM}(Q^2)$ are given by\footnote {Soffer and Teryaev \cite
{Soffer93,Soffer95} apparently use a somewhat peculiar definition of
Sachs form factors, since they do not include the factor 
$(1+Q^2/4M^2)^{-1}$ in \Eq {BC}.}
\begin{subequations}
\begin{align}
 G_\tE(Q^2) & = F_1(Q^2) - \frac {Q^2}{4M^2}\, F_2(Q^2), \\
 G_\tM(Q^2) & = F_1(Q^2) + F_2(Q^2).
\end{align}
\end{subequations}
At $Q^2=0$, \Eq {BC} reduces to
$I_2(0)=\frac14\kappa(Z+\kappa)$. Hence,
\begin{equation} \label{GDH+BC}
 I_{1+2}(0) = \frac {Z\kappa}4.
\end{equation}

\subsubsection{Bjorken sum rule and $J=1$ fixed pole}
At this point, I would like to comment on the question whether a
$Q^2\neq0$ generalization of the $J=1$ fixed pole in angular momentum
plane, which I discussed in \Sect {fixed-pole} of this thesis, would
spoil the Bjorken sum rule.  I mention in advance that \emph {it
woudn't}, although at first sight, this might be expected.

Consider the forward virtual Compton amplitude $A_1(\nu,Q^2)$ as
defined in \Eq {TA-dec}.  Its Born contribution is given by \Eq
{Born.A1}.  In order to formulate a fixed-$Q^2$
dispersion relation, the Born-\emph{pole} part must be subtracted:
\begin{equation}
 \bar A_1(\nu,Q^2)
 := A_1(\nu,Q^2) - \frac {4M^2Q^2\,F_1(Q^2)G_\tM(Q^2)}{Q^4-4M^2\nu^2}.
\end{equation}
Recall that at $Q^2=0$, one has
\begin{equation}
 f_2(\nu) = \frac{\alpha}{2M^2}\, A_1(\nu,0)
 = \frac{\alpha}{2M^2}\, \bar A_1(\nu,0).
\end{equation}

A $J=1$ fixed pole would manifest itself in a non-vanishing real part
of amplitude $A_1(\nu,Q^2)$ at $\nu\to\infty$, i.e.,
\begin{equation}
 A_1(\infty,Q^2) = \bar A_1(\infty,Q^2) \neq 0.
\end{equation}
(Except for continuity, I do not imply any assumption on the $Q^2$
dependence of $A_1(\infty,Q^2)$.)  In this case, the
associated dispersion relation would require a subtraction at
infinity.  Thus,
\begin{equation} \label{Q2-SDR}
 I_1(Q^2) = \!\int_{\nu_0}^\infty\! \frac {\td\nu}{\nu}\, G_1(\nu,Q^2)
 = \frac14 \bigl( \bar A_1(0,Q^2) - A_1(\infty,Q^2) \bigr),
\end{equation}
since $\im A_1=2\pi G_1$.

In view of \Eq {Q2-SDR}, one might be tempted to think that the
Bjorken sum rule would receive a modification if $A_1(\infty,Q^2)$
were non-zero. However, while this is true for the GDH sum rule (cf.\ \Eq
{GDH+f2(infty)}),
\begin{equation}
 I_1(0) = \frac14 \bigl( -\kappa^2 - A_1(\infty,Q^2) \bigr),
\end{equation}
the Bjorken sum rule is left unchanged,
\begin{equation} \label{Bjorken-SR'}
 I_1^\tp(Q^2) - I_1^\tn(Q^2)
 \xrightarrow[Q^2\to\infty]{} \frac{2M^2}{Q^2}\, \frac{g_\tA}6,
\end{equation}
due to the following reason.  The dispersion-theoretic derivation of
the GDH sum rule (see \Sect {DT}) rests upon the
Low-Gell-Mann-Goldberger low-energy theorem for the polarized forward
Compton amplitude, which presents a statement on the \emph {right-hand
side} of the dispersion relation \eq {Q2-SDR}.  The dispersion
relation itself enters additionally.  In contrast, modern derivation of
DIS sum rules (as well as their radiative or higher-twist corrections)
by means of operator-product expansion (see, e.g., Cheng and Li \cite
{Cheng84} or Roberts \cite {Roberts90}) yields a direct statement on
the integral on the \emph {left-hand side} of the dispersion relation.
In terms of the parton model, which represents the lowest order of the
operator-product expansion, depicted by the handbag\footnote {See
footnote \ref {animals} on p.~\pageref {animals} of this thesis.}
diagram
\begin{equation}
 \graph(3,2){
  \fmfleft{i1,i2} \fmfright{o1,o2}
  \fmf{fermion,width=thick}{i1,v1} \fmf{fermion,width=thick}{v2,o1}
  \fmf{photon}{i2,v5} \fmf{photon}{v6,o2}
  \fmf{fermion}{v4,v5,v6,v3}
  \fmfpolyn{shade,smooth,pull=1.3}{v}{4}
 }
\end{equation}
one may put it this way: The handbag diagram yields the value \eq
{Bjorken-SR'} for the proton-neutron difference of the integrals \eq
{Q2-SDR}.  It does not tell how this number is distributed among
$A_1(0,Q^2)$ and $A_1(\infty,Q^2)$ on the right-hand side of \Eq
{Q2-SDR}.

It should be noted that Bjorken's original derivation \cite
{Bjorken66,Bjorken70} of the sum rule is not exactly convenient for
the discussion of this point, because it is founded on the validity of
the naive current commutator \eq {naive-curr-comm1} \emph {and} on an
unsubtracted dispersion relation.  As a matter of fact, neither of
these assumptions is vital for the validity of the Bjorken sum rule.
Moreover, it can be proved that they are highly correlated, in such a
way that the modification introduced by a non-naive commutator is
exactly cancelled%
\footnote {I considered the general non-naive charge-density
commutator \eq {a-comm}, straightforwardly generalized \cite [Eq.\
(4.38)] {Jackiw85} to all Lorentz components, because the full
current-current commutator is needed.  Then, in addition to the
expression found by Bjorken \cite {Bjorken66}, the non-naive part of
the commutator contributes a certain \emph {constant} to the
$Q^2\to\infty$ limit of Compton amplitude $A(0,Q^2)$ occurring in \Eq
{Q2-SDR}.  On the other hand, it implies a sea\-gull \cite {Jackiw85},
which is uniquely determined by the requirement of Lorentz invariance
and gauge invariance of the amplitude.  This seagull is known to
dominate the Compton amplitude in the BJL limit (see \Sect {BJL}).
For large $Q^2$, it also dominates in the Regge limit (see discussion
on p.~\pageref {BJL+Regge} of this thesis), so that it necessitates a
subtraction at infinity in \Eq {Q2-SDR}.  The subtraction constant
$A_1(\infty,Q^2)$ turns out to exactly cancel the correction due to
the non-naive commutator, leading back to the naive result \eq
{Bjorken-SR'}.} by virtue of its accompanying seagull amplitude.  This
very simple result of a quite intricate calculation signalizes that
the current-algebra derivation of the Bjorken sum rule gets
unnecessarily awkward if a non-naive commutator is allowed for.

\subsection{The significance of $G_2(\nu,Q^2)$}
\label{Sect:gen.G2}

Owing to the explicit factor of $Q^2$ preceding structure function
$G_2(\nu,Q^2)$ in \Eqs {I-GDH},\footnote {Don't be puzzled by the
explicit factor of $1/Q^2$ preceding structure function $g_2(x,Q^2)$
in \Eqs {I-GDH}.  The small-$g$ structure functions are meaningless at
$Q^2=0$.  Observe that not even Bjorken-$x$ itself is defined for real
photons.  For a nice re-definition of Bjorken-$x$ that is applicable to
both the scaling limit and the real-photon point, see Li and Li \cite
{Li94}.} all three integrals $I_1(Q^2)$, $I_\tGDH(Q^2)$, and $\tilde
I_{\tGDH}(Q^2)$ exhibit the same real-photon limit, namely
\begin{equation} \label{I1(0)}
 I_1(0) = I_{\tGDH}(0) = \tilde I_\tGDH(0)
   = \!\int_{\nu_0}^\infty\! \frac {\td\nu}{\nu}\, G_1(\nu,0).
\end{equation}

Notably, the three integrals also have the same scaling limit, on
account of the explicit factor of $1/Q^2$ accompanying function
$g_2(x,Q^2)$ in \Eqs {I-GDH}:
\begin{equation} \label{I1(infty)}
 \lim_{Q^2\to\infty} I_{1}(Q^2) = \lim_{Q^2\to\infty} I_{\tGDH}(Q^2) =
 \lim_{Q^2\to\infty} \tilde I_{\tGDH}(Q^2) =
  \frac{2M^2}{Q^2} \!\int_0^1\! \td x\, g_1(x).
\end{equation}

So, the integrals $I_1(Q^2)$, $I_\tGDH(Q^2)$, and $\tilde
I_{\tGDH}(Q^2)$ coincide at both ends of the $Q^2$ range under consideration.
Nevertheless, I expect them to differ conspicuously at intermediate $Q^2$.
In my opinion, this fact is not appropriately dealt with in the literature.

\subsubsection{Beyond leading terms}
From definitions \eq {I-GDH}, the generalized GDH integrals
$I_\tGDH(Q^2)$ and $\tilde I_\tGDH(Q^2)$ can be expanded about $Q^2=0$
and $Q^2\to\infty$, giving
\begin{subequations}
\begin{align}
 I_\tGDH(Q^2) - \tilde I_\tGDH(Q^2) & \xrightarrow[Q^2\to0]{}
 - \frac {Q^2}2 \!\int_{\nu_0}^\infty\! \frac{\td\nu}{\nu^3}\, G_1(\nu,0),
 \label{low-Q2.GDH} \\*
 \tilde I_\tGDH(Q^2) - I_1(Q^2) & \xrightarrow[Q^2\to0]{}
 - \frac {Q^2}M \!\int_{\nu_0}^\infty\! \frac{\td\nu}{\nu^2}\, G_2(\nu,0),
 \label{low-Q2.1}
\end{align} \label{low-Q2}%
\end{subequations}
and
\begin{subequations}
\begin{align}
 I_\tGDH(Q^2) & \xrightarrow[Q^2\to\infty]{} \frac {2M^2}{Q^2}
  \left( \int_0^1\! \td x\, g_1(x,Q^2) - \frac{2M^2}{Q^2} \!\int_0^1\! \td x\,
  x^2\bigl(g_1(x)+2g_2(x)\bigr) \right), \label{high-Q2.GDH} \\*
 \tilde I_\tGDH(Q^2) & \xrightarrow[Q^2\to\infty]{} \frac {2M^2}{Q^2}
  \left( \int_0^1\! \td x\, g_1(x,Q^2) - \frac{4M^2}{Q^2} \!\int_0^1\! \td x\,
  x^2g_2(x) \right), \label{high-Q2.tilde}
\end{align} \label{high-Q2}%
\end{subequations}
respectively.  Note that the high-$Q^2$ suppression of the $g_2$
contribution to integrals \Eq {high-Q2} traces back solely to the scaling
property \eq {scaling} and the existence of the  second moment
$\int\!\td{x}\,x^2g_2(x)$.  This should not be mistaken with the
Burkhardt-Cottingham sum rule, which constrains only the \emph
{zeroth} moment of $g_2(x)$ and does not have any effect on the other
moments.  In \Eqs {high-Q2}, the $Q^2$  dependence of structure function
$g_1(x,Q^2)$ has been kept for consistency, since the $1/Q^2$ correction
to the zeroth moment of $g_1(x,Q^2)$ is of the same order than the terms
involving second moments.

\subsubsection{Historical survey}
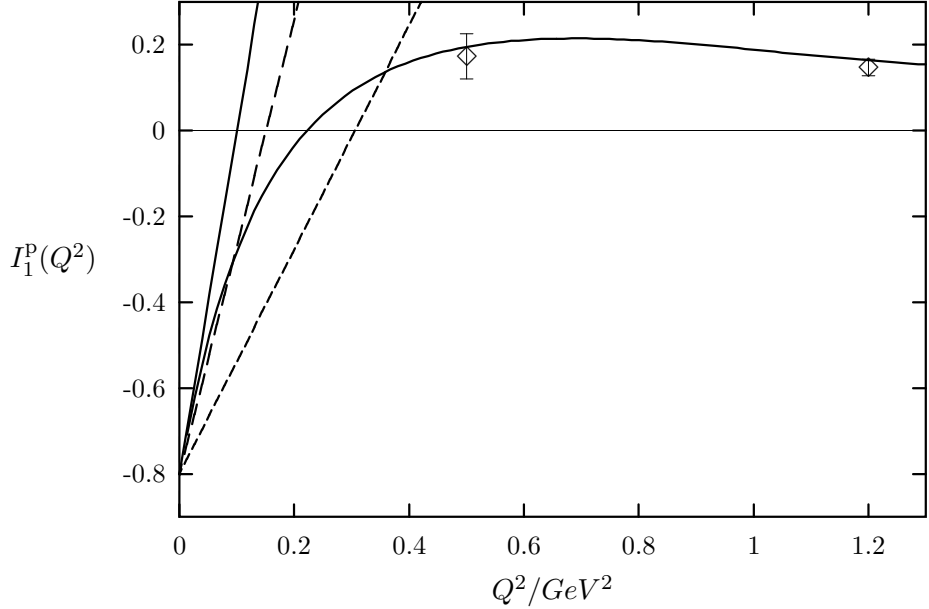
\begin{figure}[tb]
 \begin{center}
  \input{soffer}
 \end{center}
 \caption[]{
  The integral $I_1(Q^2)$, \Eq {I.1}, for the proton.
  Solid lines represent the form given by Soffer and Teryaev
  \cite {Soffer93,Soffer95}, as well as its slope at $Q^2=0$.
  Long-dashed and short-dashed lines represent estimates of the slopes
  of integrals $I_\tGDH^\tp(Q^2)$ and $\tilde I_\tGDH^\tp(Q^2)$,
  respectively.  Data are from SLAC experiment E143 \cite{Abe97a}.
  Details are given in the text.
  \label{Fig:soffer}}
\end{figure}

The literature provides a number of attempts to connect the limits
$Q^2\to0$ and $Q^2\to\infty$, i.e., \Eqs {I1(0)} and \Eq {I1(infty)}.
The former limit is related to the GDH sum rule as reflected by \Eq
{I1(0)=GDH}, while the latter is governed by the Bjorken \cite
{Bjorken66} and Ellis-Jaffe \cite {Ellis74} sum rules and has been
measured at SLAC [\plaincite {Baum83}\nocite {Anthony93}--\plaincite
{Abe97b}] and in the EMC \cite {Ashman89} and SMC \cite {Adeva93}
experiments at CERN.  Intermediate momentum transfer
($Q^2=0.5$~GeV$^2$ and 1.2~GeV$^2$) was recently explored in the SLAC
E143 experiment \cite {Abe97a}  (see \Fig {soffer}).

Anselmino, Ioffe, and Leader \cite {Anselmino89} (see also \Ref
{Anselmino95}) proposed a first-guess parametrization of integral
$I_1(Q^2)$ inspired by vector-meson dominance, namely a sum of a
single and a double pole at the $\rho$-meson mass $Q^2=-m_\rho^2$.
The residues of these poles are fitted to the GDH value at $Q^2=0$ and
the EMC data at high $Q^2$.  Burkert and Ioffe [\plaincite
{Burkert92}\nocite {Burkert94b}--\plaincite {Ioffe97}] refined this
model by adding an estimate of the resonance contribution\footnote
{Note that there is an error \cite {Ioffe97priv} in the resonance
contribution depicted in Fig.~1 of \Ref {Burkert94b}. Ioffe \cite
{Ioffe97} used the correct resonance contribution and predicted a sign
change of $I_1(Q^2)$ at $Q^2\approx0.2\ldots0.3$~GeV$^2$, whereas \Ref
{Burkert94b} predicted $Q^2\approx0.6$~GeV$^2$.} at very low $Q^2$.

Within the framework of heavy-baryon chiral perturbation theory,
Bernard, Kaiser, and Mei{\ss}ner \cite {Bernard93} investigated yet
another generalized GDH integral, distinguishing from my definition
\eq {I-GDH.flux} by the use of a different photon flux.  

Li \cite{Li93} employed a constituent-quark model including
relativistic corrections.

Soffer and Teryaev \cite {Soffer93,Soffer95} suggested a
parametrization of the proton integral $I_1^\tp(Q^2)$ founded on the
requirement that function $I_{1+2}^\tp(Q^2)$, \Eq {I.1+2}, should be
as smooth and simple as possible.  This assumption is
motivated\footnote {To me, the smoothness assumption of Soffer and
Teryaev is quite plausible also with regard to resonance saturation,
because the strong $Q^2$ dependence of the GDH integral can be
attributed mainly to the $\Delta$ resonance, which is known to die out
quickly with increasing $Q^2$ \cite {Burkert93}.  The $\Delta$
resonance is dominantly an $M_{1+}$ transition, i.e., it ``favours''
transverse photons.  On the other hand, the linear combination of
structure functions that occurs in $I_{1+2}(Q^2)$ represents an
interference between longitudinal and transverse virtual photons, as
reflected by \Eq {sigma-G.LT}.  Thus, the strongly $Q^2$ dependent
$\Delta$ resonance will only weakly affect $I_{1+2}(Q^2)$.} from QCD
sum rules.  Integral $I_{1+2}^\tp(Q^2)$ is parametrized such that it
reproduces the GDH/Burkhardt-Cottingham value \eq {GDH+BC} at $Q^2=0$
as well as the asymptotic EMC result.  Next, the Burkhardt-Cottingham
integral $I_2^\tp(Q^2)$ is subtracted (employing a reasonable
parametrization of the form factors) to gain the integral $I_1^\tp(Q^2)$.
The result is depicted in \Fig {soffer}.  Note that nothing can be
inferred on $I_\tGDH(Q^2)$ or $\tilde I_\tGDH(Q^2)$ by means of the
method of Soffer and Teryaev.  Moreover, it should be noted that the
main purpose of \Refs {Soffer93} and \plaincite {Soffer95} was the
\emph {qualitative} statement that the drastic low-$Q^2$ behavior of
the generalized GDH integral $I_1(Q^2)$ can be understood in terms of
the strong $Q^2$ dependence of the form factors occurring in the
Burkhardt-Cottingham sum rule \eq {BC}.

\subsubsection{Low $Q^2$}
I adopt the parametrization of Soffer and Teryaev \cite {Soffer93} for
the proton\footnote {Soffer and Teryaev also presented a
parametrization for the neutron \cite {Soffer97}, but in my opinion,
too much handwaving is done in this case.  (In contrast to the proton
case, $I_{1+2}^\tn(0)=0$ due to \Eq {GDH+BC}, so that at least one
extremum has to be built in, which somewhat overcharges the method.)}
to illustrate some ideas concerning different $Q^2$ evolutions of the
GDH integral.  Considering the low-$Q^2$ expansion \eq {low-Q2.GDH}
together with identity \eq {G1(0)}, one has that\footnote
{Together with the optical theorem \eq {f2-opt} and the Taylor series \eq
{gamma-def}, \Eq {I-GDH-0} represents the $\nu\to0$ limit of the
subtracted dispersion relation \eq {SDR-f2}.}
\begin{equation} \label{I-GDH-0}
 I_\tGDH(Q^2) - \tilde I_\tGDH(Q^2) \xrightarrow[Q^2\to0]{}
 - \frac {M^2Q^2}{16\pi^2\alpha} \!\int_{\nu_0}^\infty\! \frac{\td\nu}{\nu^3}\,
   \bigl( \sigma_{1/2}(\nu) - \sigma_{3/2}(\nu) \bigr)
 = - \frac {M^2Q^2}{4\alpha}\, \gamma,
\end{equation}
where $\gamma$, referred to as the spin polarizability \cite
{Bernard92b}, is defined as the second coefficient in the Taylor
series of forward Compton amplitude $f_2(\nu)$,
\begin{equation} \label{gamma-def}
 f_2(\nu) = - \frac {\alpha\kappa^2}{2M^2} + \gamma\,\nu^2 + \mathcal O(\nu^4).
\end{equation}
Taking the value
\begin{equation} \label{gamma-p}
 \gamma_\tp = -134 \ntimes 10^{-6} \text{ fm}^4,
\end{equation}
obtained from a recent saturation by pion-production multipoles \cite
{Sandorfi94}, yields
\begin{subequations}
\begin{equation} \label{low-Q2-num.GDH}
 \boxed{ I_\tGDH^{\tp}(Q^2) - \tilde I_\tGDH^{\tp}(Q^2)
  \xrightarrow[Q^2\to0]{} 2.68\, \frac {Q^2}{\text{GeV}^2} }
\end{equation}
A rough estimate can be found also for the difference $\tilde
I_\tGDH^\tp(Q^2) - I_1^\tp(Q^2)$ if one recalls that the combination
$G_1(\nu,Q^2)+\frac{\nu}{M}G_2(\nu,Q^2)$ is expected to be small.
Letting $G_2(\nu,Q^2)\approx-\frac{M}{\nu}G_1(\nu,Q^2)$ as a first
guess, \Eq {low-Q2.1} becomes
\begin{equation}
 \tilde I_\tGDH^\tp(Q^2) - I_1^\tp(Q^2) \underset{Q^2\to0}{\approx}
 Q^2 \!\int_{\nu_0}^\infty\! \frac{\td\nu}{\nu^3}\, G_1^\tp(\nu,0)
 = \frac{M^2Q^2}{2\alpha}\, \gamma_\tp.
\end{equation}
Again, inserting the value \eq {gamma-p} for the spin polarizability
$\gamma_\tp$ yields
\begin{equation}
 \boxed{ \tilde I_\tGDH^\tp(Q^2) - I_1^\tp(Q^2) \xrightarrow[Q^2\to0]{}
  -5.37\, \frac {Q^2}{\text{GeV}^2} }
\end{equation} \label{low-Q2-num}%
\end{subequations}

\Eqs {low-Q2-num} are to be confronted with the Soffer-Teryaev result
\begin{equation} \label{Soffer-num}
 \boxed{ I_1^\tp(Q^2)
  \xrightarrow[Q^2\to0]{} -0.80 + 7.98\, \frac {Q^2}{\text{GeV}^2} }
\end{equation}
The slope 7.98~GeV$^{-2}$ is composed of $-0.23$~GeV$^{-2}$ coming
from the smooth parametrization of $I_{1+2}^\tp(Q^2)$ discussed above,
plus 8.21~GeV$^{-2}$ due to $-I_2^\tp(Q^2)$, i.e., the proton
Burkhardt-Cottingham sum rule.  (Throughout, I employ the form-factor
parametrization of Mergell, Mei{\ss}ner, and Drechsel \cite
{Mergell96}.)  Thus, the sharp rise of integral $I_1^\tp(Q^2)$ around
$Q^2=0$ is attributed mainly to the Burkhardt-Cottingham sum rule.

As can be seen from \Eqs {low-Q2-num} and \eq {Soffer-num}, the
integrals $I_1(Q^2)$, $I_\tGDH(Q^2)$, and $\tilde I_\tGDH(Q^2)$ must
indeed be expected to considerably differ at low $Q^2$.  This state of
affairs is illustrated in \Fig {soffer}.

\subsubsection{High $Q^2$}
As for the high-$Q^2$ region, structure function $G_2(\nu,Q^2)$ may
conspicuously affect $1/Q^2$ higher-twist corrections.  To illustrate this, I
adopt the notation of Ehrnsperger, Mankiewicz, and Sch\"afer \cite
{Ehrnsperger94} and wright down zeroth and second moments of $g_{1,2}(x,Q^2)$:
\begin{subequations}
\begin{align}
 \int_0^1\! \td x\, g_1(x,Q^2)
  & = \frac12\,a_0 + \frac {M^2}{9Q^2}\, \bigl( a_2 + 4d_2 + 4f_2 \bigr)
    + \mathcal O(Q^{-4}), \\
 \int_0^1\! \td x\, g_2(x,Q^2)
  & = \mathcal O(Q^{-4}), \label{Ehrnsperger.g2} \\
 \int_0^1\! \td x\, x^2\, g_1(x,Q^2)
  & = \frac12\,a_2 + \mathcal O(Q^{-2}), \\
 \int_0^1\! \td x\, x^2\, g_2(x,Q^2)
  & = -\frac13\,a_2 + \frac13\,d_2 + \mathcal O(Q^{-2}),
\end{align} \label{Ehrnsperger}%
\end{subequations}
where $a_{0,2}$, $d_2$, and $f_2$ represent matrix elements of
operators of twist two, three, and four, respectively \cite
{Ehrnsperger94}.  \Eq {Ehrnsperger.g2} states the validity of the
Burkhardt-Cottingham sum rule up to twist four.  The Bjorken sum rule
states that $a_0^\tp-a_0^\tn=g_\tA/3$.  Falling back on \Eqs
{high-Q2}, I infer
\begin{equation}
 \boxed{ \begin{split}
  I_1(Q^2) & = \frac {2M^2}{Q^2}\, \left(
   \frac12\,a_0 + \frac {M^2}{9Q^2}\, \bigl( a_2 + 4d_2 + 4f_2 \bigr)
   + \mathcal O(Q^{-4}) \right) \\
  I_\tGDH(Q^2) & = \frac {2M^2}{Q^2}\, \left(
   \frac12\,a_0 + \frac {M^2}{9Q^2}\, \bigl( 4a_2 - 8d_2 + 4f_2 \bigr)
   + \mathcal O(Q^{-4}) \right) \\
  \tilde I_\tGDH(Q^2) & = \frac {2M^2}{Q^2}\, \left(
   \frac12\,a_0 + \frac {M^2}{9Q^2}\, \bigl( 13a_2 - 8d_2 + 4f_2 \bigr)
   + \mathcal O(Q^{-4}) \right)
 \end{split} } \label{twist}
\end{equation}
Observe that $1/Q^2$ corrections may considerably differ among
different generalizations of the GDH integral, depending on the
relative strength of twist-four matrix elements compared to twist-two
and -three matrix elements.  At present, estimates on $d_2$ and $f_2$
from polarized DIS data \cite {Abe96,Ji97} or from QCD sum rules \cite
{Balitsky90} are far too vague to numerically specify the differences
between \Eqs {twist}.

\subsubsection{Conclusion}
I conclude that one ought to be very conscientious about the question,
which of the integrals $I_1(Q^2)$, $I_\tGDH(Q^2)$, or $\tilde
I_\tGDH(Q^2)$ should be considered.  Different choices may be
appropriate from different points of view.  The crucial questions are:
does one want to describe properties of structure function
$G_1(\nu,Q^2)$ (corresponding to integral $I_1$), or properties of the
nucleon's spin structure as probed with transversely polarized virtual
photons (corresponding to $I_\tGDH$ or $\tilde I_\tGDH$)?  And: what
is the phenomenological input?  Experimental setups with longitudinal
target polarization generally access \emph {neither} structure
function $G_1(\nu,Q^2)$ \emph {nor} the linear combination
$G_1(\nu,Q^2)-\frac{Q^2}{M\nu}G_2(\nu,Q^2)$ (corresponding to
asymmetry $A_1$), but rather some other linear combination of
$G_1(\nu,Q^2)$ and $G_2(\nu,Q^2)$, whose coefficients involve the beam
energy $E$ in addition to $\nu$ and $Q^2$ or $E'$ and $\theta_\te$
(see, e.g., Roberts \cite {Roberts90}).  Additional assumptions about
structure function $G_2(\nu,Q^2)$ (or assymetry $A_2$) have to be put
in to extract one of the integrals under consideration.  There may be
some choice that is least affected by the arbitrariness within these
assumptions.  For instance, the intermediate-$Q^2$ E143 experiment is
capable to extract $G_1(\nu,Q^2)$ with much higher accuracy than
asymmetry $A_1$, because the former quantity is much less affected by
the lack of knowledge on the other structure function or asymmetry
\cite {Abe97a}.

On the other hand, there are estimates of the low-$Q^2$ evolution
based on resonance saturation \cite {Burkert93}, as well as
theoretical investigations within the framework of constituent-quark
models \cite {Li93} and chiral perturbation theory \cite {Bernard93}.
These investigations rather predict the behavior of integrals
$I_\tGDH(Q^2)$ or $\tilde I_\tGDH(Q^2)$ than that of $I_1(Q^2)$, due
to the lack of information on longitudinal-photon coupling.  (This
point has already been stressed by Li and Li \cite {Li94}.)  

As I demonstrated in this section, one should at any rate desist from
connecting the low-$Q^2$ range of one of these integrals with the
high-$Q^2$ range of one of the others.

Further theoretical and experimental investigation of longitudinal
photon coupling is badly needed.  As far as resonance saturation is
concerned, the following section aims at providing a basis for these
studies.

\section{The \piN\ contribution}
\label{Sect:piN}

\begin{figure}[tb]
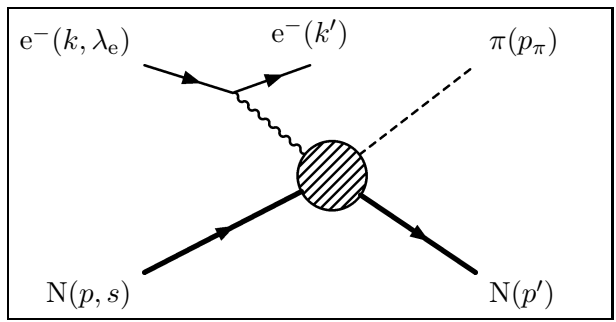

 \begin{displaymath}
  \boxedgraph(6,3){
%   \fmfstraight
   \fmfleft{i1,i2} \fmfright{o1,o2} \fmftop{t}
   \fmf{fermion,width=thick}{i1,b}
   \fmf{fermion,tension=2}{i2,f}
   \fmf{fermion}{f,t}
   \fmf{photon}{f,b}
   \fmf{fermion,width=thick}{b,o1}
   \fmf{dashes}{b,o2}
   \fmfblob{1\ul}{b}
   \fmflabel{N$(p,s)$}{i1} \fmflabel{N$(p')$}{o1} \fmflabel{$\pi(p_\pi)$}{o2}
   \fmflabel{e$^-(k,\lambda_{\rm{e}})$}{i2} \fmflabel{e$^-(k')$}{t}
  }
 \end{displaymath}
 \caption[]{
  Polarized pion electroproduction on the nucleon
  in the one-photon-exchange approximation.
  \label{Fig:pielprod}}
\end{figure}

In view of the saturation of integrals \eq {I} and \eq {I-GDH}, I now
investigate the \piN\ contribution to the inclusive electroproduction
cross sections \eq {incl1}.  That is to say, the sum over all hadronic
final states X in \Fig {incl} is approximated by the \piN\ state as depicted
in \Fig {pielprod}.%
More precisely, one has $\pi^0$p and $\pi^+$n final states for a proton in the
initial state, and $\pi^0$n and $\pi^-$p final states for a neutron.
(The discussion of different isospin channels is presented in \Sect
{piN.iso}.)

The \piN\ contribution is written in terms of helicity amplitudes,
which yield the most concise and plausible expressions, and in terms
of so-called Chew-Goldberger-Low-Nambu (CGLN) amplitudes, being best
suited for investigating Born parts.\footnote {The Born contribution
to pion electroproduction amplitudes have been given by several
authors.  I recommend von Gehlen \cite {vonGehlen69}.  The article of
Dennery \cite {Dennery61} should be corrected for a sign error: The
$\pm$ sign in Eq.\ (13) must be removed.} The partial-wave expansion
of diverse quantities of interest is computed and their isospin
decomposition is presented.

\subsection{Pion electroproduction}

I endeavor to comprehensively treat all aspects of pion
electroproduction that are needed for the purpose of saturating the
integrals \eq {I}.  For any question possibly left open, I recommend
the classical review articles of Lyth\footnote {Although Lyth's report
is very detailed and instructive, it unfortunately containes a number
of misprints.  Explicit formulae should therefore not be adopted
without verification.} \cite {Lyth78} and of Amaldi, Fubini, and
Furlan \cite {Amaldi79}.  The inclusion of polarization degrees of
freedom has been discussed in much detail by von Gehlen \cite
{vonGehlen69} and by Raskin and Donnelly \cite {Raskin89} (see also
Drechsel and Tiator \cite {Drechsel92}).

\subsubsection{Kinematics}
\begin{figure}[tb]
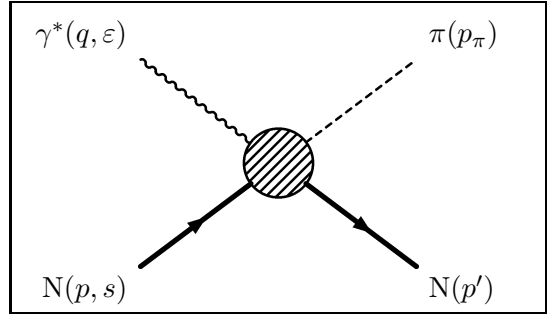

 \begin{displaymath}
  \boxedgraph(5,3){
   \fmfleft{i1,i2} \fmfright{o1,o2}
   \fmf{fermion,width=thick}{i1,v,o1}
   \fmf{boson}{i2,v}
   \fmf{dashes}{v,o2}
   \fmfblob{1\ul}{v}
   \fmflabel{N$(p,s)$}{i1} \fmflabel{$\gamma^*(q,\varepsilon)$}{i2}
   \fmflabel{N$(p')$}{o1} \fmflabel{$\pi(p_\pi)$}{o2}
  }
 \end{displaymath}
 \caption[]{
  ``Virtual photoproduction'' of pions on the nucleon.
  \label{Fig:pivirt}}
\end{figure}

Exclusive processes like pion electroproduction are conveniently
described in the center-of-mass frame of the hadronic final
state, in our case \piN$'$, where one has
\begin{equation}
 \bq+\bp = \bp_{\pi}+\bp' = 0.
\end{equation}
Note that this is \emph {not} the center-of-mass frame of the actual
reaction, eN $\to$ e$'\piN'$ (where, instead, $\bk+\bp$ would be
vanishing).  The one-photon-exchange approximation enables me to employ the
notion of ``virtual photoproduction'' of pions,
analogously to the reduction of inclusive electroproduction to
``virtual photoabsorption'' presented in \Sect {kin+X.virt}
and expressed by \Eqs {sigma-W} and \eq {incl2}.  Pictorially, this
corresponds to talking about \Fig {pielprod} in terms of
\Fig {pivirt}.%

\begin{figure}[tb]
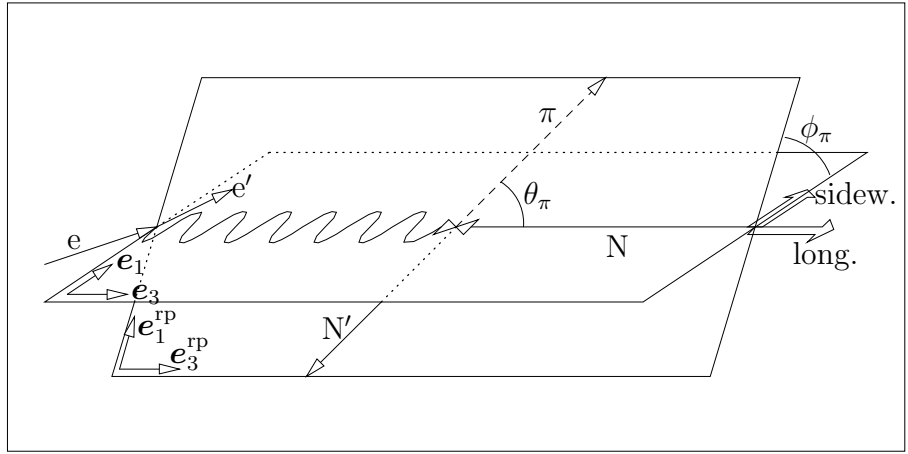

 \begin{center}
  \input kin.eepic
 \end{center}
 \caption[]{
  Kinematics of pion electroproduction in the center-of-mass frame.
  The electron momenta are scaled down in magnitude in order not to
  overshoot the dimensions of the plot.  Apart from the dreibein $\be_k$
  related to the scattering plane, the dreibein $\be^\trp_k$ related
  to the reaction plane is shown. (The respective second vectors
  are such that both dreibeine are orthonormal and right-handed.)
  Sideways and longitudinal polarizations of the incident nucleon
  are indicated.
  \label{Fig:kin}}
\end{figure}

The full kinematics of the process is presented in \Fig {kin}.%
I adopt the usual distinction between \emph {scattering plane}, defined
by the electron momenta, and \emph {reaction plane}, defined by the
momenta of the outgoing hadrons.  The angle enclosed by scattering
plane and reaction plane is called the pion azimuthal scattering angle
$\phi_\pi$.  In the center-of-mass frame, it clearly has the same
value as in the lab.  The angle between photon and pion momenta is the
frame-dependent zenith scattering angle $\theta_\pi$, which is taken to be
measured in the center-of-mass frame.  The right-handed coordinate
dreibein $\be_k$ is defined such that the photon propagates in the
direction of $\be_3$, the scattering plane is perpendicular to
$\be_2$, and the electron momenta have positive first components.  For
later use, an auxiliary right-handed coordinate dreibein $\be^\trp_k$
is defined such that $\be^\trp_3=\be_3$, the \emph {reaction} plane is
perpendicular to $\be^\trp_2$, and the first component of the \emph
{pion} momentum is positive.  These coordinate systems are related by
\begin{equation}
\begin{split}
 \be^\trp_1 &= \be_1\cos\phi_\pi + \be_2\sin\phi_\pi, \\
 \be^\trp_2 &= \be_2\cos\phi_\pi - \be_1\sin\phi_\pi.
\end{split}
\end{equation}

\subsubsection{Cross sections}
In pion electroproduction one measures triple differential cross sections
$\td\sigma_{\lambda_\te}/\td{E'}\td\Omega_\te\td\Omega_\pi$.
Here I am interested in the double differential cross section
\begin{equation}
 \frac {\td\sigma_{\lambda_\te}^{(\piN)}} {\td E'\,\td\Omega_\te} =
 \int\!\td\Omega_\pi \,
 \frac {\td\sigma_{\lambda_\te}^{(\piN)}}
  {\td{E'}\td\Omega_\te\td\Omega_\pi},
\end{equation}
where electron scattering angles are measured in the lab frame and
pion scattering angles are measured in the center-of-mass frame.  Just
like in the case of inclusive electroproduction, the difference of
electron helicities $\lambda_\te=\pm\frac12$ is considered.  The
analoga of \Eqs {sigma-W}--\eq {incl2} read
\begin{subequations}
\begin{align}
 \sigma_{1/2}^{(\piN)}(\nu,Q^2) - \sigma_{3/2}^{(\piN)}(\nu,Q^2)
  &= \frac {4\pi^2\alpha} {M\sqrt{\nu^2+Q^2}}\, \bigl[
  \eps_\mu^*(-1)\eps_\nu(-1) - \eps_\mu^*(+1)\eps_\nu(+1)
 \bigr]\, W^{\mu\nu}_{(\piN)\tlong}, \label{sigma-piN-W.GDH} \\*
 \sigma_{\tLT}^{(\piN)}(\nu,Q^2) &=
  \frac {4\pi^2\alpha} {M\sqrt{\nu^2+Q^2}}
  \im\left( \eps_\mu^*(0)\, \eps_\nu(\perp)\, W^{\mu\nu}_{(\piN)\tside}
 \right), \label{sigma-piN-W.LT}
\end{align} \label{sigma-piN-W}%
\end{subequations}
\begin{subequations}
\begin{align}
 \sigma_{1/2}^{(\piN)}(\nu,Q^2) - \sigma_{3/2}^{(\piN)}(\nu,Q^2)
 &= \frac {8\pi^2\alpha} {M^2}\, \frac {\nu} {\sqrt{\nu^2+Q^2}}
  \left[ G_1^{(\piN)}(\nu,Q^2) - \frac{Q^2}{M\nu} G_2^{(\piN)}(\nu,Q^2)
  \right], \label{sigma-piN-G.GDH} \\*
 \sigma_{\tLT}^{(\piN)}(\nu,Q^2)
 &= \frac {4\pi^2\alpha} {M^2}\, \frac {Q} {\sqrt{\nu^2+Q^2}}
  \left[ G_1^{(\piN)}(\nu,Q^2) + \frac \nu M G_2^{(\piN)}(\nu,Q^2)
  \right], \label{sigma-piN-G.LT}
\end{align} \label{sigma-piN-G}%
\end{subequations}
\begin{subequations}
\begin{align}
 \left[
  \frac {\td\sigma_{-1/2}^{(\piN)}} {\td E'\,\td\Omega_{\te}} -
  \frac {\td\sigma_{+1/2}^{(\piN)}} {\td E'\,\td\Omega_{\te}}
 \right]_{\text{long}} &=
 \Gamma \sqrt{1-\eps^2}\,
  \bigl(\sigma_{1/2}^{(\piN)}(\nu,Q^2) - \sigma_{3/2}^{(\piN)}(\nu,Q^2)\bigr),
  \label{piN.long} \\*
 \left[
  \frac {\td\sigma_{-1/2}^{(\piN)}} {\td E'\,\td\Omega_{\te}} -
  \frac {\td\sigma_{+1/2}^{(\piN)}} {\td E'\,\td\Omega_{\te}}
 \right]_{\text{side}} &=
 2\Gamma \sqrt{2\eps(1-\eps)}\, \sigma_{\tLT}^{(\piN)}(\nu,Q^2).
  \label{piN.side}
\end{align} \label{piN}%
\end{subequations}
\Eqs {sigma-piN-G} define the \piN\ contribution to structure
functions $G_{1,2}(\nu,Q^2)$.  The \piN\ contribution
$W^{\mu\nu}_{(\piN)}$ to the hadronic tensor can be read off from
\Eq {Wmunu},
\begin{align} \label{W-def-piN}
 W^{\mu\nu}_{(\piN)} &= \frac1{4\pi} \sum_{\lambda'=\pm\frac12}
  \int\! \frac{\td^3p_\pi}{(2\pi)^32p_\pi^0}
  \!\int\! \frac{\td^3p'}{(2\pi)^32p^{\prime0}} \,
  (2\pi)^4\dirac4(p+q-p'-p_{\pi}) \notag\\*
 &\quad\times \langle p,s|J^\mu(0)|p_\pi;p',\lambda'\rangle\,
  \langle p_\pi;p',\lambda'|J^\nu(0)|p,s\rangle,
\end{align}
where, as before, the incident nucleon's spin $s$ is to be taken
longitudinal (negative-helicity eigenstate) in \Eq
{sigma-piN-W.GDH} and sideways in \Eq {sigma-piN-W.LT}.  Of
course, the  helicity $\lambda'$ of the outgoing nucleon is summed over.  The
integrals run over \emph {all} pion and nucleon momenta, while the
four-dimensional delta function attends to energy and momentum
conservation.  The integration over $\bp'$ is trivial,
\begin{align}
 W^{\mu\nu}_{(\piN)} & = \frac1{64\pi^3} \sum_{\lambda'=\pm\frac12}
  \int\! \frac{\td^3p_\pi}{p_\pi^0\,p^{\prime0}} \,
  \dirac{}(p^0+q^0-p^{\prime0}-p_{\pi}^0) \notag\\*
 & \quad\times \langle p,s|J^\mu(0)|p_\pi;p',\lambda'\rangle\,
  \langle p_\pi;p',\lambda'|J^\nu(0)|p,s\rangle,
\end{align}
whereas the $\bp_\pi$ integration requires some caution.  I first
write
\begin{equation}
 \td^3p_\pi = \td\Omega_\pi\,\td|\bp_\pi|\,\bp_\pi^2
\end{equation}
in the center-of-mass frame.  Now observe that
$p^{\prime0}=\sqrt{M^2+\bp_\pi^2}$ and
$p_\pi^0=\sqrt{m_\pi^2+\bp_\pi^2}$ are functions of $|\bp_\pi|$.
Hence the argument of the delta function is quite intricate if written
in terms of $|\bp_\pi|$.  The trick is to substitute $|\bp_\pi|$ for
$W':=p^{\prime0}+p_\pi^0$.  Then the argument of the delta function is
simply $W-W'$, where the center-of-mass energy $W$ of the hadronic
final state obeys
\begin{equation}
 W^2 = \bigl(p^0+q^0\bigr)^2 = (p+q)^2 = M^2+2M\nu-Q^2.
\end{equation}
One has
$p^{\prime0}\,\td{p}^{\prime0}=p_\pi^0\,\td{p_\pi}^0=|\bp_\pi|\,\td|\bp_\pi|$,
which implies $p_\pi^0\,p^{\prime0}\td{W'}=W'\,|\bp_\pi|\,\td|\bp_\pi|$,
eventually leading to
\begin{equation} \label{Wmunu-piN}
 W^{\mu\nu}_{(\piN)} = \frac{|\bp_\pi|}{64\pi^3W}
  \sum_{\lambda'=\pm\frac12} \int\! \td\Omega_\pi \,
  \me {p,s} {J^\mu(0)} {p_\pi;p',\lambda'}\,
  \me {p_\pi;p',\lambda'} {J^\nu(0)} {p,s}.
\end{equation}
The magnitude of the pion momentum can be expressed in terms of the total
energy $W$,
\begin{equation}
 |\bp_\pi| = \frac {\sqrt{(W^2-M^2-m_\pi^2)^2-4M^2m_\pi^2}} {2W}.
\end{equation}
For later use, I also give the virtual photon's momentum,
\begin{equation}
 |\bq| = \frac {\sqrt{(W^2-M^2+Q^2)^2+4M^2Q^2}} {2W} =
 \frac MW \sqrt{\nu^2+Q^2}.
\end{equation}

\subsection{Helicity amplitudes}
The \piN\ contribution \eq {Wmunu-piN} to the hadronic tensor can
now be inserted into the pion production cross sections \eq
{sigma-piN-W}.  Strikingly simple expressions are obtained if one
introduces helicity amplitudes \cite {Jones65} defined by
\begin{equation} \label{hel-amp}
 \boxed{
 f_{\lambda',\lambda\lambda_\gamma}(W,Q^2,\theta_\pi) =
 \frac {ie}{8\pi W}\, \Bigl[ \eps_\mu(\lambda_\gamma)\,
 \me {p_\pi;p',\lambda'} {J^\mu(0)} {p,\lambda} \Bigr]_\trp}
\end{equation}
where the subscript ``rp'' means that the phases of the states are fixed
with respect to the reaction plane.  Since an angle-$\phi_\pi$ rotation
about $\bq$ transforms from scattering plane to reaction plane, the relation
between photon polarization vectors incorporated in \Eqs {sigma-piN-W}
and those of \Eq {hel-amp} is simply
\begin{equation}
 \eps(\lambda_\gamma) = e^{i\lambda_\gamma\phi_\pi} \eps_\trp(\lambda_\gamma),
\end{equation}
and the incident-nucleon's state transforms as
\begin{equation}
 |p,\lambda\rangle = e^{-i\lambda\phi_\pi} |p,\lambda\rangle_\trp.
\end{equation}
This will be relevant to the case of sideways polarization.\footnote
{Fortunately, the transformation of the \piN\ state need not be
inspected since only the phase-independent projection operator
$|p_\pi;p',\lambda'\rangle\langle p_\pi;p',\lambda'|$ appears in the
hadronic tensor \eq {Wmunu-piN}.}

From parity conservation, one has
\begin{equation}
 f_{-\lambda',-\lambda-\lambda_\gamma} =
 - (-1)^{\lambda_\gamma-\lambda+\lambda'}
 f_{\lambda',\lambda\lambda_\gamma},
\end{equation}
so that six of the twelve amplitudes
$f_{\lambda',\lambda\lambda_\gamma}$ are mutually independent.  One
commonly replaces the indices $\pm\frac12$ and $\pm1$ by simple
$\pm$'s and chooses the four transverse amplitudes
$f_{\lambda',\lambda+}$ and the two longitudinal amplitudes
$f_{\lambda',+0}$ to be the independent ones occurring explicitly.

For longitudinal nucleon polarization, \Eq {sigma-piN-W.GDH},
one lets $\lambda=-\frac12$, sums over $\lambda'$, and subtracts the
result for $\lambda_\gamma=+1$ from the result for $\lambda_\gamma=-1$,
arriving at
\begin{equation} \label{GDH-hel}
 \boxed{
 \sigma_{1/2}^{(\piN)} - \sigma_{3/2}^{(\piN)} =
 \frac{|\bp_\pi|}{|\bq|} \!\int\! \td\Omega_\pi \,
 \bigl( |f_{+,++}|^2 + |f_{-,++}|^2 - |f_{+,-+}|^2 - |f_{-,-+}|^2 \bigr).}
\end{equation}
The case of sideways nucleon polarization, \Eq {sigma-piN-W.LT},
requires some more caution with respect to phases.  Let $\bS$ denote
the spin operator.  Then, according to Jacob and Wick \cite {Jacob59},
the phase relation between different-helicity\footnote {Again, the
reader may recall that nucleon polarization in \emph {positive} $x^3$
direction corresponds to \emph {negative} helicity in the
center-of-mass frame.  Moreover, the boost transforming from the lab
to the center of mass leaves \Eqs {ladder} unaltered.} nucleon
states is such that the operators $S^1\pm iS^2$ switch between states
without giving rise to an additional phase,
\begin{equation} \begin{split} \label{ladder}
 (S^1 \pm iS^2)\, |p,\mp\tfrac12\rangle &= 0, \\*
 (S^1 \pm iS^2)\, |p,\pm\tfrac12\rangle &= |p,\mp\tfrac12\rangle.
\end{split} \end{equation}
The sideways-polarized state is an eigenstate of $S^1$ corresponding
to eigenvalue $+\frac12$, which is easily constructed from relations
\eq {ladder} to be
\begin{align}
 |p,\text{side}\rangle &= \frac1{\sqrt2}\,
  \bigl( |p,+\tfrac12\rangle + |p,-\tfrac12\rangle \bigr) \notag\\
 &= \frac1{\sqrt2}\, \bigl(
  e^{-i\phi_\pi/2}\, |p,+\tfrac12\rangle_\trp +
  e^{+i\phi_\pi/2}\, |p,-\tfrac12\rangle_\trp
 \bigr).
\end{align}
Inserting this into \Eq {sigma-piN-W.LT} using \Eq {Wmunu-piN},
one obtains
\begin{align} \label{LT-hel1}
 \sigma_{\tLT}^{(\piN)}
 &= \frac1{2\sqrt2}\, \frac{|\bp_\pi|}{|\bq|}
  \sum_{\lambda',\lambda_1,\lambda_2 = \pm\frac12} \sum_{\lambda_\gamma = \pm1}
  \!\int\! \td\Omega_\pi
  \im \bigl( i f_{\lambda',\lambda_10}^* f_{\lambda',\lambda_2\lambda_\gamma}
  e^{i(\lambda_1-\lambda_2+\lambda_\gamma)\phi_\pi} \bigr) \notag\\
 &= \frac1{\sqrt2}\, \frac{|\bp_\pi|}{|\bq|} \!\int\! \td\Omega_\pi
  \Bigl[ \re\bigl(f_{-,+0}^*f_{+,++}-f_{+,+0}^*f_{-,++}\bigr) \notag\\*
 &\qquad\qquad\qquad\qquad
  -\im\bigl(f_{-,+0}^*(f_{-,++}+f_{+,-+})+f_{+,+0}^*(f_{+,++}-f_{-,-+})
  \bigr) \sin\phi_\pi\bigg. \notag\\*
 &\qquad\qquad\qquad\qquad
  +\re\bigl(f_{-,+0}^*f_{-,-+}+f_{+,+0}^*f_{+,-+}\bigr)
  \cos(2\phi_\pi) \Bigr],
\end{align}
where $\phi_\pi$ dependent terms have been kept to render the
possibility of expressing the triple differential pion
electroproduction cross section in terms of helicity amplitudes.
Owing to the fact that both $\sin\phi_\pi$ and $\cos(2\phi_\pi)$
integrate to zero, \Eq {LT-hel1} simplifies to
\begin{equation} \label{LT-hel2}
 \boxed{ \sigma_{\tLT}^{(\piN)} =
  \frac1{\sqrt2}\, \frac{|\bp_\pi|}{|\bq|} \!\int\! \td\Omega_\pi
  \re\bigl(f_{-,+0}^*f_{+,++}-f_{+,+0}^*f_{-,++}\bigr).}
\end{equation}
 \Eqs {GDH-hel} and \eq {LT-hel2} may be inserted into \Eqs {sigma-piN-G}
to obtain the \piN\ contribution to structure functions $G_{1,2}(\nu,Q^2)$
in terms of helicity amplitudes.

\subsection{Chew-Goldberger-Low-Nambu amplitudes}
\label{Sect:piN.CGLN}

In view of the calculation of the Born contribution to cross
sections $\sigma_{1/2}^{(\piN)}-\sigma_{3/2}^{(\piN)}$ and
$\sigma_{\tLT}^{(\piN)}$, a different set of
pion-electroproduction amplitudes suggests itself, namely the
amplitudes $\mF_{1\ldots6}(W,Q^2,\theta_\pi)$ defined by \cite
{Dennery61}
\begin{align} \label{CGLN-def}
 & \frac {-ie}{8\pi W}\, \me {p_\pi;p',\lambda'}{\bJ(0)}{p,\lambda} = \notag\\*
 & \qquad \chi^\dagger(p',\lambda')\, \bigl[ \bsigma^\perp \mF_1
   -i(\bsigma\ndot\hat\bp_\pi)\,(\bsigma\ntimes\hat\bq)\, \mF_2
   +\hat\bp_\pi^\perp\, (\bsigma\ndot\hat\bq)\, \mF_3
   +\hat\bp_\pi^\perp\, (\bsigma\ndot\hat\bp_\pi)\, \mF_4 \notag\\*
 & \qquad\qquad\qquad + \hat\bq\, (\bsigma\ndot\hat\bq)\, \mF_5
   + \hat\bq\, (\bsigma\ndot\hat\bp_\pi)\, \mF_6 \bigr]\, \chi(p,\lambda)\Big.,
\end{align}
where the circumflex accent denotes a unit vector, e.g.\
$\hat\bq=\bq/|\bq|$, and the superscript $\perp$ stands for projection
onto the plane perpendicular to the photon momentum, e.g.\
$\bsigma^\perp=\bsigma-(\bsigma\ndot\hat\bq)\,\hat\bq$.
Current conservation $q\ndot J=0$ then fixes the charge-density matrix
element
\begin{equation}
 \frac {-ie}{8\pi W}\, \me {p_\pi;p',\lambda'} {J^0(0)} {p,\lambda} =
 \frac {|\bq|}{q^0}\, \chi^\dagger(p',\lambda')\,
 \bigl[ (\bsigma\ndot\hat\bq)\, \mF_5 + (\bsigma\ndot\hat\bp_\pi)\, \mF_6
 \bigr]\,
 \chi(p,\lambda).
\end{equation}
This indicates the purely longitudinal nature\footnote {Some authors
omit the projection indicated by superscript $\perp$ in \Eq
{CGLN-def}, which has the effect of contaminating amplitudes $\mF_5$ and
$\mF_6$ with transverse pieces $-\mF_1-\cos\theta_\pi \mF_3$ and
$-\cos\theta_\pi \mF_4$, respectively.  Most of these authors
subsequently define primed amplitudes $\mF_{5,6}'$, which have the
transverse portions subtracted and which coincide with my unprimed
ones.  Definition \eq {CGLN-def} is adopted from Drechsel and
Tiator \cite {Drechsel92}.} of amplitudes $\mF_{5,6}$.

The six electroproduction amplitudes defined by \Eq {CGLN-def} are
commonly called CGLN amplitudes, although Chew et al.\ \cite {Chew57}
merely considered photoproduction, where only four amplitudes
$\mF_{1\ldots4}$ occur. Dennery \cite {Dennery61} was actually the first
to treat electroproduction, where also longitudinal degrees of freedom
come in.

Introducing explicit photon-polarization vectors
$\eps(\lambda_\gamma)$ and Pauli spinors $\chi^\dagger(p',\lambda')$,
$\chi(p,\lambda)$ into definitions \eq {hel-amp} and \eq {CGLN-def},
one obtains the following connection between helicity amplitudes and CGLN
amplitudes \cite {Jones65}:
\begin{subequations}
\begin{align}
 f_{+,++} &=  f_{-,--} =  \sqrt2\, \sin\frac{\theta_\pi}2\,
  \left( \mF_1+\mF_2 + \cos^2\frac{\theta_\pi}2\, (\mF_3+\mF_4) \right), \\
 f_{-,++} &= -f_{+,--} = -\sqrt2\, \cos\frac{\theta_\pi}2\,
  \left( \mF_1-\mF_2 - \sin^2\frac{\theta_\pi}2\, (\mF_3-\mF_4) \right), \\
 f_{+,-+} &= -f_{-,+-} =  \sqrt2\, \cos\frac{\theta_\pi}2\,
  \sin^2\frac{\theta_\pi}2\, (\mF_3-\mF_4), \\
 f_{-,-+} &=  f_{+,+-} = -\sqrt2\, \sin\frac{\theta_\pi}2\,
  \cos^2\frac{\theta_\pi}2\, (\mF_3+\mF_4), \\
 f_{+,+0} &=-f_{-,-0} = \frac Q{q^0}\, \cos\frac{\theta_\pi}2\, (\mF_5+\mF_6), \\
 f_{-,+0} &= f_{+,-0} = \frac Q{q^0}\, \sin\frac{\theta_\pi}2\, (\mF_5-\mF_6).
\end{align} \label{hel-CGLN}%
\end{subequations}
Employing these relations, the pion-production cross sections \eq
{GDH-hel} and \eq {LT-hel2} can be written
\begin{subequations}
\begin{equation} \label{GDH-CGLN}
 \boxed{ \begin{split}
  \sigma_{1/2}^{(\piN)} - \sigma_{3/2}^{(\piN)} =
   2\, \frac {|\bp_\pi|} {|\bq|} \!\int\!\td\Omega_{\pi}\, \Bigl[
  & |\mF_1|^2 + |\mF_2|^2 - 2\re(\mF_1^*\mF_2)\cos\theta_\pi \\
  & + \re(\mF_1^*\mF_4+\mF_2^*\mF_3)\sin^2\theta_\pi \Bigr]
 \end{split} }
\end{equation}
and
\begin{equation}
 \boxed{ \begin{split}
  \sigma_{\tLT}^{(\piN)} =
   \frac {|\bp_\pi|} {|\bq|}\, \frac Q{q^0} \!\int\!\td\Omega_{\pi}\, \Bigl[
 & \re(\mF_5^*\mF_1-\mF_6^*\mF_2) - \re(\mF_5^*\mF_2-\mF_6^*\mF_1)
   \cos\theta_\pi \\
 & + \tfrac12 \re(\mF_5^*\mF_4-\mF_6^*\mF_3)\sin^2\theta_\pi \Bigr]
 \end{split} }
\end{equation} \label{sigma-CGLN}%
\end{subequations}
respectively.  \Eqs {sigma-CGLN} may be inserted into \Eqs
{sigma-piN-G} to obtain the \piN\ contribution to structure functions
$G_{1,2}(\nu,Q^2)$ in terms of CGLN amplitudes.

\subsection{Partial-wave expansion}
Partial-wave expansion of amplitudes corresponds to transforming the
pion scattering-angle dependence into the dependence on the discrete
total angular-momentum quantum number.  It is most concisely written
for helicity amplitudes \cite {Jones65}:
\begin{subequations}
\begin{align}
 f_{\pm,++} &= -\sqrt2 \sum_{J\ge\frac12} (J+\tfrac12)\,
  \bigl(A_{(J-\frac12)+} \pm A_{(J+\frac12)-}\bigr)\,
  d^J_{\frac12\mp\frac12}(\theta_\pi), \\
 f_{\pm,-+} &= \frac1{\sqrt2} \sum_{J\ge\frac32} (J+\tfrac12)
  \sqrt{(J-\tfrac12)(J+\tfrac32)}\,
  \bigl(B_{(J-\frac12)+} \pm B_{(J+\frac12)-}\bigr)\,
  d^J_{\frac32\mp\frac12}(\theta_\pi), \\
 f_{\pm,++} &= \sum_{J\ge\frac12} (J+\tfrac12)\,
  \bigl(C_{(J-\frac12)+} \mp C_{(J+\frac12)-}\bigr)\,
  d^J_{-\frac12\mp\frac12}(\theta_\pi).
\end{align} \label{par-wave}%
\end{subequations}
The half-integer number $J$ corresponds to the total angular momentum
of the \piN\ state.  The \emph {helicity multipoles} occurring in
\Eq {par-wave} are related to conventional electric, magnetic, and
scalar (or longitudinal) multipoles via
\begin{subequations}
\begin{align}
 A_{l+}(W,Q^2) &= \frac{l+2}2\, E_{l+} + \frac l2\, M_{l+}, \\
 A_{l-}(W,Q^2) &= -\frac{l-1}2\, E_{l-} + \frac{l+1}2\, M_{l-}, \\
 B_{l+}(W,Q^2) &= E_{l+} - M_{l+}, \bigg.\\
 B_{l-}(W,Q^2) &= E_{l-} + M_{l-}, \bigg.\\
 C_{l+}(W,Q^2) &= \frac Q{|\bq|}\, (l+1)\, S_{l+}
                = \frac Q{q^0}\, (l+1)\, L_{l+}, \\
 C_{l-}(W,Q^2) &= -\frac Q{|\bq|}\, l\, S_{l-} = -\frac Q{q^0}\, l\, L_{l-},
\end{align}
\end{subequations}
where $l$ represents the \emph {orbital} angular momentum of the \piN\
system.  In the real-photon limit, multipoles $A_{l\pm}$ and
$B_{l\pm}$ match the notation of Walker \cite {Walker69}.

As far as the rotation matrix elements $d^J_{MM'}$
are concerned, I adopt the conventions of Brink and Satchler \cite
{Brink62}.  In particular, the orthogonality condition reads (indices
$M,M'$ \emph {not summed})
\begin{equation}
 \int\!\td\Omega\, d^J_{MM'}(\theta)\, d^{J'}_{MM'}(\theta) =
 \frac{4\pi}{2J+1} \, \delta_{JJ'}.
\end{equation}
Inserting expansions \eq {par-wave} into cross sections \eq
{GDH-hel} and \eq {LT-hel2}, the integration over pion scattering
angles is easily performed, yielding
\begin{subequations}
\begin{equation}
 \boxed{ \begin{split}
 \sigma_{1/2}^{(\piN)} - \sigma_{3/2}^{(\piN)}
  = 8\pi\, \frac {|\bp_\pi|} {|\bq|} \sum_{J\ge\frac12} \biggl[
 & (J+\tfrac12)\bigl(|A_{(J-\frac12)+}|^2 + |A_{(J+\frac12)-}|^2\bigr) \\
 & - \frac14 (J-\tfrac12)(J+\tfrac12)(J+\tfrac32)
  \bigl(|B_{(J-\frac12)+}|^2 + |B_{(J+\frac12)-}|^2\bigr) \biggr]
 \end{split} }
\end{equation}
and
\begin{equation}
 \boxed{
 \sigma_{\tLT}^{(\piN)} = 4\pi\, \frac {|\bp_\pi|} {|\bq|} \sum_{J\ge\frac12}
 (J+\tfrac12) \re\bigl(C_{(J-\frac12)+}^*A_{(J-\frac12)+}
 + C_{(J+\frac12)-}^*A_{(J+\frac12)-}\bigr) }
\end{equation} \label{sigma-mult}%
\end{subequations}
 \Eqs {sigma-mult} may be inserted into \Eqs
{sigma-piN-G} to obtain the \piN\ contribution to structure functions
$G_{1,2}(\nu,Q^2)$ in terms of multipoles.

\subsection{Isospin decomposition}
\label{Sect:piN.iso}

There are four isospin channels of the pion-electroproduction process, viz.
\begin{equation} \label{piN-channels}
 \begin{split}
  \gamma^*\tp & \to \begin{cases} \pi^0\tp \\ \pi^+\tn \end{cases} \\
  \gamma^*\tn & \to \begin{cases} \pi^0\tn \\ \pi^-\tp \end{cases}
 \end{split}
\end{equation}
Assuming isospin symmetry, it is convenient to define amplitudes for
the absorption of isoscalar and isovector photons (see, e.g., Lyth
\cite {Lyth78}).  For instance, the CGLN amplitudes
$\mF_{1\ldots6}(W,Q^2,\theta_\pi)$ defined in \Sect {piN.CGLN} are
decomposed according to
\begin{equation} \label{CGLN-channels}
 \begin{split}
  \mF_i^{(\pi^0\tp)} & = \mF_i^{(0)} + \mF_i^{(+)}, \\
  \mF_i^{(\pi^+\tn)} & = \sqrt2 \bigl( \mF_i^{(0)} + \mF_i^{(-)} \bigr), \\
  \mF_i^{(\pi^0\tn)} & = - \mF_i^{(0)} + \mF_i^{(+)}, \\
  \mF_i^{(\pi^-\tp)} & = \sqrt2 \bigl( \mF_i^{(0)} - \mF_i^{(-)} \bigr),
 \end{split}
\end{equation}
where superscripts $(\pm)$ refer to isovector photons, and (0) corresponds
to an isoscalar photon.
Analogous decompositions apply to helicity amplitudes and partial-wave
multipoles.  Observe that isospin symmetry reduces the number of independent
channels from four to three.

Since I am concerned with virtual photoabsorption cross sections,
amplitudes are squared and subsequently added in pairs.  For example,
the cross-section difference \eq {GDH-CGLN} can be cast into the compact
form
\begin{equation}
 \sigma_{1/2}^{(\piN)} - \sigma_{3/2}^{(\piN)} =
  2\, \frac {|\bp_\pi|} {|\bq|} \!\int\!\td\Omega_{\pi}\,
  \sum_{i,j=1}^4 K_{ij} \mF_i^* \mF_j,
\end{equation}
where I have introduced the real symmetric matrix
\begin{equation}
 K =
 \begin{pmatrix}
  1 & -\cos\theta_\pi & 0 & \tfrac12\sin^2\theta_\pi \\
  -\cos\theta_\pi & 1 & \tfrac12\sin^2\theta_\pi & 0 \\
  0 & \tfrac12\sin^2\theta_\pi & 0 & 0 \\
  \tfrac12\sin^2\theta_\pi & 0 & 0 & 0
 \end{pmatrix}.
\end{equation}
For pion electroproduction on the proton, one has final states $\pi^0\tp$
and $\pi^+\tn$.  Accordingly,
\begin{subequations}
\begin{align}
 \Bigl[ \sigma_{1/2}^{(\piN)} - \sigma_{3/2}^{(\piN)} \Bigr]_\tp
 & = 2\, \frac {|\bp_\pi|} {|\bq|} \!\int\!\td\Omega_{\pi}\,
  \sum_{i,j=1}^4 K_{ij}
  \Bigl( \mF_i^{(\pi^0\tp)*} \mF_j^{(\pi^0\tp)}
  + \mF_i^{(\pi^+\tn)*} \mF_j^{(\pi^+\tn)} \Bigr] \notag \\
 & = 2\, \frac {|\bp_\pi|} {|\bq|} \!\int\!\td\Omega_{\pi}\,
  \sum_{i,j=1}^4 K_{ij}
  \Bigl[ 3 \mF_i^{(0)*} \mF_j^{(0)} + 2 \mF_i^{(-)*} \mF_j^{(-)}
  + \mF_i^{(+)*} \mF_j^{(+)} \notag \\*
 & \qquad\qquad\qquad\qquad\qquad
  + 2\re \Bigl( \mF_i^{(0)*} \bigl( 2\mF_j^{(-)}+\mF_j^{(+)} \bigr) \Bigr)
  \Bigr],
\end{align}
while for the neutron one has
\begin{align}
 \Bigl[ \sigma_{1/2}^{(\piN)} - \sigma_{3/2}^{(\piN)} \Bigr]_\tn
 & = 2\, \frac {|\bp_\pi|} {|\bq|} \!\int\!\td\Omega_{\pi}\,
  \sum_{i,j=1}^4 K_{ij}
  \Bigl[ \mF_i^{(\pi^0\tn)*} \mF_j^{(\pi^0\tn)}
  + \mF_i^{(\pi^-\tp)*} \mF_j^{(\pi^-\tp)} \Bigr] \notag \\
 & = 2\, \frac {|\bp_\pi|} {|\bq|} \!\int\!\td\Omega_{\pi}\,
  \sum_{i,j=1}^4 K_{ij}
  \Bigl[ 3 \mF_i^{(0)*} \mF_j^{(0)} + 2 \mF_i^{(-)*} \mF_j^{(-)}
  + \mF_i^{(+)*} \mF_j^{(+)} \notag \\*
 & \qquad\qquad\qquad\qquad\qquad
  - 2\re \Bigl( \mF_i^{(0)*} \bigl( 2\mF_j^{(-)}+\mF_j^{(+)} \bigr) \Bigr)
  \Bigr].
\end{align}
\end{subequations}
Analogous formulae apply to the decompositions of $\sigma_\tLT^{(\piN)}$
and to the expressions of virtual-photoabsorption cross sections in terms
of helicity amplitudes or multipoles.

\subsection{Outlook}
As noted in the introduction to this chapter, the present section aims at
providing formulae for the resonance saturation of generalized GDH integrals.
In order to reliably perform the saturation, a few additional investigations
are desirable.  The high-energy behavior of structure functions
$G_{1,2}(\nu,Q^2)$ should be analyzed by means of Reggeization.  Moreover,
the influence of final states other than \piN\ ought to be estimated.  In
particular, the role of double-pion production should be clarified.

\clearpage{\pagestyle{empty}\cleardoublepage}
\chapter{Summary and conclusion}

This thesis presents an attempt to comprehensively treat the
theoretical aspects of the GDH sum rule, in such a way that all
proposed modifications can rigorously be checked.  The ingredients of
the current-algebra derivation of the sum rule, in particular the
infinite-momentum limit, turn out to be most important.

\subsubsection{Derivations of the GDH sum rule}
In \Ch {derivations}, I presented all derivations of the GDH sum rule
that can be found in the literature [\plaincite {Gerasimov65}\nocite
{Drell66}--\plaincite {Hosoda66a,Kawarabayashi66a,Dicus72}].  I
commented on the impact of the lowest-order consideration of
electromagnetic coupling (see pp.~\pageref {lowest-order-story1},
\pageref {lowest-order-story2}, and \pageref {lowest-order-story3}).
As far as equal-times current algebra is concerned, I presented a
derivation that exhibits the infinite-momentum limit as its last step
-- in contrast to \Refs {Hosoda66a} and \plaincite {Kawarabayashi66a},
where this limit is employed in between.
The \emph {finite-momentum GDH sum rule} \eq {FMGDH} presents the form
that the sum rule takes before the infinite-momentum limit is taken.
At finite momenta, the GDH integration path is a parabola in
$(\nu,q^2)$ plane as depicted in \Fig {q2nu}, p.~\pageref {Fig:q2nu}.
Since the legitimacy of the infinite-momentum limit cannot be
questioned within the scope of current algebra, some perturbative
model has to be adopted as regards this problem.  The finite-momentum
GDH sum rule is well suited for this investigation.  It re-appears at
several other places within this thesis.

\subsubsection{The GDH sum rule within perturbative models}
In \Ch {pert}, I presented some important tests of the GDH sum rule,
concerning its validity within three renormalizable field theoretical
models: a pion-nucleon model with pseudoscalar \piN\ coupling and
minimal electromagnetic interaction \cite {Gerasimov75}, quantum
electrodynamics \cite {Altarelli72,Tsai75}, and the Weinberg-Salam
model of electro-weak interactions of leptons \cite {Altarelli72}.  In
all three instances, both sides of the GDH sum rule are considered to
lowest non-trivial order in the respective coupling constant $g$ and
$e$.  Since the fermion's anomalous magnetic moment is of the order
$g^2$ in the pion-nucleon model and $e^2$ in QED and the
Weinberg-Salam model, the left-hand side of the sum rule vanishes to
lowest order in all these models.  Then, I generalized the
electromagnetic coupling of the pion-nucleon model by introducing an
anomalous magnetic moment $\kappa_0$ to the nucleon on the vertex
level, which gives rise to a divergent GDH integral, because due to
the derivative coupling of the anomalous magnetic moment, the
polarized photoabsorption cross section approaches a non-zero constant
at large photon energy.

In QED and in the Weinberg-Salam model, there is no hadronic
interaction.  Hence, the lowest non-trivial order is $\alpha^2$.  (In
the pion-nucleon model, it is $\alpha g^2$.) On the other hand, the
GDH sum rule of the physical nucleon is explicitly considered to order
$\alpha$ only. That is to say, electromagnetism is treated to lowest
order, while hadronic interactions are, of course, non-perturbative.
The reader may note that this does not reduce the instructive power of
testing the GDH sum rule by means of electro-weak processes.  Rather,
the following conclusion ought to be kept in mind.  In all models
considered, the GDH integral either adopts the predicted value (viz.,
zero), or both sides of the sum rule diverge due to anomalous magnetic
moments.  No instance of a ``subtraction at infinity'', which would
spoil the GDH sum rule (see \Sect {fixed-pole}), is detected.
Moreover, the confirmation of the sum rule at next-to-leading order
($\alpha^2$) supports the idea that in principle, the forward Compton
amplitude could be considered to all orders in $\alpha$.  This is
relevant to the discussion of a possible $J=1$ fixed pole in
angular-momentum plane.

To order $\alpha g^4$ within the pion-nucleon model, or order
$\alpha^3$ within QED and the Weinberg-Salam model, the left-hand side
of the sum rule would be non-vanishing.  Thus, a study of the
respective next orders within these models would in principle be
interesting, but quite painstaking, too.

\subsubsection{Possible sources of modifications}

 \Ch {mod} has been devoted to a thorough discussion of all possible sources of
modifications of the GDH sum rule.  Particular attention has been paid to
claimed modifications within the current-algebra approach.

But first, in \Sect {fixed-pole}, I discussed the effect of a fixed
Regge singularity \cite {Abarbanel68} at $J=1$ in the complex
angular-momentum plane, emphasizing the crucial role of spin.  Such a
singularity would give rise to a ``subtraction at infinity'', where
the subtraction constant is essentially the residue of the fixed pole.
I showed that the subtraction constant is directly related to the
high-energy behaviour of a certain Compton-scattering cross section,
see \Eq {sigma-flip}.  However, I stressed that neither of the perturbative
models considered in \Ch {pert} yields a fixed pole.  Assuming that
the forward Compton amplitude can be defined to all orders in the
electromagnetic coupling and that each order obeys a dispersion
relation possibly subtracted at infinity, I argued that a very
peculiar dependence among fundamental parameters of the standard model
would emerge if not \emph {all} subtraction constants were
simultaneously vanishing (see \Eq {sum=0}).

In \Sect {anomcomm}, I thoroughly discussed the possibility of a
correction to the GDH sum rule due to an anomalous commutator of
electric charge densities.  After an introduction into general
features of Schwinger terms and anomalous commutators, I presented the
vital tools for their computation.  I particularly focussed on the
investigation presented by Chang, Liang, and Workman [\plaincite
{Chang94a}\nocite {Chang91}--\plaincite {Chang92}], which, at the
second glance, turns out to be intimately related to a theory
containing spin-1 mesons as gauge fields of chiral transformations.
Finally, I presented the most general non-naive charge-density
commutator relevant to the GDH sum rule.  I calculated the
modification of the \emph {finite-momentum} GDH sum rule that is
brought about by the non-naive charge-density commutator, emphasizing
the fact that the legitimacy of the infinite-momentum limit remains to
be checked in each concrete case under consideration.  This legitimacy
corresponds to the possibility of interchanging the infinite-momentum
limit $p^0\to\infty$ with the photon-energy integration.

In \Sect {IML}, I illustrated the possibility of a finite modification
due to the infinite-momentum limit used in the current-algebra
derivation of the GDH sum rule.  I stressed that the only instance of
such a correction is found in the Weinberg-Salam model, where it \emph
{compensates} a modification formerly brought about by an anomalous
charge-density algebra.  This has to be regarded as a very strong
evidence for the validity of the unmodified GDH sum rule.

 \Sect {PRP} comprises most of the original work of this thesis.  I
presented the calculation of the anomalous equal-times commutator of
electric charge-densities within the Weinberg-Salam model.  The
anomaly is due to the fermion-triangle loop incorporated in the
Z$^0$-exchange graphs displayed in \Fig {additionalgraphs}(d),
p.~\pageref {Fig:additionalgraphs}.  The one-electron matrix element
of the charge-density commutator is given in \Eq {Z0-comm3}.  It
translates into the electric-dipole-moment commutator presented in
\Eq {WSM-D-comm}.  Thereupon, the modified finite-momentum GDH sum
rule \eq {WSM-mod-FMGDH} arises.  

On the other hand, the model considered allows to investigate the
infinite-momentum limit by calculating the forward virtual Compton
amplitude $f_2(\nu,q^2)$ for all values of photon energy $\nu$ and
virtuality $q^2$.  That is to say, I inspected the GDH integral at
\emph {finite} electron energies $p^0$, where the integration runs
along the parabolae shown in \Fig {WSM-q2nu} (p.~\pageref
{Fig:WSM-q2nu}), and took the $p^0\to\infty$ limit subsequently.  Then,
I examined whether this limit deviates from the \emph {genuine} GDH
integral, which is taken along the straight line at $q^2=0$.  This is
indeed the case, due to the Z$^0$-exchange graphs.  The contribution
from these graphs to amplitude $f_2(\nu,q^2)$ is given by \Eq
{WSM-f2}.  I explicitly showed that the finite-momentum GDH integral
picks up a certain constant from the part of the integration path that
lies above the two-electron threshold $q^2=4m^2$.  On the other hand,
below this threshold, the contribution of Z$^0$ exchange to the GDH
integrand vanishes identically (cf.\ \Fig {fq2}, p.~\pageref
{Fig:fq2}).  This illustrates that indeed, the $p^0\to\infty$ limit
cannot be interchanged with the $\nu$ integration, i.e., the
finite-momentum GDH integral is related in a non-naive way to the
genuine GDH integral with integrand
$\sigma_{1/2}(\nu)-\sigma_{3/2}(\nu)$, as expressed by \Eq {WSM-IML}.
Inserting then the modified finite-momentum sum rule \eq
{WSM-mod-FMGDH}, one recovers the unmodified version \eq {WSM-GDH}.

This calculation has revealed the very important fact that the naive
infinite-momentum limit can\emph{not} be applied if anomalous
charge-density commutators occur.  Rather, it has to be expected that
the infinite-momentum limit \emph {compensates} any modification
obtained from anomalous charge-density commutators.

In \Sect {a1}, I returned to hadron physics and investigated the
exchange of axial-vector mesons in the $t$ channel.  The purpose of
this section has been twofold.  Firstly, I reminded to the fact that a
possible $J=1$ fixed pole in complex angular-momentum plane cannot be
attributed to $t$-channel exchange of spin-1 particles.  Secondly, I
pointed out that axial-vector exchange is analogous to Z$^0$ exchange
considered in the Salam-Weinberg model.  Therefore, the same
compensation mechanism is found.  That is to say, one obtaines an
anomalous-commutator correction at finite nucleon momenta, which is
compensated by the infinite-momentum limit.  On the other hand, I
proved in \Sect {a1} that the result of Chang, Liang, and Workman
[\plaincite {Chang94a}\nocite {Chang91}--\plaincite {Chang92}]
obtained from anomalous charge-density commutators can also be
understood by means of axial-vector exchange.  This again shows that
the modification of the GDH sum rule that is claimed in \Refs
{Chang94a}\nocite {Chang91}--\plaincite {Chang92} actually appears
only at finite momenta and does not survive the infinite-momentum
limit.  In particular, \Sect {a1} proves that the central result of
\Sect {PRP} does not crucially depend on the fact that higher orders
of $\alpha$ have to be considered in the Weinberg-Salam model.

In \Sect {quark}, I investigated the role of possible anomalous
magnetic moments $\kappa_q$ of quarks.  I calculated the modified
charge-density algebra \eq {kappa-comm} that is brought about by non-vanishing
$\kappa_q$.  Applying the current-algebra derivation,
this gives rise to a definite modification of the GDH sum rule, expressed
by \Eq {kappa-GDH} with \Eqs {kappa-gA}.  I argued that in this
instance, the modification must be expected to survive the infinite-momentum
limit.  

I emphasized that, for fundamental reasons,
 anomalous magnetic moments of \emph {current} quarks
vanish, while \emph {constituent} quarks may have
$\kappa_q\neq0$.  In view of the GDH sum rule, this leads to an apparent
inconsistency between constituent-quark models and QCD.  I argued that
this paradox can be remedied by recalling that the non-resonant
background is missing from the GDH integral in constituent-quark
models.  Thus, the putative modification constant is just this missing
portion of the integral, and an actual modification of the sum rule
must not be expected.

In \Sect {Ying}, I examined a modification of the GDH sum rule claimed
in the literature \cite {Ying96a} due to ``the presence of a localized
region inside the nucleon, in which the electromagnetic gauge symmetry
is spontaneously broken down'' \cite {Ying96a}.  I argued that the
proposed mechanism does actually \emph {not} affect the GDH sum rule.

In \Sect {modLET}, I pointed out that non-naive current algebras have \emph
{no} effect on the low-energy theorem of Low, Gell-Mann, and
Goldberger, although this may be expected upon considering Low's
derivation \cite {Low54} of the theorem.  In fact, the putative
modification of the low-energy theorem is remedied by the contribution
from the sea\-gull amplitude that is induced by the non-naive
commutator.  I want to stress that caution is advisable as soon as
statements on \emph {amplitudes} are derived from current commutators.
Very often, a commutator correction is cancelled by the
corresponding sea\-gull contribution.  Generally, this signalizes that
there is a less awkward approach.  (The Bjorken sum rule \cite
{Bjorken66} presents another example of this rule of thumb.)

\subsubsection{The state of the art}

The results of \Ch {mod} may be summarized by stating that none of the
hitherto proposed modifications of the GDH sum rule survives a close
inspection.  This is the state of the art after three decades of hard
thinking!  Thus, I conclude that a modification appears highly
unlikely.

\subsubsection{$Q^2$ evolution of the GDH sum rule}

In \Ch {Q2}, I investigated the $Q^2$ evolution of the GDH sum rule,
i.e., the transition from polarized photoabsorption to polarized
inclusive electroproduction in the one-photon-exchange approximation.
After introducing the pertinent kinematics and cross sections, I
discussed different generalizations of the GDH integral, which
coincide both at the real-photon point $Q^2=0$ and in the scaling
limit $Q^2\to\infty$.  In particular, I considered the standard
integral $I_1(Q^2)$ defined in \Eq {I.1} in terms of structure
function $G_1(\nu,Q^2)$, as well as the more physically motivated
integrals $I_\tGDH(Q^2)$ and $\tilde I_\tGDH(Q^2)$ defined in \Eqs
{I-GDH}, which involve the linear combination
$G_1(\nu,Q^2)-\frac{Q^2}{M\nu}G_2(\nu,Q^2)$.  As demonstrated in \Sect
{kin+X.virt}, this combination is directly related to the absorption
of transversely polarized virtual photons, while structure function
$G_1(\nu,Q^2)$ also incorporates the interference between longitudinal
and transverse photon polarizations.

At $Q^2=0$, the so-defined integrals coincide due to explicit
factors of $Q^2$.  At $Q^2\to\infty$, they
coincide due to the scaling behavior of $G_2(\nu,Q^2)$.  I emphasized
that at intermediate $Q^2$, substantial deviation must be expected.
Therefore, in \Sect {gen.G2}, I studied the limits $Q^2\to0$ and
$Q^2\to\infty$ beyond leading terms.  Using phenomenologically
plausible assumptions, I showed that indeed the \emph {slopes} at
$Q^2=0$, i.e., $I_1'(0)$, $I_\tGDH'(0)$ and $\tilde I_\tGDH'(0)$,
considerably differ in the case of the proton (presently, data are to poor
to consider the neutron case), see \Eqs {low-Q2-num}, \eq {Soffer-num}
and \Fig {soffer}.

As for the high-$Q^2$ region, I considered higher-twist corrections to
generalized GDH integrals $I_1(Q^2)$, $I_\tGDH(Q^2)$, and $\tilde
I_\tGDH(Q^2)$.  In \Eqs {twist}, these corrections are expressed in
terms of coefficients $a_2$, $d_2$, and $f_2$ \cite {Ehrnsperger94},
which represent nucleon matrix elements of operators of twist two,
three, and four, respectively.  Thus, depending on the relative
strength of these coefficients, conspicuous deviations among the
generalized GDH integrals have to be expected also in the high-$Q^2$
domain.

I conclude that due to the vital role of structure function
$G_2(\nu,Q^2)$, one should pay more  attention to the differences between
the generalized GDH integrals $I_1(Q^2)$, $I_\tGDH(Q^2)$, and $\tilde
I_\tGDH(Q^2)$.  Further theoretical and experimental
investigation of longitudinal photon coupling is desirable.

In \Sect {piN}, the contribution from single-pion production to
polarized inclusive electroproduction has been investigated.  Expressions
in terms of different sets of amplitudes and multipoles have been derived.
The isospin decomposition has been reviewed.  \Sect {piN} gives a
collection of formulae as a basis for investigating the saturation of
diverse generalizations of the GDH integral.

\clearpage{\pagestyle{empty}\cleardoublepage}
\addcontentsline{toc}{chapter}{Bibliography}
\bibliographystyle{prsty}
\bibliography{particle}

\clearpage{\pagestyle{empty}\cleardoublepage}
\addcontentsline{toc}{chapter}{Acknowledgments}
\chapter*{Acknowledgments}

I am grateful to Prof.\ Horst Rollnik for suggesting the subject of this
thesis and for various helpful discussions.

I wish to thank the second referee Prof.\ Herbert Petry for examining the
manuscript.

It is a pleasure to thank Dr.\ Walter Pfeil for his continuous encouragement
and numerous valuable suggestions.

This work has been supported by the Deutsche Forschungsgemeinschaft.
In particular, I wish to acknowledge long-term support by the
Graduiertenkolleg ``Die Erforschung der subnuklearen Strukturen der
Materie'' (Bonn/J\"ulich).  For their painstaking efforts, I am very
grateful to the spokesmen of the Graduiertenkolleg, Prof.\ Reinhard
Maschuw and Prof.\ Fritz Klein, as well as their assistants Dr.\ Udo
Idschok, Ralf Ziegler, and Karl-Heinz Glander, whose work enabled
fruitful and inspiring co-operation among all members -- graduates and
professors.

Dr.\ Bernard Metsch and Prof.\ Herbert Petry made available an
employment at the Institut f\"ur Theoretische Kernphysik (Bonn), which
has been very helpful for the completion of this work.

I thank my colleague Hagen Storch, with whom I shared a room at the institute,
for the pleasant working atmosphere.

Several lifely and challenging discussions with Dr.\ Klaus Helbing of
the GDH collaboration are gratefully acknowledged.

I would like to thank  Dr.\ J\"org Bonekamp, who -- among other things -- helped
me in getting familiar with the computing devices of our group.

For an instructive discussion on an aspect of the derivation of the
GDH sum rule, I am grateful to Dr.\ Anatoly L'vov.

I am greatly obliged to my colleague and dear friend Dr.\ Roland Puntigam
for conscientiously reading the manuscript and for always having an ear for
me.

\end{fmffile}
\end{document}

%% file: title.tex
Investigations on the Foundation\\
and Possible Modifications of the\\
Gerasimov-Drell-Hearn Sum Rule\\

%% file: abstract.tex
All derivations of the GDH sum rule are presented and discussed in
detail, focussing particularly on the validity of the underlying
assumptions.  Several tests of the sum rule within perturbative models
are reviewed.  Various possible sources of modifications of the sum
rule are examined.  With respect to the current-algebra derivation,
the crucial role of the infinite-momentum limit is pointed out.  A
derivation is presented that exhibits the infinite-momentum limit as
its last step, opening the prospect of studying its legitimacy within
perturbative models.  Adopting the Weinberg-Salam model as a testing
ground, it is shown that a modification due to an anomalous
charge-density commutator is remedied owing to the infinite-momentum
limit.  This finding is confirmed upon considering $t$-channel exchange
of axial-vector mesons.  Several other aspects of possible sources of
modifications are investigated.  Evolving the GDH sum rule to non-zero
photon virtualities $Q^2$, it is argued that different generalizations
of the GDH integral, which coincide both at $Q^2=0$ and at large
$Q^2$, may considerably deviate from one another at intermediate
$Q^2$.  The limits $Q^2\to0$ and $Q^2\to\infty$ are investigated
beyond leading terms.  The contribution of the pion-nucleon final
state to inclusive electroproduction is analyzed.

%% file: cauchy1.eepic
\setlength{\unitlength}{0.00087489in}
\begingroup\makeatletter\ifx\SetFigFont\undefined%
\gdef\SetFigFont#1#2#3#4#5{%
  \reset@font\fontsize{#1}{#2pt}%
  \fontfamily{#3}\fontseries{#4}\fontshape{#5}%
  \selectfont}%
\fi\endgroup%
{\renewcommand{\dashlinestretch}{30}
\begin{picture}(5424,5439)(0,-10)
\path(2712,5412)(2712,12)
\path(3387,2712)(3387,2667)(5412,2667)
\path(5142,5412)(5142,5142)(5412,5142)
\path(2037,2712)(2037,2667)(12,2667)
\path(12,2712)(5412,2712)
\path(5412,5412)(12,5412)(12,12)
	(5412,12)(5412,5412)
\thicklines
\path(3207,2712)	(3230.606,2782.261)
	(3280.388,2849.178)
	(3346.315,2914.833)
	(3382.197,2947.837)
	(3418.354,2981.306)
	(3486.474,3050.679)
	(3515.929,3087.102)
	(3540.643,3125.031)
	(3559.361,3164.725)
	(3570.829,3206.445)
	(3573.793,3250.449)
	(3567.000,3297.000)

\path(3567,3297)	(3545.358,3363.087)
	(3515.743,3423.652)
	(3478.956,3478.874)
	(3435.795,3528.931)
	(3387.062,3574.004)
	(3333.556,3614.270)
	(3276.078,3649.910)
	(3215.426,3681.101)
	(3152.402,3708.024)
	(3087.805,3730.857)
	(3022.436,3749.780)
	(2957.094,3764.970)
	(2892.579,3776.609)
	(2829.692,3784.874)
	(2769.232,3789.945)
	(2712.000,3792.000)

\path(2712,3792)	(2652.141,3790.951)
	(2588.296,3786.464)
	(2521.422,3778.421)
	(2452.478,3766.704)
	(2382.419,3751.198)
	(2312.204,3731.784)
	(2242.789,3708.347)
	(2175.131,3680.767)
	(2110.188,3648.930)
	(2048.917,3612.717)
	(1992.275,3572.012)
	(1941.219,3526.697)
	(1896.706,3476.655)
	(1859.694,3421.770)
	(1831.140,3361.924)
	(1812.000,3297.000)

\path(1812,3297)	(1807.588,3249.932)
	(1813.744,3205.554)
	(1829.134,3163.588)
	(1852.423,3123.760)
	(1882.275,3085.792)
	(1917.355,3049.409)
	(1956.329,3014.335)
	(1997.861,2980.294)
	(2040.616,2947.009)
	(2083.259,2914.204)
	(2124.455,2881.603)
	(2162.869,2848.931)
	(2226.010,2782.266)
	(2262.000,2712.000)

\path(2262,2712)	(2268.535,2665.546)
	(2265.913,2616.182)
	(2255.667,2564.254)
	(2239.334,2510.103)
	(2218.447,2454.074)
	(2194.540,2396.510)
	(2169.149,2337.755)
	(2143.807,2278.151)
	(2120.050,2218.043)
	(2099.411,2157.773)
	(2083.425,2097.686)
	(2073.627,2038.124)
	(2071.551,1979.431)
	(2078.732,1921.950)
	(2096.703,1866.025)
	(2127.000,1812.000)

\path(2127,1812)	(2160.396,1767.543)
	(2196.974,1725.093)
	(2236.509,1684.670)
	(2278.777,1646.299)
	(2323.550,1610.002)
	(2370.606,1575.801)
	(2419.717,1543.720)
	(2470.659,1513.781)
	(2523.208,1486.007)
	(2577.136,1460.420)
	(2632.220,1437.044)
	(2688.234,1415.901)
	(2744.953,1397.013)
	(2802.152,1380.405)
	(2859.605,1366.097)
	(2917.087,1354.114)
	(2974.374,1344.477)
	(3031.239,1337.210)
	(3087.458,1332.335)
	(3142.805,1329.875)
	(3197.055,1329.853)
	(3249.983,1332.291)
	(3301.364,1337.213)
	(3350.972,1344.640)
	(3398.582,1354.596)
	(3443.970,1367.103)
	(3486.909,1382.185)
	(3527.175,1399.863)
	(3564.542,1420.161)
	(3598.786,1443.101)
	(3657.000,1497.000)

\path(3657,1497)	(3688.668,1549.854)
	(3700.852,1609.526)
	(3696.019,1675.215)
	(3676.635,1746.120)
	(3662.258,1783.278)
	(3645.168,1821.438)
	(3625.674,1860.502)
	(3604.084,1900.369)
	(3580.706,1940.938)
	(3555.849,1982.110)
	(3529.821,2023.784)
	(3502.931,2065.860)
	(3475.487,2108.238)
	(3447.797,2150.818)
	(3420.169,2193.499)
	(3392.912,2236.181)
	(3366.335,2278.765)
	(3340.745,2321.149)
	(3316.451,2363.234)
	(3293.761,2404.920)
	(3272.984,2446.106)
	(3254.428,2486.692)
	(3238.402,2526.578)
	(3225.213,2565.664)
	(3215.169,2603.849)
	(3208.581,2641.034)
	(3207.000,2712.000)

\put(5232,5187){\makebox(0,0)[lb]{\smash{{{\SetFigFont{12}{14.4}{\rmdefault}{\mddefault}{\updefault}$\nu$}}}}}
\put(3342,2487){\makebox(0,0)[lb]{\smash{{{\SetFigFont{12}{14.4}{\rmdefault}{\mddefault}{\updefault}$\nu_0$}}}}}
\put(2892,2982){\makebox(0,0)[lb]{\smash{{{\SetFigFont{12}{14.4}{\rmdefault}{\mddefault}{\updefault}$\bullet\,\nu$}}}}}
\put(1812,2487){\makebox(0,0)[lb]{\smash{{{\SetFigFont{12}{14.4}{\rmdefault}{\mddefault}{\updefault}$-\nu_0$}}}}}
\put(1857,3612){\makebox(0,0)[lb]{\smash{{{\SetFigFont{12}{14.4}{\rmdefault}{\mddefault}{\updefault}$\mathcal{C}$}}}}}
\end{picture}
}

%% file: cauchy2.eepic
\setlength{\unitlength}{0.00087489in}
\begingroup\makeatletter\ifx\SetFigFont\undefined%
\gdef\SetFigFont#1#2#3#4#5{%
  \reset@font\fontsize{#1}{#2pt}%
  \fontfamily{#3}\fontseries{#4}\fontshape{#5}%
  \selectfont}%
\fi\endgroup%
{\renewcommand{\dashlinestretch}{30}
\begin{picture}(5424,5439)(0,-10)
\thicklines
\put(2037.000,2667.000){\arc{90.000}{4.7124}{7.8540}}
\put(3387.000,2667.000){\arc{90.000}{1.5708}{4.7124}}
\put(2712.000,2712.000){\arc{5220.000}{3.1416}{6.2832}}
\put(2712.000,2713.607){\arc{5223.214}{0.0351}{3.1065}}
\thinlines
\path(2712,5412)(2712,12)
\path(3387,2712)(3387,2667)(5412,2667)
\path(5142,5412)(5142,5142)(5412,5142)
\path(2037,2712)(2037,2667)(12,2667)
\path(12,2712)(5412,2712)
\thicklines
\path(102,2622)(2037,2622)
\path(102,2712)(2037,2712)
\path(3387,2712)(5322,2712)
\path(3387,2622)(5322,2622)
\thinlines
\path(5412,5412)(12,5412)(12,12)
	(5412,12)(5412,5412)
\put(5232,5187){\makebox(0,0)[lb]{\smash{{{\SetFigFont{12}{14.4}{\rmdefault}{\mddefault}{\updefault}$\nu$}}}}}
\put(3342,2487){\makebox(0,0)[lb]{\smash{{{\SetFigFont{12}{14.4}{\rmdefault}{\mddefault}{\updefault}$\nu_0$}}}}}
\put(2892,2982){\makebox(0,0)[lb]{\smash{{{\SetFigFont{12}{14.4}{\rmdefault}{\mddefault}{\updefault}$\bullet\,\nu$}}}}}
\put(1812,2487){\makebox(0,0)[lb]{\smash{{{\SetFigFont{12}{14.4}{\rmdefault}{\mddefault}{\updefault}$-\nu_0$}}}}}
\put(417,4242){\makebox(0,0)[lb]{\smash{{{\SetFigFont{12}{14.4}{\rmdefault}{\mddefault}{\updefault}$\mathcal{C}$}}}}}
\end{picture}
}

%% file: q2nu.tex
% GNUPLOT: LaTeX picture using EEPIC macros
\setlength{\unitlength}{0.240900pt}
\begin{picture}(1500,900)(0,0)
\tenrm
\thinlines \drawline[-50](264,158)(1436,158)
\thinlines \drawline[-50](264,158)(264,787)
\thicklines \path(264,158)(284,158)
\thicklines \path(1436,158)(1416,158)
\put(242,158){\makebox(0,0)[r]{$0$}}
\thicklines \path(264,284)(284,284)
\thicklines \path(1436,284)(1416,284)
\put(242,284){\makebox(0,0)[r]{$0.2$}}
\thicklines \path(264,410)(284,410)
\thicklines \path(1436,410)(1416,410)
\put(242,410){\makebox(0,0)[r]{$0.4$}}
\thicklines \path(264,535)(284,535)
\thicklines \path(1436,535)(1416,535)
\put(242,535){\makebox(0,0)[r]{$0.6$}}
\thicklines \path(264,661)(284,661)
\thicklines \path(1436,661)(1416,661)
\put(242,661){\makebox(0,0)[r]{$0.8$}}
\thicklines \path(264,787)(284,787)
\thicklines \path(1436,787)(1416,787)
\put(242,787){\makebox(0,0)[r]{$1$}}
\thicklines \path(264,158)(264,178)
\thicklines \path(264,787)(264,767)
\put(264,113){\makebox(0,0){$0$}}
\thicklines \path(498,158)(498,178)
\thicklines \path(498,787)(498,767)
\put(498,113){\makebox(0,0){$0.2$}}
\thicklines \path(733,158)(733,178)
\thicklines \path(733,787)(733,767)
\put(733,113){\makebox(0,0){$0.4$}}
\thicklines \path(967,158)(967,178)
\thicklines \path(967,787)(967,767)
\put(967,113){\makebox(0,0){$0.6$}}
\thicklines \path(1202,158)(1202,178)
\thicklines \path(1202,787)(1202,767)
\put(1202,113){\makebox(0,0){$0.8$}}
\thicklines \path(1436,158)(1436,178)
\thicklines \path(1436,787)(1436,767)
\put(1436,113){\makebox(0,0){$1$}}
\thicklines \path(264,158)(1436,158)(1436,787)(264,787)(264,158)
\put(45,472){\makebox(0,0)[l]{\shortstack{$q^2/M^2$}}}
\put(850,68){\makebox(0,0){$\nu/M$}}
\put(1084,504){\makebox(0,0)[r]{$p^0=M$}}
\put(1202,284){\makebox(0,0)[r]{$p^0=2M$}}
\put(1407,189){\makebox(0,0)[r]{$p^0\to\infty$}}
\thinlines \path(264,158)(264,158)(265,158)(265,158)(266,158)(266,158)(267,158)(268,158)(268,158)(269,158)(269,158)(270,158)(270,158)(271,158)(272,158)(272,158)(273,158)(273,158)(274,158)(275,158)(275,158)(276,158)(276,158)(277,158)(277,158)(278,158)(279,158)(279,158)(280,158)(280,158)(281,158)(282,158)(282,158)(283,158)(283,158)(284,158)(285,158)(285,158)(286,158)(286,158)(287,158)(287,158)(288,158)(289,158)(289,158)(290,158)(290,158)(291,158)(292,158)(292,158)(293,158)
\thinlines \path(293,158)(293,158)(294,158)(294,158)(295,158)(296,158)(296,158)(297,158)(297,159)(298,159)(299,159)(299,159)(300,159)(300,159)(301,159)(302,159)(302,159)(303,159)(303,159)(304,159)(304,159)(305,159)(306,159)(306,159)(307,159)(307,159)(308,159)(309,159)(309,159)(310,159)(310,159)(311,159)(311,159)(312,159)(313,159)(313,159)(314,159)(314,159)(315,159)(316,159)(316,159)(317,159)(317,159)(318,159)(319,159)(319,159)(320,159)(320,159)(321,159)(321,160)(322,160)
\thinlines \path(322,160)(323,160)(323,160)(324,160)(324,160)(325,160)(326,160)(326,160)(327,160)(327,160)(328,160)(328,160)(329,160)(330,160)(330,160)(331,160)(331,160)(332,160)(333,160)(333,160)(334,160)(334,160)(335,160)(336,160)(336,160)(337,160)(337,160)(338,160)(338,161)(339,161)(340,161)(340,161)(341,161)(341,161)(342,161)(343,161)(343,161)(344,161)(344,161)(345,161)(345,161)(346,161)(347,161)(347,161)(348,161)(348,161)(349,161)(350,161)(350,161)(351,161)(351,161)
\thinlines \path(351,161)(352,162)(353,162)(353,162)(354,162)(354,162)(355,162)(355,162)(356,162)(357,162)(357,162)(358,162)(358,162)(359,162)(360,162)(360,162)(361,162)(361,162)(362,162)(362,162)(363,162)(364,163)(364,163)(365,163)(365,163)(366,163)(367,163)(367,163)(368,163)(368,163)(369,163)(370,163)(370,163)(371,163)(371,163)(372,163)(372,163)(373,163)(374,164)(374,164)(375,164)(375,164)(376,164)(377,164)(377,164)(378,164)(378,164)(379,164)(379,164)(380,164)(381,164)
\thinlines \path(381,164)(381,164)(382,164)(382,164)(383,164)(384,165)(384,165)(385,165)(385,165)(386,165)(387,165)(387,165)(388,165)(388,165)(389,165)(389,165)(390,165)(391,165)(391,165)(392,165)(392,166)(393,166)(394,166)(394,166)(395,166)(395,166)(396,166)(397,166)(397,166)(398,166)(398,166)(399,166)(399,166)(400,166)(401,167)(401,167)(402,167)(402,167)(403,167)(404,167)(404,167)(405,167)(405,167)(406,167)(406,167)(407,167)(408,167)(408,168)(409,168)(409,168)(410,168)
\thinlines \path(410,168)(411,168)(411,168)(412,168)(412,168)(413,168)(414,168)(414,168)(415,168)(415,168)(416,169)(416,169)(417,169)(418,169)(418,169)(419,169)(419,169)(420,169)(421,169)(421,169)(422,169)(422,169)(423,170)(423,170)(424,170)(425,170)(425,170)(426,170)(426,170)(427,170)(428,170)(428,170)(429,170)(429,171)(430,171)(431,171)(431,171)(432,171)(432,171)(433,171)(433,171)(434,171)(435,171)(435,171)(436,172)(436,172)(437,172)(438,172)(438,172)(439,172)(439,172)
\thinlines \path(439,172)(440,172)(440,172)(441,172)(442,172)(442,173)(443,173)(443,173)(444,173)(445,173)(445,173)(446,173)(446,173)(447,173)(448,173)(448,174)(449,174)(449,174)(450,174)(450,174)(451,174)(452,174)(452,174)(453,174)(453,174)(454,175)(455,175)(455,175)(456,175)(456,175)(457,175)(457,175)(458,175)(459,175)(459,175)(460,176)(460,176)(461,176)(462,176)(462,176)(463,176)(463,176)(464,176)(465,176)(465,177)(466,177)(466,177)(467,177)(467,177)(468,177)(469,177)
\thinlines \path(469,177)(469,177)(470,177)(470,178)(471,178)(472,178)(472,178)(473,178)(473,178)(474,178)(474,178)(475,178)(476,179)(476,179)(477,179)(477,179)(478,179)(479,179)(479,179)(480,179)(480,179)(481,180)(482,180)(482,180)(483,180)(483,180)(484,180)(484,180)(485,180)(486,180)(486,181)(487,181)(487,181)(488,181)(489,181)(489,181)(490,181)(490,181)(491,182)(491,182)(492,182)(493,182)(493,182)(494,182)(494,182)(495,182)(496,183)(496,183)(497,183)(497,183)(498,183)
\thinlines \path(498,183)(499,183)(499,183)(500,183)(500,184)(501,184)(501,184)(502,184)(503,184)(503,184)(504,184)(504,184)(505,185)(506,185)(506,185)(507,185)(507,185)(508,185)(508,185)(509,186)(510,186)(510,186)(511,186)(511,186)(512,186)(513,186)(513,186)(514,187)(514,187)(515,187)(516,187)(516,187)(517,187)(517,187)(518,188)(518,188)(519,188)(520,188)(520,188)(521,188)(521,188)(522,188)(523,189)(523,189)(524,189)(524,189)(525,189)(525,189)(526,189)(527,190)(527,190)
\thinlines \path(527,190)(528,190)(528,190)(529,190)(530,190)(530,190)(531,191)(531,191)(532,191)(533,191)(533,191)(534,191)(534,191)(535,192)(535,192)(536,192)(537,192)(537,192)(538,192)(538,192)(539,193)(540,193)(540,193)(541,193)(541,193)(542,193)(542,194)(543,194)(544,194)(544,194)(545,194)(545,194)(546,194)(547,195)(547,195)(548,195)(548,195)(549,195)(550,195)(550,195)(551,196)(551,196)(552,196)(552,196)(553,196)(554,196)(554,197)(555,197)(555,197)(556,197)(557,197)
\thinlines \path(557,197)(557,197)(558,198)(558,198)(559,198)(559,198)(560,198)(561,198)(561,198)(562,199)(562,199)(563,199)(564,199)(564,199)(565,199)(565,200)(566,200)(567,200)(567,200)(568,200)(568,200)(569,201)(569,201)(570,201)(571,201)(571,201)(572,201)(572,202)(573,202)(574,202)(574,202)(575,202)(575,202)(576,203)(576,203)(577,203)(578,203)(578,203)(579,203)(579,204)(580,204)(581,204)(581,204)(582,204)(582,204)(583,205)(584,205)(584,205)(585,205)(585,205)(586,205)
\thinlines \path(586,205)(586,206)(587,206)(588,206)(588,206)(589,206)(589,206)(590,207)(591,207)(591,207)(592,207)(592,207)(593,208)(593,208)(594,208)(595,208)(595,208)(596,208)(596,209)(597,209)(598,209)(598,209)(599,209)(599,210)(600,210)(601,210)(601,210)(602,210)(602,210)(603,211)(603,211)(604,211)(605,211)(605,211)(606,212)(606,212)(607,212)(608,212)(608,212)(609,212)(609,213)(610,213)(610,213)(611,213)(612,213)(612,214)(613,214)(613,214)(614,214)(615,214)(615,214)
\thinlines \path(615,214)(616,215)(616,215)(617,215)(618,215)(618,215)(619,216)(619,216)(620,216)(620,216)(621,216)(622,217)(622,217)(623,217)(623,217)(624,217)(625,218)(625,218)(626,218)(626,218)(627,218)(628,219)(628,219)(629,219)(629,219)(630,219)(630,219)(631,220)(632,220)(632,220)(633,220)(633,220)(634,221)(635,221)(635,221)(636,221)(636,221)(637,222)(637,222)(638,222)(639,222)(639,222)(640,223)(640,223)(641,223)(642,223)(642,223)(643,224)(643,224)(644,224)(645,224)
\thinlines \path(645,224)(645,225)(646,225)(646,225)(647,225)(647,225)(648,226)(649,226)(649,226)(650,226)(650,226)(651,227)(652,227)(652,227)(653,227)(653,227)(654,228)(654,228)(655,228)(656,228)(656,228)(657,229)(657,229)(658,229)(659,229)(659,230)(660,230)(660,230)(661,230)(662,230)(662,231)(663,231)(663,231)(664,231)(664,231)(665,232)(666,232)(666,232)(667,232)(667,233)(668,233)(669,233)(669,233)(670,233)(670,234)(671,234)(671,234)(672,234)(673,234)(673,235)(674,235)
\thinlines \path(674,235)(674,235)(675,235)(676,236)(676,236)(677,236)(677,236)(678,236)(679,237)(679,237)(680,237)(680,237)(681,238)(681,238)(682,238)(683,238)(683,238)(684,239)(684,239)(685,239)(686,239)(686,240)(687,240)(687,240)(688,240)(688,241)(689,241)(690,241)(690,241)(691,241)(691,242)(692,242)(693,242)(693,242)(694,243)(694,243)(695,243)(696,243)(696,243)(697,244)(697,244)(698,244)(698,244)(699,245)(700,245)(700,245)(701,245)(701,246)(702,246)(703,246)(703,246)
\thinlines \path(703,246)(704,247)(704,247)(705,247)(705,247)(706,247)(707,248)(707,248)(708,248)(708,248)(709,249)(710,249)(710,249)(711,249)(711,250)(712,250)(713,250)(713,250)(714,251)(714,251)(715,251)(715,251)(716,252)(717,252)(717,252)(718,252)(718,253)(719,253)(720,253)(720,253)(721,254)(721,254)(722,254)(722,254)(723,255)(724,255)(724,255)(725,255)(725,255)(726,256)(727,256)(727,256)(728,256)(728,257)(729,257)(730,257)(730,257)(731,258)(731,258)(732,258)(732,258)
\thinlines \path(732,258)(733,259)(734,259)(734,259)(735,259)(735,260)(736,260)(737,260)(737,261)(738,261)(738,261)(739,261)(739,262)(740,262)(741,262)(741,262)(742,263)(742,263)(743,263)(744,263)(744,264)(745,264)(745,264)(746,264)(747,265)(747,265)(748,265)(748,265)(749,266)(749,266)(750,266)(751,266)(751,267)(752,267)(752,267)(753,267)(754,268)(754,268)(755,268)(755,269)(756,269)(756,269)(757,269)(758,270)(758,270)(759,270)(759,270)(760,271)(761,271)(761,271)(762,271)
\thinlines \path(762,271)(762,272)(763,272)(764,272)(764,273)(765,273)(765,273)(766,273)(766,274)(767,274)(768,274)(768,274)(769,275)(769,275)(770,275)(771,276)(771,276)(772,276)(772,276)(773,277)(773,277)(774,277)(775,277)(775,278)(776,278)(776,278)(777,279)(778,279)(778,279)(779,279)(779,280)(780,280)(781,280)(781,280)(782,281)(782,281)(783,281)(783,282)(784,282)(785,282)(785,282)(786,283)(786,283)(787,283)(788,284)(788,284)(789,284)(789,284)(790,285)(790,285)(791,285)
\thinlines \path(791,285)(792,285)(792,286)(793,286)(793,286)(794,287)(795,287)(795,287)(796,287)(796,288)(797,288)(798,288)(798,289)(799,289)(799,289)(800,289)(800,290)(801,290)(802,290)(802,291)(803,291)(803,291)(804,292)(805,292)(805,292)(806,292)(806,293)(807,293)(807,293)(808,294)(809,294)(809,294)(810,294)(810,295)(811,295)(812,295)(812,296)(813,296)(813,296)(814,296)(815,297)(815,297)(816,297)(816,298)(817,298)(817,298)(818,299)(819,299)(819,299)(820,299)(820,300)
\thinlines \path(820,300)(821,300)(822,300)(822,301)(823,301)(823,301)(824,302)(824,302)(825,302)(826,302)(826,303)(827,303)(827,303)(828,304)(829,304)(829,304)(830,305)(830,305)(831,305)(832,305)(832,306)(833,306)(833,306)(834,307)(834,307)(835,307)(836,308)(836,308)(837,308)(837,309)(838,309)(839,309)(839,309)(840,310)(840,310)(841,310)(841,311)(842,311)(843,311)(843,312)(844,312)(844,312)(845,313)(846,313)(846,313)(847,314)(847,314)(848,314)(849,314)(849,315)(850,315)
\thinlines \path(850,315)(850,315)(851,316)(851,316)(852,316)(853,317)(853,317)(854,317)(854,318)(855,318)(856,318)(856,319)(857,319)(857,319)(858,320)(859,320)(859,320)(860,320)(860,321)(861,321)(861,321)(862,322)(863,322)(863,322)(864,323)(864,323)(865,323)(866,324)(866,324)(867,324)(867,325)(868,325)(868,325)(869,326)(870,326)(870,326)(871,327)(871,327)(872,327)(873,328)(873,328)(874,328)(874,329)(875,329)(876,329)(876,330)(877,330)(877,330)(878,331)(878,331)(879,331)
\thinlines \path(879,331)(880,332)(880,332)(881,332)(881,333)(882,333)(883,333)(883,334)(884,334)(884,334)(885,335)(885,335)(886,335)(887,336)(887,336)(888,336)(888,337)(889,337)(890,337)(890,338)(891,338)(891,338)(892,339)(893,339)(893,339)(894,340)(894,340)(895,340)(895,341)(896,341)(897,341)(897,342)(898,342)(898,342)(899,343)(900,343)(900,343)(901,344)(901,344)(902,344)(902,345)(903,345)(904,345)(904,346)(905,346)(905,346)(906,347)(907,347)(907,347)(908,348)(908,348)
\thinlines \path(908,348)(909,348)(910,349)(910,349)(911,350)(911,350)(912,350)(912,351)(913,351)(914,351)(914,352)(915,352)(915,352)(916,353)(917,353)(917,353)(918,354)(918,354)(919,354)(919,355)(920,355)(921,355)(921,356)(922,356)(922,357)(923,357)(924,357)(924,358)(925,358)(925,358)(926,359)(927,359)(927,359)(928,360)(928,360)(929,360)(929,361)(930,361)(931,361)(931,362)(932,362)(932,363)(933,363)(934,363)(934,364)(935,364)(935,364)(936,365)(936,365)(937,365)(938,366)
\thinlines \path(938,366)(938,366)(939,367)(939,367)(940,367)(941,368)(941,368)(942,368)(942,369)(943,369)(944,369)(944,370)(945,370)(945,371)(946,371)(946,371)(947,372)(948,372)(948,372)(949,373)(949,373)(950,373)(951,374)(951,374)(952,375)(952,375)(953,375)(953,376)(954,376)(955,376)(955,377)(956,377)(956,378)(957,378)(958,378)(958,379)(959,379)(959,379)(960,380)(961,380)(961,381)(962,381)(962,381)(963,382)(963,382)(964,382)(965,383)(965,383)(966,384)(966,384)(967,384)
\thinlines \path(967,384)(968,385)(968,385)(969,385)(969,386)(970,386)(970,387)(971,387)(972,387)(972,388)(973,388)(973,388)(974,389)(975,389)(975,390)(976,390)(976,390)(977,391)(978,391)(978,392)(979,392)(979,392)(980,393)(980,393)(981,393)(982,394)(982,394)(983,395)(983,395)(984,395)(985,396)(985,396)(986,397)(986,397)(987,397)(987,398)(988,398)(989,398)(989,399)(990,399)(990,400)(991,400)(992,400)(992,401)(993,401)(993,402)(994,402)(995,402)(995,403)(996,403)(996,404)
\thinlines \path(996,404)(997,404)(997,404)(998,405)(999,405)(999,406)(1000,406)(1000,406)(1001,407)(1002,407)(1002,408)(1003,408)(1003,408)(1004,409)(1004,409)(1005,409)(1006,410)(1006,410)(1007,411)(1007,411)(1008,411)(1009,412)(1009,412)(1010,413)(1010,413)(1011,413)(1012,414)(1012,414)(1013,415)(1013,415)(1014,415)(1014,416)(1015,416)(1016,417)(1016,417)(1017,418)(1017,418)(1018,418)(1019,419)(1019,419)(1020,420)(1020,420)(1021,420)(1021,421)(1022,421)(1023,422)(1023,422)(1024,422)(1024,423)(1025,423)(1026,424)
\thinlines \path(1026,424)(1026,424)(1027,424)(1027,425)(1028,425)(1029,426)(1029,426)(1030,426)(1030,427)(1031,427)(1031,428)(1032,428)(1033,429)(1033,429)(1034,429)(1034,430)(1035,430)(1036,431)(1036,431)(1037,431)(1037,432)(1038,432)(1038,433)(1039,433)(1040,434)(1040,434)(1041,434)(1041,435)(1042,435)(1043,436)(1043,436)(1044,436)(1044,437)(1045,437)(1046,438)(1046,438)(1047,439)(1047,439)(1048,439)(1048,440)(1049,440)(1050,441)(1050,441)(1051,441)(1051,442)(1052,442)(1053,443)(1053,443)(1054,444)(1054,444)(1055,444)
\thinlines \path(1055,444)(1055,445)(1056,445)(1057,446)(1057,446)(1058,447)(1058,447)(1059,447)(1060,448)(1060,448)(1061,449)(1061,449)(1062,450)(1063,450)(1063,450)(1064,451)(1064,451)(1065,452)(1065,452)(1066,453)(1067,453)(1067,453)(1068,454)(1068,454)(1069,455)(1070,455)(1070,456)(1071,456)(1071,456)(1072,457)(1072,457)(1073,458)(1074,458)(1074,459)(1075,459)(1075,460)(1076,460)(1077,460)(1077,461)(1078,461)(1078,462)(1079,462)(1080,463)(1080,463)(1081,463)(1081,464)(1082,464)(1082,465)(1083,465)(1084,466)(1084,466)
\thinlines \path(1084,466)(1085,467)(1085,467)(1086,467)(1087,468)(1087,468)(1088,469)(1088,469)(1089,470)(1090,470)(1090,470)(1091,471)(1091,471)(1092,472)(1092,472)(1093,473)(1094,473)(1094,474)(1095,474)(1095,475)(1096,475)(1097,475)(1097,476)(1098,476)(1098,477)(1099,477)(1099,478)(1100,478)(1101,479)(1101,479)(1102,479)(1102,480)(1103,480)(1104,481)(1104,481)(1105,482)(1105,482)(1106,483)(1107,483)(1107,483)(1108,484)(1108,484)(1109,485)(1109,485)(1110,486)(1111,486)(1111,487)(1112,487)(1112,488)(1113,488)(1114,488)
\thinlines \path(1114,488)(1114,489)(1115,489)(1115,490)(1116,490)(1116,491)(1117,491)(1118,492)(1118,492)(1119,493)(1119,493)(1120,494)(1121,494)(1121,494)(1122,495)(1122,495)(1123,496)(1124,496)(1124,497)(1125,497)(1125,498)(1126,498)(1126,499)(1127,499)(1128,500)(1128,500)(1129,500)(1129,501)(1130,501)(1131,502)(1131,502)(1132,503)(1132,503)(1133,504)(1133,504)(1134,505)(1135,505)(1135,506)(1136,506)(1136,507)(1137,507)(1138,507)(1138,508)(1139,508)(1139,509)(1140,509)(1141,510)(1141,510)(1142,511)(1142,511)(1143,512)
\thinlines \path(1143,512)(1143,512)(1144,513)(1145,513)(1145,514)(1146,514)(1146,515)(1147,515)(1148,515)(1148,516)(1149,516)(1149,517)(1150,517)(1150,518)(1151,518)(1152,519)(1152,519)(1153,520)(1153,520)(1154,521)(1155,521)(1155,522)(1156,522)(1156,523)(1157,523)(1158,524)(1158,524)(1159,525)(1159,525)(1160,526)(1160,526)(1161,526)(1162,527)(1162,527)(1163,528)(1163,528)(1164,529)(1165,529)(1165,530)(1166,530)(1166,531)(1167,531)(1167,532)(1168,532)(1169,533)(1169,533)(1170,534)(1170,534)(1171,535)(1172,535)(1172,536)
\thinlines \path(1172,536)(1173,536)(1173,537)(1174,537)(1175,538)(1175,538)(1176,539)(1176,539)(1177,540)(1177,540)(1178,541)(1179,541)(1179,542)(1180,542)(1180,543)(1181,543)(1182,544)(1182,544)(1183,545)(1183,545)(1184,545)(1184,546)(1185,546)(1186,547)(1186,547)(1187,548)(1187,548)(1188,549)(1189,549)(1189,550)(1190,550)(1190,551)(1191,551)(1192,552)(1192,552)(1193,553)(1193,553)(1194,554)(1194,554)(1195,555)(1196,555)(1196,556)(1197,556)(1197,557)(1198,557)(1199,558)(1199,558)(1200,559)(1200,559)(1201,560)(1201,560)
\thinlines \path(1201,560)(1202,561)(1203,561)(1203,562)(1204,562)(1204,563)(1205,563)(1206,564)(1206,564)(1207,565)(1207,566)(1208,566)(1209,567)(1209,567)(1210,568)(1210,568)(1211,569)(1211,569)(1212,570)(1213,570)(1213,571)(1214,571)(1214,572)(1215,572)(1216,573)(1216,573)(1217,574)(1217,574)(1218,575)(1218,575)(1219,576)(1220,576)(1220,577)(1221,577)(1221,578)(1222,578)(1223,579)(1223,579)(1224,580)(1224,580)(1225,581)(1226,581)(1226,582)(1227,582)(1227,583)(1228,583)(1228,584)(1229,584)(1230,585)(1230,586)(1231,586)
\thinlines \path(1231,586)(1231,587)(1232,587)(1233,588)(1233,588)(1234,589)(1234,589)(1235,590)(1235,590)(1236,591)(1237,591)(1237,592)(1238,592)(1238,593)(1239,593)(1240,594)(1240,594)(1241,595)(1241,595)(1242,596)(1243,596)(1243,597)(1244,598)(1244,598)(1245,599)(1245,599)(1246,600)(1247,600)(1247,601)(1248,601)(1248,602)(1249,602)(1250,603)(1250,603)(1251,604)(1251,604)(1252,605)(1252,605)(1253,606)(1254,607)(1254,607)(1255,608)(1255,608)(1256,609)(1257,609)(1257,610)(1258,610)(1258,611)(1259,611)(1260,612)(1260,612)
\thinlines \path(1260,612)(1261,613)(1261,613)(1262,614)(1262,615)(1263,615)(1264,616)(1264,616)(1265,617)(1265,617)(1266,618)(1267,618)(1267,619)(1268,619)(1268,620)(1269,620)(1269,621)(1270,622)(1271,622)(1271,623)(1272,623)(1272,624)(1273,624)(1274,625)(1274,625)(1275,626)(1275,626)(1276,627)(1277,627)(1277,628)(1278,629)(1278,629)(1279,630)(1279,630)(1280,631)(1281,631)(1281,632)(1282,632)(1282,633)(1283,633)(1284,634)(1284,635)(1285,635)(1285,636)(1286,636)(1286,637)(1287,637)(1288,638)(1288,638)(1289,639)(1289,640)
\thinlines \path(1289,640)(1290,640)(1291,641)(1291,641)(1292,642)(1292,642)(1293,643)(1294,643)(1294,644)(1295,644)(1295,645)(1296,646)(1296,646)(1297,647)(1298,647)(1298,648)(1299,648)(1299,649)(1300,649)(1301,650)(1301,651)(1302,651)(1302,652)(1303,652)(1303,653)(1304,653)(1305,654)(1305,654)(1306,655)(1306,656)(1307,656)(1308,657)(1308,657)(1309,658)(1309,658)(1310,659)(1311,660)(1311,660)(1312,661)(1312,661)(1313,662)(1313,662)(1314,663)(1315,663)(1315,664)(1316,665)(1316,665)(1317,666)(1318,666)(1318,667)(1319,667)
\thinlines \path(1319,667)(1319,668)(1320,669)(1321,669)(1321,670)(1322,670)(1322,671)(1323,671)(1323,672)(1324,673)(1325,673)(1325,674)(1326,674)(1326,675)(1327,675)(1328,676)(1328,677)(1329,677)(1329,678)(1330,678)(1330,679)(1331,679)(1332,680)(1332,681)(1333,681)(1333,682)(1334,682)(1335,683)(1335,683)(1336,684)(1336,685)(1337,685)(1338,686)(1338,686)(1339,687)(1339,687)(1340,688)(1340,689)(1341,689)(1342,690)(1342,690)(1343,691)(1343,691)(1344,692)(1345,693)(1345,693)(1346,694)(1346,694)(1347,695)(1347,696)(1348,696)
\thinlines \path(1348,696)(1349,697)(1349,697)(1350,698)(1350,698)(1351,699)(1352,700)(1352,700)(1353,701)(1353,701)(1354,702)(1355,703)(1355,703)(1356,704)(1356,704)(1357,705)(1357,705)(1358,706)(1359,707)(1359,707)(1360,708)(1360,708)(1361,709)(1362,710)(1362,710)(1363,711)(1363,711)(1364,712)(1364,713)(1365,713)(1366,714)(1366,714)(1367,715)(1367,716)(1368,716)(1369,717)(1369,717)(1370,718)(1370,718)(1371,719)(1372,720)(1372,720)(1373,721)(1373,721)(1374,722)(1374,723)(1375,723)(1376,724)(1376,724)(1377,725)(1377,726)
\thinlines \path(1377,726)(1378,726)(1379,727)(1379,727)(1380,728)(1380,729)(1381,729)(1381,730)(1382,730)(1383,731)(1383,732)(1384,732)(1384,733)(1385,733)(1386,734)(1386,735)(1387,735)(1387,736)(1388,736)(1389,737)(1389,738)(1390,738)(1390,739)(1391,739)(1391,740)(1392,741)(1393,741)(1393,742)(1394,743)(1394,743)(1395,744)(1396,744)(1396,745)(1397,746)(1397,746)(1398,747)(1398,747)(1399,748)(1400,749)(1400,749)(1401,750)(1401,750)(1402,751)(1403,752)(1403,752)(1404,753)(1404,753)(1405,754)(1406,755)(1406,755)(1407,756)
\thinlines \path(1407,756)(1407,757)(1408,757)(1408,758)(1409,758)(1410,759)(1410,760)(1411,760)(1411,761)(1412,761)(1413,762)(1413,763)(1414,763)(1414,764)(1415,765)(1415,765)(1416,766)(1417,766)(1417,767)(1418,768)(1418,768)(1419,769)(1420,770)(1420,770)(1421,771)(1421,771)(1422,772)(1423,773)(1423,773)(1424,774)(1424,774)(1425,775)(1425,776)(1426,776)(1427,777)(1427,778)(1428,778)(1428,779)(1429,779)(1430,780)(1430,781)(1431,781)(1431,782)(1432,783)(1432,783)(1433,784)(1434,784)(1434,785)(1435,786)(1435,786)(1436,787)
\thinlines \path(264,158)(264,158)(265,158)(265,158)(266,158)(266,158)(267,158)(268,158)(268,158)(269,158)(269,158)(270,158)(270,158)(271,158)(272,158)(272,158)(273,158)(273,158)(274,158)(275,158)(275,158)(276,158)(276,158)(277,158)(277,158)(278,158)(279,158)(279,158)(280,158)(280,158)(281,158)(282,158)(282,158)(283,158)(283,158)(284,158)(285,158)(285,158)(286,158)(286,158)(287,158)(287,158)(288,158)(289,158)(289,158)(290,158)(290,158)(291,158)(292,158)(292,158)(293,158)
\thinlines \path(293,158)(293,158)(294,158)(294,158)(295,158)(296,158)(296,158)(297,158)(297,158)(298,158)(299,158)(299,158)(300,158)(300,158)(301,158)(302,158)(302,158)(303,158)(303,158)(304,158)(304,158)(305,158)(306,158)(306,158)(307,158)(307,158)(308,158)(309,158)(309,158)(310,158)(310,158)(311,158)(311,158)(312,158)(313,158)(313,158)(314,158)(314,158)(315,158)(316,158)(316,158)(317,158)(317,158)(318,158)(319,158)(319,158)(320,158)(320,158)(321,158)(321,158)(322,158)
\thinlines \path(322,158)(323,158)(323,158)(324,158)(324,158)(325,158)(326,158)(326,158)(327,158)(327,158)(328,158)(328,158)(329,158)(330,158)(330,159)(331,159)(331,159)(332,159)(333,159)(333,159)(334,159)(334,159)(335,159)(336,159)(336,159)(337,159)(337,159)(338,159)(338,159)(339,159)(340,159)(340,159)(341,159)(341,159)(342,159)(343,159)(343,159)(344,159)(344,159)(345,159)(345,159)(346,159)(347,159)(347,159)(348,159)(348,159)(349,159)(350,159)(350,159)(351,159)(351,159)
\thinlines \path(351,159)(352,159)(353,159)(353,159)(354,159)(354,159)(355,159)(355,159)(356,159)(357,159)(357,159)(358,159)(358,159)(359,159)(360,159)(360,159)(361,159)(361,159)(362,159)(362,159)(363,159)(364,159)(364,159)(365,159)(365,159)(366,159)(367,159)(367,159)(368,159)(368,159)(369,159)(370,159)(370,159)(371,159)(371,159)(372,159)(372,159)(373,159)(374,159)(374,159)(375,159)(375,159)(376,159)(377,159)(377,159)(378,159)(378,159)(379,160)(379,160)(380,160)(381,160)
\thinlines \path(381,160)(381,160)(382,160)(382,160)(383,160)(384,160)(384,160)(385,160)(385,160)(386,160)(387,160)(387,160)(388,160)(388,160)(389,160)(389,160)(390,160)(391,160)(391,160)(392,160)(392,160)(393,160)(394,160)(394,160)(395,160)(395,160)(396,160)(397,160)(397,160)(398,160)(398,160)(399,160)(399,160)(400,160)(401,160)(401,160)(402,160)(402,160)(403,160)(404,160)(404,160)(405,160)(405,160)(406,160)(406,160)(407,160)(408,160)(408,160)(409,160)(409,160)(410,160)
\thinlines \path(410,160)(411,160)(411,160)(412,160)(412,161)(413,161)(414,161)(414,161)(415,161)(415,161)(416,161)(416,161)(417,161)(418,161)(418,161)(419,161)(419,161)(420,161)(421,161)(421,161)(422,161)(422,161)(423,161)(423,161)(424,161)(425,161)(425,161)(426,161)(426,161)(427,161)(428,161)(428,161)(429,161)(429,161)(430,161)(431,161)(431,161)(432,161)(432,161)(433,161)(433,161)(434,161)(435,161)(435,161)(436,161)(436,161)(437,161)(438,161)(438,161)(439,161)(439,162)
\thinlines \path(439,162)(440,162)(440,162)(441,162)(442,162)(442,162)(443,162)(443,162)(444,162)(445,162)(445,162)(446,162)(446,162)(447,162)(448,162)(448,162)(449,162)(449,162)(450,162)(450,162)(451,162)(452,162)(452,162)(453,162)(453,162)(454,162)(455,162)(455,162)(456,162)(456,162)(457,162)(457,162)(458,162)(459,162)(459,162)(460,162)(460,162)(461,162)(462,162)(462,162)(463,163)(463,163)(464,163)(465,163)(465,163)(466,163)(466,163)(467,163)(467,163)(468,163)(469,163)
\thinlines \path(469,163)(469,163)(470,163)(470,163)(471,163)(472,163)(472,163)(473,163)(473,163)(474,163)(474,163)(475,163)(476,163)(476,163)(477,163)(477,163)(478,163)(479,163)(479,163)(480,163)(480,163)(481,163)(482,163)(482,163)(483,163)(483,164)(484,164)(484,164)(485,164)(486,164)(486,164)(487,164)(487,164)(488,164)(489,164)(489,164)(490,164)(490,164)(491,164)(491,164)(492,164)(493,164)(493,164)(494,164)(494,164)(495,164)(496,164)(496,164)(497,164)(497,164)(498,164)
\thinlines \path(498,164)(499,164)(499,164)(500,164)(500,164)(501,164)(501,164)(502,164)(503,165)(503,165)(504,165)(504,165)(505,165)(506,165)(506,165)(507,165)(507,165)(508,165)(508,165)(509,165)(510,165)(510,165)(511,165)(511,165)(512,165)(513,165)(513,165)(514,165)(514,165)(515,165)(516,165)(516,165)(517,165)(517,165)(518,165)(518,165)(519,165)(520,165)(520,166)(521,166)(521,166)(522,166)(523,166)(523,166)(524,166)(524,166)(525,166)(525,166)(526,166)(527,166)(527,166)
\thinlines \path(527,166)(528,166)(528,166)(529,166)(530,166)(530,166)(531,166)(531,166)(532,166)(533,166)(533,166)(534,166)(534,166)(535,166)(535,166)(536,166)(537,167)(537,167)(538,167)(538,167)(539,167)(540,167)(540,167)(541,167)(541,167)(542,167)(542,167)(543,167)(544,167)(544,167)(545,167)(545,167)(546,167)(547,167)(547,167)(548,167)(548,167)(549,167)(550,167)(550,167)(551,167)(551,167)(552,167)(552,168)(553,168)(554,168)(554,168)(555,168)(555,168)(556,168)(557,168)
\thinlines \path(557,168)(557,168)(558,168)(558,168)(559,168)(559,168)(560,168)(561,168)(561,168)(562,168)(562,168)(563,168)(564,168)(564,168)(565,168)(565,168)(566,168)(567,168)(567,169)(568,169)(568,169)(569,169)(569,169)(570,169)(571,169)(571,169)(572,169)(572,169)(573,169)(574,169)(574,169)(575,169)(575,169)(576,169)(576,169)(577,169)(578,169)(578,169)(579,169)(579,169)(580,169)(581,169)(581,170)(582,170)(582,170)(583,170)(584,170)(584,170)(585,170)(585,170)(586,170)
\thinlines \path(586,170)(586,170)(587,170)(588,170)(588,170)(589,170)(589,170)(590,170)(591,170)(591,170)(592,170)(592,170)(593,170)(593,170)(594,170)(595,171)(595,171)(596,171)(596,171)(597,171)(598,171)(598,171)(599,171)(599,171)(600,171)(601,171)(601,171)(602,171)(602,171)(603,171)(603,171)(604,171)(605,171)(605,171)(606,171)(606,171)(607,171)(608,172)(608,172)(609,172)(609,172)(610,172)(610,172)(611,172)(612,172)(612,172)(613,172)(613,172)(614,172)(615,172)(615,172)
\thinlines \path(615,172)(616,172)(616,172)(617,172)(618,172)(618,172)(619,172)(619,172)(620,172)(620,173)(621,173)(622,173)(622,173)(623,173)(623,173)(624,173)(625,173)(625,173)(626,173)(626,173)(627,173)(628,173)(628,173)(629,173)(629,173)(630,173)(630,173)(631,173)(632,173)(632,174)(633,174)(633,174)(634,174)(635,174)(635,174)(636,174)(636,174)(637,174)(637,174)(638,174)(639,174)(639,174)(640,174)(640,174)(641,174)(642,174)(642,174)(643,174)(643,174)(644,175)(645,175)
\thinlines \path(645,175)(645,175)(646,175)(646,175)(647,175)(647,175)(648,175)(649,175)(649,175)(650,175)(650,175)(651,175)(652,175)(652,175)(653,175)(653,175)(654,175)(654,175)(655,176)(656,176)(656,176)(657,176)(657,176)(658,176)(659,176)(659,176)(660,176)(660,176)(661,176)(662,176)(662,176)(663,176)(663,176)(664,176)(664,176)(665,176)(666,176)(666,177)(667,177)(667,177)(668,177)(669,177)(669,177)(670,177)(670,177)(671,177)(671,177)(672,177)(673,177)(673,177)(674,177)
\thinlines \path(674,177)(674,177)(675,177)(676,177)(676,177)(677,178)(677,178)(678,178)(679,178)(679,178)(680,178)(680,178)(681,178)(681,178)(682,178)(683,178)(683,178)(684,178)(684,178)(685,178)(686,178)(686,178)(687,178)(687,179)(688,179)(688,179)(689,179)(690,179)(690,179)(691,179)(691,179)(692,179)(693,179)(693,179)(694,179)(694,179)(695,179)(696,179)(696,179)(697,179)(697,179)(698,180)(698,180)(699,180)(700,180)(700,180)(701,180)(701,180)(702,180)(703,180)(703,180)
\thinlines \path(703,180)(704,180)(704,180)(705,180)(705,180)(706,180)(707,180)(707,180)(708,181)(708,181)(709,181)(710,181)(710,181)(711,181)(711,181)(712,181)(713,181)(713,181)(714,181)(714,181)(715,181)(715,181)(716,181)(717,181)(717,182)(718,182)(718,182)(719,182)(720,182)(720,182)(721,182)(721,182)(722,182)(722,182)(723,182)(724,182)(724,182)(725,182)(725,182)(726,182)(727,182)(727,183)(728,183)(728,183)(729,183)(730,183)(730,183)(731,183)(731,183)(732,183)(732,183)
\thinlines \path(732,183)(733,183)(734,183)(734,183)(735,183)(735,183)(736,184)(737,184)(737,184)(738,184)(738,184)(739,184)(739,184)(740,184)(741,184)(741,184)(742,184)(742,184)(743,184)(744,184)(744,184)(745,184)(745,185)(746,185)(747,185)(747,185)(748,185)(748,185)(749,185)(749,185)(750,185)(751,185)(751,185)(752,185)(752,185)(753,185)(754,185)(754,186)(755,186)(755,186)(756,186)(756,186)(757,186)(758,186)(758,186)(759,186)(759,186)(760,186)(761,186)(761,186)(762,186)
\thinlines \path(762,186)(762,186)(763,186)(764,187)(764,187)(765,187)(765,187)(766,187)(766,187)(767,187)(768,187)(768,187)(769,187)(769,187)(770,187)(771,187)(771,187)(772,188)(772,188)(773,188)(773,188)(774,188)(775,188)(775,188)(776,188)(776,188)(777,188)(778,188)(778,188)(779,188)(779,188)(780,188)(781,189)(781,189)(782,189)(782,189)(783,189)(783,189)(784,189)(785,189)(785,189)(786,189)(786,189)(787,189)(788,189)(788,189)(789,190)(789,190)(790,190)(790,190)(791,190)
\thinlines \path(791,190)(792,190)(792,190)(793,190)(793,190)(794,190)(795,190)(795,190)(796,190)(796,190)(797,191)(798,191)(798,191)(799,191)(799,191)(800,191)(800,191)(801,191)(802,191)(802,191)(803,191)(803,191)(804,191)(805,191)(805,192)(806,192)(806,192)(807,192)(807,192)(808,192)(809,192)(809,192)(810,192)(810,192)(811,192)(812,192)(812,192)(813,192)(813,193)(814,193)(815,193)(815,193)(816,193)(816,193)(817,193)(817,193)(818,193)(819,193)(819,193)(820,193)(820,193)
\thinlines \path(820,193)(821,194)(822,194)(822,194)(823,194)(823,194)(824,194)(824,194)(825,194)(826,194)(826,194)(827,194)(827,194)(828,194)(829,194)(829,195)(830,195)(830,195)(831,195)(832,195)(832,195)(833,195)(833,195)(834,195)(834,195)(835,195)(836,195)(836,195)(837,196)(837,196)(838,196)(839,196)(839,196)(840,196)(840,196)(841,196)(841,196)(842,196)(843,196)(843,196)(844,196)(844,197)(845,197)(846,197)(846,197)(847,197)(847,197)(848,197)(849,197)(849,197)(850,197)
\thinlines \path(850,197)(850,197)(851,197)(851,198)(852,198)(853,198)(853,198)(854,198)(854,198)(855,198)(856,198)(856,198)(857,198)(857,198)(858,198)(859,198)(859,199)(860,199)(860,199)(861,199)(861,199)(862,199)(863,199)(863,199)(864,199)(864,199)(865,199)(866,199)(866,200)(867,200)(867,200)(868,200)(868,200)(869,200)(870,200)(870,200)(871,200)(871,200)(872,200)(873,200)(873,200)(874,201)(874,201)(875,201)(876,201)(876,201)(877,201)(877,201)(878,201)(878,201)(879,201)
\thinlines \path(879,201)(880,201)(880,201)(881,202)(881,202)(882,202)(883,202)(883,202)(884,202)(884,202)(885,202)(885,202)(886,202)(887,202)(887,202)(888,203)(888,203)(889,203)(890,203)(890,203)(891,203)(891,203)(892,203)(893,203)(893,203)(894,203)(894,203)(895,204)(895,204)(896,204)(897,204)(897,204)(898,204)(898,204)(899,204)(900,204)(900,204)(901,204)(901,204)(902,205)(902,205)(903,205)(904,205)(904,205)(905,205)(905,205)(906,205)(907,205)(907,205)(908,205)(908,206)
\thinlines \path(908,206)(909,206)(910,206)(910,206)(911,206)(911,206)(912,206)(912,206)(913,206)(914,206)(914,206)(915,206)(915,207)(916,207)(917,207)(917,207)(918,207)(918,207)(919,207)(919,207)(920,207)(921,207)(921,207)(922,208)(922,208)(923,208)(924,208)(924,208)(925,208)(925,208)(926,208)(927,208)(927,208)(928,208)(928,209)(929,209)(929,209)(930,209)(931,209)(931,209)(932,209)(932,209)(933,209)(934,209)(934,209)(935,210)(935,210)(936,210)(936,210)(937,210)(938,210)
\thinlines \path(938,210)(938,210)(939,210)(939,210)(940,210)(941,210)(941,210)(942,211)(942,211)(943,211)(944,211)(944,211)(945,211)(945,211)(946,211)(946,211)(947,211)(948,212)(948,212)(949,212)(949,212)(950,212)(951,212)(951,212)(952,212)(952,212)(953,212)(953,212)(954,213)(955,213)(955,213)(956,213)(956,213)(957,213)(958,213)(958,213)(959,213)(959,213)(960,213)(961,214)(961,214)(962,214)(962,214)(963,214)(963,214)(964,214)(965,214)(965,214)(966,214)(966,214)(967,215)
\thinlines \path(967,215)(968,215)(968,215)(969,215)(969,215)(970,215)(970,215)(971,215)(972,215)(972,215)(973,216)(973,216)(974,216)(975,216)(975,216)(976,216)(976,216)(977,216)(978,216)(978,216)(979,216)(979,217)(980,217)(980,217)(981,217)(982,217)(982,217)(983,217)(983,217)(984,217)(985,217)(985,218)(986,218)(986,218)(987,218)(987,218)(988,218)(989,218)(989,218)(990,218)(990,218)(991,219)(992,219)(992,219)(993,219)(993,219)(994,219)(995,219)(995,219)(996,219)(996,219)
\thinlines \path(996,219)(997,219)(997,220)(998,220)(999,220)(999,220)(1000,220)(1000,220)(1001,220)(1002,220)(1002,220)(1003,220)(1003,221)(1004,221)(1004,221)(1005,221)(1006,221)(1006,221)(1007,221)(1007,221)(1008,221)(1009,221)(1009,222)(1010,222)(1010,222)(1011,222)(1012,222)(1012,222)(1013,222)(1013,222)(1014,222)(1014,222)(1015,223)(1016,223)(1016,223)(1017,223)(1017,223)(1018,223)(1019,223)(1019,223)(1020,223)(1020,223)(1021,224)(1021,224)(1022,224)(1023,224)(1023,224)(1024,224)(1024,224)(1025,224)(1026,224)
\thinlines \path(1026,224)(1026,225)(1027,225)(1027,225)(1028,225)(1029,225)(1029,225)(1030,225)(1030,225)(1031,225)(1031,225)(1032,226)(1033,226)(1033,226)(1034,226)(1034,226)(1035,226)(1036,226)(1036,226)(1037,226)(1037,226)(1038,227)(1038,227)(1039,227)(1040,227)(1040,227)(1041,227)(1041,227)(1042,227)(1043,227)(1043,228)(1044,228)(1044,228)(1045,228)(1046,228)(1046,228)(1047,228)(1047,228)(1048,228)(1048,228)(1049,229)(1050,229)(1050,229)(1051,229)(1051,229)(1052,229)(1053,229)(1053,229)(1054,229)(1054,230)(1055,230)
\thinlines \path(1055,230)(1055,230)(1056,230)(1057,230)(1057,230)(1058,230)(1058,230)(1059,230)(1060,230)(1060,231)(1061,231)(1061,231)(1062,231)(1063,231)(1063,231)(1064,231)(1064,231)(1065,231)(1065,232)(1066,232)(1067,232)(1067,232)(1068,232)(1068,232)(1069,232)(1070,232)(1070,232)(1071,233)(1071,233)(1072,233)(1072,233)(1073,233)(1074,233)(1074,233)(1075,233)(1075,233)(1076,233)(1077,234)(1077,234)(1078,234)(1078,234)(1079,234)(1080,234)(1080,234)(1081,234)(1081,234)(1082,235)(1082,235)(1083,235)(1084,235)(1084,235)
\thinlines \path(1084,235)(1085,235)(1085,235)(1086,235)(1087,235)(1087,236)(1088,236)(1088,236)(1089,236)(1090,236)(1090,236)(1091,236)(1091,236)(1092,236)(1092,237)(1093,237)(1094,237)(1094,237)(1095,237)(1095,237)(1096,237)(1097,237)(1097,237)(1098,238)(1098,238)(1099,238)(1099,238)(1100,238)(1101,238)(1101,238)(1102,238)(1102,238)(1103,239)(1104,239)(1104,239)(1105,239)(1105,239)(1106,239)(1107,239)(1107,239)(1108,239)(1108,240)(1109,240)(1109,240)(1110,240)(1111,240)(1111,240)(1112,240)(1112,240)(1113,241)(1114,241)
\thinlines \path(1114,241)(1114,241)(1115,241)(1115,241)(1116,241)(1116,241)(1117,241)(1118,241)(1118,242)(1119,242)(1119,242)(1120,242)(1121,242)(1121,242)(1122,242)(1122,242)(1123,242)(1124,243)(1124,243)(1125,243)(1125,243)(1126,243)(1126,243)(1127,243)(1128,243)(1128,243)(1129,244)(1129,244)(1130,244)(1131,244)(1131,244)(1132,244)(1132,244)(1133,244)(1133,245)(1134,245)(1135,245)(1135,245)(1136,245)(1136,245)(1137,245)(1138,245)(1138,245)(1139,246)(1139,246)(1140,246)(1141,246)(1141,246)(1142,246)(1142,246)(1143,246)
\thinlines \path(1143,246)(1143,247)(1144,247)(1145,247)(1145,247)(1146,247)(1146,247)(1147,247)(1148,247)(1148,247)(1149,248)(1149,248)(1150,248)(1150,248)(1151,248)(1152,248)(1152,248)(1153,248)(1153,249)(1154,249)(1155,249)(1155,249)(1156,249)(1156,249)(1157,249)(1158,249)(1158,250)(1159,250)(1159,250)(1160,250)(1160,250)(1161,250)(1162,250)(1162,250)(1163,250)(1163,251)(1164,251)(1165,251)(1165,251)(1166,251)(1166,251)(1167,251)(1167,251)(1168,252)(1169,252)(1169,252)(1170,252)(1170,252)(1171,252)(1172,252)(1172,252)
\thinlines \path(1172,252)(1173,253)(1173,253)(1174,253)(1175,253)(1175,253)(1176,253)(1176,253)(1177,253)(1177,254)(1178,254)(1179,254)(1179,254)(1180,254)(1180,254)(1181,254)(1182,254)(1182,255)(1183,255)(1183,255)(1184,255)(1184,255)(1185,255)(1186,255)(1186,255)(1187,255)(1187,256)(1188,256)(1189,256)(1189,256)(1190,256)(1190,256)(1191,256)(1192,256)(1192,257)(1193,257)(1193,257)(1194,257)(1194,257)(1195,257)(1196,257)(1196,257)(1197,258)(1197,258)(1198,258)(1199,258)(1199,258)(1200,258)(1200,258)(1201,258)(1201,259)
\thinlines \path(1201,259)(1202,259)(1203,259)(1203,259)(1204,259)(1204,259)(1205,259)(1206,259)(1206,260)(1207,260)(1207,260)(1208,260)(1209,260)(1209,260)(1210,260)(1210,261)(1211,261)(1211,261)(1212,261)(1213,261)(1213,261)(1214,261)(1214,261)(1215,262)(1216,262)(1216,262)(1217,262)(1217,262)(1218,262)(1218,262)(1219,262)(1220,263)(1220,263)(1221,263)(1221,263)(1222,263)(1223,263)(1223,263)(1224,263)(1224,264)(1225,264)(1226,264)(1226,264)(1227,264)(1227,264)(1228,264)(1228,264)(1229,265)(1230,265)(1230,265)(1231,265)
\thinlines \path(1231,265)(1231,265)(1232,265)(1233,265)(1233,266)(1234,266)(1234,266)(1235,266)(1235,266)(1236,266)(1237,266)(1237,266)(1238,267)(1238,267)(1239,267)(1240,267)(1240,267)(1241,267)(1241,267)(1242,267)(1243,268)(1243,268)(1244,268)(1244,268)(1245,268)(1245,268)(1246,268)(1247,269)(1247,269)(1248,269)(1248,269)(1249,269)(1250,269)(1250,269)(1251,269)(1251,270)(1252,270)(1252,270)(1253,270)(1254,270)(1254,270)(1255,270)(1255,271)(1256,271)(1257,271)(1257,271)(1258,271)(1258,271)(1259,271)(1260,271)(1260,272)
\thinlines \path(1260,272)(1261,272)(1261,272)(1262,272)(1262,272)(1263,272)(1264,272)(1264,273)(1265,273)(1265,273)(1266,273)(1267,273)(1267,273)(1268,273)(1268,273)(1269,274)(1269,274)(1270,274)(1271,274)(1271,274)(1272,274)(1272,274)(1273,275)(1274,275)(1274,275)(1275,275)(1275,275)(1276,275)(1277,275)(1277,276)(1278,276)(1278,276)(1279,276)(1279,276)(1280,276)(1281,276)(1281,276)(1282,277)(1282,277)(1283,277)(1284,277)(1284,277)(1285,277)(1285,277)(1286,278)(1286,278)(1287,278)(1288,278)(1288,278)(1289,278)(1289,278)
\thinlines \path(1289,278)(1290,279)(1291,279)(1291,279)(1292,279)(1292,279)(1293,279)(1294,279)(1294,279)(1295,280)(1295,280)(1296,280)(1296,280)(1297,280)(1298,280)(1298,280)(1299,281)(1299,281)(1300,281)(1301,281)(1301,281)(1302,281)(1302,281)(1303,282)(1303,282)(1304,282)(1305,282)(1305,282)(1306,282)(1306,282)(1307,283)(1308,283)(1308,283)(1309,283)(1309,283)(1310,283)(1311,283)(1311,284)(1312,284)(1312,284)(1313,284)(1313,284)(1314,284)(1315,284)(1315,285)(1316,285)(1316,285)(1317,285)(1318,285)(1318,285)(1319,285)
\thinlines \path(1319,285)(1319,285)(1320,286)(1321,286)(1321,286)(1322,286)(1322,286)(1323,286)(1323,286)(1324,287)(1325,287)(1325,287)(1326,287)(1326,287)(1327,287)(1328,287)(1328,288)(1329,288)(1329,288)(1330,288)(1330,288)(1331,288)(1332,288)(1332,289)(1333,289)(1333,289)(1334,289)(1335,289)(1335,289)(1336,289)(1336,290)(1337,290)(1338,290)(1338,290)(1339,290)(1339,290)(1340,291)(1340,291)(1341,291)(1342,291)(1342,291)(1343,291)(1343,291)(1344,292)(1345,292)(1345,292)(1346,292)(1346,292)(1347,292)(1347,292)(1348,293)
\thinlines \path(1348,293)(1349,293)(1349,293)(1350,293)(1350,293)(1351,293)(1352,293)(1352,294)(1353,294)(1353,294)(1354,294)(1355,294)(1355,294)(1356,294)(1356,295)(1357,295)(1357,295)(1358,295)(1359,295)(1359,295)(1360,295)(1360,296)(1361,296)(1362,296)(1362,296)(1363,296)(1363,296)(1364,296)(1364,297)(1365,297)(1366,297)(1366,297)(1367,297)(1367,297)(1368,298)(1369,298)(1369,298)(1370,298)(1370,298)(1371,298)(1372,298)(1372,299)(1373,299)(1373,299)(1374,299)(1374,299)(1375,299)(1376,299)(1376,300)(1377,300)(1377,300)
\thinlines \path(1377,300)(1378,300)(1379,300)(1379,300)(1380,301)(1380,301)(1381,301)(1381,301)(1382,301)(1383,301)(1383,301)(1384,302)(1384,302)(1385,302)(1386,302)(1386,302)(1387,302)(1387,302)(1388,303)(1389,303)(1389,303)(1390,303)(1390,303)(1391,303)(1391,304)(1392,304)(1393,304)(1393,304)(1394,304)(1394,304)(1395,304)(1396,305)(1396,305)(1397,305)(1397,305)(1398,305)(1398,305)(1399,305)(1400,306)(1400,306)(1401,306)(1401,306)(1402,306)(1403,306)(1403,307)(1404,307)(1404,307)(1405,307)(1406,307)(1406,307)(1407,307)
\thinlines \path(1407,307)(1407,308)(1408,308)(1408,308)(1409,308)(1410,308)(1410,308)(1411,309)(1411,309)(1412,309)(1413,309)(1413,309)(1414,309)(1414,309)(1415,310)(1415,310)(1416,310)(1417,310)(1417,310)(1418,310)(1418,311)(1419,311)(1420,311)(1420,311)(1421,311)(1421,311)(1422,311)(1423,312)(1423,312)(1424,312)(1424,312)(1425,312)(1425,312)(1426,313)(1427,313)(1427,313)(1428,313)(1428,313)(1429,313)(1430,314)(1430,314)(1431,314)(1431,314)(1432,314)(1432,314)(1433,314)(1434,315)(1434,315)(1435,315)(1435,315)(1436,315)
\Thicklines \path(264,353)(264,353)(265,352)(265,352)(266,351)(266,350)(267,350)(268,349)(268,349)(269,348)(269,347)(270,347)(270,346)(271,345)(272,345)(272,344)(273,344)(273,343)(274,342)(275,342)(275,341)(276,340)(276,340)(277,339)(277,339)(278,338)(279,337)(279,337)(280,336)(280,335)(281,335)(282,334)(282,333)(283,333)(283,332)(284,332)(285,331)(285,330)(286,330)(286,329)(287,328)(287,328)(288,327)(289,327)(289,326)(290,325)(290,325)(291,324)(292,323)(292,323)(293,322)
\Thicklines \path(293,322)(293,322)(294,321)(294,320)(295,320)(296,319)(296,318)(297,318)(297,317)(298,316)(299,316)(299,315)(300,315)(300,314)(301,313)(302,313)(302,312)(303,311)(303,311)(304,310)(304,310)(305,309)(306,308)(306,308)(307,307)(307,306)(308,306)(309,305)(309,305)(310,304)(310,303)(311,303)(311,302)(312,301)(313,301)(313,300)(314,299)(314,299)(315,298)(316,298)(316,297)(317,296)(317,296)(318,295)(319,294)(319,294)(320,293)(320,293)(321,292)(321,291)(322,291)
\Thicklines \path(322,291)(323,290)(323,289)(324,289)(324,288)(325,288)(326,287)(326,286)(327,286)(327,285)(328,284)(328,284)(329,283)(330,283)(330,282)(331,281)(331,281)(332,280)(333,279)(333,279)(334,278)(334,277)(335,277)(336,276)(336,276)(337,275)(337,274)(338,274)(338,273)(339,272)(340,272)(340,271)(341,271)(341,270)(342,269)(343,269)(343,268)(344,267)(344,267)(345,266)(345,266)(346,265)(347,264)(347,264)(348,263)(348,262)(349,262)(350,261)(350,260)(351,260)(351,259)
\Thicklines \path(351,259)(352,259)(353,258)(353,257)(354,257)(354,256)(355,255)(355,255)(356,254)(357,254)(357,253)(358,252)(358,252)(359,251)(360,250)(360,250)(361,249)(361,249)(362,248)(362,247)(363,247)(364,246)(364,245)(365,245)(365,244)(366,243)(367,243)(367,242)(368,242)(368,241)(369,240)(370,240)(370,239)(371,238)(371,238)(372,237)(372,237)(373,236)(374,235)(374,235)(375,234)(375,233)(376,233)(377,232)(377,232)(378,231)(378,230)(379,230)(379,229)(380,228)(381,228)
\Thicklines \path(381,228)(381,227)(382,226)(382,226)(383,225)(384,225)(384,224)(385,223)(385,223)(386,222)(387,221)(387,221)(388,220)(388,220)(389,219)(389,218)(390,218)(391,217)(391,216)(392,216)(392,215)(393,215)(394,214)(394,213)(395,213)(395,212)(396,211)(397,211)(397,210)(398,210)(398,209)(399,208)(399,208)(400,207)(401,206)(401,206)(402,205)(402,204)(403,204)(404,203)(404,203)(405,202)(405,201)(406,201)(406,200)(407,199)(408,199)(408,198)(409,198)(409,197)(410,196)
\Thicklines \path(410,196)(411,196)(411,195)(412,194)(412,194)(413,193)(414,193)(414,192)(415,191)(415,191)(416,190)(416,189)(417,189)(418,188)(418,187)(419,187)(419,186)(420,186)(421,185)(421,184)(422,184)(422,183)(423,182)(423,182)(424,181)(425,181)(425,180)(426,179)(426,179)(427,178)(428,177)(428,177)(429,176)(429,176)(430,175)(431,174)(431,174)(432,173)(432,172)(433,172)(433,171)(434,170)(435,170)(435,169)(436,169)(436,168)(437,167)(438,167)(438,166)(439,165)(439,165)
\Thicklines \path(439,165)(440,164)(440,164)(441,163)(442,162)(442,162)(443,161)(443,160)(444,160)(445,159)(445,159)(446,158)
\end{picture}

%% file: gm-xsect.tex
% GNUPLOT: LaTeX picture using EEPIC macros
\setlength{\unitlength}{0.240900pt}
\begin{picture}(1500,900)(0,0)
\tenrm
\thinlines \drawline(264,338)(1436,338)
\thicklines \path(264,158)(284,158)
\thicklines \path(1436,158)(1416,158)
\put(242,158){\makebox(0,0)[r]{$-100$}}
\thicklines \path(264,248)(284,248)
\thicklines \path(1436,248)(1416,248)
\put(242,248){\makebox(0,0)[r]{$-50$}}
\thicklines \path(264,338)(284,338)
\thicklines \path(1436,338)(1416,338)
\put(242,338){\makebox(0,0)[r]{$0$}}
\thicklines \path(264,428)(284,428)
\thicklines \path(1436,428)(1416,428)
\put(242,428){\makebox(0,0)[r]{$50$}}
\thicklines \path(264,517)(284,517)
\thicklines \path(1436,517)(1416,517)
\put(242,517){\makebox(0,0)[r]{$100$}}
\thicklines \path(264,607)(284,607)
\thicklines \path(1436,607)(1416,607)
\put(242,607){\makebox(0,0)[r]{$150$}}
\thicklines \path(264,697)(284,697)
\thicklines \path(1436,697)(1416,697)
\put(242,697){\makebox(0,0)[r]{$200$}}
\thicklines \path(264,787)(284,787)
\thicklines \path(1436,787)(1416,787)
\put(242,787){\makebox(0,0)[r]{$250$}}
\thicklines \path(264,158)(264,178)
\thicklines \path(264,787)(264,767)
\put(264,113){\makebox(0,0){$0.1$}}
\thicklines \path(382,158)(382,168)
\thicklines \path(382,787)(382,777)
\put(382,113){\makebox(0,0){}}
\thicklines \path(450,158)(450,168)
\thicklines \path(450,787)(450,777)
\put(450,113){\makebox(0,0){}}
\thicklines \path(499,158)(499,168)
\thicklines \path(499,787)(499,777)
\put(499,113){\makebox(0,0){}}
\thicklines \path(537,158)(537,168)
\thicklines \path(537,787)(537,777)
\put(537,113){\makebox(0,0){}}
\thicklines \path(568,158)(568,168)
\thicklines \path(568,787)(568,777)
\put(568,113){\makebox(0,0){}}
\thicklines \path(594,158)(594,168)
\thicklines \path(594,787)(594,777)
\put(594,113){\makebox(0,0){}}
\thicklines \path(617,158)(617,168)
\thicklines \path(617,787)(617,777)
\put(617,113){\makebox(0,0){}}
\thicklines \path(637,158)(637,168)
\thicklines \path(637,787)(637,777)
\put(637,113){\makebox(0,0){}}
\thicklines \path(655,158)(655,178)
\thicklines \path(655,787)(655,767)
\put(655,113){\makebox(0,0){$1$}}
\thicklines \path(772,158)(772,168)
\thicklines \path(772,787)(772,777)
\put(772,113){\makebox(0,0){}}
\thicklines \path(841,158)(841,168)
\thicklines \path(841,787)(841,777)
\put(841,113){\makebox(0,0){}}
\thicklines \path(890,158)(890,168)
\thicklines \path(890,787)(890,777)
\put(890,113){\makebox(0,0){}}
\thicklines \path(928,158)(928,168)
\thicklines \path(928,787)(928,777)
\put(928,113){\makebox(0,0){}}
\thicklines \path(959,158)(959,168)
\thicklines \path(959,787)(959,777)
\put(959,113){\makebox(0,0){}}
\thicklines \path(985,158)(985,168)
\thicklines \path(985,787)(985,777)
\put(985,113){\makebox(0,0){}}
\thicklines \path(1007,158)(1007,168)
\thicklines \path(1007,787)(1007,777)
\put(1007,113){\makebox(0,0){}}
\thicklines \path(1027,158)(1027,168)
\thicklines \path(1027,787)(1027,777)
\put(1027,113){\makebox(0,0){}}
\thicklines \path(1045,158)(1045,178)
\thicklines \path(1045,787)(1045,767)
\put(1045,113){\makebox(0,0){$10$}}
\thicklines \path(1163,158)(1163,168)
\thicklines \path(1163,787)(1163,777)
\put(1163,113){\makebox(0,0){}}
\thicklines \path(1232,158)(1232,168)
\thicklines \path(1232,787)(1232,777)
\put(1232,113){\makebox(0,0){}}
\thicklines \path(1281,158)(1281,168)
\thicklines \path(1281,787)(1281,777)
\put(1281,113){\makebox(0,0){}}
\thicklines \path(1318,158)(1318,168)
\thicklines \path(1318,787)(1318,777)
\put(1318,113){\makebox(0,0){}}
\thicklines \path(1349,158)(1349,168)
\thicklines \path(1349,787)(1349,777)
\put(1349,113){\makebox(0,0){}}
\thicklines \path(1375,158)(1375,168)
\thicklines \path(1375,787)(1375,777)
\put(1375,113){\makebox(0,0){}}
\thicklines \path(1398,158)(1398,168)
\thicklines \path(1398,787)(1398,777)
\put(1398,113){\makebox(0,0){}}
\thicklines \path(1418,158)(1418,168)
\thicklines \path(1418,787)(1418,777)
\put(1418,113){\makebox(0,0){}}
\thicklines \path(1436,158)(1436,178)
\thicklines \path(1436,787)(1436,767)
\put(1436,113){\makebox(0,0){$100$}}
\thicklines \path(264,158)(1436,158)(1436,787)(264,787)(264,158)
\put(45,472){\makebox(0,0)[l]{\shortstack{\begin{sideways} 		$\big(\sigma_{1/2}(\nu) - \sigma_{3/2}(\nu)\big)/\mu$b 		\end{sideways}}}}
\put(850,68){\makebox(0,0){$\nu/$GeV}}
\put(1306,722){\makebox(0,0)[r]{neutron}}
\thinlines \path(1328,722)(1394,722)
\thinlines \path(333,338)(333,338)(335,450)(336,494)(339,551)(340,573)(342,591)(345,621)(348,644)(351,662)(353,676)(356,688)(359,697)(362,704)(365,709)(366,711)(368,713)(369,714)(371,715)(372,716)(373,716)(375,716)(376,716)(378,716)(379,715)(381,715)(382,714)(385,712)(391,706)(396,697)(402,688)(425,640)(471,535)(517,444)(563,373)(586,346)(609,323)(632,303)(655,288)(678,275)(701,265)(724,257)(735,254)(747,251)(758,249)(770,247)(781,246)(787,245)(793,245)(798,245)(804,244)
\thinlines \path(804,244)(807,244)(810,244)(813,244)(816,244)(819,244)(820,244)(821,244)(823,244)(824,244)(826,244)(827,244)(829,244)(830,244)(832,244)(833,244)(834,244)(836,244)(839,244)(844,244)(850,244)(862,245)(873,245)(885,246)(931,252)(977,258)(1022,266)(1068,274)(1114,282)(1160,290)(1206,297)(1252,303)(1298,308)(1344,313)(1390,317)(1436,321)
\put(1306,677){\makebox(0,0)[r]{proton}}
\thicklines \path(1328,677)(1394,677)
\thicklines \path(333,338)(333,338)(335,424)(336,457)(339,500)(340,516)(342,530)(345,551)(348,568)(351,581)(353,591)(356,598)(359,604)(361,606)(362,608)(363,609)(365,610)(366,611)(368,612)(369,612)(371,612)(372,612)(373,612)(375,611)(376,610)(379,609)(382,606)(385,603)(391,596)(402,577)(425,533)(471,442)(494,403)(517,370)(540,343)(563,322)(586,305)(609,293)(620,288)(632,284)(643,281)(655,278)(666,276)(678,275)(684,275)(689,274)(692,274)(695,274)(698,274)(701,274)(704,274)
\thicklines \path(704,274)(705,274)(707,273)(708,273)(709,273)(711,273)(712,273)(714,273)(715,273)(717,273)(718,273)(720,273)(721,274)(724,274)(730,274)(735,274)(747,275)(758,275)(770,276)(793,279)(839,284)(885,290)(931,295)(977,301)(1022,305)(1068,310)(1114,314)(1160,317)(1206,320)(1252,323)(1298,325)(1344,327)(1390,329)(1436,330)
\end{picture}

%% file: piN-p.tex
% GNUPLOT: LaTeX picture using EEPIC macros
\setlength{\unitlength}{0.240900pt}
\begin{picture}(1500,900)(0,0)
\tenrm
\thinlines \drawline(220,400)(1436,400)
\thicklines \path(220,113)(240,113)
\thicklines \path(1436,113)(1416,113)
\put(198,113){\makebox(0,0)[r]{$-150$}}
\thicklines \path(220,209)(240,209)
\thicklines \path(1436,209)(1416,209)
\put(198,209){\makebox(0,0)[r]{$-100$}}
\thicklines \path(220,304)(240,304)
\thicklines \path(1436,304)(1416,304)
\put(198,304){\makebox(0,0)[r]{$-50$}}
\thicklines \path(220,400)(240,400)
\thicklines \path(1436,400)(1416,400)
\put(198,400){\makebox(0,0)[r]{$0$}}
\thicklines \path(220,495)(240,495)
\thicklines \path(1436,495)(1416,495)
\put(198,495){\makebox(0,0)[r]{$50$}}
\thicklines \path(220,591)(240,591)
\thicklines \path(1436,591)(1416,591)
\put(198,591){\makebox(0,0)[r]{$100$}}
\thicklines \path(220,686)(240,686)
\thicklines \path(1436,686)(1416,686)
\put(198,686){\makebox(0,0)[r]{$150$}}
\thicklines \path(220,782)(240,782)
\thicklines \path(1436,782)(1416,782)
\put(198,782){\makebox(0,0)[r]{$200$}}
\thicklines \path(220,877)(240,877)
\thicklines \path(1436,877)(1416,877)
\put(198,877){\makebox(0,0)[r]{$250$}}
\thicklines \path(220,113)(220,133)
\thicklines \path(220,877)(220,857)
\put(220,68){\makebox(0,0){$0.1$}}
\thicklines \path(312,113)(312,123)
\thicklines \path(312,877)(312,867)
\put(312,68){\makebox(0,0){}}
\thicklines \path(365,113)(365,123)
\thicklines \path(365,877)(365,867)
\put(365,68){\makebox(0,0){}}
\thicklines \path(403,113)(403,123)
\thicklines \path(403,877)(403,867)
\put(403,68){\makebox(0,0){}}
\thicklines \path(432,113)(432,123)
\thicklines \path(432,877)(432,867)
\put(432,68){\makebox(0,0){}}
\thicklines \path(457,113)(457,123)
\thicklines \path(457,877)(457,867)
\put(457,68){\makebox(0,0){}}
\thicklines \path(477,113)(477,123)
\thicklines \path(477,877)(477,867)
\put(477,68){\makebox(0,0){}}
\thicklines \path(495,113)(495,123)
\thicklines \path(495,877)(495,867)
\put(495,68){\makebox(0,0){}}
\thicklines \path(510,113)(510,123)
\thicklines \path(510,877)(510,867)
\put(510,68){\makebox(0,0){}}
\thicklines \path(524,113)(524,133)
\thicklines \path(524,877)(524,857)
\put(524,68){\makebox(0,0){$1$}}
\thicklines \path(616,113)(616,123)
\thicklines \path(616,877)(616,867)
\put(616,68){\makebox(0,0){}}
\thicklines \path(669,113)(669,123)
\thicklines \path(669,877)(669,867)
\put(669,68){\makebox(0,0){}}
\thicklines \path(707,113)(707,123)
\thicklines \path(707,877)(707,867)
\put(707,68){\makebox(0,0){}}
\thicklines \path(736,113)(736,123)
\thicklines \path(736,877)(736,867)
\put(736,68){\makebox(0,0){}}
\thicklines \path(761,113)(761,123)
\thicklines \path(761,877)(761,867)
\put(761,68){\makebox(0,0){}}
\thicklines \path(781,113)(781,123)
\thicklines \path(781,877)(781,867)
\put(781,68){\makebox(0,0){}}
\thicklines \path(799,113)(799,123)
\thicklines \path(799,877)(799,867)
\put(799,68){\makebox(0,0){}}
\thicklines \path(814,113)(814,123)
\thicklines \path(814,877)(814,867)
\put(814,68){\makebox(0,0){}}
\thicklines \path(828,113)(828,133)
\thicklines \path(828,877)(828,857)
\put(828,68){\makebox(0,0){$10$}}
\thicklines \path(920,113)(920,123)
\thicklines \path(920,877)(920,867)
\put(920,68){\makebox(0,0){}}
\thicklines \path(973,113)(973,123)
\thicklines \path(973,877)(973,867)
\put(973,68){\makebox(0,0){}}
\thicklines \path(1011,113)(1011,123)
\thicklines \path(1011,877)(1011,867)
\put(1011,68){\makebox(0,0){}}
\thicklines \path(1040,113)(1040,123)
\thicklines \path(1040,877)(1040,867)
\put(1040,68){\makebox(0,0){}}
\thicklines \path(1065,113)(1065,123)
\thicklines \path(1065,877)(1065,867)
\put(1065,68){\makebox(0,0){}}
\thicklines \path(1085,113)(1085,123)
\thicklines \path(1085,877)(1085,867)
\put(1085,68){\makebox(0,0){}}
\thicklines \path(1103,113)(1103,123)
\thicklines \path(1103,877)(1103,867)
\put(1103,68){\makebox(0,0){}}
\thicklines \path(1118,113)(1118,123)
\thicklines \path(1118,877)(1118,867)
\put(1118,68){\makebox(0,0){}}
\thicklines \path(1132,113)(1132,133)
\thicklines \path(1132,877)(1132,857)
\put(1132,68){\makebox(0,0){$100$}}
\thicklines \path(1224,113)(1224,123)
\thicklines \path(1224,877)(1224,867)
\put(1224,68){\makebox(0,0){}}
\thicklines \path(1277,113)(1277,123)
\thicklines \path(1277,877)(1277,867)
\put(1277,68){\makebox(0,0){}}
\thicklines \path(1315,113)(1315,123)
\thicklines \path(1315,877)(1315,867)
\put(1315,68){\makebox(0,0){}}
\thicklines \path(1344,113)(1344,123)
\thicklines \path(1344,877)(1344,867)
\put(1344,68){\makebox(0,0){}}
\thicklines \path(1369,113)(1369,123)
\thicklines \path(1369,877)(1369,867)
\put(1369,68){\makebox(0,0){}}
\thicklines \path(1389,113)(1389,123)
\thicklines \path(1389,877)(1389,867)
\put(1389,68){\makebox(0,0){}}
\thicklines \path(1407,113)(1407,123)
\thicklines \path(1407,877)(1407,867)
\put(1407,68){\makebox(0,0){}}
\thicklines \path(1422,113)(1422,123)
\thicklines \path(1422,877)(1422,867)
\put(1422,68){\makebox(0,0){}}
\thicklines \path(1436,113)(1436,133)
\thicklines \path(1436,877)(1436,857)
\put(1436,68){\makebox(0,0){$1000$}}
\thicklines \path(220,113)(1436,113)(1436,877)(220,877)(220,113)
\put(45,495){\makebox(0,0)[l]{\shortstack{\begin{sideways} 		$\big(\sigma_{1/2}(\nu) - \sigma_{3/2}(\nu)\big)/\mu$b 		\end{sideways}}}}
\put(828,23){\makebox(0,0){$\nu/$GeV}}
\put(1306,812){\makebox(0,0)[r]{proton, $\kappa_0=0$}}
\thinlines \path(1328,812)(1394,812)
\thinlines \path(274,401)(274,401)(275,505)(277,545)(278,574)(280,596)(281,613)(283,628)(286,650)(288,659)(289,666)(291,672)(292,677)(294,681)(295,685)(297,687)(298,689)(300,690)(301,691)(303,691)(304,691)(306,690)(307,690)(310,687)(313,683)(316,678)(322,666)(371,537)(395,478)(419,430)(443,394)(468,367)(480,358)(492,350)(504,344)(516,339)(522,337)(528,335)(534,334)(540,333)(546,332)(549,332)(552,332)(555,332)(558,331)(560,331)(561,331)(563,331)(564,331)(566,331)(567,331)
\thinlines \path(567,331)(569,331)(570,331)(572,331)(574,331)(575,331)(577,331)(580,331)(583,332)(589,332)(595,332)(601,333)(613,334)(661,341)(710,350)(758,358)(807,365)(855,371)(903,376)(952,381)(1000,384)(1049,388)(1097,390)(1145,392)(1194,394)(1242,395)(1291,396)(1339,397)(1388,398)(1436,398)
\put(1306,767){\makebox(0,0)[r]{proton, $\kappa_0\neq0$}}
\thicklines \path(1328,767)(1394,767)
\thicklines \path(274,402)(274,402)(275,515)(277,558)(278,589)(280,613)(281,632)(283,647)(284,660)(286,671)(288,680)(289,688)(291,694)(292,699)(294,703)(295,706)(297,708)(298,710)(300,711)(301,711)(303,711)(304,710)(306,709)(307,708)(310,704)(313,699)(316,693)(322,679)(371,538)(395,481)(407,459)(419,442)(431,428)(437,423)(443,418)(449,415)(455,412)(459,411)(462,410)(465,409)(466,409)(468,409)(469,409)(471,409)(472,408)(474,408)(475,408)(477,408)(478,408)(480,408)(481,409)
\thicklines \path(481,409)(483,409)(486,409)(489,410)(492,410)(498,412)(504,414)(516,419)(564,446)(589,460)(613,472)(625,476)(637,480)(649,484)(661,486)(667,487)(670,487)(673,488)(676,488)(679,488)(681,488)(682,488)(684,488)(685,488)(687,488)(689,488)(690,488)(692,488)(693,488)(695,488)(696,488)(698,488)(701,488)(704,488)(710,487)(716,487)(722,486)(734,484)(758,478)(807,463)(855,444)(903,426)(952,408)(1000,393)(1049,381)(1097,371)(1145,363)(1194,357)(1242,352)(1291,348)(1339,346)
\thicklines \path(1339,346)(1388,343)(1436,342)
\end{picture}

%% file: piN-n.tex
% GNUPLOT: LaTeX picture using EEPIC macros
\setlength{\unitlength}{0.240900pt}
\begin{picture}(1500,900)(0,0)
\tenrm
\thinlines \drawline(220,400)(1436,400)
\thicklines \path(220,113)(240,113)
\thicklines \path(1436,113)(1416,113)
\put(198,113){\makebox(0,0)[r]{$-150$}}
\thicklines \path(220,209)(240,209)
\thicklines \path(1436,209)(1416,209)
\put(198,209){\makebox(0,0)[r]{$-100$}}
\thicklines \path(220,304)(240,304)
\thicklines \path(1436,304)(1416,304)
\put(198,304){\makebox(0,0)[r]{$-50$}}
\thicklines \path(220,400)(240,400)
\thicklines \path(1436,400)(1416,400)
\put(198,400){\makebox(0,0)[r]{$0$}}
\thicklines \path(220,495)(240,495)
\thicklines \path(1436,495)(1416,495)
\put(198,495){\makebox(0,0)[r]{$50$}}
\thicklines \path(220,591)(240,591)
\thicklines \path(1436,591)(1416,591)
\put(198,591){\makebox(0,0)[r]{$100$}}
\thicklines \path(220,686)(240,686)
\thicklines \path(1436,686)(1416,686)
\put(198,686){\makebox(0,0)[r]{$150$}}
\thicklines \path(220,782)(240,782)
\thicklines \path(1436,782)(1416,782)
\put(198,782){\makebox(0,0)[r]{$200$}}
\thicklines \path(220,877)(240,877)
\thicklines \path(1436,877)(1416,877)
\put(198,877){\makebox(0,0)[r]{$250$}}
\thicklines \path(220,113)(220,133)
\thicklines \path(220,877)(220,857)
\put(220,68){\makebox(0,0){$0.1$}}
\thicklines \path(312,113)(312,123)
\thicklines \path(312,877)(312,867)
\put(312,68){\makebox(0,0){}}
\thicklines \path(365,113)(365,123)
\thicklines \path(365,877)(365,867)
\put(365,68){\makebox(0,0){}}
\thicklines \path(403,113)(403,123)
\thicklines \path(403,877)(403,867)
\put(403,68){\makebox(0,0){}}
\thicklines \path(432,113)(432,123)
\thicklines \path(432,877)(432,867)
\put(432,68){\makebox(0,0){}}
\thicklines \path(457,113)(457,123)
\thicklines \path(457,877)(457,867)
\put(457,68){\makebox(0,0){}}
\thicklines \path(477,113)(477,123)
\thicklines \path(477,877)(477,867)
\put(477,68){\makebox(0,0){}}
\thicklines \path(495,113)(495,123)
\thicklines \path(495,877)(495,867)
\put(495,68){\makebox(0,0){}}
\thicklines \path(510,113)(510,123)
\thicklines \path(510,877)(510,867)
\put(510,68){\makebox(0,0){}}
\thicklines \path(524,113)(524,133)
\thicklines \path(524,877)(524,857)
\put(524,68){\makebox(0,0){$1$}}
\thicklines \path(616,113)(616,123)
\thicklines \path(616,877)(616,867)
\put(616,68){\makebox(0,0){}}
\thicklines \path(669,113)(669,123)
\thicklines \path(669,877)(669,867)
\put(669,68){\makebox(0,0){}}
\thicklines \path(707,113)(707,123)
\thicklines \path(707,877)(707,867)
\put(707,68){\makebox(0,0){}}
\thicklines \path(736,113)(736,123)
\thicklines \path(736,877)(736,867)
\put(736,68){\makebox(0,0){}}
\thicklines \path(761,113)(761,123)
\thicklines \path(761,877)(761,867)
\put(761,68){\makebox(0,0){}}
\thicklines \path(781,113)(781,123)
\thicklines \path(781,877)(781,867)
\put(781,68){\makebox(0,0){}}
\thicklines \path(799,113)(799,123)
\thicklines \path(799,877)(799,867)
\put(799,68){\makebox(0,0){}}
\thicklines \path(814,113)(814,123)
\thicklines \path(814,877)(814,867)
\put(814,68){\makebox(0,0){}}
\thicklines \path(828,113)(828,133)
\thicklines \path(828,877)(828,857)
\put(828,68){\makebox(0,0){$10$}}
\thicklines \path(920,113)(920,123)
\thicklines \path(920,877)(920,867)
\put(920,68){\makebox(0,0){}}
\thicklines \path(973,113)(973,123)
\thicklines \path(973,877)(973,867)
\put(973,68){\makebox(0,0){}}
\thicklines \path(1011,113)(1011,123)
\thicklines \path(1011,877)(1011,867)
\put(1011,68){\makebox(0,0){}}
\thicklines \path(1040,113)(1040,123)
\thicklines \path(1040,877)(1040,867)
\put(1040,68){\makebox(0,0){}}
\thicklines \path(1065,113)(1065,123)
\thicklines \path(1065,877)(1065,867)
\put(1065,68){\makebox(0,0){}}
\thicklines \path(1085,113)(1085,123)
\thicklines \path(1085,877)(1085,867)
\put(1085,68){\makebox(0,0){}}
\thicklines \path(1103,113)(1103,123)
\thicklines \path(1103,877)(1103,867)
\put(1103,68){\makebox(0,0){}}
\thicklines \path(1118,113)(1118,123)
\thicklines \path(1118,877)(1118,867)
\put(1118,68){\makebox(0,0){}}
\thicklines \path(1132,113)(1132,133)
\thicklines \path(1132,877)(1132,857)
\put(1132,68){\makebox(0,0){$100$}}
\thicklines \path(1224,113)(1224,123)
\thicklines \path(1224,877)(1224,867)
\put(1224,68){\makebox(0,0){}}
\thicklines \path(1277,113)(1277,123)
\thicklines \path(1277,877)(1277,867)
\put(1277,68){\makebox(0,0){}}
\thicklines \path(1315,113)(1315,123)
\thicklines \path(1315,877)(1315,867)
\put(1315,68){\makebox(0,0){}}
\thicklines \path(1344,113)(1344,123)
\thicklines \path(1344,877)(1344,867)
\put(1344,68){\makebox(0,0){}}
\thicklines \path(1369,113)(1369,123)
\thicklines \path(1369,877)(1369,867)
\put(1369,68){\makebox(0,0){}}
\thicklines \path(1389,113)(1389,123)
\thicklines \path(1389,877)(1389,867)
\put(1389,68){\makebox(0,0){}}
\thicklines \path(1407,113)(1407,123)
\thicklines \path(1407,877)(1407,867)
\put(1407,68){\makebox(0,0){}}
\thicklines \path(1422,113)(1422,123)
\thicklines \path(1422,877)(1422,867)
\put(1422,68){\makebox(0,0){}}
\thicklines \path(1436,113)(1436,133)
\thicklines \path(1436,877)(1436,857)
\put(1436,68){\makebox(0,0){$1000$}}
\thicklines \path(220,113)(1436,113)(1436,877)(220,877)(220,113)
\put(45,495){\makebox(0,0)[l]{\shortstack{\begin{sideways} 		$\big(\sigma_{1/2}(\nu) - \sigma_{3/2}(\nu)\big)/\mu$b 		\end{sideways}}}}
\put(828,23){\makebox(0,0){$\nu/$GeV}}
\put(1306,812){\makebox(0,0)[r]{neutron, $\kappa_0=0$}}
\thinlines \path(1328,812)(1394,812)
\thinlines \path(274,402)(274,402)(275,538)(277,590)(278,628)(280,658)(283,702)(286,733)(289,756)(292,773)(294,780)(295,785)(297,790)(298,793)(300,796)(301,799)(303,800)(304,801)(306,802)(307,802)(309,801)(310,801)(312,799)(313,798)(316,794)(322,783)(334,754)(347,718)(371,641)(395,569)(419,507)(443,455)(468,413)(492,379)(516,353)(540,334)(552,326)(564,320)(577,314)(589,310)(601,306)(613,304)(619,303)(625,302)(631,301)(637,300)(640,300)(643,300)(646,300)(649,300)(651,300)
\thinlines \path(651,300)(652,300)(654,300)(655,300)(657,300)(658,300)(660,300)(661,300)(663,300)(664,300)(667,300)(670,300)(673,300)(679,300)(685,301)(698,302)(710,303)(758,312)(807,323)(855,334)(903,345)(952,356)(1000,364)(1049,372)(1097,378)(1145,383)(1194,387)(1242,390)(1291,392)(1339,394)(1388,396)(1436,397)
\put(1306,767){\makebox(0,0)[r]{neutron, $\kappa_0\neq0$}}
\thicklines \path(1328,767)(1394,767)
\thicklines \path(274,402)(274,402)(275,539)(277,592)(278,630)(280,659)(281,682)(283,701)(284,717)(286,731)(288,742)(289,751)(291,759)(292,765)(294,770)(295,774)(297,777)(298,779)(300,780)(301,780)(303,780)(304,779)(306,778)(307,776)(310,771)(313,765)(316,757)(322,739)(371,538)(395,443)(419,362)(443,296)(468,245)(492,206)(504,191)(516,179)(528,168)(540,160)(552,154)(558,152)(564,150)(570,148)(577,147)(583,146)(586,146)(587,145)(589,145)(590,145)(592,145)(593,145)(595,145)
\thicklines \path(595,145)(596,145)(598,145)(599,145)(601,145)(602,145)(604,145)(607,146)(610,146)(613,146)(619,147)(625,148)(637,151)(661,160)(685,171)(710,184)(758,214)(807,247)(855,280)(903,311)(952,339)(1000,363)(1049,384)(1097,401)(1145,415)(1194,426)(1242,434)(1291,441)(1339,446)(1388,450)(1436,453)
\end{picture}

%% file: qed-xsect.tex
% GNUPLOT: LaTeX picture using EEPIC macros
\setlength{\unitlength}{0.240900pt}
\begin{picture}(1500,900)(0,0)
\tenrm
\thinlines \drawline(220,495)(1436,495)
\thicklines \path(220,113)(240,113)
\thicklines \path(1436,113)(1416,113)
\put(198,113){\makebox(0,0)[r]{$-60$}}
\thicklines \path(220,240)(240,240)
\thicklines \path(1436,240)(1416,240)
\put(198,240){\makebox(0,0)[r]{$-40$}}
\thicklines \path(220,368)(240,368)
\thicklines \path(1436,368)(1416,368)
\put(198,368){\makebox(0,0)[r]{$-20$}}
\thicklines \path(220,495)(240,495)
\thicklines \path(1436,495)(1416,495)
\put(198,495){\makebox(0,0)[r]{$0$}}
\thicklines \path(220,622)(240,622)
\thicklines \path(1436,622)(1416,622)
\put(198,622){\makebox(0,0)[r]{$20$}}
\thicklines \path(220,750)(240,750)
\thicklines \path(1436,750)(1416,750)
\put(198,750){\makebox(0,0)[r]{$40$}}
\thicklines \path(220,877)(240,877)
\thicklines \path(1436,877)(1416,877)
\put(198,877){\makebox(0,0)[r]{$60$}}
\thicklines \path(220,113)(220,133)
\thicklines \path(220,877)(220,857)
\put(220,68){\makebox(0,0){$0.001$}}
\thicklines \path(423,113)(423,133)
\thicklines \path(423,877)(423,857)
\put(423,68){\makebox(0,0){$0.01$}}
\thicklines \path(625,113)(625,133)
\thicklines \path(625,877)(625,857)
\put(625,68){\makebox(0,0){$0.1$}}
\thicklines \path(828,113)(828,133)
\thicklines \path(828,877)(828,857)
\put(828,68){\makebox(0,0){$1$}}
\thicklines \path(1031,113)(1031,133)
\thicklines \path(1031,877)(1031,857)
\put(1031,68){\makebox(0,0){$10$}}
\thicklines \path(1233,113)(1233,133)
\thicklines \path(1233,877)(1233,857)
\put(1233,68){\makebox(0,0){$100$}}
\thicklines \path(1436,113)(1436,133)
\thicklines \path(1436,877)(1436,857)
\put(1436,68){\makebox(0,0){$1000$}}
\thicklines \path(220,113)(1436,113)(1436,877)(220,877)(220,113)
\put(45,495){\makebox(0,0)[l]{\shortstack{ \begin{sideways} 		$\big(\sigma_{1/2}(\nu) - \sigma_{3/2}(\nu)\big)/$mb 		\end{sideways}}}}
\put(828,23){\makebox(0,0){$\nu/m$}}
\thicklines \path(220,491)(220,491)(232,490)(245,489)(257,489)(269,488)(281,487)(294,485)(306,484)(318,482)(331,480)(343,478)(355,476)(367,473)(380,470)(392,466)(404,462)(417,457)(429,452)(441,446)(453,439)(466,431)(478,423)(490,413)(503,402)(515,390)(527,377)(539,362)(552,346)(564,329)(576,311)(588,291)(601,271)(613,251)(625,230)(638,209)(650,190)(662,172)(674,157)(687,145)(699,138)(711,136)(724,140)(736,150)(748,167)(760,190)(773,220)(785,257)(797,298)(810,343)(822,391)
\thicklines \path(822,391)(834,440)(846,490)(859,538)(871,584)(883,626)(896,664)(908,698)(920,726)(932,750)(945,768)(957,781)(969,790)(982,794)(994,795)(1006,792)(1018,787)(1031,779)(1043,770)(1055,759)(1068,747)(1080,734)(1092,721)(1104,707)(1117,694)(1129,680)(1141,667)(1153,654)(1166,642)(1178,631)(1190,620)(1203,609)(1215,599)(1227,590)(1239,582)(1252,574)(1264,567)(1276,560)(1289,554)(1301,548)(1313,543)(1325,538)(1338,534)(1350,530)(1362,526)(1375,523)(1387,520)(1399,518)(1411,515)(1424,513)(1436,511)
\thicklines \path(220,495)(220,491)
\thicklines \path(232,495)(232,490)
\thicklines \path(245,495)(245,489)
\thicklines \path(257,495)(257,489)
\thicklines \path(269,495)(269,488)
\thicklines \path(281,495)(281,487)
\thicklines \path(294,495)(294,485)
\thicklines \path(306,495)(306,484)
\thicklines \path(318,495)(318,482)
\thicklines \path(331,495)(331,480)
\thicklines \path(343,495)(343,478)
\thicklines \path(355,495)(355,476)
\thicklines \path(367,495)(367,473)
\thicklines \path(380,495)(380,470)
\thicklines \path(392,495)(392,466)
\thicklines \path(404,495)(404,462)
\thicklines \path(417,495)(417,457)
\thicklines \path(429,495)(429,452)
\thicklines \path(441,495)(441,446)
\thicklines \path(453,495)(453,439)
\thicklines \path(466,495)(466,431)
\thicklines \path(478,495)(478,423)
\thicklines \path(490,495)(490,413)
\thicklines \path(503,495)(503,402)
\thicklines \path(515,495)(515,390)
\thicklines \path(527,495)(527,377)
\thicklines \path(539,495)(539,362)
\thicklines \path(552,495)(552,346)
\thicklines \path(564,495)(564,329)
\thicklines \path(576,495)(576,311)
\thicklines \path(588,495)(588,291)
\thicklines \path(601,495)(601,271)
\thicklines \path(613,495)(613,251)
\thicklines \path(625,495)(625,230)
\thicklines \path(638,495)(638,209)
\thicklines \path(650,495)(650,190)
\thicklines \path(662,495)(662,172)
\thicklines \path(674,495)(674,157)
\thicklines \path(687,495)(687,145)
\thicklines \path(699,495)(699,138)
\thicklines \path(711,495)(711,136)
\thicklines \path(724,495)(724,140)
\thicklines \path(736,495)(736,150)
\thicklines \path(748,495)(748,167)
\thicklines \path(760,495)(760,190)
\thicklines \path(773,495)(773,220)
\thicklines \path(785,495)(785,257)
\thicklines \path(797,495)(797,298)
\thicklines \path(810,495)(810,343)
\thicklines \path(822,495)(822,391)
\thicklines \path(834,495)(834,440)
\thicklines \path(846,495)(846,490)
\thicklines \path(859,495)(859,538)
\thicklines \path(871,495)(871,584)
\thicklines \path(883,495)(883,626)
\thicklines \path(896,495)(896,664)
\thicklines \path(908,495)(908,698)
\thicklines \path(920,495)(920,726)
\thicklines \path(932,495)(932,750)
\thicklines \path(945,495)(945,768)
\thicklines \path(957,495)(957,781)
\thicklines \path(969,495)(969,790)
\thicklines \path(982,495)(982,794)
\thicklines \path(994,495)(994,795)
\thicklines \path(1006,495)(1006,792)
\thicklines \path(1018,495)(1018,787)
\thicklines \path(1031,495)(1031,779)
\thicklines \path(1043,495)(1043,770)
\thicklines \path(1055,495)(1055,759)
\thicklines \path(1068,495)(1068,747)
\thicklines \path(1080,495)(1080,734)
\thicklines \path(1092,495)(1092,721)
\thicklines \path(1104,495)(1104,707)
\thicklines \path(1117,495)(1117,694)
\thicklines \path(1129,495)(1129,680)
\thicklines \path(1141,495)(1141,667)
\thicklines \path(1153,495)(1153,654)
\thicklines \path(1166,495)(1166,642)
\thicklines \path(1178,495)(1178,631)
\thicklines \path(1190,495)(1190,620)
\thicklines \path(1203,495)(1203,609)
\thicklines \path(1215,495)(1215,599)
\thicklines \path(1227,495)(1227,590)
\thicklines \path(1239,495)(1239,582)
\thicklines \path(1252,495)(1252,574)
\thicklines \path(1264,495)(1264,567)
\thicklines \path(1276,495)(1276,560)
\thicklines \path(1289,495)(1289,554)
\thicklines \path(1301,495)(1301,548)
\thicklines \path(1313,495)(1313,543)
\thicklines \path(1325,495)(1325,538)
\thicklines \path(1338,495)(1338,534)
\thicklines \path(1350,495)(1350,530)
\thicklines \path(1362,495)(1362,526)
\thicklines \path(1375,495)(1375,523)
\thicklines \path(1387,495)(1387,520)
\thicklines \path(1399,495)(1399,518)
\thicklines \path(1411,495)(1411,515)
\thicklines \path(1424,495)(1424,513)
\thicklines \path(1436,495)(1436,511)
\end{picture}

%% file: wsm-q2nu.tex
% GNUPLOT: LaTeX picture using EEPIC macros
\setlength{\unitlength}{0.240900pt}
\begin{picture}(1500,900)(0,0)
\tenrm
\thinlines \drawline[-50](264,158)(1436,158)
\thinlines \drawline[-50](264,158)(264,787)
\thicklines \path(264,158)(284,158)
\thicklines \path(1436,158)(1416,158)
\put(242,158){\makebox(0,0)[r]{$0$}}
\thicklines \path(264,284)(284,284)
\thicklines \path(1436,284)(1416,284)
\put(242,284){\makebox(0,0)[r]{$2$}}
\thicklines \path(264,410)(284,410)
\thicklines \path(1436,410)(1416,410)
\put(242,410){\makebox(0,0)[r]{$4$}}
\thicklines \path(264,535)(284,535)
\thicklines \path(1436,535)(1416,535)
\put(242,535){\makebox(0,0)[r]{$6$}}
\thicklines \path(264,661)(284,661)
\thicklines \path(1436,661)(1416,661)
\put(242,661){\makebox(0,0)[r]{$8$}}
\thicklines \path(264,787)(284,787)
\thicklines \path(1436,787)(1416,787)
\put(242,787){\makebox(0,0)[r]{$10$}}
\thicklines \path(264,158)(264,178)
\thicklines \path(264,787)(264,767)
\put(264,113){\makebox(0,0){$0$}}
\thicklines \path(498,158)(498,178)
\thicklines \path(498,787)(498,767)
\put(498,113){\makebox(0,0){$2$}}
\thicklines \path(733,158)(733,178)
\thicklines \path(733,787)(733,767)
\put(733,113){\makebox(0,0){$4$}}
\thicklines \path(967,158)(967,178)
\thicklines \path(967,787)(967,767)
\put(967,113){\makebox(0,0){$6$}}
\thicklines \path(1202,158)(1202,178)
\thicklines \path(1202,787)(1202,767)
\put(1202,113){\makebox(0,0){$8$}}
\thicklines \path(1436,158)(1436,178)
\thicklines \path(1436,787)(1436,767)
\put(1436,113){\makebox(0,0){$10$}}
\thicklines \path(264,158)(1436,158)(1436,787)(264,787)(264,158)
\put(45,472){\makebox(0,0)[l]{\shortstack{$q^2/m^2$}}}
\put(850,68){\makebox(0,0){$\nu/m$}}
\put(557,661){\makebox(0,0)[r]{$p^0=m$}}
\put(1084,535){\makebox(0,0)[r]{$p^0=3m$}}
\put(1319,315){\makebox(0,0)[r]{$p^0=6m$}}
\put(1407,189){\makebox(0,0)[r]{$p^0\to\infty$}}
\put(850,441){\makebox(0,0)[r]{$q^2=4m^2$}}
\thinlines \path(264,158)(264,158)(266,158)(269,158)(271,158)(273,158)(276,159)(278,159)(280,159)(283,160)(285,160)(287,161)(290,161)(292,162)(295,162)(297,163)(299,164)(302,164)(304,165)(306,166)(309,167)(311,168)(313,169)(316,170)(318,171)(320,173)(323,174)(325,175)(327,176)(330,178)(332,179)(334,181)(337,182)(339,184)(342,186)(344,187)(346,189)(349,191)(351,193)(353,194)(356,196)(358,198)(360,200)(363,203)(365,205)(367,207)(370,209)(372,211)(374,214)(377,216)(379,219)
\thinlines \path(379,219)(381,221)(384,224)(386,226)(388,229)(391,232)(393,234)(396,237)(398,240)(400,243)(403,246)(405,249)(407,252)(410,255)(412,258)(414,261)(417,265)(419,268)(421,271)(424,275)(426,278)(428,282)(431,285)(433,289)(435,293)(438,296)(440,300)(443,304)(445,308)(447,312)(450,316)(452,320)(454,324)(457,328)(459,332)(461,336)(464,341)(466,345)(468,349)(471,354)(473,358)(475,363)(478,367)(480,372)(482,376)(485,381)(487,386)(489,391)(492,396)(494,401)(497,406)
\thinlines \path(497,406)(499,411)(501,416)(504,421)(506,426)(508,431)(511,437)(513,442)(515,447)(518,453)(520,458)(522,464)(525,469)(527,475)(529,481)(532,486)(534,492)(536,498)(539,504)(541,510)(543,516)(546,522)(548,528)(551,534)(553,540)(555,546)(558,553)(560,559)(562,565)(565,572)(567,578)(569,585)(572,592)(574,598)(576,605)(579,612)(581,618)(583,625)(586,632)(588,639)(590,646)(593,653)(595,660)(598,667)(600,675)(602,682)(605,689)(607,696)(609,704)(612,711)(614,719)
\thinlines \path(614,719)(616,726)(619,734)(621,742)(623,749)(626,757)(628,765)(630,773)(633,781)(635,787)
\thinlines \path(264,158)(264,158)(266,158)(269,158)(271,158)(273,158)(276,158)(278,158)(280,158)(283,158)(285,158)(287,158)(290,158)(292,158)(295,158)(297,159)(299,159)(302,159)(304,159)(306,159)(309,159)(311,159)(313,159)(316,159)(318,159)(320,160)(323,160)(325,160)(327,160)(330,160)(332,160)(334,161)(337,161)(339,161)(342,161)(344,161)(346,161)(349,162)(351,162)(353,162)(356,162)(358,162)(360,163)(363,163)(365,163)(367,163)(370,164)(372,164)(374,164)(377,164)(379,165)
\thinlines \path(379,165)(381,165)(384,165)(386,166)(388,166)(391,166)(393,166)(396,167)(398,167)(400,167)(403,168)(405,168)(407,168)(410,169)(412,169)(414,169)(417,170)(419,170)(421,171)(424,171)(426,171)(428,172)(431,172)(433,173)(435,173)(438,173)(440,174)(443,174)(445,175)(447,175)(450,176)(452,176)(454,176)(457,177)(459,177)(461,178)(464,178)(466,179)(468,179)(471,180)(473,180)(475,181)(478,181)(480,182)(482,182)(485,183)(487,183)(489,184)(492,184)(494,185)(497,186)
\thinlines \path(497,186)(499,186)(501,187)(504,187)(506,188)(508,188)(511,189)(513,190)(515,190)(518,191)(520,191)(522,192)(525,193)(527,193)(529,194)(532,194)(534,195)(536,196)(539,196)(541,197)(543,198)(546,198)(548,199)(551,200)(553,200)(555,201)(558,202)(560,203)(562,203)(565,204)(567,205)(569,205)(572,206)(574,207)(576,208)(579,208)(581,209)(583,210)(586,211)(588,211)(590,212)(593,213)(595,214)(598,215)(600,215)(602,216)(605,217)(607,218)(609,219)(612,219)(614,220)
\thinlines \path(614,220)(616,221)(619,222)(621,223)(623,224)(626,225)(628,225)(630,226)(633,227)(635,228)(637,229)(640,230)(642,231)(644,232)(647,233)(649,233)(652,234)(654,235)(656,236)(659,237)(661,238)(663,239)(666,240)(668,241)(670,242)(673,243)(675,244)(677,245)(680,246)(682,247)(684,248)(687,249)(689,250)(691,251)(694,252)(696,253)(699,254)(701,255)(703,256)(706,257)(708,258)(710,259)(713,260)(715,261)(717,263)(720,264)(722,265)(724,266)(727,267)(729,268)(731,269)
\thinlines \path(731,269)(734,270)(736,271)(738,273)(741,274)(743,275)(745,276)(748,277)(750,278)(753,279)(755,281)(757,282)(760,283)(762,284)(764,285)(767,287)(769,288)(771,289)(774,290)(776,291)(778,293)(781,294)(783,295)(785,296)(788,298)(790,299)(792,300)(795,301)(797,303)(800,304)(802,305)(804,306)(807,308)(809,309)(811,310)(814,312)(816,313)(818,314)(821,316)(823,317)(825,318)(828,320)(830,321)(832,322)(835,324)(837,325)(839,326)(842,328)(844,329)(846,331)(849,332)
\thinlines \path(849,332)(851,333)(854,335)(856,336)(858,338)(861,339)(863,341)(865,342)(868,343)(870,345)(872,346)(875,348)(877,349)(879,351)(882,352)(884,354)(886,355)(889,357)(891,358)(893,360)(896,361)(898,363)(900,364)(903,366)(905,367)(908,369)(910,370)(912,372)(915,373)(917,375)(919,376)(922,378)(924,380)(926,381)(929,383)(931,384)(933,386)(936,388)(938,389)(940,391)(943,392)(945,394)(947,396)(950,397)(952,399)(955,401)(957,402)(959,404)(962,406)(964,407)(966,409)
\thinlines \path(966,409)(969,411)(971,412)(973,414)(976,416)(978,417)(980,419)(983,421)(985,423)(987,424)(990,426)(992,428)(994,429)(997,431)(999,433)(1001,435)(1004,437)(1006,438)(1009,440)(1011,442)(1013,444)(1016,445)(1018,447)(1020,449)(1023,451)(1025,453)(1027,454)(1030,456)(1032,458)(1034,460)(1037,462)(1039,464)(1041,466)(1044,467)(1046,469)(1048,471)(1051,473)(1053,475)(1056,477)(1058,479)(1060,481)(1063,482)(1065,484)(1067,486)(1070,488)(1072,490)(1074,492)(1077,494)(1079,496)(1081,498)(1084,500)
\thinlines \path(1084,500)(1086,502)(1088,504)(1091,506)(1093,508)(1095,510)(1098,512)(1100,514)(1102,516)(1105,518)(1107,520)(1110,522)(1112,524)(1114,526)(1117,528)(1119,530)(1121,532)(1124,534)(1126,536)(1128,538)(1131,540)(1133,542)(1135,544)(1138,546)(1140,549)(1142,551)(1145,553)(1147,555)(1149,557)(1152,559)(1154,561)(1157,563)(1159,565)(1161,568)(1164,570)(1166,572)(1168,574)(1171,576)(1173,578)(1175,581)(1178,583)(1180,585)(1182,587)(1185,589)(1187,592)(1189,594)(1192,596)(1194,598)(1196,600)(1199,603)(1201,605)
\thinlines \path(1201,605)(1203,607)(1206,609)(1208,612)(1211,614)(1213,616)(1215,618)(1218,621)(1220,623)(1222,625)(1225,628)(1227,630)(1229,632)(1232,634)(1234,637)(1236,639)(1239,641)(1241,644)(1243,646)(1246,648)(1248,651)(1250,653)(1253,655)(1255,658)(1257,660)(1260,663)(1262,665)(1265,667)(1267,670)(1269,672)(1272,675)(1274,677)(1276,679)(1279,682)(1281,684)(1283,687)(1286,689)(1288,692)(1290,694)(1293,696)(1295,699)(1297,701)(1300,704)(1302,706)(1304,709)(1307,711)(1309,714)(1312,716)(1314,719)(1316,721)(1319,724)
\thinlines \path(1319,724)(1321,726)(1323,729)(1326,731)(1328,734)(1330,737)(1333,739)(1335,742)(1337,744)(1340,747)(1342,749)(1344,752)(1347,754)(1349,757)(1351,760)(1354,762)(1356,765)(1358,768)(1361,770)(1363,773)(1366,775)(1368,778)(1370,781)(1373,783)(1375,786)(1376,787)
\thinlines \path(264,158)(264,158)(266,158)(269,158)(271,158)(273,158)(276,158)(278,158)(280,158)(283,158)(285,158)(287,158)(290,158)(292,158)(295,158)(297,158)(299,158)(302,158)(304,158)(306,158)(309,158)(311,158)(313,158)(316,158)(318,158)(320,158)(323,158)(325,158)(327,159)(330,159)(332,159)(334,159)(337,159)(339,159)(342,159)(344,159)(346,159)(349,159)(351,159)(353,159)(356,159)(358,159)(360,159)(363,159)(365,159)(367,159)(370,159)(372,159)(374,160)(377,160)(379,160)
\thinlines \path(379,160)(381,160)(384,160)(386,160)(388,160)(391,160)(393,160)(396,160)(398,160)(400,160)(403,160)(405,161)(407,161)(410,161)(412,161)(414,161)(417,161)(419,161)(421,161)(424,161)(426,161)(428,161)(431,162)(433,162)(435,162)(438,162)(440,162)(443,162)(445,162)(447,162)(450,162)(452,162)(454,163)(457,163)(459,163)(461,163)(464,163)(466,163)(468,163)(471,163)(473,164)(475,164)(478,164)(480,164)(482,164)(485,164)(487,164)(489,164)(492,165)(494,165)(497,165)
\thinlines \path(497,165)(499,165)(501,165)(504,165)(506,165)(508,166)(511,166)(513,166)(515,166)(518,166)(520,166)(522,166)(525,167)(527,167)(529,167)(532,167)(534,167)(536,167)(539,168)(541,168)(543,168)(546,168)(548,168)(551,168)(553,169)(555,169)(558,169)(560,169)(562,169)(565,169)(567,170)(569,170)(572,170)(574,170)(576,170)(579,171)(581,171)(583,171)(586,171)(588,171)(590,172)(593,172)(595,172)(598,172)(600,172)(602,173)(605,173)(607,173)(609,173)(612,173)(614,174)
\thinlines \path(614,174)(616,174)(619,174)(621,174)(623,174)(626,175)(628,175)(630,175)(633,175)(635,176)(637,176)(640,176)(642,176)(644,176)(647,177)(649,177)(652,177)(654,177)(656,178)(659,178)(661,178)(663,178)(666,179)(668,179)(670,179)(673,179)(675,179)(677,180)(680,180)(682,180)(684,180)(687,181)(689,181)(691,181)(694,181)(696,182)(699,182)(701,182)(703,183)(706,183)(708,183)(710,183)(713,184)(715,184)(717,184)(720,184)(722,185)(724,185)(727,185)(729,186)(731,186)
\thinlines \path(731,186)(734,186)(736,186)(738,187)(741,187)(743,187)(745,187)(748,188)(750,188)(753,188)(755,189)(757,189)(760,189)(762,190)(764,190)(767,190)(769,190)(771,191)(774,191)(776,191)(778,192)(781,192)(783,192)(785,193)(788,193)(790,193)(792,194)(795,194)(797,194)(800,194)(802,195)(804,195)(807,195)(809,196)(811,196)(814,196)(816,197)(818,197)(821,197)(823,198)(825,198)(828,198)(830,199)(832,199)(835,199)(837,200)(839,200)(842,200)(844,201)(846,201)(849,202)
\thinlines \path(849,202)(851,202)(854,202)(856,203)(858,203)(861,203)(863,204)(865,204)(868,204)(870,205)(872,205)(875,205)(877,206)(879,206)(882,207)(884,207)(886,207)(889,208)(891,208)(893,208)(896,209)(898,209)(900,210)(903,210)(905,210)(908,211)(910,211)(912,211)(915,212)(917,212)(919,213)(922,213)(924,213)(926,214)(929,214)(931,215)(933,215)(936,215)(938,216)(940,216)(943,217)(945,217)(947,217)(950,218)(952,218)(955,219)(957,219)(959,219)(962,220)(964,220)(966,221)
\thinlines \path(966,221)(969,221)(971,222)(973,222)(976,222)(978,223)(980,223)(983,224)(985,224)(987,225)(990,225)(992,225)(994,226)(997,226)(999,227)(1001,227)(1004,228)(1006,228)(1009,229)(1011,229)(1013,229)(1016,230)(1018,230)(1020,231)(1023,231)(1025,232)(1027,232)(1030,233)(1032,233)(1034,233)(1037,234)(1039,234)(1041,235)(1044,235)(1046,236)(1048,236)(1051,237)(1053,237)(1056,238)(1058,238)(1060,239)(1063,239)(1065,240)(1067,240)(1070,241)(1072,241)(1074,242)(1077,242)(1079,242)(1081,243)(1084,243)
\thinlines \path(1084,243)(1086,244)(1088,244)(1091,245)(1093,245)(1095,246)(1098,246)(1100,247)(1102,247)(1105,248)(1107,248)(1110,249)(1112,249)(1114,250)(1117,250)(1119,251)(1121,251)(1124,252)(1126,253)(1128,253)(1131,254)(1133,254)(1135,255)(1138,255)(1140,256)(1142,256)(1145,257)(1147,257)(1149,258)(1152,258)(1154,259)(1157,259)(1159,260)(1161,260)(1164,261)(1166,261)(1168,262)(1171,263)(1173,263)(1175,264)(1178,264)(1180,265)(1182,265)(1185,266)(1187,266)(1189,267)(1192,267)(1194,268)(1196,269)(1199,269)(1201,270)
\thinlines \path(1201,270)(1203,270)(1206,271)(1208,271)(1211,272)(1213,273)(1215,273)(1218,274)(1220,274)(1222,275)(1225,275)(1227,276)(1229,277)(1232,277)(1234,278)(1236,278)(1239,279)(1241,279)(1243,280)(1246,281)(1248,281)(1250,282)(1253,282)(1255,283)(1257,284)(1260,284)(1262,285)(1265,285)(1267,286)(1269,287)(1272,287)(1274,288)(1276,288)(1279,289)(1281,290)(1283,290)(1286,291)(1288,291)(1290,292)(1293,293)(1295,293)(1297,294)(1300,294)(1302,295)(1304,296)(1307,296)(1309,297)(1312,298)(1314,298)(1316,299)(1319,299)
\thinlines \path(1319,299)(1321,300)(1323,301)(1326,301)(1328,302)(1330,303)(1333,303)(1335,304)(1337,305)(1340,305)(1342,306)(1344,306)(1347,307)(1349,308)(1351,308)(1354,309)(1356,310)(1358,310)(1361,311)(1363,312)(1366,312)(1368,313)(1370,314)(1373,314)(1375,315)(1377,316)(1380,316)(1382,317)(1384,318)(1387,318)(1389,319)(1391,320)(1394,320)(1396,321)(1398,322)(1401,322)(1403,323)(1405,324)(1408,324)(1410,325)(1413,326)(1415,326)(1417,327)(1420,328)(1422,329)(1424,329)(1427,330)(1429,331)(1431,331)(1434,332)(1436,333)
\Thicklines \path(264,410)(264,410)(266,410)(269,410)(271,410)(273,410)(276,410)(278,410)(280,410)(283,410)(285,410)(287,410)(290,410)(292,410)(295,410)(297,410)(299,410)(302,410)(304,410)(306,410)(309,410)(311,410)(313,410)(316,410)(318,410)(320,410)(323,410)(325,410)(327,410)(330,410)(332,410)(334,410)(337,410)(339,410)(342,410)(344,410)(346,410)(349,410)(351,410)(353,410)(356,410)(358,410)(360,410)(363,410)(365,410)(367,410)(370,410)(372,410)(374,410)(377,410)(379,410)
\Thicklines \path(379,410)(381,410)(384,410)(386,410)(388,410)(391,410)(393,410)(396,410)(398,410)(400,410)(403,410)(405,410)(407,410)(410,410)(412,410)(414,410)(417,410)(419,410)(421,410)(424,410)(426,410)(428,410)(431,410)(433,410)(435,410)(438,410)(440,410)(443,410)(445,410)(447,410)(450,410)(452,410)(454,410)(457,410)(459,410)(461,410)(464,410)(466,410)(468,410)(471,410)(473,410)(475,410)(478,410)(480,410)(482,410)(485,410)(487,410)(489,410)(492,410)(494,410)(497,410)
\Thicklines \path(497,410)(499,410)(501,410)(504,410)(506,410)(508,410)(511,410)(513,410)(515,410)(518,410)(520,410)(522,410)(525,410)(527,410)(529,410)(532,410)(534,410)(536,410)(539,410)(541,410)(543,410)(546,410)(548,410)(551,410)(553,410)(555,410)(558,410)(560,410)(562,410)(565,410)(567,410)(569,410)(572,410)(574,410)(576,410)(579,410)(581,410)(583,410)(586,410)(588,410)(590,410)(593,410)(595,410)(598,410)(600,410)(602,410)(605,410)(607,410)(609,410)(612,410)(614,410)
\Thicklines \path(614,410)(616,410)(619,410)(621,410)(623,410)(626,410)(628,410)(630,410)(633,410)(635,410)(637,410)(640,410)(642,410)(644,410)(647,410)(649,410)(652,410)(654,410)(656,410)(659,410)(661,410)(663,410)(666,410)(668,410)(670,410)(673,410)(675,410)(677,410)(680,410)(682,410)(684,410)(687,410)(689,410)(691,410)(694,410)(696,410)(699,410)(701,410)(703,410)(706,410)(708,410)(710,410)(713,410)(715,410)(717,410)(720,410)(722,410)(724,410)(727,410)(729,410)(731,410)
\Thicklines \path(731,410)(734,410)(736,410)(738,410)(741,410)(743,410)(745,410)(748,410)(750,410)(753,410)(755,410)(757,410)(760,410)(762,410)(764,410)(767,410)(769,410)(771,410)(774,410)(776,410)(778,410)(781,410)(783,410)(785,410)(788,410)(790,410)(792,410)(795,410)(797,410)(800,410)(802,410)(804,410)(807,410)(809,410)(811,410)(814,410)(816,410)(818,410)(821,410)(823,410)(825,410)(828,410)(830,410)(832,410)(835,410)(837,410)(839,410)(842,410)(844,410)(846,410)(849,410)
\Thicklines \path(849,410)(851,410)(854,410)(856,410)(858,410)(861,410)(863,410)(865,410)(868,410)(870,410)(872,410)(875,410)(877,410)(879,410)(882,410)(884,410)(886,410)(889,410)(891,410)(893,410)(896,410)(898,410)(900,410)(903,410)(905,410)(908,410)(910,410)(912,410)(915,410)(917,410)(919,410)(922,410)(924,410)(926,410)(929,410)(931,410)(933,410)(936,410)(938,410)(940,410)(943,410)(945,410)(947,410)(950,410)(952,410)(955,410)(957,410)(959,410)(962,410)(964,410)(966,410)
\Thicklines \path(966,410)(969,410)(971,410)(973,410)(976,410)(978,410)(980,410)(983,410)(985,410)(987,410)(990,410)(992,410)(994,410)(997,410)(999,410)(1001,410)(1004,410)(1006,410)(1009,410)(1011,410)(1013,410)(1016,410)(1018,410)(1020,410)(1023,410)(1025,410)(1027,410)(1030,410)(1032,410)(1034,410)(1037,410)(1039,410)(1041,410)(1044,410)(1046,410)(1048,410)(1051,410)(1053,410)(1056,410)(1058,410)(1060,410)(1063,410)(1065,410)(1067,410)(1070,410)(1072,410)(1074,410)(1077,410)(1079,410)(1081,410)(1084,410)
\Thicklines \path(1084,410)(1086,410)(1088,410)(1091,410)(1093,410)(1095,410)(1098,410)(1100,410)(1102,410)(1105,410)(1107,410)(1110,410)(1112,410)(1114,410)(1117,410)(1119,410)(1121,410)(1124,410)(1126,410)(1128,410)(1131,410)(1133,410)(1135,410)(1138,410)(1140,410)(1142,410)(1145,410)(1147,410)(1149,410)(1152,410)(1154,410)(1157,410)(1159,410)(1161,410)(1164,410)(1166,410)(1168,410)(1171,410)(1173,410)(1175,410)(1178,410)(1180,410)(1182,410)(1185,410)(1187,410)(1189,410)(1192,410)(1194,410)(1196,410)(1199,410)(1201,410)
\Thicklines \path(1201,410)(1203,410)(1206,410)(1208,410)(1211,410)(1213,410)(1215,410)(1218,410)(1220,410)(1222,410)(1225,410)(1227,410)(1229,410)(1232,410)(1234,410)(1236,410)(1239,410)(1241,410)(1243,410)(1246,410)(1248,410)(1250,410)(1253,410)(1255,410)(1257,410)(1260,410)(1262,410)(1265,410)(1267,410)(1269,410)(1272,410)(1274,410)(1276,410)(1279,410)(1281,410)(1283,410)(1286,410)(1288,410)(1290,410)(1293,410)(1295,410)(1297,410)(1300,410)(1302,410)(1304,410)(1307,410)(1309,410)(1312,410)(1314,410)(1316,410)(1319,410)
\Thicklines \path(1319,410)(1321,410)(1323,410)(1326,410)(1328,410)(1330,410)(1333,410)(1335,410)(1337,410)(1340,410)(1342,410)(1344,410)(1347,410)(1349,410)(1351,410)(1354,410)(1356,410)(1358,410)(1361,410)(1363,410)(1366,410)(1368,410)(1370,410)(1373,410)(1375,410)(1377,410)(1380,410)(1382,410)(1384,410)(1387,410)(1389,410)(1391,410)(1394,410)(1396,410)(1398,410)(1401,410)(1403,410)(1405,410)(1408,410)(1410,410)(1413,410)(1415,410)(1417,410)(1420,410)(1422,410)(1424,410)(1427,410)(1429,410)(1431,410)(1434,410)(1436,410)
\end{picture}

%% file: fq2-lin.tex
% GNUPLOT: LaTeX picture using EEPIC macros
\setlength{\unitlength}{0.240900pt}
\begin{picture}(1500,900)(0,0)
\tenrm
\thinlines \drawline[-50](264,473)(1436,473)
\thinlines \drawline[-50](850,158)(850,787)
\thicklines \path(264,158)(284,158)
\thicklines \path(1436,158)(1416,158)
\put(242,158){\makebox(0,0)[r]{$-1$}}
\thicklines \path(264,473)(284,473)
\thicklines \path(1436,473)(1416,473)
\put(242,473){\makebox(0,0)[r]{$0$}}
\thicklines \path(264,787)(284,787)
\thicklines \path(1436,787)(1416,787)
\put(242,787){\makebox(0,0)[r]{$1$}}
\thicklines \path(264,158)(264,178)
\thicklines \path(264,787)(264,767)
\put(264,113){\makebox(0,0){$-10$}}
\thicklines \path(381,158)(381,178)
\thicklines \path(381,787)(381,767)
\put(381,113){\makebox(0,0){$-8$}}
\thicklines \path(498,158)(498,178)
\thicklines \path(498,787)(498,767)
\put(498,113){\makebox(0,0){$-6$}}
\thicklines \path(616,158)(616,178)
\thicklines \path(616,787)(616,767)
\put(616,113){\makebox(0,0){$-4$}}
\thicklines \path(733,158)(733,178)
\thicklines \path(733,787)(733,767)
\put(733,113){\makebox(0,0){$-2$}}
\thicklines \path(850,158)(850,178)
\thicklines \path(850,787)(850,767)
\put(850,113){\makebox(0,0){$0$}}
\thicklines \path(967,158)(967,178)
\thicklines \path(967,787)(967,767)
\put(967,113){\makebox(0,0){$2$}}
\thicklines \path(1084,158)(1084,178)
\thicklines \path(1084,787)(1084,767)
\put(1084,113){\makebox(0,0){$4$}}
\thicklines \path(1202,158)(1202,178)
\thicklines \path(1202,787)(1202,767)
\put(1202,113){\makebox(0,0){$6$}}
\thicklines \path(1319,158)(1319,178)
\thicklines \path(1319,787)(1319,767)
\put(1319,113){\makebox(0,0){$8$}}
\thicklines \path(1436,158)(1436,178)
\thicklines \path(1436,787)(1436,767)
\put(1436,113){\makebox(0,0){$10$}}
\thicklines \path(264,158)(1436,158)(1436,787)(264,787)(264,158)
\put(850,68){\makebox(0,0){$q^2/4m^2$}}
\put(1306,722){\makebox(0,0)[r]{\small $\re f(q^2)$}}
\thinlines \path(1328,722)(1394,722)
\thinlines \path(264,214)(264,214)(265,214)(265,214)(266,214)(266,214)(267,214)(268,214)(268,214)(269,214)(269,214)(270,214)(270,214)(271,215)(272,215)(272,215)(273,215)(273,215)(274,215)(275,215)(275,215)(276,215)(276,215)(277,215)(277,215)(278,215)(279,215)(279,215)(280,215)(280,215)(281,215)(282,215)(282,215)(283,215)(283,215)(284,215)(285,215)(285,215)(286,216)(286,216)(287,216)(287,216)(288,216)(289,216)(289,216)(290,216)(290,216)(291,216)(292,216)(292,216)(293,216)
\thinlines \path(293,216)(293,216)(294,216)(294,216)(295,216)(296,216)(296,216)(297,216)(297,216)(298,216)(299,216)(299,216)(300,217)(300,217)(301,217)(302,217)(302,217)(303,217)(303,217)(304,217)(304,217)(305,217)(306,217)(306,217)(307,217)(307,217)(308,217)(309,217)(309,217)(310,217)(310,217)(311,217)(311,217)(312,217)(313,218)(313,218)(314,218)(314,218)(315,218)(316,218)(316,218)(317,218)(317,218)(318,218)(319,218)(319,218)(320,218)(320,218)(321,218)(321,218)(322,218)
\thinlines \path(322,218)(323,218)(323,218)(324,218)(324,218)(325,218)(326,219)(326,219)(327,219)(327,219)(328,219)(328,219)(329,219)(330,219)(330,219)(331,219)(331,219)(332,219)(333,219)(333,219)(334,219)(334,219)(335,219)(336,219)(336,219)(337,219)(337,219)(338,220)(338,220)(339,220)(340,220)(340,220)(341,220)(341,220)(342,220)(343,220)(343,220)(344,220)(344,220)(345,220)(345,220)(346,220)(347,220)(347,220)(348,220)(348,220)(349,220)(350,220)(350,221)(351,221)(351,221)
\thinlines \path(351,221)(352,221)(353,221)(353,221)(354,221)(354,221)(355,221)(355,221)(356,221)(357,221)(357,221)(358,221)(358,221)(359,221)(360,221)(360,221)(361,221)(361,222)(362,222)(362,222)(363,222)(364,222)(364,222)(365,222)(365,222)(366,222)(367,222)(367,222)(368,222)(368,222)(369,222)(370,222)(370,222)(371,222)(371,222)(372,222)(372,223)(373,223)(374,223)(374,223)(375,223)(375,223)(376,223)(377,223)(377,223)(378,223)(378,223)(379,223)(379,223)(380,223)(381,223)
\thinlines \path(381,223)(381,223)(382,223)(382,223)(383,224)(384,224)(384,224)(385,224)(385,224)(386,224)(387,224)(387,224)(388,224)(388,224)(389,224)(389,224)(390,224)(391,224)(391,224)(392,224)(392,224)(393,224)(394,225)(394,225)(395,225)(395,225)(396,225)(397,225)(397,225)(398,225)(398,225)(399,225)(399,225)(400,225)(401,225)(401,225)(402,225)(402,225)(403,225)(404,226)(404,226)(405,226)(405,226)(406,226)(406,226)(407,226)(408,226)(408,226)(409,226)(409,226)(410,226)
\thinlines \path(410,226)(411,226)(411,226)(412,226)(412,226)(413,226)(414,227)(414,227)(415,227)(415,227)(416,227)(416,227)(417,227)(418,227)(418,227)(419,227)(419,227)(420,227)(421,227)(421,227)(422,227)(422,227)(423,228)(423,228)(424,228)(425,228)(425,228)(426,228)(426,228)(427,228)(428,228)(428,228)(429,228)(429,228)(430,228)(431,228)(431,228)(432,229)(432,229)(433,229)(433,229)(434,229)(435,229)(435,229)(436,229)(436,229)(437,229)(438,229)(438,229)(439,229)(439,229)
\thinlines \path(439,229)(440,229)(440,230)(441,230)(442,230)(442,230)(443,230)(443,230)(444,230)(445,230)(445,230)(446,230)(446,230)(447,230)(448,230)(448,230)(449,230)(449,231)(450,231)(450,231)(451,231)(452,231)(452,231)(453,231)(453,231)(454,231)(455,231)(455,231)(456,231)(456,231)(457,231)(457,232)(458,232)(459,232)(459,232)(460,232)(460,232)(461,232)(462,232)(462,232)(463,232)(463,232)(464,232)(465,232)(465,233)(466,233)(466,233)(467,233)(467,233)(468,233)(469,233)
\thinlines \path(469,233)(469,233)(470,233)(470,233)(471,233)(472,233)(472,233)(473,233)(473,234)(474,234)(474,234)(475,234)(476,234)(476,234)(477,234)(477,234)(478,234)(479,234)(479,234)(480,234)(480,234)(481,235)(482,235)(482,235)(483,235)(483,235)(484,235)(484,235)(485,235)(486,235)(486,235)(487,235)(487,235)(488,236)(489,236)(489,236)(490,236)(490,236)(491,236)(491,236)(492,236)(493,236)(493,236)(494,236)(494,236)(495,237)(496,237)(496,237)(497,237)(497,237)(498,237)
\thinlines \path(498,237)(499,237)(499,237)(500,237)(500,237)(501,237)(501,237)(502,238)(503,238)(503,238)(504,238)(504,238)(505,238)(506,238)(506,238)(507,238)(507,238)(508,238)(508,239)(509,239)(510,239)(510,239)(511,239)(511,239)(512,239)(513,239)(513,239)(514,239)(514,239)(515,239)(516,240)(516,240)(517,240)(517,240)(518,240)(518,240)(519,240)(520,240)(520,240)(521,240)(521,241)(522,241)(523,241)(523,241)(524,241)(524,241)(525,241)(525,241)(526,241)(527,241)(527,241)
\thinlines \path(527,241)(528,242)(528,242)(529,242)(530,242)(530,242)(531,242)(531,242)(532,242)(533,242)(533,242)(534,243)(534,243)(535,243)(535,243)(536,243)(537,243)(537,243)(538,243)(538,243)(539,243)(540,244)(540,244)(541,244)(541,244)(542,244)(542,244)(543,244)(544,244)(544,244)(545,244)(545,245)(546,245)(547,245)(547,245)(548,245)(548,245)(549,245)(550,245)(550,245)(551,246)(551,246)(552,246)(552,246)(553,246)(554,246)(554,246)(555,246)(555,246)(556,246)(557,247)
\thinlines \path(557,247)(557,247)(558,247)(558,247)(559,247)(559,247)(560,247)(561,247)(561,247)(562,248)(562,248)(563,248)(564,248)(564,248)(565,248)(565,248)(566,248)(567,249)(567,249)(568,249)(568,249)(569,249)(569,249)(570,249)(571,249)(571,249)(572,250)(572,250)(573,250)(574,250)(574,250)(575,250)(575,250)(576,250)(576,251)(577,251)(578,251)(578,251)(579,251)(579,251)(580,251)(581,251)(581,252)(582,252)(582,252)(583,252)(584,252)(584,252)(585,252)(585,252)(586,253)
\thinlines \path(586,253)(586,253)(587,253)(588,253)(588,253)(589,253)(589,253)(590,253)(591,254)(591,254)(592,254)(592,254)(593,254)(593,254)(594,254)(595,254)(595,255)(596,255)(596,255)(597,255)(598,255)(598,255)(599,255)(599,256)(600,256)(601,256)(601,256)(602,256)(602,256)(603,256)(603,257)(604,257)(605,257)(605,257)(606,257)(606,257)(607,257)(608,258)(608,258)(609,258)(609,258)(610,258)(610,258)(611,258)(612,259)(612,259)(613,259)(613,259)(614,259)(615,259)(615,259)
\thinlines \path(615,259)(616,260)(616,260)(617,260)(618,260)(618,260)(619,260)(619,260)(620,261)(620,261)(621,261)(622,261)(622,261)(623,261)(623,262)(624,262)(625,262)(625,262)(626,262)(626,262)(627,263)(628,263)(628,263)(629,263)(629,263)(630,263)(630,263)(631,264)(632,264)(632,264)(633,264)(633,264)(634,264)(635,265)(635,265)(636,265)(636,265)(637,265)(637,265)(638,266)(639,266)(639,266)(640,266)(640,266)(641,267)(642,267)(642,267)(643,267)(643,267)(644,267)(645,268)
\thinlines \path(645,268)(645,268)(646,268)(646,268)(647,268)(647,268)(648,269)(649,269)(649,269)(650,269)(650,269)(651,270)(652,270)(652,270)(653,270)(653,270)(654,271)(654,271)(655,271)(656,271)(656,271)(657,271)(657,272)(658,272)(659,272)(659,272)(660,272)(660,273)(661,273)(662,273)(662,273)(663,273)(663,274)(664,274)(664,274)(665,274)(666,274)(666,275)(667,275)(667,275)(668,275)(669,276)(669,276)(670,276)(670,276)(671,276)(671,277)(672,277)(673,277)(673,277)(674,277)
\thinlines \path(674,277)(674,278)(675,278)(676,278)(676,278)(677,279)(677,279)(678,279)(679,279)(679,279)(680,280)(680,280)(681,280)(681,280)(682,281)(683,281)(683,281)(684,281)(684,281)(685,282)(686,282)(686,282)(687,282)(687,283)(688,283)(688,283)(689,283)(690,284)(690,284)(691,284)(691,284)(692,285)(693,285)(693,285)(694,285)(694,286)(695,286)(696,286)(696,286)(697,287)(697,287)(698,287)(698,287)(699,288)(700,288)(700,288)(701,288)(701,289)(702,289)(703,289)(703,290)
\thinlines \path(703,290)(704,290)(704,290)(705,290)(705,291)(706,291)(707,291)(707,291)(708,292)(708,292)(709,292)(710,293)(710,293)(711,293)(711,293)(712,294)(713,294)(713,294)(714,295)(714,295)(715,295)(715,296)(716,296)(717,296)(717,296)(718,297)(718,297)(719,297)(720,298)(720,298)(721,298)(721,299)(722,299)(722,299)(723,300)(724,300)(724,300)(725,301)(725,301)(726,301)(727,302)(727,302)(728,302)(728,303)(729,303)(730,303)(730,304)(731,304)(731,304)(732,305)(732,305)
\thinlines \path(732,305)(733,305)(734,306)(734,306)(735,306)(735,307)(736,307)(737,307)(737,308)(738,308)(738,309)(739,309)(739,309)(740,310)(741,310)(741,310)(742,311)(742,311)(743,312)(744,312)(744,312)(745,313)(745,313)(746,314)(747,314)(747,314)(748,315)(748,315)(749,316)(749,316)(750,316)(751,317)(751,317)(752,318)(752,318)(753,319)(754,319)(754,319)(755,320)(755,320)(756,321)(756,321)(757,322)(758,322)(758,323)(759,323)(759,323)(760,324)(761,324)(761,325)(762,325)
\thinlines \path(762,325)(762,326)(763,326)(764,327)(764,327)(765,328)(765,328)(766,329)(766,329)(767,330)(768,330)(768,331)(769,331)(769,332)(770,332)(771,333)(771,333)(772,334)(772,334)(773,335)(773,335)(774,336)(775,337)(775,337)(776,338)(776,338)(777,339)(778,339)(778,340)(779,340)(779,341)(780,342)(781,342)(781,343)(782,343)(782,344)(783,345)(783,345)(784,346)(785,346)(785,347)(786,348)(786,348)(787,349)(788,350)(788,350)(789,351)(789,352)(790,352)(790,353)(791,354)
\thinlines \path(791,354)(792,354)(792,355)(793,356)(793,356)(794,357)(795,358)(795,359)(796,359)(796,360)(797,361)(798,361)(798,362)(799,363)(799,364)(800,365)(800,365)(801,366)(802,367)(802,368)(803,368)(803,369)(804,370)(805,371)(805,372)(806,373)(806,373)(807,374)(807,375)(808,376)(809,377)(809,378)(810,379)(810,380)(811,381)(812,381)(812,382)(813,383)(813,384)(814,385)(815,386)(815,387)(816,388)(816,389)(817,390)(817,391)(818,392)(819,393)(819,394)(820,395)(820,397)
\thinlines \path(820,397)(821,398)(822,399)(822,400)(823,401)(823,402)(824,403)(824,404)(825,406)(826,407)(826,408)(827,409)(827,410)(828,412)(829,413)(829,414)(830,416)(830,417)(831,418)(832,420)(832,421)(833,422)(833,424)(834,425)(834,427)(835,428)(836,429)(836,431)(837,432)(837,434)(838,436)(839,437)(839,439)(840,440)(840,442)(841,444)(841,445)(842,447)(843,449)(843,450)(844,452)(844,454)(845,456)(846,458)(846,460)(847,461)(847,463)(848,465)(849,467)(849,469)(850,471)
\thinlines \path(850,471)(850,474)(851,476)(851,478)(852,480)(853,482)(853,485)(854,487)(854,489)(855,492)(856,494)(856,497)(857,499)(857,502)(858,504)(859,507)(859,510)(860,512)(860,515)(861,518)(861,521)(862,524)(863,527)(863,530)(864,533)(864,537)(865,540)(866,543)(866,547)(867,550)(867,554)(868,558)(868,562)(869,565)(870,569)(870,573)(871,578)(871,582)(872,586)(873,591)(873,595)(874,600)(874,605)(875,610)(876,615)(876,620)(877,626)(877,631)(878,637)(878,643)(879,649)
\thinlines \path(879,649)(880,655)(880,662)(881,669)(881,676)(882,683)(883,690)(883,698)(884,706)(884,714)(885,723)(885,732)(886,741)(887,751)(887,761)(888,771)(888,782)
\thinlines \path(909,471)(910,469)(910,467)(911,465)(911,463)(912,461)(912,459)(913,458)(914,456)(914,454)(915,452)(915,450)(916,449)(917,447)(917,445)(918,443)(918,442)(919,440)(919,439)(920,437)(921,435)(921,434)(922,432)(922,431)(923,429)(924,428)(924,426)(925,425)(925,424)(926,422)(927,421)(927,419)(928,418)(928,417)(929,416)(929,414)(930,413)(931,412)(931,410)(932,409)(932,408)(933,407)(934,406)(934,404)(935,403)(935,402)(936,401)(936,400)(937,399)(938,398)(938,397)
\thinlines \path(938,397)(939,395)(939,394)(940,393)(941,392)(941,391)(942,390)(942,389)(943,388)(944,387)(944,386)(945,385)(945,384)(946,383)(946,382)(947,381)(948,381)(948,380)(949,379)(949,378)(950,377)(951,376)(951,375)(952,374)(952,373)(953,373)(953,372)(954,371)(955,370)(955,369)(956,368)(956,368)(957,367)(958,366)(958,365)(959,364)(959,364)(960,363)(961,362)(961,361)(962,361)(962,360)(963,359)(963,359)(964,358)(965,357)(965,356)(966,356)(966,355)(967,354)(968,354)
\thinlines \path(968,354)(968,353)(969,352)(969,352)(970,351)(970,350)(971,350)(972,349)(972,348)(973,348)(973,347)(974,346)(975,346)(975,345)(976,345)(976,344)(977,343)(978,343)(978,342)(979,342)(979,341)(980,340)(980,340)(981,339)(982,339)(982,338)(983,338)(983,337)(984,336)(985,336)(985,335)(986,335)(986,334)(987,334)(987,333)(988,333)(989,332)(989,332)(990,331)(990,331)(991,330)(992,330)(992,329)(993,329)(993,328)(994,328)(995,327)(995,327)(996,326)(996,326)(997,325)
\thinlines \path(997,325)(997,325)(998,324)(999,324)(999,323)(1000,323)(1000,322)(1001,322)(1002,322)(1002,321)(1003,321)(1003,320)(1004,320)(1004,319)(1005,319)(1006,319)(1006,318)(1007,318)(1007,317)(1008,317)(1009,316)(1009,316)(1010,316)(1010,315)(1011,315)(1012,314)(1012,314)(1013,314)(1013,313)(1014,313)(1014,312)(1015,312)(1016,312)(1016,311)(1017,311)(1017,310)(1018,310)(1019,310)(1019,309)(1020,309)(1020,309)(1021,308)(1021,308)(1022,307)(1023,307)(1023,307)(1024,306)(1024,306)(1025,306)(1026,305)(1026,305)
\thinlines \path(1026,305)(1027,305)(1027,304)(1028,304)(1029,304)(1029,303)(1030,303)(1030,303)(1031,302)(1031,302)(1032,302)(1033,301)(1033,301)(1034,301)(1034,300)(1035,300)(1036,300)(1036,299)(1037,299)(1037,299)(1038,298)(1038,298)(1039,298)(1040,297)(1040,297)(1041,297)(1041,296)(1042,296)(1043,296)(1043,296)(1044,295)(1044,295)(1045,295)(1046,294)(1046,294)(1047,294)(1047,293)(1048,293)(1048,293)(1049,293)(1050,292)(1050,292)(1051,292)(1051,291)(1052,291)(1053,291)(1053,291)(1054,290)(1054,290)(1055,290)(1055,290)
\thinlines \path(1055,290)(1056,289)(1057,289)(1057,289)(1058,288)(1058,288)(1059,288)(1060,288)(1060,287)(1061,287)(1061,287)(1062,287)(1063,286)(1063,286)(1064,286)(1064,286)(1065,285)(1065,285)(1066,285)(1067,285)(1067,284)(1068,284)(1068,284)(1069,284)(1070,283)(1070,283)(1071,283)(1071,283)(1072,282)(1072,282)(1073,282)(1074,282)(1074,281)(1075,281)(1075,281)(1076,281)(1077,281)(1077,280)(1078,280)(1078,280)(1079,280)(1080,279)(1080,279)(1081,279)(1081,279)(1082,279)(1082,278)(1083,278)(1084,278)(1084,278)(1085,277)
\thinlines \path(1085,277)(1085,277)(1086,277)(1087,277)(1087,277)(1088,276)(1088,276)(1089,276)(1090,276)(1090,275)(1091,275)(1091,275)(1092,275)(1092,275)(1093,274)(1094,274)(1094,274)(1095,274)(1095,274)(1096,273)(1097,273)(1097,273)(1098,273)(1098,273)(1099,272)(1099,272)(1100,272)(1101,272)(1101,272)(1102,271)(1102,271)(1103,271)(1104,271)(1104,271)(1105,271)(1105,270)(1106,270)(1107,270)(1107,270)(1108,270)(1108,269)(1109,269)(1109,269)(1110,269)(1111,269)(1111,268)(1112,268)(1112,268)(1113,268)(1114,268)(1114,268)
\thinlines \path(1114,268)(1115,267)(1115,267)(1116,267)(1116,267)(1117,267)(1118,267)(1118,266)(1119,266)(1119,266)(1120,266)(1121,266)(1121,265)(1122,265)(1122,265)(1123,265)(1124,265)(1124,265)(1125,264)(1125,264)(1126,264)(1126,264)(1127,264)(1128,264)(1128,263)(1129,263)(1129,263)(1130,263)(1131,263)(1131,263)(1132,263)(1132,262)(1133,262)(1133,262)(1134,262)(1135,262)(1135,262)(1136,261)(1136,261)(1137,261)(1138,261)(1138,261)(1139,261)(1139,260)(1140,260)(1141,260)(1141,260)(1142,260)(1142,260)(1143,260)(1143,259)
\thinlines \path(1143,259)(1144,259)(1145,259)(1145,259)(1146,259)(1146,259)(1147,259)(1148,258)(1148,258)(1149,258)(1149,258)(1150,258)(1150,258)(1151,258)(1152,257)(1152,257)(1153,257)(1153,257)(1154,257)(1155,257)(1155,257)(1156,256)(1156,256)(1157,256)(1158,256)(1158,256)(1159,256)(1159,256)(1160,255)(1160,255)(1161,255)(1162,255)(1162,255)(1163,255)(1163,255)(1164,254)(1165,254)(1165,254)(1166,254)(1166,254)(1167,254)(1167,254)(1168,254)(1169,253)(1169,253)(1170,253)(1170,253)(1171,253)(1172,253)(1172,253)(1173,253)
\thinlines \path(1173,253)(1173,252)(1174,252)(1175,252)(1175,252)(1176,252)(1176,252)(1177,252)(1177,252)(1178,251)(1179,251)(1179,251)(1180,251)(1180,251)(1181,251)(1182,251)(1182,251)(1183,250)(1183,250)(1184,250)(1184,250)(1185,250)(1186,250)(1186,250)(1187,250)(1187,249)(1188,249)(1189,249)(1189,249)(1190,249)(1190,249)(1191,249)(1192,249)(1192,249)(1193,248)(1193,248)(1194,248)(1194,248)(1195,248)(1196,248)(1196,248)(1197,248)(1197,247)(1198,247)(1199,247)(1199,247)(1200,247)(1200,247)(1201,247)(1201,247)(1202,247)
\thinlines \path(1202,247)(1203,246)(1203,246)(1204,246)(1204,246)(1205,246)(1206,246)(1206,246)(1207,246)(1207,246)(1208,246)(1209,245)(1209,245)(1210,245)(1210,245)(1211,245)(1211,245)(1212,245)(1213,245)(1213,245)(1214,244)(1214,244)(1215,244)(1216,244)(1216,244)(1217,244)(1217,244)(1218,244)(1218,244)(1219,244)(1220,243)(1220,243)(1221,243)(1221,243)(1222,243)(1223,243)(1223,243)(1224,243)(1224,243)(1225,243)(1226,242)(1226,242)(1227,242)(1227,242)(1228,242)(1228,242)(1229,242)(1230,242)(1230,242)(1231,242)(1231,241)
\thinlines \path(1231,241)(1232,241)(1233,241)(1233,241)(1234,241)(1234,241)(1235,241)(1235,241)(1236,241)(1237,241)(1237,241)(1238,240)(1238,240)(1239,240)(1240,240)(1240,240)(1241,240)(1241,240)(1242,240)(1243,240)(1243,240)(1244,239)(1244,239)(1245,239)(1245,239)(1246,239)(1247,239)(1247,239)(1248,239)(1248,239)(1249,239)(1250,239)(1250,239)(1251,238)(1251,238)(1252,238)(1252,238)(1253,238)(1254,238)(1254,238)(1255,238)(1255,238)(1256,238)(1257,238)(1257,237)(1258,237)(1258,237)(1259,237)(1260,237)(1260,237)(1261,237)
\thinlines \path(1261,237)(1261,237)(1262,237)(1262,237)(1263,237)(1264,237)(1264,236)(1265,236)(1265,236)(1266,236)(1267,236)(1267,236)(1268,236)(1268,236)(1269,236)(1269,236)(1270,236)(1271,236)(1271,235)(1272,235)(1272,235)(1273,235)(1274,235)(1274,235)(1275,235)(1275,235)(1276,235)(1277,235)(1277,235)(1278,235)(1278,234)(1279,234)(1279,234)(1280,234)(1281,234)(1281,234)(1282,234)(1282,234)(1283,234)(1284,234)(1284,234)(1285,234)(1285,234)(1286,233)(1286,233)(1287,233)(1288,233)(1288,233)(1289,233)(1289,233)(1290,233)
\thinlines \path(1290,233)(1291,233)(1291,233)(1292,233)(1292,233)(1293,233)(1294,233)(1294,232)(1295,232)(1295,232)(1296,232)(1296,232)(1297,232)(1298,232)(1298,232)(1299,232)(1299,232)(1300,232)(1301,232)(1301,232)(1302,231)(1302,231)(1303,231)(1303,231)(1304,231)(1305,231)(1305,231)(1306,231)(1306,231)(1307,231)(1308,231)(1308,231)(1309,231)(1309,231)(1310,230)(1311,230)(1311,230)(1312,230)(1312,230)(1313,230)(1313,230)(1314,230)(1315,230)(1315,230)(1316,230)(1316,230)(1317,230)(1318,230)(1318,230)(1319,229)(1319,229)
\thinlines \path(1319,229)(1320,229)(1321,229)(1321,229)(1322,229)(1322,229)(1323,229)(1323,229)(1324,229)(1325,229)(1325,229)(1326,229)(1326,229)(1327,229)(1328,228)(1328,228)(1329,228)(1329,228)(1330,228)(1330,228)(1331,228)(1332,228)(1332,228)(1333,228)(1333,228)(1334,228)(1335,228)(1335,228)(1336,228)(1336,227)(1337,227)(1338,227)(1338,227)(1339,227)(1339,227)(1340,227)(1340,227)(1341,227)(1342,227)(1342,227)(1343,227)(1343,227)(1344,227)(1345,227)(1345,227)(1346,226)(1346,226)(1347,226)(1347,226)(1348,226)(1349,226)
\thinlines \path(1349,226)(1349,226)(1350,226)(1350,226)(1351,226)(1352,226)(1352,226)(1353,226)(1353,226)(1354,226)(1355,226)(1355,226)(1356,225)(1356,225)(1357,225)(1357,225)(1358,225)(1359,225)(1359,225)(1360,225)(1360,225)(1361,225)(1362,225)(1362,225)(1363,225)(1363,225)(1364,225)(1364,225)(1365,225)(1366,224)(1366,224)(1367,224)(1367,224)(1368,224)(1369,224)(1369,224)(1370,224)(1370,224)(1371,224)(1372,224)(1372,224)(1373,224)(1373,224)(1374,224)(1374,224)(1375,224)(1376,224)(1376,223)(1377,223)(1377,223)(1378,223)
\thinlines \path(1378,223)(1379,223)(1379,223)(1380,223)(1380,223)(1381,223)(1381,223)(1382,223)(1383,223)(1383,223)(1384,223)(1384,223)(1385,223)(1386,223)(1386,223)(1387,222)(1387,222)(1388,222)(1389,222)(1389,222)(1390,222)(1390,222)(1391,222)(1391,222)(1392,222)(1393,222)(1393,222)(1394,222)(1394,222)(1395,222)(1396,222)(1396,222)(1397,222)(1397,222)(1398,221)(1398,221)(1399,221)(1400,221)(1400,221)(1401,221)(1401,221)(1402,221)(1403,221)(1403,221)(1404,221)(1404,221)(1405,221)(1406,221)(1406,221)(1407,221)(1407,221)
\thinlines \path(1407,221)(1408,221)(1408,221)(1409,220)(1410,220)(1410,220)(1411,220)(1411,220)(1412,220)(1413,220)(1413,220)(1414,220)(1414,220)(1415,220)(1415,220)(1416,220)(1417,220)(1417,220)(1418,220)(1418,220)(1419,220)(1420,220)(1420,220)(1421,220)(1421,219)(1422,219)(1423,219)(1423,219)(1424,219)(1424,219)(1425,219)(1425,219)(1426,219)(1427,219)(1427,219)(1428,219)(1428,219)(1429,219)(1430,219)(1430,219)(1431,219)(1431,219)(1432,219)(1432,219)(1433,219)(1434,218)(1434,218)(1435,218)(1435,218)(1436,218)
\put(1306,677){\makebox(0,0)[r]{\small $\im f(q^2)$}}
\Thicklines \path(1328,677)(1394,677)
\Thicklines \path(264,473)(264,473)(265,473)(265,473)(266,473)(266,473)(267,473)(268,473)(268,473)(269,473)(269,473)(270,473)(270,473)(271,473)(272,473)(272,473)(273,473)(273,473)(274,473)(275,473)(275,473)(276,473)(276,473)(277,473)(277,473)(278,473)(279,473)(279,473)(280,473)(280,473)(281,473)(282,473)(282,473)(283,473)(283,473)(284,473)(285,473)(285,473)(286,473)(286,473)(287,473)(287,473)(288,473)(289,473)(289,473)(290,473)(290,473)(291,473)(292,473)(292,473)(293,473)
\Thicklines \path(293,473)(293,473)(294,473)(294,473)(295,473)(296,473)(296,473)(297,473)(297,473)(298,473)(299,473)(299,473)(300,473)(300,473)(301,473)(302,473)(302,473)(303,473)(303,473)(304,473)(304,473)(305,473)(306,473)(306,473)(307,473)(307,473)(308,473)(309,473)(309,473)(310,473)(310,473)(311,473)(311,473)(312,473)(313,473)(313,473)(314,473)(314,473)(315,473)(316,473)(316,473)(317,473)(317,473)(318,473)(319,473)(319,473)(320,473)(320,473)(321,473)(321,473)(322,473)
\Thicklines \path(322,473)(323,473)(323,473)(324,473)(324,473)(325,473)(326,473)(326,473)(327,473)(327,473)(328,473)(328,473)(329,473)(330,473)(330,473)(331,473)(331,473)(332,473)(333,473)(333,473)(334,473)(334,473)(335,473)(336,473)(336,473)(337,473)(337,473)(338,473)(338,473)(339,473)(340,473)(340,473)(341,473)(341,473)(342,473)(343,473)(343,473)(344,473)(344,473)(345,473)(345,473)(346,473)(347,473)(347,473)(348,473)(348,473)(349,473)(350,473)(350,473)(351,473)(351,473)
\Thicklines \path(351,473)(352,473)(353,473)(353,473)(354,473)(354,473)(355,473)(355,473)(356,473)(357,473)(357,473)(358,473)(358,473)(359,473)(360,473)(360,473)(361,473)(361,473)(362,473)(362,473)(363,473)(364,473)(364,473)(365,473)(365,473)(366,473)(367,473)(367,473)(368,473)(368,473)(369,473)(370,473)(370,473)(371,473)(371,473)(372,473)(372,473)(373,473)(374,473)(374,473)(375,473)(375,473)(376,473)(377,473)(377,473)(378,473)(378,473)(379,473)(379,473)(380,473)(381,473)
\Thicklines \path(381,473)(381,473)(382,473)(382,473)(383,473)(384,473)(384,473)(385,473)(385,473)(386,473)(387,473)(387,473)(388,473)(388,473)(389,473)(389,473)(390,473)(391,473)(391,473)(392,473)(392,473)(393,473)(394,473)(394,473)(395,473)(395,473)(396,473)(397,473)(397,473)(398,473)(398,473)(399,473)(399,473)(400,473)(401,473)(401,473)(402,473)(402,473)(403,473)(404,473)(404,473)(405,473)(405,473)(406,473)(406,473)(407,473)(408,473)(408,473)(409,473)(409,473)(410,473)
\Thicklines \path(410,473)(411,473)(411,473)(412,473)(412,473)(413,473)(414,473)(414,473)(415,473)(415,473)(416,473)(416,473)(417,473)(418,473)(418,473)(419,473)(419,473)(420,473)(421,473)(421,473)(422,473)(422,473)(423,473)(423,473)(424,473)(425,473)(425,473)(426,473)(426,473)(427,473)(428,473)(428,473)(429,473)(429,473)(430,473)(431,473)(431,473)(432,473)(432,473)(433,473)(433,473)(434,473)(435,473)(435,473)(436,473)(436,473)(437,473)(438,473)(438,473)(439,473)(439,473)
\Thicklines \path(439,473)(440,473)(440,473)(441,473)(442,473)(442,473)(443,473)(443,473)(444,473)(445,473)(445,473)(446,473)(446,473)(447,473)(448,473)(448,473)(449,473)(449,473)(450,473)(450,473)(451,473)(452,473)(452,473)(453,473)(453,473)(454,473)(455,473)(455,473)(456,473)(456,473)(457,473)(457,473)(458,473)(459,473)(459,473)(460,473)(460,473)(461,473)(462,473)(462,473)(463,473)(463,473)(464,473)(465,473)(465,473)(466,473)(466,473)(467,473)(467,473)(468,473)(469,473)
\Thicklines \path(469,473)(469,473)(470,473)(470,473)(471,473)(472,473)(472,473)(473,473)(473,473)(474,473)(474,473)(475,473)(476,473)(476,473)(477,473)(477,473)(478,473)(479,473)(479,473)(480,473)(480,473)(481,473)(482,473)(482,473)(483,473)(483,473)(484,473)(484,473)(485,473)(486,473)(486,473)(487,473)(487,473)(488,473)(489,473)(489,473)(490,473)(490,473)(491,473)(491,473)(492,473)(493,473)(493,473)(494,473)(494,473)(495,473)(496,473)(496,473)(497,473)(497,473)(498,473)
\Thicklines \path(498,473)(499,473)(499,473)(500,473)(500,473)(501,473)(501,473)(502,473)(503,473)(503,473)(504,473)(504,473)(505,473)(506,473)(506,473)(507,473)(507,473)(508,473)(508,473)(509,473)(510,473)(510,473)(511,473)(511,473)(512,473)(513,473)(513,473)(514,473)(514,473)(515,473)(516,473)(516,473)(517,473)(517,473)(518,473)(518,473)(519,473)(520,473)(520,473)(521,473)(521,473)(522,473)(523,473)(523,473)(524,473)(524,473)(525,473)(525,473)(526,473)(527,473)(527,473)
\Thicklines \path(527,473)(528,473)(528,473)(529,473)(530,473)(530,473)(531,473)(531,473)(532,473)(533,473)(533,473)(534,473)(534,473)(535,473)(535,473)(536,473)(537,473)(537,473)(538,473)(538,473)(539,473)(540,473)(540,473)(541,473)(541,473)(542,473)(542,473)(543,473)(544,473)(544,473)(545,473)(545,473)(546,473)(547,473)(547,473)(548,473)(548,473)(549,473)(550,473)(550,473)(551,473)(551,473)(552,473)(552,473)(553,473)(554,473)(554,473)(555,473)(555,473)(556,473)(557,473)
\Thicklines \path(557,473)(557,473)(558,473)(558,473)(559,473)(559,473)(560,473)(561,473)(561,473)(562,473)(562,473)(563,473)(564,473)(564,473)(565,473)(565,473)(566,473)(567,473)(567,473)(568,473)(568,473)(569,473)(569,473)(570,473)(571,473)(571,473)(572,473)(572,473)(573,473)(574,473)(574,473)(575,473)(575,473)(576,473)(576,473)(577,473)(578,473)(578,473)(579,473)(579,473)(580,473)(581,473)(581,473)(582,473)(582,473)(583,473)(584,473)(584,473)(585,473)(585,473)(586,473)
\Thicklines \path(586,473)(586,473)(587,473)(588,473)(588,473)(589,473)(589,473)(590,473)(591,473)(591,473)(592,473)(592,473)(593,473)(593,473)(594,473)(595,473)(595,473)(596,473)(596,473)(597,473)(598,473)(598,473)(599,473)(599,473)(600,473)(601,473)(601,473)(602,473)(602,473)(603,473)(603,473)(604,473)(605,473)(605,473)(606,473)(606,473)(607,473)(608,473)(608,473)(609,473)(609,473)(610,473)(610,473)(611,473)(612,473)(612,473)(613,473)(613,473)(614,473)(615,473)(615,473)
\Thicklines \path(615,473)(616,473)(616,473)(617,473)(618,473)(618,473)(619,473)(619,473)(620,473)(620,473)(621,473)(622,473)(622,473)(623,473)(623,473)(624,473)(625,473)(625,473)(626,473)(626,473)(627,473)(628,473)(628,473)(629,473)(629,473)(630,473)(630,473)(631,473)(632,473)(632,473)(633,473)(633,473)(634,473)(635,473)(635,473)(636,473)(636,473)(637,473)(637,473)(638,473)(639,473)(639,473)(640,473)(640,473)(641,473)(642,473)(642,473)(643,473)(643,473)(644,473)(645,473)
\Thicklines \path(645,473)(645,473)(646,473)(646,473)(647,473)(647,473)(648,473)(649,473)(649,473)(650,473)(650,473)(651,473)(652,473)(652,473)(653,473)(653,473)(654,473)(654,473)(655,473)(656,473)(656,473)(657,473)(657,473)(658,473)(659,473)(659,473)(660,473)(660,473)(661,473)(662,473)(662,473)(663,473)(663,473)(664,473)(664,473)(665,473)(666,473)(666,473)(667,473)(667,473)(668,473)(669,473)(669,473)(670,473)(670,473)(671,473)(671,473)(672,473)(673,473)(673,473)(674,473)
\Thicklines \path(674,473)(674,473)(675,473)(676,473)(676,473)(677,473)(677,473)(678,473)(679,473)(679,473)(680,473)(680,473)(681,473)(681,473)(682,473)(683,473)(683,473)(684,473)(684,473)(685,473)(686,473)(686,473)(687,473)(687,473)(688,473)(688,473)(689,473)(690,473)(690,473)(691,473)(691,473)(692,473)(693,473)(693,473)(694,473)(694,473)(695,473)(696,473)(696,473)(697,473)(697,473)(698,473)(698,473)(699,473)(700,473)(700,473)(701,473)(701,473)(702,473)(703,473)(703,473)
\Thicklines \path(703,473)(704,473)(704,473)(705,473)(705,473)(706,473)(707,473)(707,473)(708,473)(708,473)(709,473)(710,473)(710,473)(711,473)(711,473)(712,473)(713,473)(713,473)(714,473)(714,473)(715,473)(715,473)(716,473)(717,473)(717,473)(718,473)(718,473)(719,473)(720,473)(720,473)(721,473)(721,473)(722,473)(722,473)(723,473)(724,473)(724,473)(725,473)(725,473)(726,473)(727,473)(727,473)(728,473)(728,473)(729,473)(730,473)(730,473)(731,473)(731,473)(732,473)(732,473)
\Thicklines \path(732,473)(733,473)(734,473)(734,473)(735,473)(735,473)(736,473)(737,473)(737,473)(738,473)(738,473)(739,473)(739,473)(740,473)(741,473)(741,473)(742,473)(742,473)(743,473)(744,473)(744,473)(745,473)(745,473)(746,473)(747,473)(747,473)(748,473)(748,473)(749,473)(749,473)(750,473)(751,473)(751,473)(752,473)(752,473)(753,473)(754,473)(754,473)(755,473)(755,473)(756,473)(756,473)(757,473)(758,473)(758,473)(759,473)(759,473)(760,473)(761,473)(761,473)(762,473)
\Thicklines \path(762,473)(762,473)(763,473)(764,473)(764,473)(765,473)(765,473)(766,473)(766,473)(767,473)(768,473)(768,473)(769,473)(769,473)(770,473)(771,473)(771,473)(772,473)(772,473)(773,473)(773,473)(774,473)(775,473)(775,473)(776,473)(776,473)(777,473)(778,473)(778,473)(779,473)(779,473)(780,473)(781,473)(781,473)(782,473)(782,473)(783,473)(783,473)(784,473)(785,473)(785,473)(786,473)(786,473)(787,473)(788,473)(788,473)(789,473)(789,473)(790,473)(790,473)(791,473)
\Thicklines \path(791,473)(792,473)(792,473)(793,473)(793,473)(794,473)(795,473)(795,473)(796,473)(796,473)(797,473)(798,473)(798,473)(799,473)(799,473)(800,473)(800,473)(801,473)(802,473)(802,473)(803,473)(803,473)(804,473)(805,473)(805,473)(806,473)(806,473)(807,473)(807,473)(808,473)(809,473)(809,473)(810,473)(810,473)(811,473)(812,473)(812,473)(813,473)(813,473)(814,473)(815,473)(815,473)(816,473)(816,473)(817,473)(817,473)(818,473)(819,473)(819,473)(820,473)(820,473)
\Thicklines \path(820,473)(821,473)(822,473)(822,473)(823,473)(823,473)(824,473)(824,473)(825,473)(826,473)(826,473)(827,473)(827,473)(828,473)(829,473)(829,473)(830,473)(830,473)(831,473)(832,473)(832,473)(833,473)(833,473)(834,473)(834,473)(835,473)(836,473)(836,473)(837,473)(837,473)(838,473)(839,473)(839,473)(840,473)(840,473)(841,473)(841,473)(842,473)(843,473)(843,473)(844,473)(844,473)(845,473)(846,473)(846,473)(847,473)(847,473)(848,473)(849,473)(849,473)(850,473)
\Thicklines \path(850,473)(850,473)(851,473)(851,473)(852,473)(853,473)(853,473)(854,473)(854,473)(855,473)(856,473)(856,473)(857,473)(857,473)(858,473)(859,473)(859,473)(860,473)(860,473)(861,473)(861,473)(862,473)(863,473)(863,473)(864,473)(864,473)(865,473)(866,473)(866,473)(867,473)(867,473)(868,473)(868,473)(869,473)(870,473)(870,473)(871,473)(871,473)(872,473)(873,473)(873,473)(874,473)(874,473)(875,473)(876,473)(876,473)(877,473)(877,473)(878,473)(878,473)(879,473)
\Thicklines \path(879,473)(880,473)(880,473)(881,473)(881,473)(882,473)(883,473)(883,473)(884,473)(884,473)(885,473)(885,473)(886,473)(887,473)(887,473)(888,473)(888,473)(889,473)(890,473)(890,473)(891,473)(891,473)(892,473)(893,473)(893,473)(894,473)(894,473)(895,473)(895,473)(896,473)(897,473)(897,473)(898,473)(898,473)(899,473)(900,473)(900,473)(901,473)(901,473)(902,473)(902,473)(903,473)(904,473)(904,473)(905,473)(905,473)(906,473)(907,473)(907,473)(908,473)(908,473)
\Thicklines \path(976,785)(977,783)(978,781)(978,779)(979,777)(979,775)(980,773)(980,772)(981,770)(982,768)(982,766)(983,764)(983,762)(984,761)(985,759)(985,757)(986,755)(986,754)(987,752)(987,750)(988,749)(989,747)(989,746)(990,744)(990,743)(991,741)(992,740)(992,738)(993,737)(993,735)(994,734)(995,732)(995,731)(996,729)(996,728)(997,727)(997,725)(998,724)(999,723)(999,721)(1000,720)(1000,719)(1001,718)(1002,716)(1002,715)(1003,714)(1003,713)(1004,712)(1004,710)(1005,709)(1006,708)
\Thicklines \path(1006,708)(1006,707)(1007,706)(1007,705)(1008,703)(1009,702)(1009,701)(1010,700)(1010,699)(1011,698)(1012,697)(1012,696)(1013,695)(1013,694)(1014,693)(1014,692)(1015,691)(1016,690)(1016,689)(1017,688)(1017,687)(1018,686)(1019,685)(1019,684)(1020,683)(1020,682)(1021,681)(1021,681)(1022,680)(1023,679)(1023,678)(1024,677)(1024,676)(1025,675)(1026,674)(1026,674)(1027,673)(1027,672)(1028,671)(1029,670)(1029,670)(1030,669)(1030,668)(1031,667)(1031,666)(1032,666)(1033,665)(1033,664)(1034,663)(1034,663)(1035,662)
\Thicklines \path(1035,662)(1036,661)(1036,660)(1037,660)(1037,659)(1038,658)(1038,658)(1039,657)(1040,656)(1040,655)(1041,655)(1041,654)(1042,653)(1043,653)(1043,652)(1044,651)(1044,651)(1045,650)(1046,649)(1046,649)(1047,648)(1047,648)(1048,647)(1048,646)(1049,646)(1050,645)(1050,644)(1051,644)(1051,643)(1052,643)(1053,642)(1053,641)(1054,641)(1054,640)(1055,640)(1055,639)(1056,639)(1057,638)(1057,637)(1058,637)(1058,636)(1059,636)(1060,635)(1060,635)(1061,634)(1061,634)(1062,633)(1063,633)(1063,632)(1064,632)(1064,631)
\Thicklines \path(1064,631)(1065,630)(1065,630)(1066,629)(1067,629)(1067,628)(1068,628)(1068,627)(1069,627)(1070,626)(1070,626)(1071,626)(1071,625)(1072,625)(1072,624)(1073,624)(1074,623)(1074,623)(1075,622)(1075,622)(1076,621)(1077,621)(1077,620)(1078,620)(1078,620)(1079,619)(1080,619)(1080,618)(1081,618)(1081,617)(1082,617)(1082,616)(1083,616)(1084,616)(1084,615)(1085,615)(1085,614)(1086,614)(1087,614)(1087,613)(1088,613)(1088,612)(1089,612)(1090,612)(1090,611)(1091,611)(1091,610)(1092,610)(1092,610)(1093,609)(1094,609)
\Thicklines \path(1094,609)(1094,608)(1095,608)(1095,608)(1096,607)(1097,607)(1097,607)(1098,606)(1098,606)(1099,606)(1099,605)(1100,605)(1101,604)(1101,604)(1102,604)(1102,603)(1103,603)(1104,603)(1104,602)(1105,602)(1105,602)(1106,601)(1107,601)(1107,601)(1108,600)(1108,600)(1109,600)(1109,599)(1110,599)(1111,599)(1111,598)(1112,598)(1112,598)(1113,597)(1114,597)(1114,597)(1115,596)(1115,596)(1116,596)(1116,596)(1117,595)(1118,595)(1118,595)(1119,594)(1119,594)(1120,594)(1121,593)(1121,593)(1122,593)(1122,592)(1123,592)
\Thicklines \path(1123,592)(1124,592)(1124,592)(1125,591)(1125,591)(1126,591)(1126,590)(1127,590)(1128,590)(1128,590)(1129,589)(1129,589)(1130,589)(1131,589)(1131,588)(1132,588)(1132,588)(1133,587)(1133,587)(1134,587)(1135,587)(1135,586)(1136,586)(1136,586)(1137,586)(1138,585)(1138,585)(1139,585)(1139,585)(1140,584)(1141,584)(1141,584)(1142,584)(1142,583)(1143,583)(1143,583)(1144,583)(1145,582)(1145,582)(1146,582)(1146,582)(1147,581)(1148,581)(1148,581)(1149,581)(1149,580)(1150,580)(1150,580)(1151,580)(1152,579)(1152,579)
\Thicklines \path(1152,579)(1153,579)(1153,579)(1154,578)(1155,578)(1155,578)(1156,578)(1156,578)(1157,577)(1158,577)(1158,577)(1159,577)(1159,576)(1160,576)(1160,576)(1161,576)(1162,576)(1162,575)(1163,575)(1163,575)(1164,575)(1165,575)(1165,574)(1166,574)(1166,574)(1167,574)(1167,573)(1168,573)(1169,573)(1169,573)(1170,573)(1170,572)(1171,572)(1172,572)(1172,572)(1173,572)(1173,571)(1174,571)(1175,571)(1175,571)(1176,571)(1176,570)(1177,570)(1177,570)(1178,570)(1179,570)(1179,569)(1180,569)(1180,569)(1181,569)(1182,569)
\Thicklines \path(1182,569)(1182,569)(1183,568)(1183,568)(1184,568)(1184,568)(1185,568)(1186,567)(1186,567)(1187,567)(1187,567)(1188,567)(1189,567)(1189,566)(1190,566)(1190,566)(1191,566)(1192,566)(1192,565)(1193,565)(1193,565)(1194,565)(1194,565)(1195,565)(1196,564)(1196,564)(1197,564)(1197,564)(1198,564)(1199,564)(1199,563)(1200,563)(1200,563)(1201,563)(1201,563)(1202,563)(1203,562)(1203,562)(1204,562)(1204,562)(1205,562)(1206,562)(1206,561)(1207,561)(1207,561)(1208,561)(1209,561)(1209,561)(1210,560)(1210,560)(1211,560)
\Thicklines \path(1211,560)(1211,560)(1212,560)(1213,560)(1213,560)(1214,559)(1214,559)(1215,559)(1216,559)(1216,559)(1217,559)(1217,558)(1218,558)(1218,558)(1219,558)(1220,558)(1220,558)(1221,558)(1221,557)(1222,557)(1223,557)(1223,557)(1224,557)(1224,557)(1225,557)(1226,556)(1226,556)(1227,556)(1227,556)(1228,556)(1228,556)(1229,556)(1230,555)(1230,555)(1231,555)(1231,555)(1232,555)(1233,555)(1233,555)(1234,554)(1234,554)(1235,554)(1235,554)(1236,554)(1237,554)(1237,554)(1238,554)(1238,553)(1239,553)(1240,553)(1240,553)
\Thicklines \path(1240,553)(1241,553)(1241,553)(1242,553)(1243,552)(1243,552)(1244,552)(1244,552)(1245,552)(1245,552)(1246,552)(1247,552)(1247,551)(1248,551)(1248,551)(1249,551)(1250,551)(1250,551)(1251,551)(1251,551)(1252,550)(1252,550)(1253,550)(1254,550)(1254,550)(1255,550)(1255,550)(1256,550)(1257,549)(1257,549)(1258,549)(1258,549)(1259,549)(1260,549)(1260,549)(1261,549)(1261,549)(1262,548)(1262,548)(1263,548)(1264,548)(1264,548)(1265,548)(1265,548)(1266,548)(1267,547)(1267,547)(1268,547)(1268,547)(1269,547)(1269,547)
\Thicklines \path(1269,547)(1270,547)(1271,547)(1271,547)(1272,546)(1272,546)(1273,546)(1274,546)(1274,546)(1275,546)(1275,546)(1276,546)(1277,546)(1277,545)(1278,545)(1278,545)(1279,545)(1279,545)(1280,545)(1281,545)(1281,545)(1282,545)(1282,545)(1283,544)(1284,544)(1284,544)(1285,544)(1285,544)(1286,544)(1286,544)(1287,544)(1288,544)(1288,543)(1289,543)(1289,543)(1290,543)(1291,543)(1291,543)(1292,543)(1292,543)(1293,543)(1294,543)(1294,542)(1295,542)(1295,542)(1296,542)(1296,542)(1297,542)(1298,542)(1298,542)(1299,542)
\Thicklines \path(1299,542)(1299,542)(1300,541)(1301,541)(1301,541)(1302,541)(1302,541)(1303,541)(1303,541)(1304,541)(1305,541)(1305,541)(1306,541)(1306,540)(1307,540)(1308,540)(1308,540)(1309,540)(1309,540)(1310,540)(1311,540)(1311,540)(1312,540)(1312,540)(1313,539)(1313,539)(1314,539)(1315,539)(1315,539)(1316,539)(1316,539)(1317,539)(1318,539)(1318,539)(1319,539)(1319,538)(1320,538)(1321,538)(1321,538)(1322,538)(1322,538)(1323,538)(1323,538)(1324,538)(1325,538)(1325,538)(1326,537)(1326,537)(1327,537)(1328,537)(1328,537)
\Thicklines \path(1328,537)(1329,537)(1329,537)(1330,537)(1330,537)(1331,537)(1332,537)(1332,537)(1333,536)(1333,536)(1334,536)(1335,536)(1335,536)(1336,536)(1336,536)(1337,536)(1338,536)(1338,536)(1339,536)(1339,536)(1340,535)(1340,535)(1341,535)(1342,535)(1342,535)(1343,535)(1343,535)(1344,535)(1345,535)(1345,535)(1346,535)(1346,535)(1347,535)(1347,534)(1348,534)(1349,534)(1349,534)(1350,534)(1350,534)(1351,534)(1352,534)(1352,534)(1353,534)(1353,534)(1354,534)(1355,534)(1355,533)(1356,533)(1356,533)(1357,533)(1357,533)
\Thicklines \path(1357,533)(1358,533)(1359,533)(1359,533)(1360,533)(1360,533)(1361,533)(1362,533)(1362,533)(1363,532)(1363,532)(1364,532)(1364,532)(1365,532)(1366,532)(1366,532)(1367,532)(1367,532)(1368,532)(1369,532)(1369,532)(1370,532)(1370,532)(1371,531)(1372,531)(1372,531)(1373,531)(1373,531)(1374,531)(1374,531)(1375,531)(1376,531)(1376,531)(1377,531)(1377,531)(1378,531)(1379,531)(1379,531)(1380,530)(1380,530)(1381,530)(1381,530)(1382,530)(1383,530)(1383,530)(1384,530)(1384,530)(1385,530)(1386,530)(1386,530)(1387,530)
\Thicklines \path(1387,530)(1387,530)(1388,530)(1389,529)(1389,529)(1390,529)(1390,529)(1391,529)(1391,529)(1392,529)(1393,529)(1393,529)(1394,529)(1394,529)(1395,529)(1396,529)(1396,529)(1397,529)(1397,528)(1398,528)(1398,528)(1399,528)(1400,528)(1400,528)(1401,528)(1401,528)(1402,528)(1403,528)(1403,528)(1404,528)(1404,528)(1405,528)(1406,528)(1406,528)(1407,527)(1407,527)(1408,527)(1408,527)(1409,527)(1410,527)(1410,527)(1411,527)(1411,527)(1412,527)(1413,527)(1413,527)(1414,527)(1414,527)(1415,527)(1415,527)(1416,527)
\Thicklines \path(1416,527)(1417,526)(1417,526)(1418,526)(1418,526)(1419,526)(1420,526)(1420,526)(1421,526)(1421,526)(1422,526)(1423,526)(1423,526)(1424,526)(1424,526)(1425,526)(1425,526)(1426,526)(1427,525)(1427,525)(1428,525)(1428,525)(1429,525)(1430,525)(1430,525)(1431,525)(1431,525)(1432,525)(1432,525)(1433,525)(1434,525)(1434,525)(1435,525)(1435,525)(1436,525)
\end{picture}

%% file: fq2-log.tex
% GNUPLOT: LaTeX picture using EEPIC macros
\setlength{\unitlength}{0.240900pt}
\begin{picture}(1500,900)(0,0)
\tenrm
\thinlines \drawline[-50](264,473)(1436,473)
\thicklines \path(264,158)(284,158)
\thicklines \path(1436,158)(1416,158)
\put(242,158){\makebox(0,0)[r]{$-1$}}
\thicklines \path(264,473)(284,473)
\thicklines \path(1436,473)(1416,473)
\put(242,473){\makebox(0,0)[r]{$0$}}
\thicklines \path(264,787)(284,787)
\thicklines \path(1436,787)(1416,787)
\put(242,787){\makebox(0,0)[r]{$1$}}
\thicklines \path(264,158)(264,178)
\thicklines \path(264,787)(264,767)
\put(264,113){\makebox(0,0){$0.01$}}
\thicklines \path(498,158)(498,178)
\thicklines \path(498,787)(498,767)
\put(498,113){\makebox(0,0){$0.1$}}
\thicklines \path(733,158)(733,178)
\thicklines \path(733,787)(733,767)
\put(733,113){\makebox(0,0){$1$}}
\thinlines \drawline[-50](733,158)(733,787)
\thicklines \path(967,158)(967,178)
\thicklines \path(967,787)(967,767)
\put(967,113){\makebox(0,0){$10$}}
\thicklines \path(1202,158)(1202,178)
\thicklines \path(1202,787)(1202,767)
\put(1202,113){\makebox(0,0){$100$}}
\thicklines \path(1436,158)(1436,178)
\thicklines \path(1436,787)(1436,767)
\put(1436,113){\makebox(0,0){$1000$}}
\thicklines \path(264,158)(1436,158)(1436,787)(264,787)(264,158)
\put(850,68){\makebox(0,0){$q^2/4m^2$}}
\put(1306,722){\makebox(0,0)[r]{\small $\re f(q^2)$}}
\thinlines \path(1328,722)(1394,722)
\thinlines \path(264,475)(264,475)(265,475)(266,475)(268,475)(269,475)(270,475)(271,475)(272,475)(273,475)(275,475)(276,475)(277,475)(278,475)(279,475)(280,475)(282,475)(283,475)(284,475)(285,475)(286,475)(287,475)(289,475)(290,475)(291,475)(292,475)(293,475)(295,475)(296,475)(297,475)(298,475)(299,475)(300,476)(302,476)(303,476)(304,476)(305,476)(306,476)(307,476)(309,476)(310,476)(311,476)(312,476)(313,476)(314,476)(316,476)(317,476)(318,476)(319,476)(320,476)(321,476)
\thinlines \path(321,476)(323,476)(324,476)(325,476)(326,476)(327,476)(329,477)(330,477)(331,477)(332,477)(333,477)(334,477)(336,477)(337,477)(338,477)(339,477)(340,477)(341,477)(343,477)(344,477)(345,477)(346,477)(347,477)(348,477)(350,477)(351,478)(352,478)(353,478)(354,478)(356,478)(357,478)(358,478)(359,478)(360,478)(361,478)(363,478)(364,478)(365,478)(366,478)(367,478)(368,478)(370,479)(371,479)(372,479)(373,479)(374,479)(375,479)(377,479)(378,479)(379,479)(380,479)
\thinlines \path(380,479)(381,479)(382,479)(384,479)(385,480)(386,480)(387,480)(388,480)(390,480)(391,480)(392,480)(393,480)(394,480)(395,480)(397,480)(398,481)(399,481)(400,481)(401,481)(402,481)(404,481)(405,481)(406,481)(407,481)(408,481)(409,482)(411,482)(412,482)(413,482)(414,482)(415,482)(417,482)(418,482)(419,482)(420,483)(421,483)(422,483)(424,483)(425,483)(426,483)(427,483)(428,483)(429,484)(431,484)(432,484)(433,484)(434,484)(435,484)(436,484)(438,485)(439,485)
\thinlines \path(439,485)(440,485)(441,485)(442,485)(443,485)(445,485)(446,486)(447,486)(448,486)(449,486)(451,486)(452,486)(453,487)(454,487)(455,487)(456,487)(458,487)(459,488)(460,488)(461,488)(462,488)(463,488)(465,488)(466,489)(467,489)(468,489)(469,489)(470,489)(472,490)(473,490)(474,490)(475,490)(476,491)(478,491)(479,491)(480,491)(481,491)(482,492)(483,492)(485,492)(486,492)(487,493)(488,493)(489,493)(490,493)(492,494)(493,494)(494,494)(495,495)(496,495)(497,495)
\thinlines \path(497,495)(499,495)(500,496)(501,496)(502,496)(503,497)(505,497)(506,497)(507,497)(508,498)(509,498)(510,498)(512,499)(513,499)(514,499)(515,500)(516,500)(517,500)(519,501)(520,501)(521,502)(522,502)(523,502)(524,503)(526,503)(527,504)(528,504)(529,504)(530,505)(531,505)(533,506)(534,506)(535,506)(536,507)(537,507)(539,508)(540,508)(541,509)(542,509)(543,510)(544,510)(546,511)(547,511)(548,512)(549,512)(550,513)(551,513)(553,514)(554,514)(555,515)(556,516)
\thinlines \path(556,516)(557,516)(558,517)(560,517)(561,518)(562,519)(563,519)(564,520)(566,521)(567,521)(568,522)(569,523)(570,523)(571,524)(573,525)(574,525)(575,526)(576,527)(577,528)(578,528)(580,529)(581,530)(582,531)(583,532)(584,533)(585,533)(587,534)(588,535)(589,536)(590,537)(591,538)(592,539)(594,540)(595,541)(596,542)(597,543)(598,544)(600,545)(601,546)(602,547)(603,548)(604,550)(605,551)(607,552)(608,553)(609,554)(610,556)(611,557)(612,558)(614,560)(615,561)
\thinlines \path(615,561)(616,562)(617,564)(618,565)(619,567)(621,568)(622,570)(623,571)(624,573)(625,575)(627,576)(628,578)(629,580)(630,582)(631,583)(632,585)(634,587)(635,589)(636,591)(637,593)(638,595)(639,597)(641,600)(642,602)(643,604)(644,606)(645,609)(646,611)(648,614)(649,617)(650,619)(651,622)(652,625)(653,628)(655,631)(656,634)(657,637)(658,640)(659,643)(661,647)(662,650)(663,654)(664,658)(665,662)(666,666)(668,670)(669,674)(670,678)(671,683)(672,687)(673,692)
\thinlines \path(673,692)(675,697)(676,702)(677,708)(678,713)(679,719)(680,725)(682,731)(683,738)(684,745)(685,752)(686,759)(688,767)(689,775)(690,783)
\thinlines \path(733,472)(734,469)(736,467)(737,464)(738,462)(739,460)(740,457)(741,455)(743,453)(744,451)(745,448)(746,446)(747,444)(749,442)(750,439)(751,437)(752,435)(753,433)(754,431)(756,429)(757,427)(758,424)(759,422)(760,420)(761,418)(763,416)(764,414)(765,412)(766,410)(767,408)(768,406)(770,404)(771,402)(772,401)(773,399)(774,397)(776,395)(777,393)(778,391)(779,389)(780,387)(781,386)(783,384)(784,382)(785,380)(786,379)(787,377)(788,375)(790,373)(791,372)(792,370)
\thinlines \path(792,370)(793,368)(794,367)(795,365)(797,363)(798,362)(799,360)(800,358)(801,357)(802,355)(804,354)(805,352)(806,350)(807,349)(808,347)(810,346)(811,344)(812,343)(813,341)(814,340)(815,338)(817,337)(818,336)(819,334)(820,333)(821,331)(822,330)(824,328)(825,327)(826,326)(827,324)(828,323)(829,322)(831,320)(832,319)(833,318)(834,316)(835,315)(837,314)(838,313)(839,311)(840,310)(841,309)(842,308)(844,306)(845,305)(846,304)(847,303)(848,301)(849,300)(851,299)
\thinlines \path(851,299)(852,298)(853,297)(854,296)(855,295)(856,293)(858,292)(859,291)(860,290)(861,289)(862,288)(863,287)(865,286)(866,285)(867,284)(868,283)(869,282)(871,280)(872,279)(873,278)(874,277)(875,276)(876,275)(878,274)(879,274)(880,273)(881,272)(882,271)(883,270)(885,269)(886,268)(887,267)(888,266)(889,265)(890,264)(892,263)(893,262)(894,261)(895,261)(896,260)(898,259)(899,258)(900,257)(901,256)(902,256)(903,255)(905,254)(906,253)(907,252)(908,251)(909,251)
\thinlines \path(909,251)(910,250)(912,249)(913,248)(914,248)(915,247)(916,246)(917,245)(919,244)(920,244)(921,243)(922,242)(923,242)(924,241)(926,240)(927,239)(928,239)(929,238)(930,237)(932,237)(933,236)(934,235)(935,235)(936,234)(937,233)(939,233)(940,232)(941,231)(942,231)(943,230)(944,229)(946,229)(947,228)(948,228)(949,227)(950,226)(951,226)(953,225)(954,225)(955,224)(956,223)(957,223)(959,222)(960,222)(961,221)(962,221)(963,220)(964,220)(966,219)(967,218)(968,218)
\thinlines \path(968,218)(969,217)(970,217)(971,216)(973,216)(974,215)(975,215)(976,214)(977,214)(978,213)(980,213)(981,212)(982,212)(983,211)(984,211)(986,210)(987,210)(988,210)(989,209)(990,209)(991,208)(993,208)(994,207)(995,207)(996,206)(997,206)(998,206)(1000,205)(1001,205)(1002,204)(1003,204)(1004,203)(1005,203)(1007,203)(1008,202)(1009,202)(1010,201)(1011,201)(1012,201)(1014,200)(1015,200)(1016,200)(1017,199)(1018,199)(1020,198)(1021,198)(1022,198)(1023,197)(1024,197)(1025,197)(1027,196)
\thinlines \path(1027,196)(1028,196)(1029,196)(1030,195)(1031,195)(1032,195)(1034,194)(1035,194)(1036,194)(1037,193)(1038,193)(1039,193)(1041,192)(1042,192)(1043,192)(1044,191)(1045,191)(1047,191)(1048,191)(1049,190)(1050,190)(1051,190)(1052,189)(1054,189)(1055,189)(1056,189)(1057,188)(1058,188)(1059,188)(1061,187)(1062,187)(1063,187)(1064,187)(1065,186)(1066,186)(1068,186)(1069,186)(1070,185)(1071,185)(1072,185)(1073,185)(1075,184)(1076,184)(1077,184)(1078,184)(1079,183)(1081,183)(1082,183)(1083,183)(1084,182)(1085,182)
\thinlines \path(1085,182)(1086,182)(1088,182)(1089,182)(1090,181)(1091,181)(1092,181)(1093,181)(1095,180)(1096,180)(1097,180)(1098,180)(1099,180)(1100,179)(1102,179)(1103,179)(1104,179)(1105,179)(1106,178)(1108,178)(1109,178)(1110,178)(1111,178)(1112,178)(1113,177)(1115,177)(1116,177)(1117,177)(1118,177)(1119,176)(1120,176)(1122,176)(1123,176)(1124,176)(1125,176)(1126,175)(1127,175)(1129,175)(1130,175)(1131,175)(1132,175)(1133,174)(1134,174)(1136,174)(1137,174)(1138,174)(1139,174)(1140,174)(1142,173)(1143,173)(1144,173)
\thinlines \path(1144,173)(1145,173)(1146,173)(1147,173)(1149,173)(1150,172)(1151,172)(1152,172)(1153,172)(1154,172)(1156,172)(1157,172)(1158,172)(1159,171)(1160,171)(1161,171)(1163,171)(1164,171)(1165,171)(1166,171)(1167,171)(1169,170)(1170,170)(1171,170)(1172,170)(1173,170)(1174,170)(1176,170)(1177,170)(1178,169)(1179,169)(1180,169)(1181,169)(1183,169)(1184,169)(1185,169)(1186,169)(1187,169)(1188,169)(1190,168)(1191,168)(1192,168)(1193,168)(1194,168)(1195,168)(1197,168)(1198,168)(1199,168)(1200,168)(1201,167)(1203,167)
\thinlines \path(1203,167)(1204,167)(1205,167)(1206,167)(1207,167)(1208,167)(1210,167)(1211,167)(1212,167)(1213,167)(1214,167)(1215,166)(1217,166)(1218,166)(1219,166)(1220,166)(1221,166)(1222,166)(1224,166)(1225,166)(1226,166)(1227,166)(1228,166)(1230,166)(1231,165)(1232,165)(1233,165)(1234,165)(1235,165)(1237,165)(1238,165)(1239,165)(1240,165)(1241,165)(1242,165)(1244,165)(1245,165)(1246,165)(1247,164)(1248,164)(1249,164)(1251,164)(1252,164)(1253,164)(1254,164)(1255,164)(1257,164)(1258,164)(1259,164)(1260,164)(1261,164)
\thinlines \path(1261,164)(1262,164)(1264,164)(1265,164)(1266,164)(1267,163)(1268,163)(1269,163)(1271,163)(1272,163)(1273,163)(1274,163)(1275,163)(1276,163)(1278,163)(1279,163)(1280,163)(1281,163)(1282,163)(1283,163)(1285,163)(1286,163)(1287,163)(1288,163)(1289,163)(1291,163)(1292,162)(1293,162)(1294,162)(1295,162)(1296,162)(1298,162)(1299,162)(1300,162)(1301,162)(1302,162)(1303,162)(1305,162)(1306,162)(1307,162)(1308,162)(1309,162)(1310,162)(1312,162)(1313,162)(1314,162)(1315,162)(1316,162)(1318,162)(1319,162)(1320,162)
\thinlines \path(1320,162)(1321,161)(1322,161)(1323,161)(1325,161)(1326,161)(1327,161)(1328,161)(1329,161)(1330,161)(1332,161)(1333,161)(1334,161)(1335,161)(1336,161)(1337,161)(1339,161)(1340,161)(1341,161)(1342,161)(1343,161)(1344,161)(1346,161)(1347,161)(1348,161)(1349,161)(1350,161)(1352,161)(1353,161)(1354,161)(1355,161)(1356,161)(1357,161)(1359,161)(1360,161)(1361,160)(1362,160)(1363,160)(1364,160)(1366,160)(1367,160)(1368,160)(1369,160)(1370,160)(1371,160)(1373,160)(1374,160)(1375,160)(1376,160)(1377,160)(1379,160)
\thinlines \path(1379,160)(1380,160)(1381,160)(1382,160)(1383,160)(1384,160)(1386,160)(1387,160)(1388,160)(1389,160)(1390,160)(1391,160)(1393,160)(1394,160)(1395,160)(1396,160)(1397,160)(1398,160)(1400,160)(1401,160)(1402,160)(1403,160)(1404,160)(1405,160)(1407,160)(1408,160)(1409,160)(1410,160)(1411,160)(1413,160)(1414,160)(1415,160)(1416,160)(1417,160)(1418,160)(1420,160)(1421,159)(1422,159)(1423,159)(1424,159)(1425,159)(1427,159)(1428,159)(1429,159)(1430,159)(1431,159)(1432,159)(1434,159)(1435,159)(1436,159)
\put(1306,677){\makebox(0,0)[r]{\small $\im f(q^2)$}}
\Thicklines \path(1328,677)(1394,677)
\Thicklines \path(264,473)(264,473)(265,473)(266,473)(268,473)(269,473)(270,473)(271,473)(272,473)(273,473)(275,473)(276,473)(277,473)(278,473)(279,473)(280,473)(282,473)(283,473)(284,473)(285,473)(286,473)(287,473)(289,473)(290,473)(291,473)(292,473)(293,473)(295,473)(296,473)(297,473)(298,473)(299,473)(300,473)(302,473)(303,473)(304,473)(305,473)(306,473)(307,473)(309,473)(310,473)(311,473)(312,473)(313,473)(314,473)(316,473)(317,473)(318,473)(319,473)(320,473)(321,473)
\Thicklines \path(321,473)(323,473)(324,473)(325,473)(326,473)(327,473)(329,473)(330,473)(331,473)(332,473)(333,473)(334,473)(336,473)(337,473)(338,473)(339,473)(340,473)(341,473)(343,473)(344,473)(345,473)(346,473)(347,473)(348,473)(350,473)(351,473)(352,473)(353,473)(354,473)(356,473)(357,473)(358,473)(359,473)(360,473)(361,473)(363,473)(364,473)(365,473)(366,473)(367,473)(368,473)(370,473)(371,473)(372,473)(373,473)(374,473)(375,473)(377,473)(378,473)(379,473)(380,473)
\Thicklines \path(380,473)(381,473)(382,473)(384,473)(385,473)(386,473)(387,473)(388,473)(390,473)(391,473)(392,473)(393,473)(394,473)(395,473)(397,473)(398,473)(399,473)(400,473)(401,473)(402,473)(404,473)(405,473)(406,473)(407,473)(408,473)(409,473)(411,473)(412,473)(413,473)(414,473)(415,473)(417,473)(418,473)(419,473)(420,473)(421,473)(422,473)(424,473)(425,473)(426,473)(427,473)(428,473)(429,473)(431,473)(432,473)(433,473)(434,473)(435,473)(436,473)(438,473)(439,473)
\Thicklines \path(439,473)(440,473)(441,473)(442,473)(443,473)(445,473)(446,473)(447,473)(448,473)(449,473)(451,473)(452,473)(453,473)(454,473)(455,473)(456,473)(458,473)(459,473)(460,473)(461,473)(462,473)(463,473)(465,473)(466,473)(467,473)(468,473)(469,473)(470,473)(472,473)(473,473)(474,473)(475,473)(476,473)(478,473)(479,473)(480,473)(481,473)(482,473)(483,473)(485,473)(486,473)(487,473)(488,473)(489,473)(490,473)(492,473)(493,473)(494,473)(495,473)(496,473)(497,473)
\Thicklines \path(497,473)(499,473)(500,473)(501,473)(502,473)(503,473)(505,473)(506,473)(507,473)(508,473)(509,473)(510,473)(512,473)(513,473)(514,473)(515,473)(516,473)(517,473)(519,473)(520,473)(521,473)(522,473)(523,473)(524,473)(526,473)(527,473)(528,473)(529,473)(530,473)(531,473)(533,473)(534,473)(535,473)(536,473)(537,473)(539,473)(540,473)(541,473)(542,473)(543,473)(544,473)(546,473)(547,473)(548,473)(549,473)(550,473)(551,473)(553,473)(554,473)(555,473)(556,473)
\Thicklines \path(556,473)(557,473)(558,473)(560,473)(561,473)(562,473)(563,473)(564,473)(566,473)(567,473)(568,473)(569,473)(570,473)(571,473)(573,473)(574,473)(575,473)(576,473)(577,473)(578,473)(580,473)(581,473)(582,473)(583,473)(584,473)(585,473)(587,473)(588,473)(589,473)(590,473)(591,473)(592,473)(594,473)(595,473)(596,473)(597,473)(598,473)(600,473)(601,473)(602,473)(603,473)(604,473)(605,473)(607,473)(608,473)(609,473)(610,473)(611,473)(612,473)(614,473)(615,473)
\Thicklines \path(615,473)(616,473)(617,473)(618,473)(619,473)(621,473)(622,473)(623,473)(624,473)(625,473)(627,473)(628,473)(629,473)(630,473)(631,473)(632,473)(634,473)(635,473)(636,473)(637,473)(638,473)(639,473)(641,473)(642,473)(643,473)(644,473)(645,473)(646,473)(648,473)(649,473)(650,473)(651,473)(652,473)(653,473)(655,473)(656,473)(657,473)(658,473)(659,473)(661,473)(662,473)(663,473)(664,473)(665,473)(666,473)(668,473)(669,473)(670,473)(671,473)(672,473)(673,473)
\Thicklines \path(673,473)(675,473)(676,473)(677,473)(678,473)(679,473)(680,473)(682,473)(683,473)(684,473)(685,473)(686,473)(688,473)(689,473)(690,473)(691,473)(692,473)(693,473)(695,473)(696,473)(697,473)(698,473)(699,473)(700,473)(702,473)(703,473)(704,473)(705,473)(706,473)(707,473)(709,473)(710,473)(711,473)(712,473)(713,473)(714,473)(716,473)(717,473)(718,473)(719,473)(720,473)(722,473)(723,473)(724,473)(725,473)(726,473)(727,473)(729,473)(730,473)(731,473)(732,473)
\Thicklines \path(811,787)(812,782)(813,777)(814,772)(815,767)(817,762)(818,758)(819,753)(820,749)(821,744)(822,740)(824,736)(825,732)(826,728)(827,724)(828,720)(829,717)(831,713)(832,709)(833,706)(834,702)(835,699)(837,696)(838,692)(839,689)(840,686)(841,683)(842,680)(844,677)(845,674)(846,671)(847,668)(848,665)(849,663)(851,660)(852,657)(853,655)(854,652)(855,650)(856,647)(858,645)(859,643)(860,640)(861,638)(862,636)(863,633)(865,631)(866,629)(867,627)(868,625)(869,623)
\Thicklines \path(869,623)(871,621)(872,619)(873,617)(874,615)(875,613)(876,611)(878,609)(879,607)(880,606)(881,604)(882,602)(883,600)(885,599)(886,597)(887,596)(888,594)(889,592)(890,591)(892,589)(893,588)(894,586)(895,585)(896,583)(898,582)(899,580)(900,579)(901,578)(902,576)(903,575)(905,574)(906,572)(907,571)(908,570)(909,569)(910,567)(912,566)(913,565)(914,564)(915,563)(916,562)(917,561)(919,559)(920,558)(921,557)(922,556)(923,555)(924,554)(926,553)(927,552)(928,551)
\Thicklines \path(928,551)(929,550)(930,549)(932,548)(933,547)(934,546)(935,545)(936,545)(937,544)(939,543)(940,542)(941,541)(942,540)(943,539)(944,539)(946,538)(947,537)(948,536)(949,535)(950,535)(951,534)(953,533)(954,532)(955,532)(956,531)(957,530)(959,529)(960,529)(961,528)(962,527)(963,527)(964,526)(966,525)(967,525)(968,524)(969,524)(970,523)(971,522)(973,522)(974,521)(975,521)(976,520)(977,519)(978,519)(980,518)(981,518)(982,517)(983,517)(984,516)(986,516)(987,515)
\Thicklines \path(987,515)(988,515)(989,514)(990,514)(991,513)(993,513)(994,512)(995,512)(996,511)(997,511)(998,510)(1000,510)(1001,509)(1002,509)(1003,509)(1004,508)(1005,508)(1007,507)(1008,507)(1009,506)(1010,506)(1011,506)(1012,505)(1014,505)(1015,504)(1016,504)(1017,504)(1018,503)(1020,503)(1021,503)(1022,502)(1023,502)(1024,502)(1025,501)(1027,501)(1028,501)(1029,500)(1030,500)(1031,500)(1032,499)(1034,499)(1035,499)(1036,498)(1037,498)(1038,498)(1039,497)(1041,497)(1042,497)(1043,497)(1044,496)(1045,496)
\Thicklines \path(1045,496)(1047,496)(1048,495)(1049,495)(1050,495)(1051,495)(1052,494)(1054,494)(1055,494)(1056,494)(1057,493)(1058,493)(1059,493)(1061,493)(1062,492)(1063,492)(1064,492)(1065,492)(1066,491)(1068,491)(1069,491)(1070,491)(1071,491)(1072,490)(1073,490)(1075,490)(1076,490)(1077,490)(1078,489)(1079,489)(1081,489)(1082,489)(1083,489)(1084,488)(1085,488)(1086,488)(1088,488)(1089,488)(1090,488)(1091,487)(1092,487)(1093,487)(1095,487)(1096,487)(1097,487)(1098,486)(1099,486)(1100,486)(1102,486)(1103,486)(1104,486)
\Thicklines \path(1104,486)(1105,485)(1106,485)(1108,485)(1109,485)(1110,485)(1111,485)(1112,485)(1113,484)(1115,484)(1116,484)(1117,484)(1118,484)(1119,484)(1120,484)(1122,483)(1123,483)(1124,483)(1125,483)(1126,483)(1127,483)(1129,483)(1130,483)(1131,482)(1132,482)(1133,482)(1134,482)(1136,482)(1137,482)(1138,482)(1139,482)(1140,482)(1142,481)(1143,481)(1144,481)(1145,481)(1146,481)(1147,481)(1149,481)(1150,481)(1151,481)(1152,481)(1153,481)(1154,480)(1156,480)(1157,480)(1158,480)(1159,480)(1160,480)(1161,480)(1163,480)
\Thicklines \path(1163,480)(1164,480)(1165,480)(1166,480)(1167,479)(1169,479)(1170,479)(1171,479)(1172,479)(1173,479)(1174,479)(1176,479)(1177,479)(1178,479)(1179,479)(1180,479)(1181,479)(1183,478)(1184,478)(1185,478)(1186,478)(1187,478)(1188,478)(1190,478)(1191,478)(1192,478)(1193,478)(1194,478)(1195,478)(1197,478)(1198,478)(1199,478)(1200,478)(1201,477)(1203,477)(1204,477)(1205,477)(1206,477)(1207,477)(1208,477)(1210,477)(1211,477)(1212,477)(1213,477)(1214,477)(1215,477)(1217,477)(1218,477)(1219,477)(1220,477)(1221,477)
\Thicklines \path(1221,477)(1222,477)(1224,476)(1225,476)(1226,476)(1227,476)(1228,476)(1230,476)(1231,476)(1232,476)(1233,476)(1234,476)(1235,476)(1237,476)(1238,476)(1239,476)(1240,476)(1241,476)(1242,476)(1244,476)(1245,476)(1246,476)(1247,476)(1248,476)(1249,476)(1251,476)(1252,476)(1253,475)(1254,475)(1255,475)(1257,475)(1258,475)(1259,475)(1260,475)(1261,475)(1262,475)(1264,475)(1265,475)(1266,475)(1267,475)(1268,475)(1269,475)(1271,475)(1272,475)(1273,475)(1274,475)(1275,475)(1276,475)(1278,475)(1279,475)(1280,475)
\Thicklines \path(1280,475)(1281,475)(1282,475)(1283,475)(1285,475)(1286,475)(1287,475)(1288,475)(1289,475)(1291,475)(1292,475)(1293,475)(1294,474)(1295,474)(1296,474)(1298,474)(1299,474)(1300,474)(1301,474)(1302,474)(1303,474)(1305,474)(1306,474)(1307,474)(1308,474)(1309,474)(1310,474)(1312,474)(1313,474)(1314,474)(1315,474)(1316,474)(1318,474)(1319,474)(1320,474)(1321,474)(1322,474)(1323,474)(1325,474)(1326,474)(1327,474)(1328,474)(1329,474)(1330,474)(1332,474)(1333,474)(1334,474)(1335,474)(1336,474)(1337,474)(1339,474)
\Thicklines \path(1339,474)(1340,474)(1341,474)(1342,474)(1343,474)(1344,474)(1346,474)(1347,474)(1348,474)(1349,474)(1350,474)(1352,474)(1353,474)(1354,474)(1355,474)(1356,474)(1357,474)(1359,474)(1360,474)(1361,474)(1362,474)(1363,474)(1364,473)(1366,473)(1367,473)(1368,473)(1369,473)(1370,473)(1371,473)(1373,473)(1374,473)(1375,473)(1376,473)(1377,473)(1379,473)(1380,473)(1381,473)(1382,473)(1383,473)(1384,473)(1386,473)(1387,473)(1388,473)(1389,473)(1390,473)(1391,473)(1393,473)(1394,473)(1395,473)(1396,473)(1397,473)
\Thicklines \path(1397,473)(1398,473)(1400,473)(1401,473)(1402,473)(1403,473)(1404,473)(1405,473)(1407,473)(1408,473)(1409,473)(1410,473)(1411,473)(1413,473)(1414,473)(1415,473)(1416,473)(1417,473)(1418,473)(1420,473)(1421,473)(1422,473)(1423,473)(1424,473)(1425,473)(1427,473)(1428,473)(1429,473)(1430,473)(1431,473)(1432,473)(1434,473)(1435,473)(1436,473)
\end{picture}

%% file: spin.eepic
%
%  File spin.eepic
%  Changes marked %%%
%
\setlength{\unitlength}{0.00087489in}
\begingroup\makeatletter\ifx\SetFigFont\undefined%
\gdef\SetFigFont#1#2#3#4#5{%
  \reset@font\fontsize{#1}{#2pt}%
  \fontfamily{#3}\fontseries{#4}\fontshape{#5}%
  \selectfont}%
\fi\endgroup%
{\renewcommand{\dashlinestretch}{30}
\begin{picture}(4524,2739)(0,-10)
\path(12,2712)(4512,2712)(4512,12)
	(12,12)(12,2712)
\thicklines %%%
\put(1660.125,1553.250){\arc{708.393}{4.8880}{5.9263}}
\put(3004,1430){\ellipse{320}{320}}
\path(2982,2262)(3117,2127)(3027,2127)
	(3027,1632)(2937,1632)(2937,2127)
	(2847,2127)(2982,2262)
\path(3837,1452)(3702,1317)(3702,1407)
	(3207,1407)(3207,1497)(3702,1497)
	(3702,1587)(3837,1452)
\path(1632,1452)(1812,2352)
\whiten\path(1817.883,2228.447)(1812.000,2352.000)(1759.049,2240.214)(1817.883,2228.447)
\path(2712,1407)(2712,1497)(2802,1452)(2712,1407)
\dashline{60.000}(1632,1452)(2307,1902)
\path(282,552)(1632,1452)
\whiten\path(1548.795,1360.474)(1632.000,1452.000)(1515.513,1410.397)(1548.795,1360.474)
\path(462,1452)(912,1452)
\whiten\path(792.000,1422.000)(912.000,1452.000)(792.000,1482.000)(792.000,1422.000)
\path(462,1452)(462,1902)
\whiten\path(492.000,1782.000)(462.000,1902.000)(432.000,1782.000)(492.000,1782.000)
\path(2712,1432)	(2691.562,1482.890)
	(2671.155,1516.467)
	(2622.000,1542.000)

\path(2622,1542)	(2568.212,1515.849)
	(2533.939,1453.755)
	(2498.696,1390.783)
	(2442.000,1362.000)

\path(2442,1362)	(2386.031,1390.125)
	(2352.000,1452.000)
	(2317.969,1513.875)
	(2262.000,1542.000)

\path(2262,1542)	(2206.031,1513.875)
	(2172.000,1452.000)
	(2137.969,1390.125)
	(2082.000,1362.000)

\path(2082,1362)	(2026.031,1390.125)
	(1992.000,1452.000)
	(1957.969,1513.875)
	(1902.000,1542.000)

\path(1902,1542)	(1846.031,1513.875)
	(1812.000,1452.000)
	(1777.969,1390.125)
	(1722.000,1362.000)

\path(1722,1362)	(1674.473,1381.688)
	(1632.000,1452.000)

\put(2937,1362){\makebox(0,0)[lb]{\smash{{{\SetFigFont{12}{14.4}{\rmdefault}{\mddefault}{\updefault}N}}}}}
\put(957,1407){\makebox(0,0)[lb]{\smash{{{\SetFigFont{12}{14.4}{\rmdefault}{\mddefault}{\updefault}$\be_3$}}}}} %%%
\put(417,1947){\makebox(0,0)[lb]{\smash{{{\SetFigFont{12}{14.4}{\rmdefault}{\mddefault}{\updefault}$\be_1$}}}}} %%%
\put(1857,1902){\makebox(0,0)[lb]{\smash{{{\SetFigFont{12}{14.4}{\rmdefault}{\mddefault}{\updefault}$\theta_{\text{e}}$}}}}} %%%
\put(3297,1137){\makebox(0,0)[lb]{\smash{{{\SetFigFont{12}{14.4}{\rmdefault}{\mddefault}{\updefault}longitudinal}}}}}
\put(3297,912){\makebox(0,0)[lb]{\smash{{{\SetFigFont{12}{14.4}{\rmdefault}{\mddefault}{\updefault}polarization}}}}}
\put(3162,2172){\makebox(0,0)[lb]{\smash{{{\SetFigFont{12}{14.4}{\rmdefault}{\mddefault}{\updefault}sideways}}}}}
\put(3162,1947){\makebox(0,0)[lb]{\smash{{{\SetFigFont{12}{14.4}{\rmdefault}{\mddefault}{\updefault}polarization}}}}}
\put(1632,2262){\makebox(0,0)[lb]{\smash{{{\SetFigFont{12}{14.4}{\rmdefault}{\mddefault}{\updefault}$\boldsymbol k'$}}}}} %%%
\put(1002,867){\makebox(0,0)[lb]{\smash{{{\SetFigFont{12}{14.4}{\rmdefault}{\mddefault}{\updefault}$\boldsymbol k$}}}}} %%%
\put(2397,1227){\makebox(0,0)[lb]{\smash{{{\SetFigFont{12}{14.4}{\rmdefault}{\mddefault}{\updefault}$\boldsymbol q$}}}}} %%%
\end{picture}
}

%% file: soffer.tex
% GNUPLOT: LaTeX picture using EEPIC macros
\setlength{\unitlength}{0.240900pt}
\begin{picture}(1500,1080)(0,0)
\tenrm
\thinlines \drawline(264,765)(1436,765)
\thinlines \drawline[-50](264,158)(264,967)
\thicklines \path(264,225)(284,225)
\thicklines \path(1436,225)(1416,225)
\put(242,225){\makebox(0,0)[r]{-0.8}}
\thicklines \path(264,360)(284,360)
\thicklines \path(1436,360)(1416,360)
\put(242,360){\makebox(0,0)[r]{-0.6}}
\thicklines \path(264,495)(284,495)
\thicklines \path(1436,495)(1416,495)
\put(242,495){\makebox(0,0)[r]{-0.4}}
\thicklines \path(264,630)(284,630)
\thicklines \path(1436,630)(1416,630)
\put(242,630){\makebox(0,0)[r]{-0.2}}
\thicklines \path(264,765)(284,765)
\thicklines \path(1436,765)(1416,765)
\put(242,765){\makebox(0,0)[r]{0}}
\thicklines \path(264,900)(284,900)
\thicklines \path(1436,900)(1416,900)
\put(242,900){\makebox(0,0)[r]{0.2}}
\thicklines \path(264,158)(264,178)
\thicklines \path(264,967)(264,947)
\put(264,113){\makebox(0,0){0}}
\thicklines \path(444,158)(444,178)
\thicklines \path(444,967)(444,947)
\put(444,113){\makebox(0,0){0.2}}
\thicklines \path(625,158)(625,178)
\thicklines \path(625,967)(625,947)
\put(625,113){\makebox(0,0){0.4}}
\thicklines \path(805,158)(805,178)
\thicklines \path(805,967)(805,947)
\put(805,113){\makebox(0,0){0.6}}
\thicklines \path(985,158)(985,178)
\thicklines \path(985,967)(985,947)
\put(985,113){\makebox(0,0){0.8}}
\thicklines \path(1166,158)(1166,178)
\thicklines \path(1166,967)(1166,947)
\put(1166,113){\makebox(0,0){1}}
\thicklines \path(1346,158)(1346,178)
\thicklines \path(1346,967)(1346,947)
\put(1346,113){\makebox(0,0){1.2}}
\thicklines \path(264,158)(1436,158)(1436,967)(264,967)(264,158)
\put(45,562){\makebox(0,0)[l]{\shortstack{\hspace{-4mm}$I_1^{\rm p}(Q^2)$}}}
\put(850,68){\makebox(0,0){$\begin{matrix}\\Q^2/GeV^2\end{matrix}$}}
\put(715,881){\raisebox{-1.2pt}{\makebox(0,0){$\Diamond$}}}
\put(1346,864){\raisebox{-1.2pt}{\makebox(0,0){$\Diamond$}}}
\thinlines \path(715,846)(715,917)
\thinlines \path(705,846)(725,846)
\thinlines \path(705,917)(725,917)
\thinlines \path(1346,851)(1346,877)
\thinlines \path(1336,851)(1356,851)
\thinlines \path(1336,877)(1356,877)
\thicklines \path(276,287)(276,287)(288,345)(300,397)(311,443)(323,485)(335,522)(347,556)(359,586)(371,614)(382,639)(394,662)(406,683)(418,703)(430,720)(442,737)(453,752)(465,765)(477,778)(489,790)(501,800)(513,810)(524,819)(536,828)(548,835)(560,842)(572,849)(584,855)(595,861)(607,866)(619,870)(631,875)(643,879)(655,882)(667,886)(678,889)(690,891)(702,894)(714,896)(726,898)(738,900)(749,902)(761,903)(773,904)(785,906)(797,906)(809,907)(820,908)(832,909)(844,909)(856,909)
\thicklines \path(856,909)(868,909)(880,910)(891,910)(903,910)(915,909)(927,909)(939,909)(951,908)(962,908)(974,907)(986,907)(998,906)(1010,905)(1022,905)(1033,904)(1045,903)(1057,902)(1069,901)(1081,900)(1093,899)(1105,898)(1116,897)(1128,896)(1140,895)(1152,893)(1164,892)(1176,891)(1187,890)(1199,889)(1211,887)(1223,886)(1235,885)(1247,884)(1258,883)(1270,882)(1282,881)(1294,880)(1306,879)(1318,878)(1329,877)(1341,876)(1353,875)(1365,874)(1377,873)(1389,872)(1400,871)(1412,870)(1424,869)(1436,869)
\thicklines \path(264,225)(264,225)(276,296)(288,367)(300,437)(311,508)(323,579)(335,649)(347,720)(359,791)(371,861)(382,932)(388,967)
\thicklines \drawline[-50](264,225)(264,225)(276,249)(288,272)(300,295)(311,318)(323,341)(335,364)(347,387)(359,410)(371,433)(382,456)(394,480)(406,503)(418,526)(430,549)(442,572)(453,595)(465,618)(477,641)(489,664)(501,688)(513,711)(524,734)(536,757)(548,780)(560,803)(572,826)(584,849)(595,872)(607,895)(619,919)(631,942)(643,965)(644,967)
\thicklines \drawline[-50](264,225)(264,225)(276,272)(288,319)(300,366)(311,413)(323,460)(335,506)(347,553)(359,600)(371,647)(382,694)(394,741)(406,787)(418,834)(430,881)(442,928)(451,967)
\end{picture}

%% file: kin.eepic
\setlength{\unitlength}{0.00087489in}
\begingroup\makeatletter\ifx\SetFigFont\undefined%
\gdef\SetFigFont#1#2#3#4#5{%
  \reset@font\fontsize{#1}{#2pt}%
  \fontfamily{#3}\fontseries{#4}\fontshape{#5}%
  \selectfont}%
\fi\endgroup%
{\renewcommand{\dashlinestretch}{30}
\begin{picture}(5424,2739)(0,-10)
\put(2824.500,1384.500){\arc{586.728}{5.2791}{6.3600}}
\put(4630.125,1553.250){\arc{708.393}{4.8880}{5.9263}}
\path(777,912)(642,462)(4242,462)
	(4782,2262)(1182,2262)(912,1362)
\path(912,1362)(237,912)(3837,912)
	(5187,1812)(4647,1812)
\path(2577,1317)(2667,1407)(2757,1317)
	(2847,1407)(2577,1317)
\path(2802,1362)(4512,1362)
\path(237,1137)(912,1362)
\whiten\path(807.645,1295.592)(912.000,1362.000)(788.671,1352.513)(807.645,1295.592)
\path(912,1362)(1362,1587)
\whiten\path(1268.085,1506.502)(1362.000,1587.000)(1241.252,1560.167)(1268.085,1506.502)
\dottedline{45}(2712,1362)(2262,912)
\path(2262,912)(1812,462)
\whiten\path(1875.640,568.066)(1812.000,462.000)(1918.066,525.640)(1875.640,568.066)
\dashline{60.000}(2712,1362)(3612,2262)
\whiten\path(3548.360,2155.934)(3612.000,2262.000)(3505.934,2198.360)(3548.360,2155.934)
\dottedline{45}(912,1362)(1587,1812)(4647,1812)
\dottedline{45}(912,1362)(777,912)
\path(687,507)(1047,507)
\whiten\path(927.000,477.000)(1047.000,507.000)(927.000,537.000)(927.000,477.000)
\path(4467,1362)(4557,1362)(4827,1542)
	(4872,1542)(4827,1587)(4692,1542)
	(4737,1542)(4467,1362)
\path(4467,1317)(4872,1317)(4827,1272)
	(4984,1347)(4962,1407)(4917,1362)
	(4512,1362)(4467,1317)
\path(12,2712)(5412,2712)(5412,12)
	(12,12)(12,2712)
\path(372,957)(732,957)
\whiten\path(612.000,927.000)(732.000,957.000)(612.000,987.000)(612.000,927.000)
\path(372,957)(642,1137)
\whiten\path(558.795,1045.474)(642.000,1137.000)(525.513,1095.397)(558.795,1045.474)
\path(687,507)(777,822)
\whiten\path(772.879,698.375)(777.000,822.000)(715.188,714.859)(772.879,698.375)
\path(912,1362)	(866.693,1323.297)
	(837.998,1296.803)
	(822.000,1272.000)

\path(822,1272)	(882.413,1293.479)
	(936.423,1326.606)
	(997.553,1365.855)
	(1059.238,1404.622)
	(1114.916,1436.303)
	(1182.000,1452.000)

\path(1182,1452)	(1165.862,1418.092)
	(1114.050,1362.952)
	(1062.463,1307.336)
	(1047.000,1272.000)

\path(1047,1272)	(1073.815,1267.922)
	(1121.039,1286.625)
	(1181.868,1320.516)
	(1249.500,1362.000)
	(1317.132,1403.484)
	(1377.961,1437.375)
	(1425.185,1456.078)
	(1452.000,1452.000)

\path(1452,1452)	(1436.368,1417.021)
	(1384.500,1362.000)
	(1332.632,1306.979)
	(1317.000,1272.000)

\path(1317,1272)	(1343.815,1267.922)
	(1391.039,1286.625)
	(1451.868,1320.516)
	(1519.500,1362.000)
	(1587.132,1403.484)
	(1647.961,1437.375)
	(1695.185,1456.078)
	(1722.000,1452.000)

\path(1722,1452)	(1706.368,1417.021)
	(1654.500,1362.000)
	(1602.632,1306.979)
	(1587.000,1272.000)

\path(1587,1272)	(1613.815,1267.922)
	(1661.039,1286.625)
	(1721.868,1320.516)
	(1789.500,1362.000)
	(1857.132,1403.484)
	(1917.961,1437.375)
	(1965.185,1456.078)
	(1992.000,1452.000)

\path(1992,1452)	(1976.368,1417.021)
	(1924.500,1362.000)
	(1872.632,1306.979)
	(1857.000,1272.000)

\path(1857,1272)	(1883.815,1267.922)
	(1931.039,1286.625)
	(1991.868,1320.516)
	(2059.500,1362.000)
	(2127.132,1403.484)
	(2187.961,1437.375)
	(2235.185,1456.078)
	(2262.000,1452.000)

\path(2262,1452)	(2246.368,1417.021)
	(2194.500,1362.000)
	(2142.632,1306.979)
	(2127.000,1272.000)

\path(2127,1272)	(2153.815,1267.922)
	(2201.039,1286.625)
	(2261.868,1320.516)
	(2329.500,1362.000)
	(2397.132,1403.484)
	(2457.961,1437.375)
	(2505.185,1456.078)
	(2532.000,1452.000)

\path(2532,1452)	(2516.240,1417.323)
	(2464.159,1362.806)
	(2412.248,1307.886)
	(2397.000,1272.000)

\path(2397,1272)	(2458.534,1275.622)
	(2524.889,1307.412)
	(2569.270,1331.724)
	(2622.000,1362.000)

\put(3117,1497){\makebox(0,0)[lb]{\smash{{{\SetFigFont{12}{14.4}{\rmdefault}{\mddefault}{\updefault}$\theta_\pi$}}}}}
\put(4782,1902){\makebox(0,0)[lb]{\smash{{{\SetFigFont{12}{14.4}{\rmdefault}{\mddefault}{\updefault}$\phi_\pi$}}}}}
\put(4872,1497){\makebox(0,0)[lb]{\smash{{{\SetFigFont{12}{14.4}{\rmdefault}{\mddefault}{\updefault}sidew.}}}}}
\put(4737,1137){\makebox(0,0)[lb]{\smash{{{\SetFigFont{12}{14.4}{\rmdefault}{\mddefault}{\updefault}long.}}}}}
\put(372,1227){\makebox(0,0)[lb]{\smash{{{\SetFigFont{12}{14.4}{\rmdefault}{\mddefault}{\updefault}e}}}}}
\put(1362,1542){\makebox(0,0)[lb]{\smash{{{\SetFigFont{12}{14.4}{\rmdefault}{\mddefault}{\updefault}e$'$}}}}}
\put(1902,687){\makebox(0,0)[lb]{\smash{{{\SetFigFont{12}{14.4}{\rmdefault}{\mddefault}{\updefault}N$'$}}}}}
\put(3612,1182){\makebox(0,0)[lb]{\smash{{{\SetFigFont{12}{14.4}{\rmdefault}{\mddefault}{\updefault}N}}}}}
\put(3207,1992){\makebox(0,0)[lb]{\smash{{{\SetFigFont{12}{14.4}{\rmdefault}{\mddefault}{\updefault}$\pi$}}}}}
\put(792,710){\makebox(0,0)[lb]{\smash{{{\SetFigFont{12}{14.4}{\rmdefault}{\mddefault}{\updefault}$\boldsymbol{e}^{\text{rp}}_1$}}}}}
\put(987,544){\makebox(0,0)[lb]{\smash{{{\SetFigFont{12}{14.4}{\rmdefault}{\mddefault}{\updefault}$\boldsymbol{e}^{\text{rp}}_3$}}}}}
\put(754,927){\makebox(0,0)[lb]{\smash{{{\SetFigFont{12}{14.4}{\rmdefault}{\mddefault}{\updefault}$\boldsymbol{e}_3$}}}}}
\put(664,1107){\makebox(0,0)[lb]{\smash{{{\SetFigFont{12}{14.4}{\rmdefault}{\mddefault}{\updefault}$\boldsymbol{e}_1$}}}}}
\end{picture}
}